\documentclass{aa} 

\usepackage{graphicx}
\usepackage{natbib}
\usepackage{lscape}
\usepackage{txfonts}
\begin{document}
\title{The Taurus Boundary of Stellar/Substellar (TBOSS) Survey I: \\
far-IR disk emission measured with {\it Herschel}\thanks{{\it Herschel} is an ESA space observatory with science instruments provided by European-led Principal Investigator consortia and with important participation from NASA}}

\author{
J. Bulger\inst{1, 2}, J. Patience\inst{1,2}, K. Ward-Duong\inst{2}, C. Pinte\inst{3}, H. Bouy\inst{4} F. M{\'e}nard\inst{3}, J.-L. Monin\inst{5}
}

\institute{School of Physics, University of Exeter, Exeter, EX4 4QL, UK
\and
School of Earth and Space Exploration, Arizona State University, Tempe, AZ USA 85281\\
\email{jbulger1@asu.edu}
\and
Universidad de Chile, Camino El Observatorio 1515, Las Condes, Santiago, Chile.
\and
Centro de Astrobiolog{\'i}a - Depto. Astrof{\'i}sica (CSIC- INTA), ESAC Campus, PO Box 78, E-28691 Villanueva de la Ca{\~n}ada, Spain
\and
UJF-Grenoble 1 / CNRS-INSU, Institut de Plan{\'e}tologie et d$'$Astrophysique de Grenoble (IPAG) UMR 5274, Grenoble, F-38041, France
}

\date{Received November 21, 2013 /  Accepted July 2, 2014 }

 
  \abstract
{With the PACS instrument on {\it Herschel}, 134 low mass members of the Taurus star-forming region spanning the M4-L0 spectral type range and covering the transition from low mass stars to brown dwarfs were observed. Combining the new {\it Herschel} results with other {\it Herschel} programs, a total of 150 of the 154 M4-L0 Taurus members members have observations, and we have added an additional 3 targets from {\it Spitzer} to form the 153-object TBOSS (Taurus Boundary of Stellar/Substellar) sample, a 99\% complete study. Among the 150 targets, 70~$\mu$m flux densities were measured for 7 of the 7 Class~I objects, 48 of the 67 Class~II objects, and 3 of the 76 Class~III objects. For the detected Class II objects, the median 70~$\mu$m flux density level declines with spectral type, however, the distribution of excess relative to central object flux density does not change across the stellar/substellar boundary in the M4-L0 range. Connecting the 70~$\mu$m TBOSS values with the results from K0-M3 Class~II members results in the first comprehensive census of far-IR emission across the full mass spectrum of the stellar and substellar population of a star-forming region, and the median flux density declines with spectral type in a trend analogous to the flux density decline expected for the central objects. Spectral energy distributions (SEDs) were constructed for all TBOSS targets covering the optical to far-IR range and extending to the submm/mm for a subset of sources with longer wavelength data. Based on an initial exploration of the impact of different physical parameters on the {\it Herschel} flux densities, geometrical factors such as inclination and structural parameters such as scale height and flaring have the largest influence on the flux densities in the PACS bands. 
From the 24~$\mu$m to 70~$\mu$m spectral index of the SEDs, 5 new candidate transition disks were identified. Considering the previously known and new candidate transition disks, the spectral indices over longer wavelengths ($\geq$70~$\mu$m) are not distinct from those of the full Class~II population, suggesting that the outer regions of the transition disks are similar to Class~II disks. The steep 24~$\mu$m to 70~$\mu$m slope for a subset of 8 TBOSS targets may be an indication of truncated disks in these systems, however additional measurements are required to establish the outer radii of these disks conclusively. 
From existing high angular resolution companion search observations, two examples of mixed pair systems that include secondaries with disks were measured in the {\it Herschel} data.
Finally, comparing the TBOSS results with a {\it Herschel} study of Ophiuchus brown dwarfs reveals a lower fraction of disks around the Taurus substellar population with flux densities comparable to the Ophiuchus disks.
}

\keywords{(stars:) brown dwarfs, stars: pre-main sequence, protoplanetary disks}

\authorrunning{Bulger, Patience, Ward-Duong, Pinte, Bouy, M\'enard, Monin}
\titlerunning{Disk emission across the stellar/substellar boundary in Taurus}

\maketitle

\section{Introduction}
Disks are critical structures in the star and planet formation process, as they provide a conduit to channel material onto the central object (e.g., \citealp{Hartmann:1997}) and supply a reservoir of dust and gas to form planets (e.g., \citealp{Pollack:1996, Boss:1997}). Variations in the frequency, lifetimes, and structure of disks as a function of central object mass have important implications for formation models (e.g., \citealp{Bate:2003,Reipurth:2001}) and the viability of future planet formation. Infrared and mid-infrared observations of young brown dwarfs (e.g., \citealp{Luhman:2006,Luhman:2010,Guieu:2007}) have enabled the identification of inner disks. This common presence of disks around stellar and substellar objects has been used to argue for a common formation mechanism \citep{Scholz:2008, Haisch:2010}, while other observations hint at possible differences in the dynamical history of low mass stars/brown dwarfs compared to higher mass stars \citep{Thies:2007}. Formation scenarios for brown dwarfs also have implications for the properties of brown dwarf disks, and a number of possible mechanisms have been proposed, including ejection of stellar embryos (e.g., \citealp{Bate:2003,Reipurth:2001}), erosion of star-forming clouds by radiation from massive stars \citep{Whitworth:2004}, gravitational instabilities in the early stages of disks (e.g., \citealp{Stamatellos:2011, Basu:2012}), and gravitational collapse of clouds analogous to the formation of stars (e.g., \citealp{Padoan:2005, Hennebelle:2008}).

Trends in the population of exoplanets orbiting low mass stars have suggested intriguing differences compared to the planetary systems around more massive stars and observations of protoplanetary disks around low mass objects are required to investigate possible origins of the distinct populations. Analysis of the {\it Kepler} planet candidate statistics shows a steep increase in the occurrence of planets with the smallest radii as host star mass declines \citep{Howard:2012}. By contrast, the frequency of hot Jupiters around M-stars is lower than that for higher mass stars (e.g., \citealp{Johnson:2010}). The detection of brown dwarfs with directly-imaged planetary mass companions \citep{Chauvin:2004, Song:2006} indicate that at least some substellar objects are associated with planet formation early in their history. The overall exoplanet population frequency remains unclear around brown dwarfs. By detecting and characterizing the disks around young M-stars and brown dwarfs, it is possible to investigate the origins of the exoplanet populations.

As one of the nearest regions of star-formation ($\sim$140 pc; \citealp{Kenyon:1994,Bertout:1999,Torres:2009}), Taurus represents an important population of young stars and brown dwarfs for which detailed investigations of disk frequencies and properties are possible. The stellar density of the Taurus region is exceptionally low compared to other star-forming regions based on a calculation from the most recent membership studies and a consistent approach \citep{King:2012}. This low density enables isolating the individual objects and avoiding confusion within the beam of longer wavelength observations. Large-scale surveys of Taurus members have identified the presence of disks based on excess emission above the photosphere out to 24~$\mu$m with sensitive {\it Spitzer} measurements (e.g., \citealp{Luhman:2010}), but these wavelengths do not probe the cooler material in the disk that covers the portions of the spectral energy distributions over which the emission transitions from optically thick to optically thin, and is strongly influenced by factors such as flaring, scale height, and settling. The {\it Spitzer} survey of Taurus at 70~$\mu$m \citep{Rebull:2010} detected a large population of the earlier spectral type members, but did not have the sensitivity to measure disk emission from the lowest mass stars and brown dwarfs.  In Taurus, $\sim$50\% of the known members have spectral types later than M3 \citep{Luhman:2010}, making the existing far-IR measurements fundamentally incomplete and resulting in a biased view of the early stages of star and planet formation. As pre-Main Sequence stars contract onto the Main Sequence, their effective temperatures increase and a spectral type of M4 in a star-forming region corresponds to $\sim$M2 at an age of 1~Gyr (e.g., \citealp{Chabrier:2000}), comparable to the ages of the stars in the Solar Neighborhood. Among the nearest stars, the majority are M-dwarfs (e.g., \citealp{Reid:1997,Henry:1998}), again highlighting the importance of understanding the environments of low mass stars to develop a comprehensive disk population study.

To obtain a nearly complete census of far-IR emission across the full population of Taurus, we have performed a large-scale survey of low mass stars and brown dwarfs with {\it Herschel}, targeting the M4-L0 members -- the Taurus Boundary of Stellar/Substellar (TBOSS) Survey -- to extend the exploration of disk mass and structure into the substellar regime with a sensitivity that was not possible prior to {\it Herschel}.  The nearly complete sample for this study is defined in Section~\ref{Sec:sample}, followed by the new {\it Herschel} PACS observations in Section~\ref{Sec:Obs}. The data analysis including the measurement of far-IR flux densities and construction of the SEDs for each target in the full sample is described in Section~\ref{Sec:Data}. The results for the members of each evolutionary class, a comparison with {\it Spitzer} data, the detection rates, and spatially extended and multiple sources are reported in Section~\ref{Sec:Results}. In Section~\ref{Sec:Discussion}, the discussion covers a number of topics including the dependence of the results on spectral type and companions, the SEDs for different types of targets such as transition disks, and a comparison of the Taurus substellar disks with the population in Ophiuchus. Finally, Section~\ref{Sec:Conclusions} summarizes the conclusions.

\begin{figure*}
\centering
\includegraphics[scale=0.8]{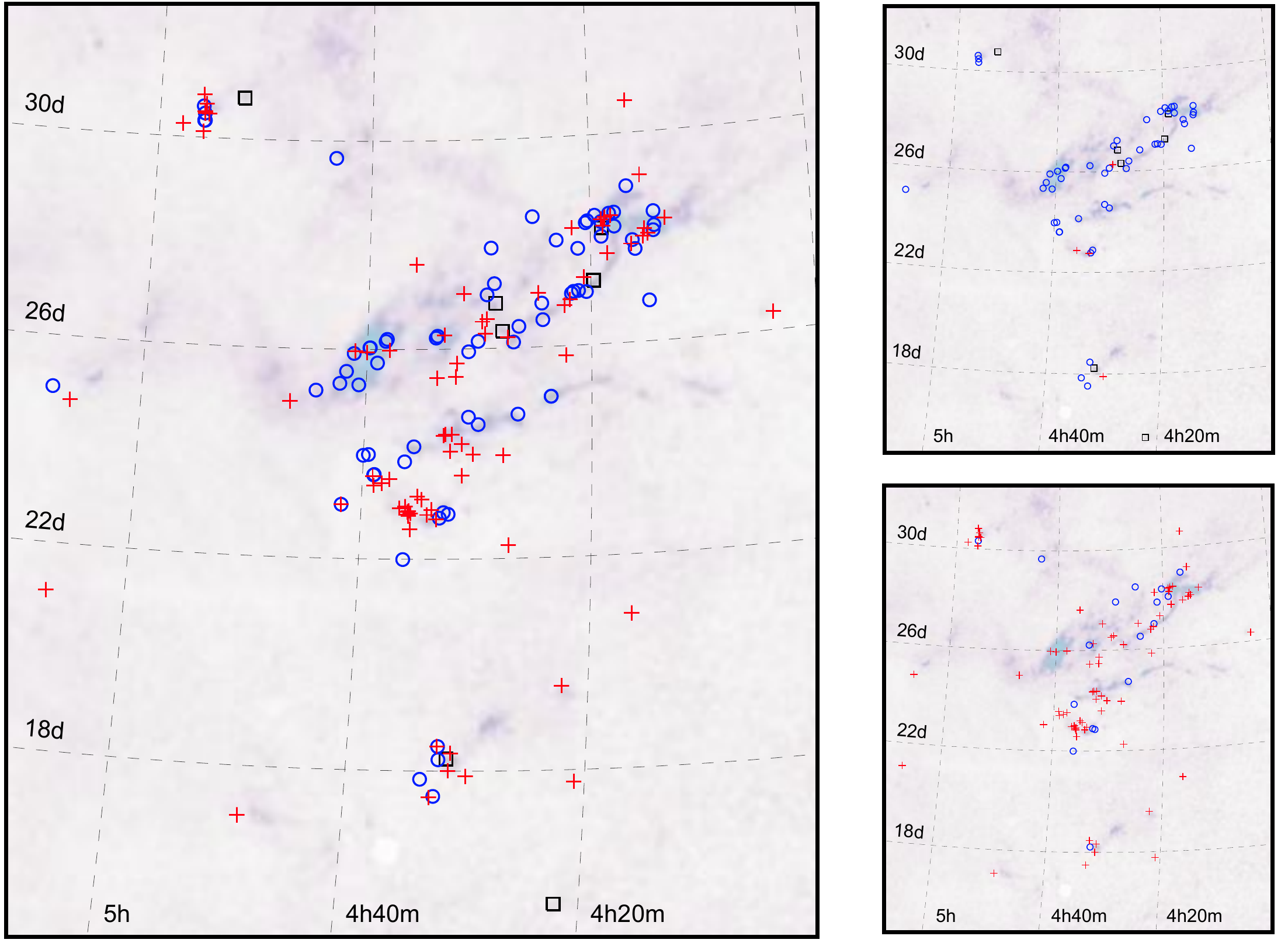}
\caption {The spatial distribution of the TBOSS sample across the Taurus molecular cloud (shown in the map of extinction from \citealp{Dobashi:2005}) is shown on the left. Black squares indicate Class I objects, blue circles indicate Class II objects, and red crosses indicate Class III objects. The distribution of the TBOSS sample with {\it Herschel} PACS detections are shown in the top, right map and the distribution of the TBOSS sample with {\it Herschel} PACS upper limits are shown in the bottom, right map.
}
\label{Av}
\end{figure*}

\section{Sample}
\label{Sec:sample}

The target sample was selected to provide a comprehensive census of far-IR disk emission from Taurus members spanning both sides of the stellar/substellar boundary. The demarcation of the substellar boundary occurs at a spectral type of M6.25 at the age of Taurus (e.g., \citealp{Luhman:2005}), and the spectral type range of the TBOSS sample is M4-L0. From a {\it Spitzer} 3.6-24 $\mu$m study of the Taurus population \citep{Luhman:2010}, 152 M4-L0 members were included, which represented 99\% of the currently known membership, with the remaining 1\% of the known Taurus population comprised of objects with spectral types earlier than M4. Of the newest Taurus members identified in a {\it Spitzer} study including 70~$\mu$m data \citep{Rebull:2010}, all but 2 with spectral types of M4 and later were included in the \citet{Luhman:2010} study. 
Of the 154 M4-L0 members listed in the combination of the {\it Spitzer} studies, {\it Herschel} Photodetector Array Camera and Spectrometer (PACS; \citealp{Poglitsch:2010}) 70~$\mu$m or 100~$\mu$m and 160~$\mu$m flux densities or detection limits are reported for 150 objects. 

The large majority of the {\it Herschel} observations, covering 134 targets, were performed as part of our program (OT1\_patienc\_1). PACS photometry for 12 targets were drawn from the literature -- 4 targets from \citet{Harvey:2012} from the program GT1\_pharve01\_2, and 8 targets from \citet{Howard:2013} from the program KPOT\_bdent\_1 as part of the Gas Survey of Protoplanetary Systems (GASPS; \citealp{Dent:2013}). Another 4 targets were observed as part of the Guaranteed Time Key Program, KPGT\_pandre\_1, and for these targets PACS photometry measurements were made from the reduced archive maps. For the 4 targets not observed by {\it Herschel}, far-IR {\it Spitzer} measurements exist for J04251550+2829275, J04355760+2253574, IRAS S04414+2506 \citep{Rebull:2010}, but no measurements exist for XEST 26-052. Table \ref{Table:sample} reports the 2MASS name, alternate name, evolutionary class and spectral type of each of the 154 M4-L0 Taurus members. 
The 153-object TBOSS sample consists of the known 154 Taurus members with either {\it Herschel} or {\it Spitzer} far-IR photometry, and is comprised of 7 Class~I, 69 Class~II, and 77 Class~III members, and 63 of the targets are brown dwarfs (M6.25-L0). The spectral types have been compiled from the original reference (noted in Table \ref{Table:sample}), and for the targets with more than one spectral type measurement reported, the most recent result was adopted for this study. Figure~\ref{Av} shows the distribution of the TBOSS sample across the Taurus molecular cloud.

\section{Observations}
\label{Sec:Obs}

Of the 153 targets in the TBOSS sample, 150 were observed with the {\it Herschel} PACS instrument. PACS records two wavelengths simultaneously, and the observations of 146 include both 70~$\mu$m data from the Blue channel and 160~$\mu$m data from the Red channel. Amongst the 146 targets, 8 targets were also observed with the PACS Blue channel operating at 100 $\mu$m. Observations of the remaining 4 targets of the 150 total were carried out with PACS operating at 100~$\mu$m and 160~$\mu$m.
For majority of the sample, the 134 targets observed under our {\it Herschel} PACS program (OT1\_patienc\_1), the mapping strategy involved obtaining two scan maps per target, with each map at a different orientation (70$\degr$ and 110$\degr$). The cross scans are designed to reduce the stripping effect of the 1/{\it f} noise and, consequently, to provide more homogeneous and higher redundancy coverage of the map area. The medium mapping scan speed (20$\arcsec$/s) was employed with scan leg lengths of 3$\arcmin$.0, cross scan steps of 4$\farcs$0, and a total of 8 scan legs per orientation. Each map was centered on the target and the field-of-view was typically 3$\arcmin$ $\times$ 7$\arcmin$. The pointing accuracy of {\it Herschel} is 0$\farcs$8\footnote{http://herschel.esac.esa.int/twiki/bin/view/Public/SummaryPointing}, and the beam size is 5$\farcs$6 in the 70 $\mu$m maps, 6$\farcs$8 in the 100 $\mu$m maps and 11$\farcs$3 in the 160~$\mu$m maps.  The integration time between the PACS channels is the same, and the total on-source integration time was 180s for all but 5 targets for which the integration time was 240s. Table~\ref{Table:obsLog} reports the {\it Herschel} scan and cross scan ID numbers along with the observation date, observation duration and total on-source integration time for each target in the OT1\_patienc\_1 program.
Table~\ref{Table:obsLog_other} summarizes the program ID, scan ID numbers, observation date, and observation duration per target for the 16 targets not observed in the OT1\_patienc\_1 program. 

For the 3 targets of the 153-object TBOSS sample with {\it Spitzer} Multiband Imaging Photometer (MIPS; \citealp{Rieke:2004}) measurements only, the observation details are reported in \citet{Rebull:2010}.

\section{Data Analysis}
\label{Sec:Data}

\subsection{Measurement of PACS flux densities}
The {\it Herschel} program OT1\_jpatienc\_1 data were reduced using the {\it Herschel} Interactive Processing Environment (HIPE; \citealp{Ott:2010}) software version 9.0.0. For each target, the scan and corresponding cross scan were pre-processed using the standardized routines given in the PACS Data Reduction Guide (version~7, June 2011), and the calibration employed the PACS calibration file set version 48. The initial map was used to identify any bright sources that needed to be masked prior to performing the Multiresolution Median Transform (MMT) de-glitching routine and applying a high pass filter to each data set. The location of any point source with a signal stronger than three times the standard deviation of the initial map was masked. 

Depending on the brightness and size of the source in the initial map, a different number of frames were used in the high pass filtering, as suggested in the Data Reduction Guide. For the 9 brightest and/or resolved targets, a high pass filter radius corresponding to 51 frames in the Blue channel and 71 frames in the Red channel was used; for the 125 fainter point sources, increments of 31 frames in the Blue channel and 51 frames in the Red channel were used to filter the thousands of frames involved in the scans of each target. After filtering, each scan and cross-scan was re-processed and the final map for each target was formed by a co-addition using the {\tt photProject()} task.

The target flux densities were measured using aperture photometry, with the aperture size determined by the observation wavelength and object size. Among the 146 targets observed at 70~$\mu$m and 160~$\mu$m, 141 are unresolved point sources and 5 are spatially resolved. For the point sources, the aperture radii of 5$\farcs$5 at 70 $\mu$m and 10$\farcs$5 at 160 $\mu$m recommended by the {\it Herschel} Science Center\footnote{Technical note: PICC-ME-TN-037} were used to measure the target flux densities. For the extended sources, the flux densities were measured within a 3$\sigma$ contour, derived from three times the standard deviation of the map noise. The source flux density error was calculated as the standard deviation of the flux density in nine comparison apertures of the same size as the target aperture and distributed in the high coverage area of each map. Upper limits were determined by three times the standard deviation of the nine comparison apertures. Finally, aperture corrections defined in a {\it Herschel} Science Center technical note$^{2}$ were applied to the source flux densities and upper limits. Additional absolute flux calibration uncertainties of 2.6\% at 70~$\mu$m and 4.2\% at 160~$\mu$m as reported in the technical note$^{2}$ are not included in the uncertainties reported here. For the 4 targets observed as part of the KPGT\_pandre\_1 program, 3 targets were unresolved point sources and 1 target was spatially resolved. For the point sources, flux densities were measured from aperture photometry of the level 2.5 processed maps\footnote{Here level 2.5 maps are the products of calibrated and combined PACS scan and cross-scan data.}; an aperture of 5$\farcs$6 was used for the 100~$\mu$m data. For the resolved target, the flux density was measured within a 3$\sigma$ contour from the level 2.5 processed map. For the 12 targets observed as part of the GT1\_pharve01\_2 and KPOT\_bdent\_1 programs, the flux densities were obtained from the literature \citep{Harvey:2012, Howard:2013}.

\begin{figure}
\centering
\includegraphics[scale=0.53]{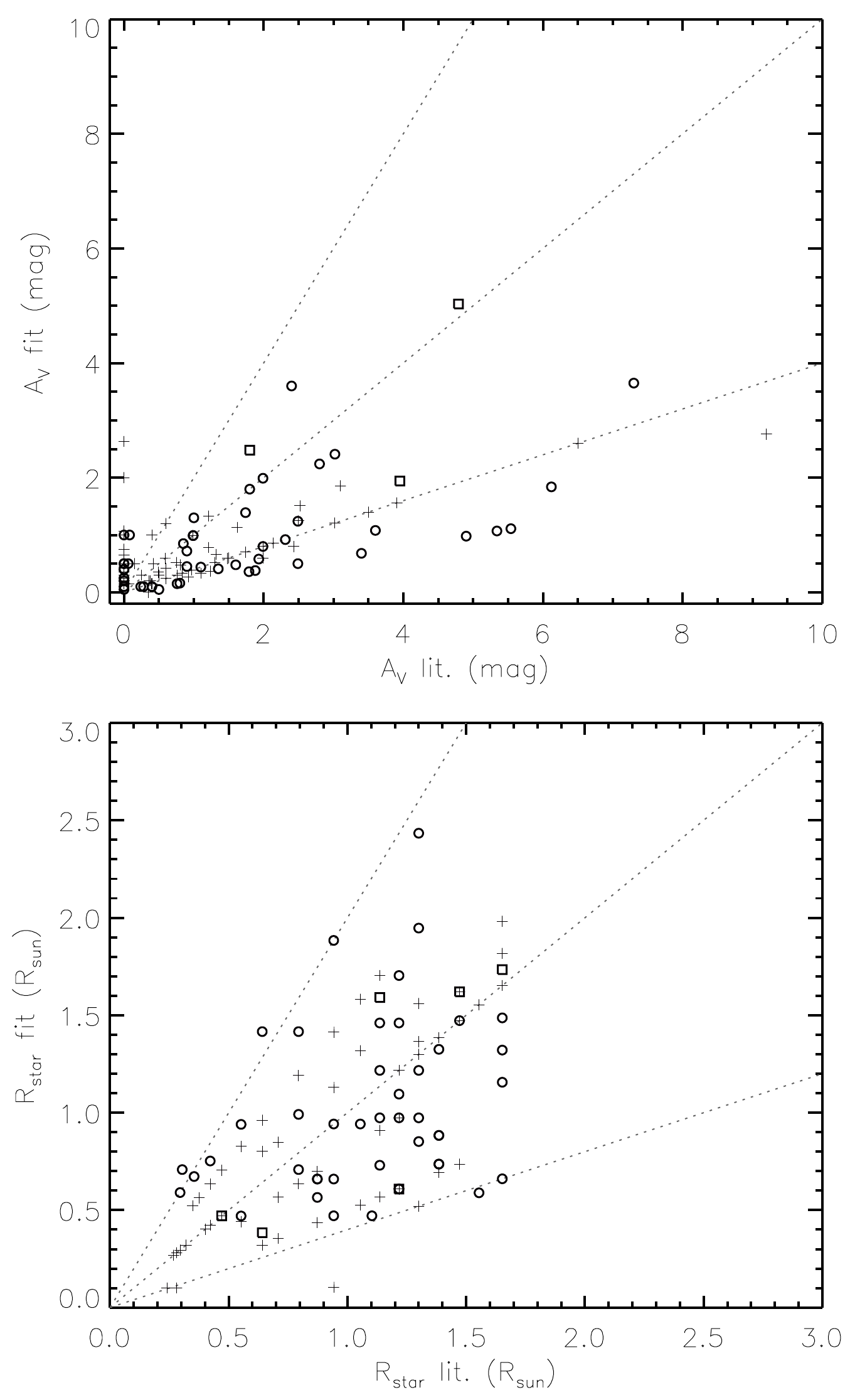}
\caption {Comparison plots of the SED best-fit values of {\it A$_{V}$} {\bf (top)} and {\it R}$_{\mathrm{star}}$ {\bf (bottom)} against those reported in the literature. Literature {\it A$_{V}$} values are compiled from the references listed in Table \ref{Table:sed}. {\it R}$_{\mathrm{star}}$ values are derived from the spectral types reported in the literature, using the temperature conversion of \citet{Luhman:2003b} and the 1 Myr evolutionary models of \citet{Baraffe:1998} and \citet{Chabrier:2000}. Squares indicate Class~I objects, circles indicate Class~II objects, and crosses indicate Class~III objects. The dotted lines represent fits that are 0.4, 1 and 2 times that of the literature values.
}
\label{Rstar}
\end{figure}

\subsection{Construction of Spectral Energy Distributions}
\label{Sec:SEDs}
Spectral energy distributions (SEDs) were constructed for all targets of the TBOSS sample. The details of the source photometry included in the SEDs from non-{\it Herschel} observations are given in this section, and the resulting SED plots are presented and discussed in Section~\ref{Sec:Discussion}. In addition to the {\it Herschel} PACS photometry, the SEDs include photometry from optical to millimeter wavelengths when available.

The optical photometry measurements consist of {\it R$_{C}$} and {\it I$_{C}$} band measurements compiled from the literature. The original measurements were in the {\it R$_{C}$} and {\it I$_{C}$} filters for those targets reported in  \citet{Kenyon:1995}, whilst all other compiled literature studies transformed the measured photometric value into the Cousins system \citep{Guieu:2007, Guieu:2006, Luhman:2004, Luhman:2003, Briceno:2002, Martin:2001, Luhman:2000, Briceno:1999}.
For targets with more than one photometric measurement, the median value was adopted. Uncertainties were not reported in up to half of the the literature sources, therefore, to account for both source variability and the uncertainties in the photometric transformation to the Cousins system, adopted common values for the {\it R$_{C}$} and {\it I$_{C}$} uncertainties were used (e.g., \citealp{Briceno:2002, Mayne:2012}). The error adopted for {\it R$_{C}$} was 0.8~mag, and the error adopted for {\it I$_{C}$} was 0.3~mag. 
Magnitudes were converted to flux densities using zero-points of 3080~Jy  and 2550~Jy for {\it R$_{C}$} and {\it I$_{C}$}, respectively \citep{Bessell:1979}. The {\it R$_{C}$} and {\it I$_{C}$} flux densities, and literature references are listed in Table~\ref{table:Optical}.

In the near-infrared, {\it JHK$_{S}$} magnitudes for all targets were extracted from the Two Micron All Sky Survey (2MASS) point source catalog \citep{Skrutskie:2006} and were converted into flux densities using zero-points of 1594 $\pm$ 28 Jy, 1024 $\pm$ 20 Jy and 667 $\pm$ 13 Jy \citep{Cohen:2003} for {\it J}, {\it H}, and {\it K$_{S}$}, respectively. The 2MASS {\it JHK$_{S}$} flux densities and uncertainties are listed in Table~\ref{table:Optical}.

In the mid-IR, with the {\it Spitzer} Infrared Array Camera (IRAC: 3.6, 4.5, 5.8 and 8.0 $\mu$m; \citealp{Fazio:2004}), and at 24~$\mu$m with the Multiband Imaging Photometer for {\it Spitzer} (MIPS; \citealp{Rieke:2004}), magnitudes were compiled from \citet{Luhman:2010} and \citet{Rebull:2010}. The mean-weighted magnitudes were calculated for targets in which multiepoch observations are reported. Magnitudes were converted into flux densities using the zero-points of 281 $\pm$ 4 Jy, 180 $\pm$ 3 Jy, 115~$\pm$~2~Jy and 65 $\pm$ 1 Jy at 3.6, 4.5, 5.8 and 8.0 $\mu$m respectively \citep{Reach:2005}, and 7.2~$\pm$~0.1~Jy at 24~$\mu$m \citep{Engelbracht:2007}.  Additionally, the Wide-field Infrared Survey Explorer ({\it WISE}) All-Sky Data Release \citep{Cutri:2012} was used to extract photometry for all targets measured in the four {\it WISE} channels. The {\it WISE} magnitudes were converted into flux densities using the zero-points of 310~$\pm$~5~Jy, 172~$\pm$~3 Jy, 31.7~$\pm$ 0.5~Jy and 8.4~$\pm$~0.1~Jy at 3.4, 4.6, 12 and 22~$\mu$m respectively \citep{Wright:2010}. The IRAC, {\it WISE} flux densities and uncertainties are given in Table~\ref{Table:midIR}. The MIPS-1 flux densities and uncertainties are given in Table~\ref{Table:MIPS}. 

In the far-IR, {\it Spitzer} MIPS-2 (70~$\mu$m) measurements were compiled from \citet{Rebull:2010}, using the zero-point of  0.78~$\pm$~0.01~Jy \citep{Gordon:2007}. In addition to the new {\it Herschel} PACS observations presented here, a subset of the TBOSS sample 70~$\mu$m, 100~$\mu$m and 160~$\mu$m PACS measurements were compiled from \citet{Harvey:2012} and \citet{Howard:2013}. All far-IR flux densities and uncertainties are given in Table~\ref{Table:MIPS}.

The submillimeter and millimeter measurements consist of observations at 350 - 1200 $\mu$m, and were compiled from the following studies: \citet{Klein:2003, Young:2003, Andrews:2005, Bourke:2006, Andrews:2008, Bouy:2008, DiFrancesco:2008, Harris:2012}, and  \citet{Mohanty:2013}. At 1.3 mm and 2.6 mm, measurements were compiled from the following studies:  \citet{Beckwith:1990, Osterloh:1995, Motte:2001, Scholz:2006, Schaefer:2009, PhanBao:2011, Ricci:2013} and \citet{Andrews:2013}. All flux densities and uncertainties for the submillimeter and millimeter data are given in Table~\ref{Table:submm}.

In addition to the observed photometry, the SEDs for each target also show the underlying best-fit stellar atmospheric model. The two parameters that define the model atmosphere are effective stellar temperature ({\it T}$\mathrm{_{eff}}$) and surface gravity (${\rm log}(g)$); metallicity was assumed to be Solar.
The atmospheric model temperature was fixed based on the adopted spectral type of each target and using the temperature-spectral type conversion from \citet{Luhman:2003b}.
The value of ${\rm log}(g)$ and the initial value of stellar radius ({\it R$_{\mathrm{star}}$}) was set from the 1~Myr evolutionary models of \citet{Baraffe:1998} for targets with {\it T}$\mathrm{_{eff}}>$2700 K ($\leq$M8) and the models from \citet{Chabrier:2000} for targets with {\it T}$\mathrm{_{eff}}<$2700 K. 
The model atmosphere flux is scaled by {\it R$_{\mathrm{star}}$} and extinction before plotting on the SED. The values of {\it R$_{\mathrm{star}}$} and visual extinction ({\it A}$_{\it V}$) were used as free parameters in the fitting process, since the atmospheric models, set by target spectral type, resulted in poor fits for some cases. 
The initial extinction estimate for each target was scaled to {\it A}$_{\it V}$ from the reported {\it A}$_{\it J}$ or {\it A}$_{\it H}$, following the extinction law of \citet{Mathis:1990}.
For targets with no reported extinction, the initial {\it A}$_{\it V}$ was estimated using $A_{V} = 9.44 \times E(J-H)$, derived from the extinction law of \citet{Mathis:1990}, where $E(J-H) = (J-H) - (J-H)_{0}$ and is the excess with respect to the expected stellar photosphere. The intrinsic ({\it J-H})$_{0}$ colors were taken from \citet{Luhman:2010}. The visual extinctions compiled from the literature and those derived in this paper are given in Table~\ref{Table:sed}. 

The best-fit atmospheric model for each target was obtained by minimizing the $\chi^{2}$ value, summed over the 2MASS {\it JHK$_{S}$} bands where detections are reported. 
Varying the value of {\it A}$_{\it V}$ also accounts for the cases in which there is a disk excess affecting the photometry.
{\it A}$_V$ and {\it R$_{\mathrm{star}}$} were typically constrained to vary from no extinction to twice the reported {\it A}$_V$ values and from 0.4 to 2 times the derived stellar radii values. The reported {\it A}$_V$ and {\it R$_{\mathrm{star}}$} values for each target, and the corresponding best-fit SED values are listed in Table~\ref{Table:sed} and are also shown in Figure \ref{Rstar}.
The SEDs show the {\text PHOENIX}-based \citep{Hauschildt:1999}, "BT-settl" models \citep{Allard:2003, Allard:2011} for targets with {\it T}$\mathrm{_{eff}}>$2700~K and the "AMES-Dusty" models \citep{Allard:2001} for targets with {\it T}$\mathrm{_{eff}}<$2700~K. The redding law from \citet{Mathis:1990} with the optical total-to-selective extinction ratio of $R_{v} = 3.1$ was applied to the stellar atmospheric models. \\

For the multiple systems with angular separations in the range $\sim$5-11$\arcsec$ -- FU~Tau~A+B, GG~Tau~Ba+Bb, IRAS~04191+1523~A+B, and IRAS~04325+2402~AB+C \citep{Luhman:2009b, Duchene:2004} -- emission long-ward of the mid-IR is unresolved. For these systems, the combined system photometry were compiled, and the underlying best-fit spectra displayed on the SEDs was combined from the spectral types of both the primary and secondary components. The best-fit extinctions and stellar radii for these four multiples systems are given in Table~\ref{Table:sed}. Furthermore, for the specific case of GG~Tau~Ba+Bb, with an angular separation of 10$\farcs$75 from the primary system Aa+Ab \citep{Kraus:2009}, emission is resolved with both {\it Herschel} PACS at 70~$\mu$m \citep{Howard:2013} and the SMA at 1.3~mm \citep{Harris:2012}. The SED for GG~Tau~Ba+Bb was constructed with photometry from the secondary system, and the underlying best-fit spectrum was combined from both the spectral types of the Ba and Bb components. 


\begin{figure}
\centering
\includegraphics[scale=0.6]{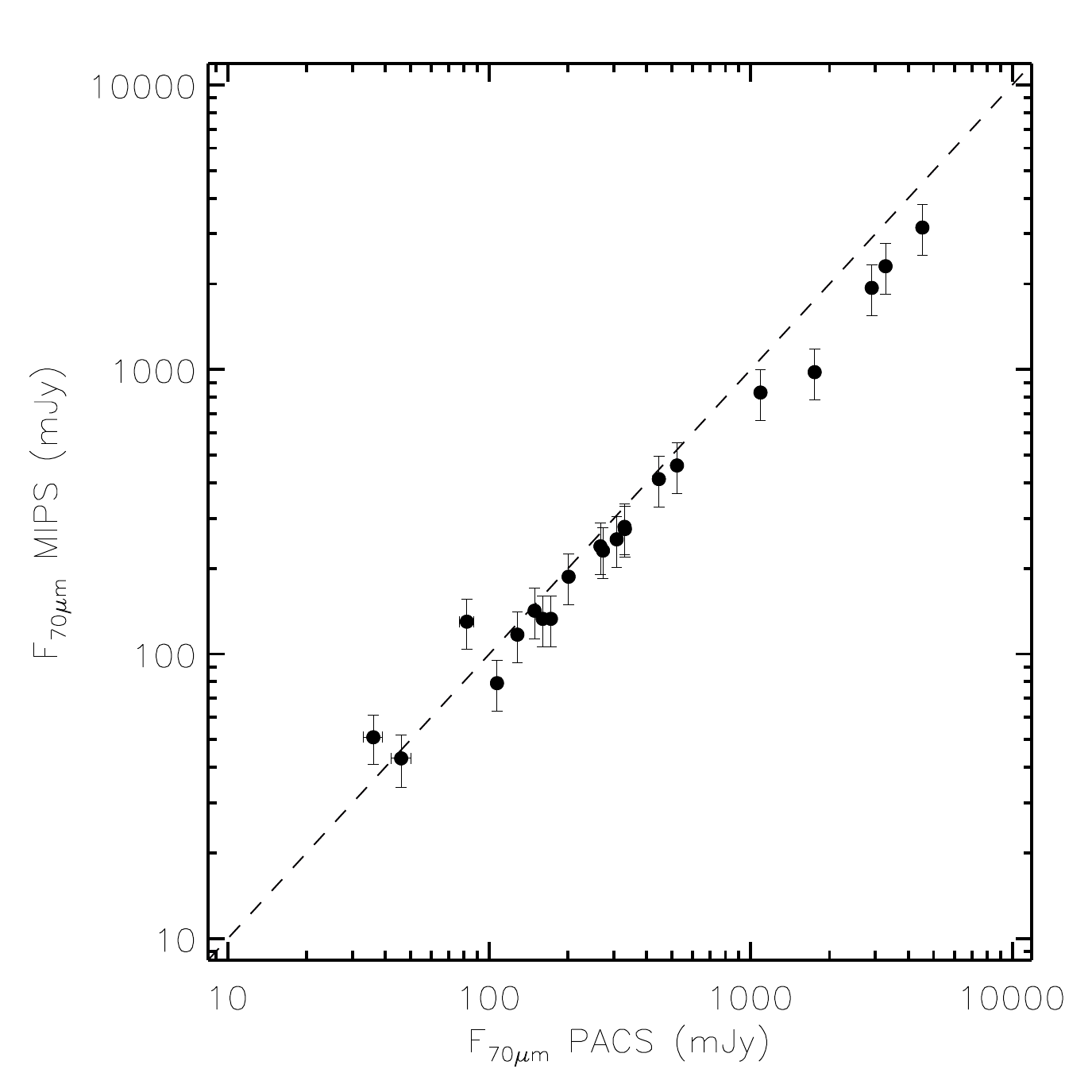}
\caption {Plot showing the relation of PACS 70 $\mu$m and corresponding MIPS 70~$\mu$m flux densities \citep{Rebull:2010} for the TBOSS targets detected with both instruments. Non-linearity of the MIPS detector at $\sim$1~Jy is the likely cause of the 25-40\% flux discrepancy seen for the targets  $>$1 Jy, measured with PACS.}
\label{mips}
\end{figure}
\begin{figure*}
\centering
\includegraphics[scale=0.75]{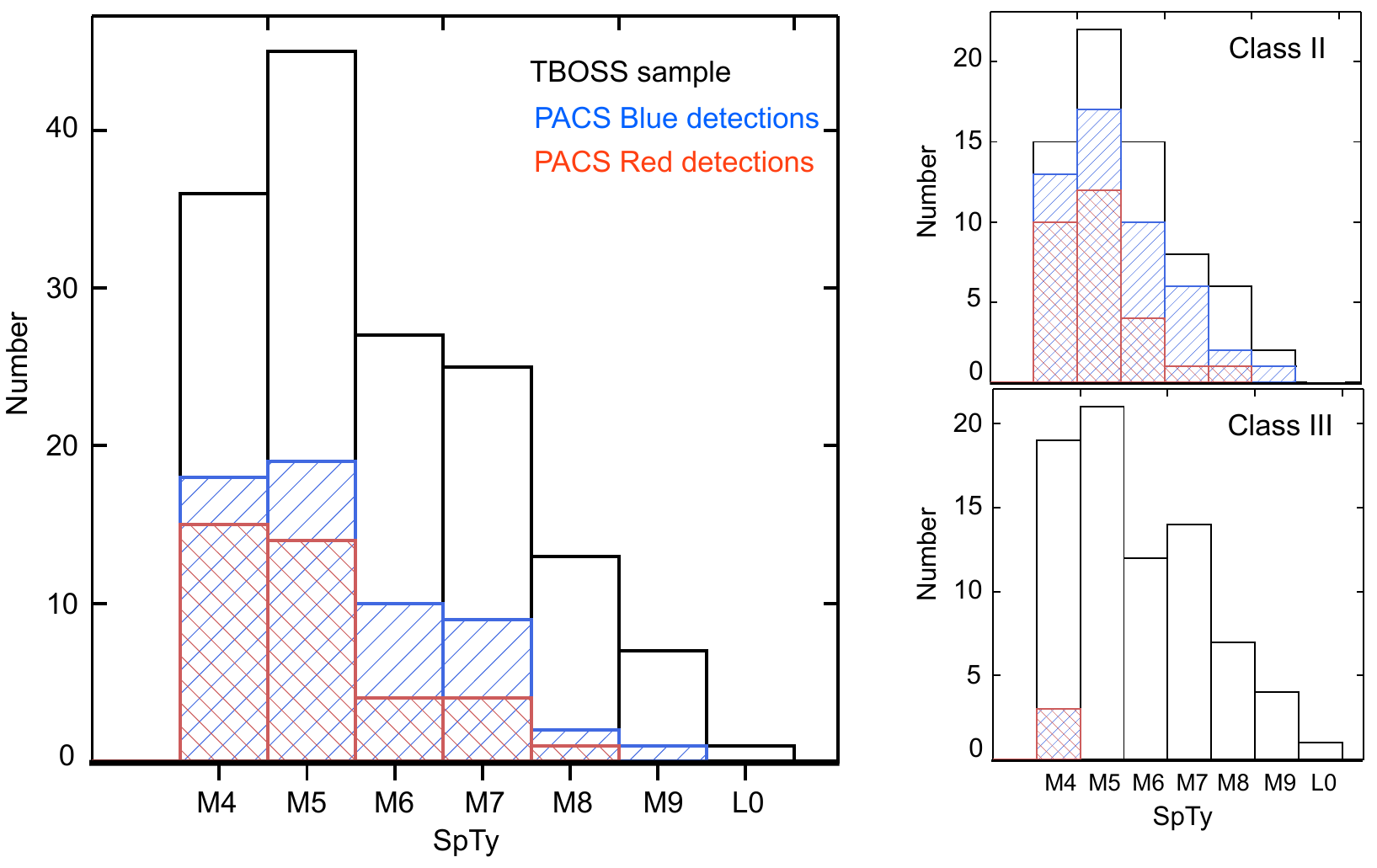}
\caption {Histogram of spectral type for the TBOSS sample observed with {\it Herschel} PACS. The full observed sample and the number of PACS Blue channel (70/100~$\mu$m) detections and Red channel (160~$\mu$m) detections are shown on the left. The Class~II and Class~III observed sample, and PACS Blue and Red detections are shown in the top and bottom plots on the right respectively. }
\label{histogram}
\end{figure*}

\section{Results}
\label{Sec:Results}

Of the 150 TBOSS targets observed with {\it Herschel}, 58 were detected, 2 have unresolved emission from earlier spectral type companions, and 90 have upper limits. The PACS photometry compilation for these 150 targets is reported in Table~\ref{table:PACS}. The spatial distribution of the full TBOSS sample is shown in Figure~\ref{Av}, along with the corresponding subset of detections and upper limits. The {\it Herschel} results for each evolutionary class are reported in Section \ref{Sec:Class1} (Class~I), \ref{Sec:Class2} (Class~II), and \ref{Sec:Class3} (Class~III). After the summary of {\it Herschel} results, the PACS flux densities are compared with {\it Spitzer} MIPS flux densities in Section~\ref{Sec:mips}. Combining the {\it Herschel} and {\it Spitzer} results, the overall 153-object TBOSS sample far-IR detection rates are calculated in Section~\ref{Sec:DetRates}. Although the {\it Herschel} beam size corresponds to >750~AU in Taurus, several TBOSS targets are spatially extended, and these cases are reported in Section~\ref{Sec:Extended}. 

\subsection{Class I population}
\label{Sec:Class1}
All of the seven Class I targets within the TBOSS sample are detected in both PACS Blue (70~$\mu$m or 100~$\mu$m) and Red (160~$\mu$m) channels.
The Class I targets include the brightest targets amongst the TBOSS sample observed with {\it Herschel}, although the two brightest Class I sources -- IRAS~04191+1523~B and IRAS~04248+2612 -- are unresolved binaries, as noted in Table~\ref{table:PACS}. The flux densities of the isolated Class~I substellar objects range from $\sim$270~mJy to >4~Jy, larger than nearly all of the Class~II sources and indicate that a substantial amount of dust is present around brown dwarfs in the early stage of evolution. 
The maps for the spatially extended Class I targets are given in Section~\ref{Sec:Extended} and the maps of the remaining unresolved Class I targets are given in Appendix~\ref{Sec:maps}.

\subsection{Class II population}
\label{Sec:Class2}
The majority (>70\%) of the Class~II TBOSS targets are detected with the PACS Blue channel at either 70~$\mu$m or 100~$\mu$m, with a large range of flux densities spanning three orders of magnitude. Approximately half of the Class~II TBOSS sample also have PACS 160~$\mu$m flux densities. 
Furthermore, several of the detected Class~II targets have flux densities comparable to the Class~I targets. Table~\ref{table:PACS} reports the flux densities and upper limits.
The angular resolution degrades at longer wavelengths, and  the source JH~112~B, is undetected at 70~$\mu$m, but a long wavelength detection is due to contamination from emission associated with an earlier spectral type primary. Additionally, the 100~$\mu$m emission witnessed for IRAS~04325+2402~C (shown in Figure~\ref{resolved}) is unresolved emission from an earlier spectral type primary, and is further discussed in Section~\ref{Sec:Extended}. 
The maps of the detected Class~II sources are given in Appendix~\ref{Sec:maps}.

\subsection{Class III population}
\label{Sec:Class3}
Only a small subset (<5\%) of Class~III TBOSS targets are detected at one or more wavelengths in the {\it Herschel} data. Although the overall frequency of detected disks is low for Class~III targets, the measured flux densities of the few Class~III detections are larger than some Class~II targets of equivalent spectral type. One of the detected Class~III targets -- XEST 08-033 -- is unusual, since it is the only source with a non-detection at 70~$\mu$m and a longer wavelength detection uncontaminated by a more massive primary. The 160~$\mu$m map for XEST~08-033 shows two sources unlike most of the detected targets. The source at the coordinates of XEST~08-033 is unresolved as expected for a circumstellar disk. 
Maps of the one detected Class~III target in which the emission is spatially extended -- XEST 17-036 -- is given in Section~\ref{Sec:Extended}, and the maps of the remaining detected Class III targets are given in Appendix~\ref{Sec:maps}.

\subsection{Comparison with {\it Spitzer} MIPS}
\label{Sec:mips}
Among the TBOSS targets detected at 70~$\mu$m with the {\it Herschel} PACS maps, 21 sources have existing 70~$\mu$m detections with {\it Spitzer} MIPS data \citep{Rebull:2010}. A plot of the PACS 70~$\mu$m flux densities as a function of the corresponding MIPS 70 $\mu$m flux densities for these 21 sources are shown in Figure~\ref{mips}. For the sources with flux densities <1~Jy, the agreement between the two values is typically within the MIPS absolute flux calibration uncertainty (<10\%). For the six sources brighter than 1~Jy, the discrepancy between the PACS and MIPS fluxes range from 25-40\%, and the {\it Herschel} PACS flux is systematically higher. As noted in a {\it Herschel} Science Center technical note\footnote{Technical note: PICC-ME-TN-037}, the processed PACS data are affected by the high pass filter width, source masking, drizzling, and aperture sizes used in the photometry. The contribution of these factors is expected to have an impact on the final flux densities by only a small percent and cannot explain the full discrepancy between the larger systematic offset. The most likely cause is due to the impact of non-linearity for MIPS data compared to PACS; the PACS detector is linear up to flux density levels well above the brightest TBOSS target, while the MIPS detector is non-linear for flux densities above $\sim$1~Jy\footnote{https://nhscsci.ipac.caltech.edu/sc/index.php/Pacs/AbsoluteCalibration}.

\subsection{Far-IR Detection rates for Taurus members}
\label{Sec:DetRates}
Histograms of the TBOSS sample observed with {\it Herschel} PACS are plotted in Figure~\ref{histogram} with detections indicated. By combining results from the TBOSS far-IR observations with similar {\it Herschel} PACS \citep{Howard:2013} and {\it Spitzer} MIPS \citep{Rebull:2010} measurements of the earlier spectral type Taurus members, a comprehensive accounting of the detection rates above and below the stellar/substellar boundary is calculated and given in Table~\ref{table:diskFraction}. The detections rates at 70~$\mu$m of Class~I and Class~II targets of any spectral type are 93\% and 84\%, respectively. 73\% of Class~I targets were detected at both 70~$\mu$m and 160~$\mu$m, and 52\% of observed Class~II targets were similarly detected at both wavelengths. 
The detection rate declines for later spectral types, however even the substellar (M6.25-L0) Class~II members exhibit disk emission at 70~$\mu$m in the majority ($>$50\%) of systems. 
For the Class III members detected at 70~$\mu$m, and both 70~$\mu$m and 160~$\mu$m, the detection rate is systematically low ranging from $\sim$5\% for K0-M6 stars to 0\% for M6.25-L0 brown dwarfs. Direct comparisons of the far-IR emission between the substellar members of Taurus and other star-forming regions are not possible, since the TBOSS survey represents a uniquely sensitive and complete study.

\subsection{Spatially extended targets}
\label{Sec:Extended}
\begin{figure}
\centering
\includegraphics[scale=0.25]{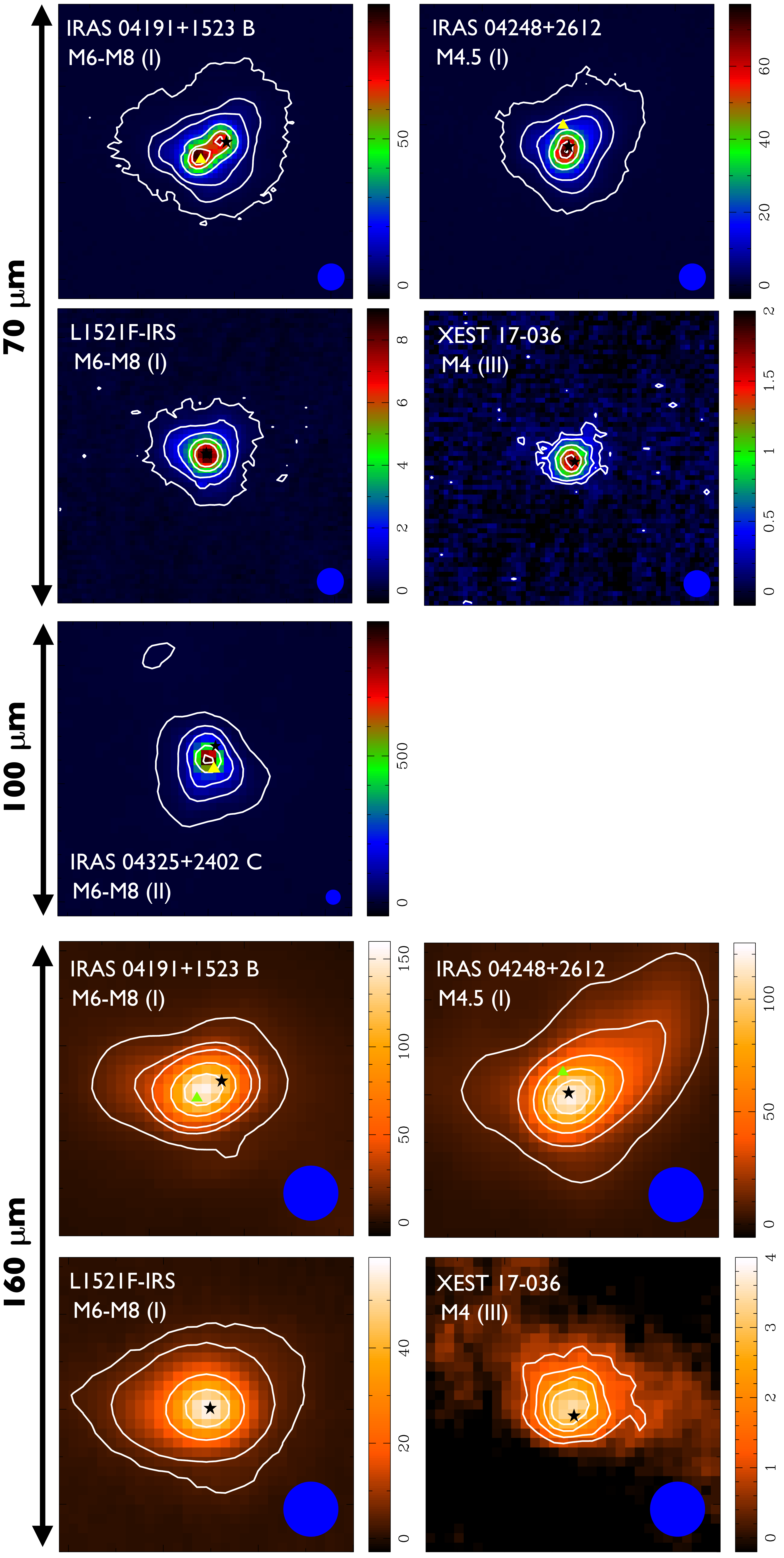}
\caption {PACS Blue channel (70~$\mu$m and 100~$\mu$m) and Red channel (160~$\mu$m) maps of the five spatially extended TBOSS targets. Target name, spectral type and evolutionary class are labeled in the top, lefthand corner of each map. The 70~$\mu$m and 160~$\mu$m maps are 60$\arcsec$ in size along each axis, and the 100~$\mu$m maps are 120$\arcsec$ in size along each axis. Beam sizes are represented by the blue circles in the bottom, right-hand corner of each map. The scale bar shows the intensity of each map in units of mJy/pixel. 2MASS coordinates of the TBOSS targets are represented by the black stars. For binary systems, the position of the companion star are indicted with yellow triangles in the 70 $\mu$m and 100~$\mu$m maps, and by green triangles in the 160~$\mu$m maps. Contours levels begin at 3$\sigma$ in all maps and extend up to 500$\sigma$, 400$\sigma$, 150$\sigma$, 40$\sigma$ and 300$\sigma$ in the Blue channel maps of IRAS~04191+1523 B, IRAS~04248+2612, L1521F-IRS, XEST~17-036 and IRAS~04325+2402~C respectively, and up to 50$\sigma$, 60$\sigma$, 70$\sigma$ and 10$\sigma$ in the Red channel maps of IRAS~04191+1523~B, IRAS~04248+2612, L1521F-IRS and XEST~17-036 respectively.}
\label{resolved}
\end{figure}

Although the {\it Herschel} PACS 70~$\mu$m and 160~$\mu$m beam sizes are 5$\farcs$6 and 11$\farcs$3 -- which translate into spatial scales of >750~AU at the distance of Taurus -- five TBOSS targets exhibit extended structure in the PACS maps shown in Figure~\ref{resolved}.
Amongst the 5 spatially extended targets 3 have been imaged for companions -- IRAS~04191+1523, IRAS~04248+2612, and IRAS~04325+2402. These three resolved sources are binary systems for the which the angular separations are listed in Table~\ref{table:Multiplicity} \citep{Duchene:2004}. The components for two of these systems are separated by an angle larger than the 70~$\mu$m beam size. IRAS~04191+1523 is extended along the binary axis and two peaks are visible in the map. IRAS~04325+2402 also extended along the binary axis and only one peak is present in the map. In these two wider systems, the TBOSS target is the secondary component. For the remaining resolved binary IRAS~04248+2612, both components are M4+, only one peak is present, and a large extension to the northwest direction $\sim$45$\degr$ to the binary axis is evident. The two additional spatially extended source are the Class~I target L1521F-IRS and the Class~III target XEST~17-036. In both systems, the emission is largely symmetric. The Class~I emission may result from an envelope of dust, however the origin of extended emission in the Class~III source is difficult to explain.

\begin{figure}
\centering
\includegraphics[scale=0.3]{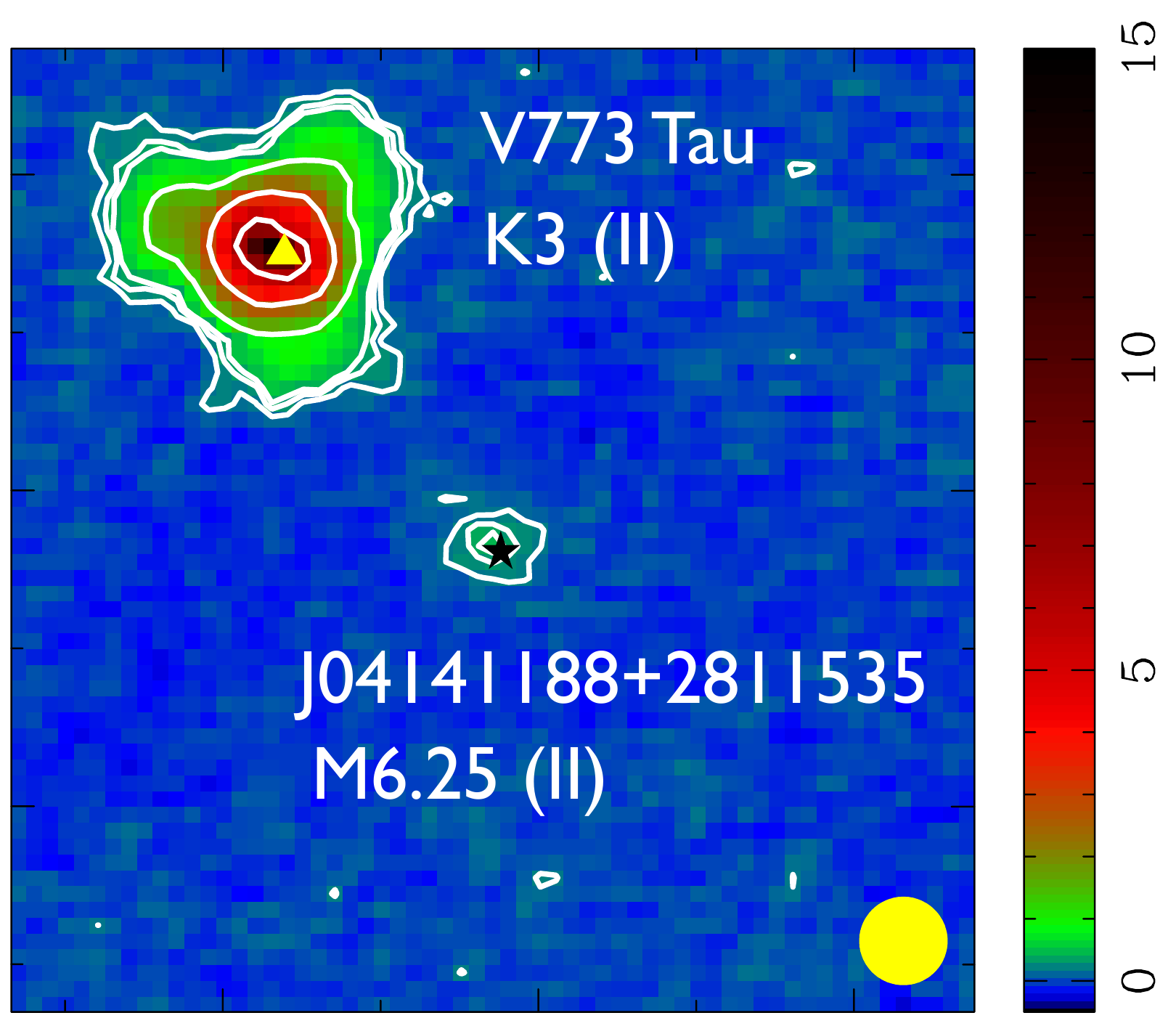}
\caption {PACS Blue channel (70~$\mu$m) map of the faintly detected brown dwarf J04141188+2811535 (black star). The earlier type companion V773~Tau (yellow triangle) is also seen within the map and is spatially extended.
Target name, spectral type and evolutionary class for each component of the system are indicated on the map. The map is 60$\arcsec$ in size along each axis. The beam size is represented by the yellow circle in the bottom, right-hand corner. Contour levels are plotted at 3, 5, 7, 20, 50 and 100$\sigma$. The scale bar shows the intensity of the map in mJy/pixel.}
\label{V773Tau}
\end{figure}

Within the 70~$\mu$m map of the faintly detected brown dwarf J04141188+2811535 is the earlier spectral type object V773 Tau which also appears to be spatially extended, as shown in Figure~\ref{V773Tau}. The FWHM of V773 Tau is 6$\farcs$5, while the FWHM of ten targets with similar brightness is 5$\farcs$5~$\pm$~0$\farcs$2. Follow-up interferometric imaging should be able to resolve the disk structure and determine which component or components harbor the disk. In addition to the wide separation TBOSS brown dwarf companion to V773 Tau, the primary is a multiple system that includes many components, within an 0$\farcs$3 radius, there are at least four stars \citep{Boden:2007,Boden:2012}. The spatially extended {\it Herschel} emission indicates that there may be a circumbinary distribution to the dust. 




\section{Discussion}
\label{Sec:Discussion}

\subsection{Far-IR emission as a function of spectral type}
\begin{figure}
\centering
\includegraphics[scale=0.63]{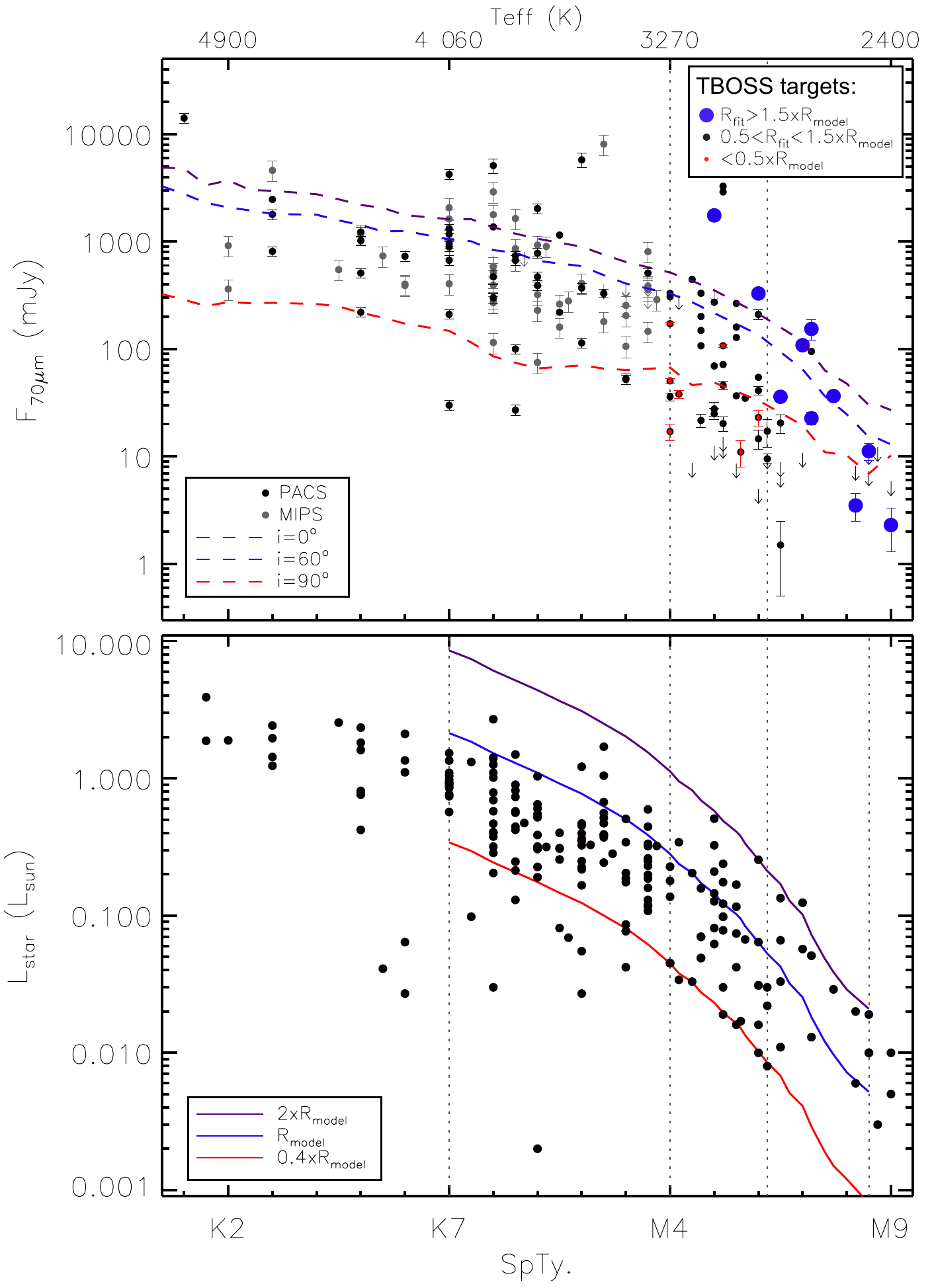}
\caption {({\bf top}) 70~$\mu$m flux density shown as a function of spectral type for the Taurus Class~II population, with spectral types spanning early K to late M. The vertical dotted lines mark the boundaries of the TBOSS sample (targets of spectral types M4+) and the stellar/substellar limit (M6.25). Flux densities measured from PACS and MIPS are represented by the black and gray points respectively. Additionally, for the detected TBOSS targets, the large blue points correspond to best-fit stellar radii of >1.5 times that of the evolutionary model values, and the small red points correspond to best-fit stellar radii of <0.5 times that of the evolutionary model values. Downwards arrows represent the 3$\sigma$ upper limits. The dashed lines represent the 70~$\mu$m flux densities extracted from a test grid of model SEDs, generated with the radiative transfer code MCFOST and represent the flux densities for disk inclinations of face-on (0$\degr$; purple dashed-line), 60$\degr$ (blue dashed-line) and edge-on (90$\degr$; red dashed-line). ({\bf bottom}) Bolometric luminosities of the Taurus Class~II population, with spectral types spanning early K to late M. The luminosities of the TBOSS sample are derived from the best-fit stellar radii. The luminosities for targets of spectral type <M4 are from those reported in \citet{Andrews:2013}. The 1~Myr isochrone from \citet{Baraffe:1998} is represented with the blue solid line and the corresponding isochrones are shown for values of 2 and 0.4 times that of the model stellar radii (purple and red solid lines respectively). 
}
\label{detRate}
\end{figure}

\begin{figure}
\centering
\includegraphics[scale=0.6]{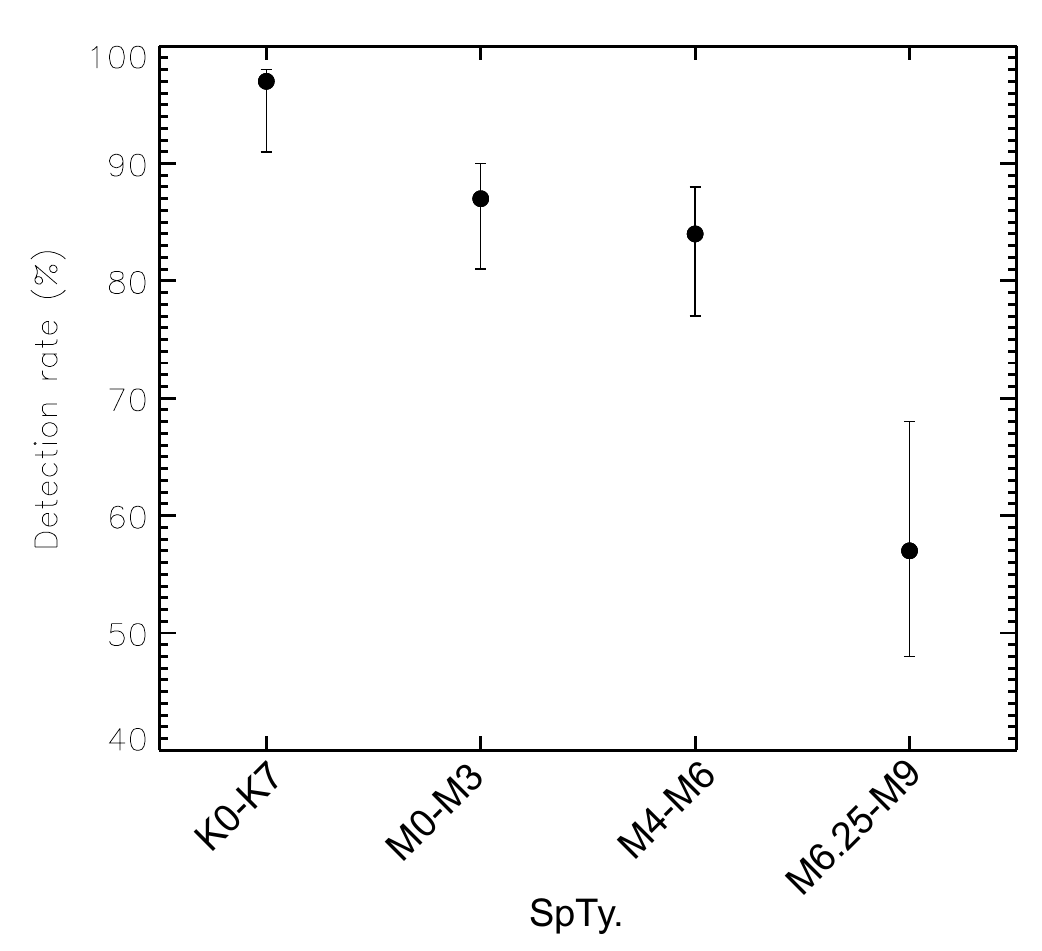}
\caption {Taurus Class~II, 70~$\mu$m detection rates for objects of spectral types K0-M9. The M4-M6 and M6.25-M9 detections rates are compiled from the TBOSS sample observed with {\it Herschel} PACS. The K0-K7 and M0-M3 detections rates are compiled form the {\it Herschel} PACS observations reported in \citet{Howard:2013} and the {\it Spitzer} MIPS observations reported in \citet{Rebull:2010}.
}
\label{detRate2}
\end{figure}

The large number of PACS 70~$\mu$m detections in the TBOSS Class~II sample defines the upper boundary of far-IR emission. Connecting the TBOSS population flux densities with previous measurements of earlier spectral type Class~II Taurus members reported in the literature \citep{Howard:2013,Rebull:2010} provides the first nearly complete census of 70~$\mu$m emission for a full population of a star-forming region. Figure~\ref{detRate} plots the 70~$\mu$m flux densities and upper limits as a function of spectral type spanning K0-M9 for 159 Taurus Class~II members, representing 89\% of all objects of this evolutionary class. Although the {\it Spitzer} and {\it Herschel} GASPS programs targeting earlier spectral types had shallower detection limits than the TBOSS observations, the detection rates for the early spectral types is nearly 100\% as shown in Figure~\ref{detRate2}, indicating that the trends seen in the flux density as a function of spectral type are not due to an observational bias. As shown in Figure~\ref{detRate2}, the Class~II disk frequency based on a 70~$\mu$m flux density detection declines moderately from nearly 100\% for the earliest spectral types (K0-K7) to $\sim$80\% to the latest stellar spectral types (M4-M6). Below the stellar limit (spectral types M6.25-M9), the frequency of Class~II disks drops significantly to $\sim$50\%.

\begin{figure}
\centering
\includegraphics[scale=0.5]{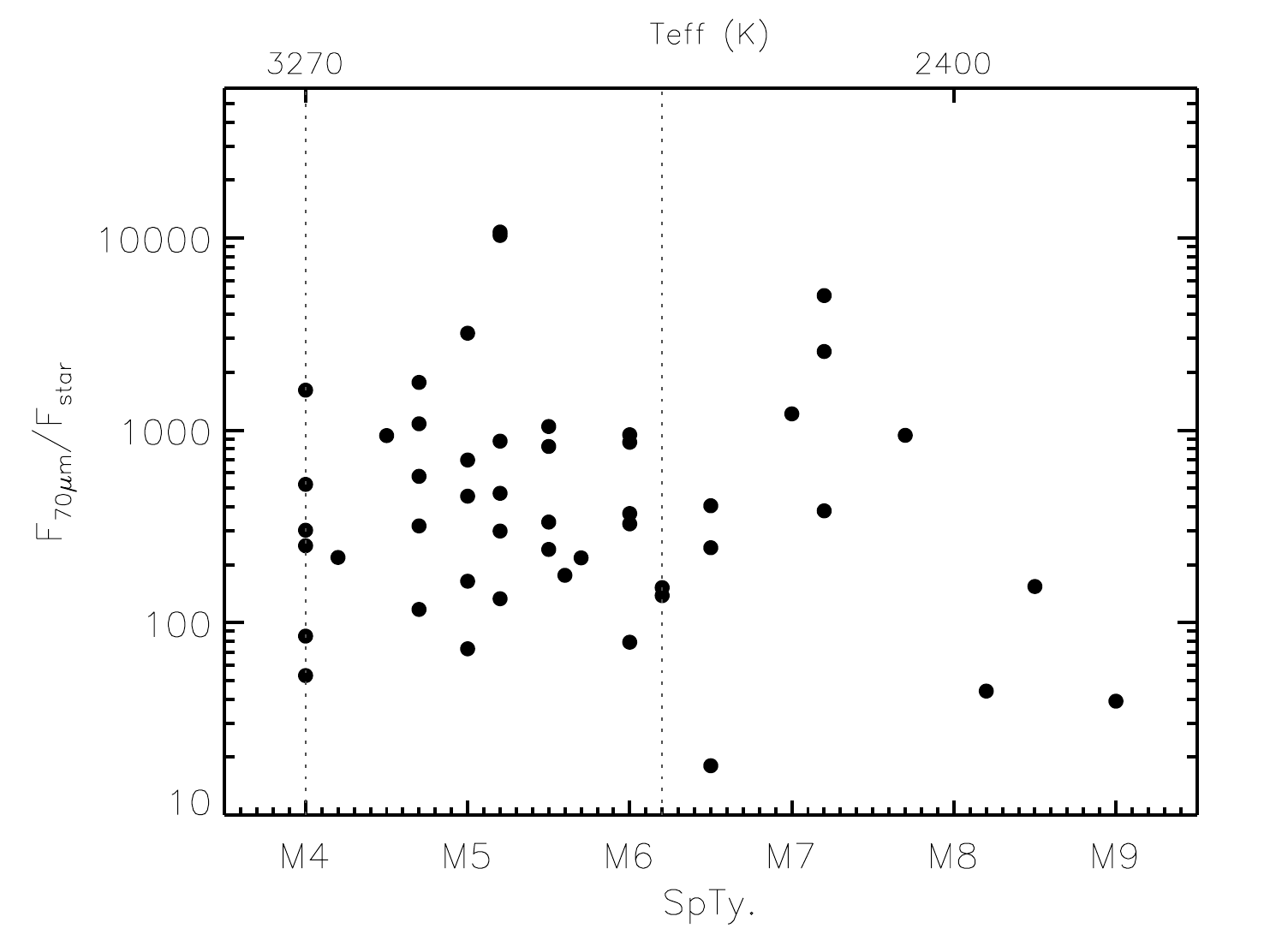}
\caption {70~$\mu$m disk excess fraction shown as a function of spectral type for the detected Class~II objects within the TBOSS sample.
}
\label{Fstar}
\end{figure}

The overall shape of the flux density distribution in Figure~\ref{detRate} exhibits a systematic decline with later spectral types, and the slope of the decline is shallower for K- and early-M stars than for mid- to late-M dwarfs. For comparison with the disk flux density trend, the stellar bolometric luminosity predicted from the 1~Myr evolutionary models of \citet{Baraffe:1998}, that span the spectral types K7 to $\sim$M8 (i.e., {\it T}$_{\mathrm{eff}}$ $\sim$4100-2300 K), is displayed in the lower panel of Figure \ref{detRate}. For the TBOSS Class~II objects detected at 70~$\mu$m, bolometric luminosities were calculated from the best-fit stellar radii derived in the SED fitting procedure described in Section~\ref{Sec:SEDs}. For the Taurus Class II population outside the TBOSS sample (<M4), bolometric luminosities were supplemented from those reported in \citet{Andrews:2013}. The similarity in the shapes of the stellar bolometric luminosity function and the disk flux density data suggests that the excess flux density fraction from the disk is similar across the full population of Taurus. For the TBOSS sample, the disk excess fraction is plotted as a function of spectral type in Figure~\ref{Fstar}. There is no distinction across the stellar/substellar boundary and a Kolmogorov-Smirnov (K-S) test indicates that the stellar and substellar TBOSS samples are drawn from the same parent distribution.


\begin{figure*}
\centering
\includegraphics[scale=0.38]{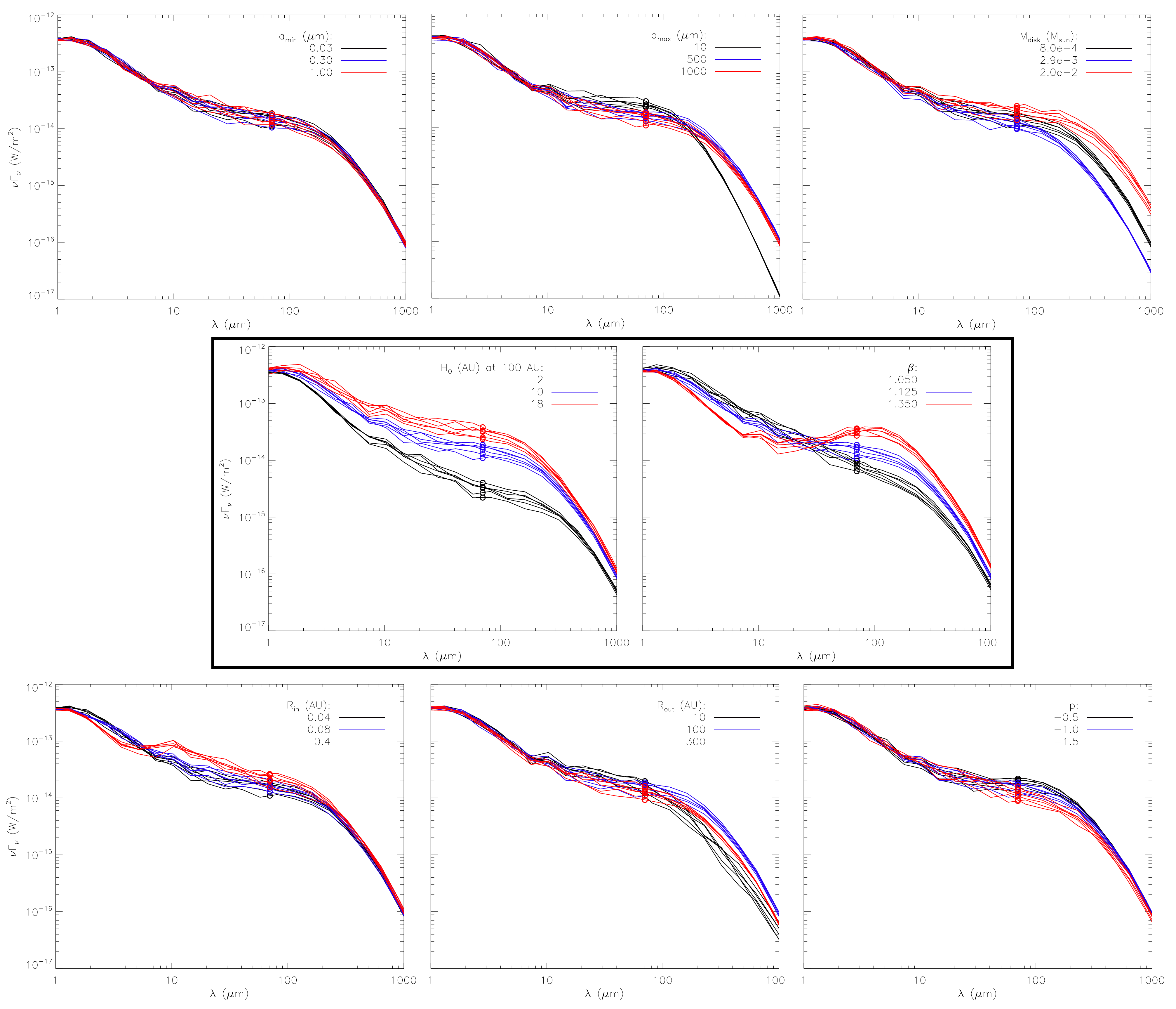}
\caption {SED visualization showing the impact that individual disk parameters have on excesses at 70 $\mu$m, the wavelength of which is highlighted on the SEDs with the open circle symbols. The SEDs were generated with the 3D radiative transfer code MCFOST \citep{Pinte:2006, Pinte:2009}. In each individual panel, 3 models of SEDs were generated (black, blue and red lines), by varying one of the the eight disk parameters with values that are displayed in each panel. From left to right and top to bottom, the disk parameters are: minimum grain size ({\it a}$_{\mathrm{min}}$), maximum grain size ({\it a}$_{\mathrm{max}}$), disk mass ({\it M}$_{\mathrm{disk}}$), scale height ({\it H}$_{0}$) at a reference disk radius of 100 AU, disk flaring exponent ($\beta$), inner radius ({\it R}$_{\mathrm{in}}$), outer radius ({\it R}$_{\mathrm{out}}$), and surface density exponent ({\it p}). The SEDs were generated for a star of spectral type M4, and the stellar properties were set based on the temperature conversion of \citet{Luhman:2003b} and the 1 Myr evolutionary models of \citet{Baraffe:1998}, i.e., {\it T}$_{\mathrm{eff}}$=3300 K, {\it R}$_{\mathrm{star}}$=1.65 R$_\sun$, {\it M}$_{\mathrm{star}}$=0.29 M$_\sun$. The model SEDs are shown at disk inclinations of 18$\degr$, 32$\degr$, 41$\degr$, 49$\degr$, 57$\degr$ and 63$\degr$ (where 0$\degr$ is for a face-on disk inclination), for each value of the varied disk parameters. Disk geometry;  inclination, {\it H}$_{0}$ and $\beta$ are seen to have the largest impact on the emission at 70 $\mu$m.
}
\label{mcfost}
\end{figure*}

At any given spectral type, there is a significant spread in the far-IR emission, typically over an order of magnitude, as seen in Figure~\ref{detRate}. To investigate possible origins of the large dispersion in disk flux density, we generated a series of model SEDs with the radiative transfer code MCFOST \citep{Pinte:2006,Pinte:2009} and considered the range of 70~$\mu$m flux densities in the model SEDs. Rather than creating an exhaustive multi-parameter grid search on any given target, we varied individual disk parameters about a reference disk defined by the following parameters: inner and outer radius ({\it R}$_{\mathrm{in}}$, {\it R}$_{\mathrm{out}}$), minimum and maximum grain size ({\it a}$_{\mathrm{min}}$, {\it a}$_{\mathrm{max}}$), surface density exponent ({\it p}), scale height ({\it H}$_{0}$), flaring exponent ($\beta$) and disk mass ({\it M}$_{\mathrm{disk}}$). As shown in Figure~\ref{mcfost}, disk properties such as {\it R}$_{\mathrm{in}}$, {\it R}$_{\mathrm{out}}$, {\it a}$_{\mathrm{min}}$, {\it a}$_{\mathrm{max}}$, {\it p}, and {\it M}$_{\mathrm{disk}}$ do not have a large impact on the 70~$\mu$m flux density relative to other wavelength regimes, and the PACS data do not constrain these parameters. In contrast, variations in geometrical and structural factors such as disk inclination, {\it H}$_{0}$ (equivalent to disk opening angle), and $\beta$, can individually account for variations up to nearly an order of magnitude. Following on from this assessment, properties for a canonical disk model were used to generate a grid of model SEDs, across spectral types ranging from K0-M9, in order to investigate the effect that disk inclination alone can have in the spread of 70~$\mu$m flux density for objects with the same spectral type. The input stellar properties ({\it T}$_{\mathrm{eff}}$, {\it R}$_{\mathrm{star}}$, and {\it M}$_{\mathrm{star}}$) were set based on the 1~Myr evolutionary models of \citet{Palla:1999} for spectral types <K7, \citet{Baraffe:1998} for spectral types K7-M8, and those from \citet{Chabrier:2000} for spectral types $\geq$M8. Six of the eight input disk parameters were fixed in accordance to typical values reported in the literature; {\it a}$_{\mathrm{min}}$ = 0.03 $\mu$m, {\it a}$_{\mathrm{max}}$~=~1000~$\mu$m, {\it H}$_{0} = 10$~AU at a reference radius of 100 AU, $\beta$ = 1.125, {\it p}~=~-1.0 and {\it R}$_{\mathrm{out}}$~=~100 AU. The two remaining disk parameters were scaled in accordance to the properties of the central object; {\it M}$_{\mathrm{disk}}$ was set at 1\% of the stellar mass (e.g \citealp{Scholz:2006, Andrews:2013}) with a standard gas to dust ratio of 100:1, and {\it R}$_{\mathrm{in}}$ was set at the sublimation radius for dust grains at {\it T}$_{\mathrm{sub}}$ = 1500 K. The SED models were generated for disk inclinations at; face-on (0$\degr$), 60$\degr$, and edge-on (90$\degr$), from which the 70~$\mu$m flux densities were extracted and are shown in Figure~\ref{detRate}. Between the face- and edge-on disk inclinations, over an order of magnitude spread in flux density is seen across the majority of the spectral range, and 73\% of the observed population is bounded between these two models. 
Direct observations of disks around brown dwarfs, and constraints from SED modeling results reported in the literature, suggest that the outer disk radius is typically in the range $\sim$10-40 AU for these low mass objects (e.g., \citealp{Scholz:2006, Luhman:2007, Ricci:2013}). In order to ensure that the canonical value {\it R}$_{\mathrm{out}}$ = 100 AU does not bias the model results for objects in the brown dwarf regime, a second grid of SEDs were generated in which {\it R}$_{\mathrm{out}}$ was scaled as {\it M}$_{\mathrm{star}}^{1/2}$ (keeping the density across the area of the disk constant across the spectral range) and bound at 10 AU for an object of spectral type M9. No significant change in the 70~$\mu$m flux density was seen between the these two grid of models.  
Whilst the disk model 70~$\mu$m flux densities from face- to edge-on exhibit a spread of over a magnitude, and encompass the majority of the observed population, an equivalent trend and spread is seen in the observed bolometric luminosities in the lower panel of Figure~\ref{detRate}. In addition to the 1 Myr isochrone shown in Figure~\ref{detRate}, the isochrones for 0.4 and 2 times that of the model stellar radii are also displayed, and selected based on the results of the SED fitting of the TBOSS sample. To distinguish whether or not the stellar properties alone are the underlying cause of the observed trend and spread in disk flux densities, the TBOSS data points in Figure~\ref{detRate} are plotted with a size corresponding to the best-fit stellar radii. The random scatter of large (>1.5 times the model radii) and small (<0.5 times the model radii) best-fit model radii throughout the population indicates that the spread seen in the 70~$\mu$m disk excess cannot be solely explained due to the range of fluxes of the central source. This further reinforces the requirement for detailed modeling of the disk properties across the full population, and is the subject of a forthcoming paper.

\subsection{Spectral Energy Distributions of the TBOSS sample}
\begin{figure}
\centering
\includegraphics[scale=0.465]{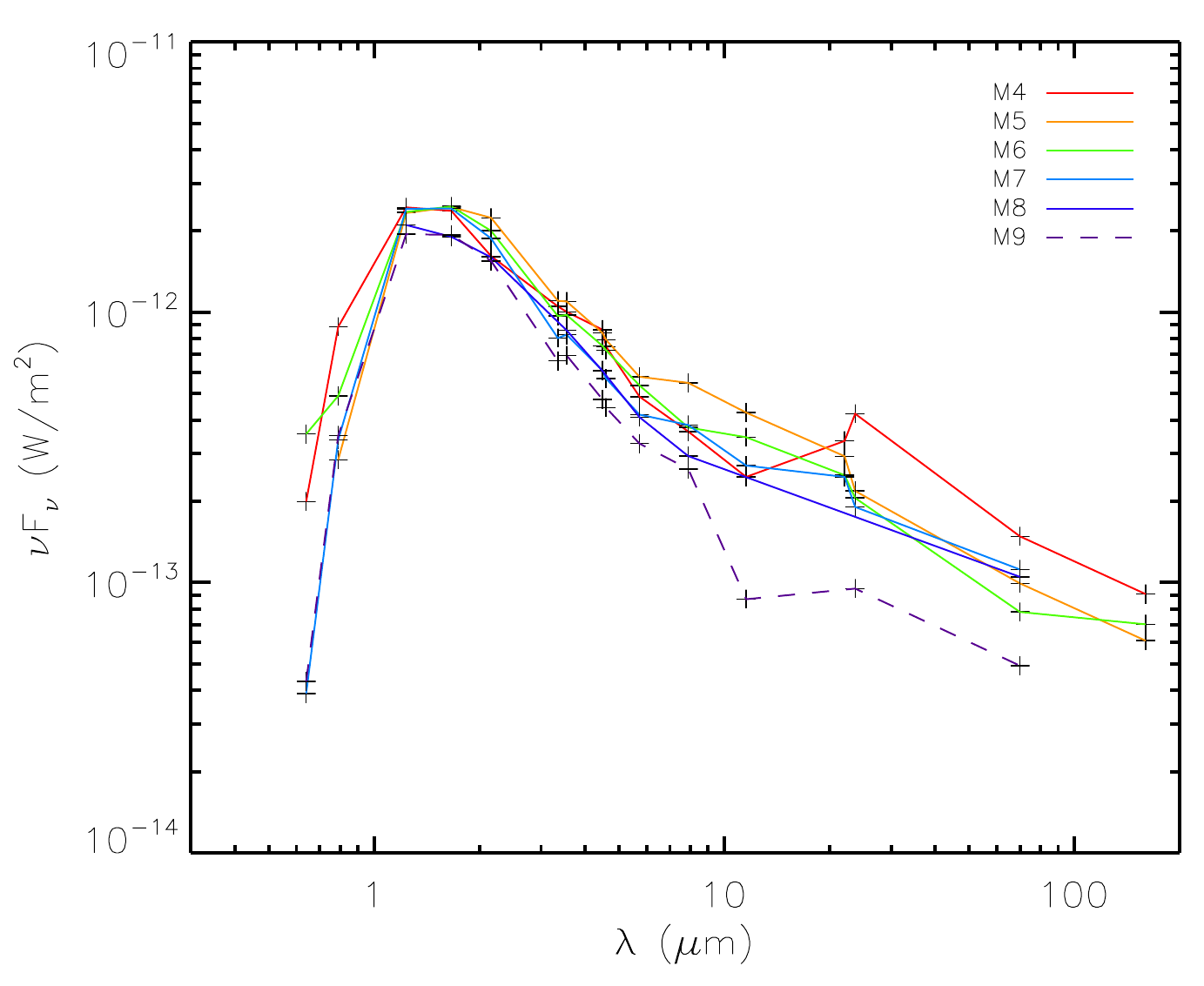}
\caption{Median SEDs of the Class~II TBOSS targets detected with {\it Herschel} PACS. The median SEDs are normalized by the {\it J}-band flux density, and shown per spectral type for M4-M8. The normalized SED of the one, M9 detected target (KPNO~6; \citealp{Harvey:2012}) is shown and represented by the purple, dashed line.
}
\label{medianSED}
\end{figure}

SEDs were constructed for each target of the TBOSS sample by combining the new {\it Herschel} far-IR photometry with literature photometry measurements from optical - submm/mm wavelengths. The underlying stellar photosphere is also plotted on each SED. For 65 of the 148 TBOSS targets, the {\it Herschel} data represents the longest wavelength measurement.

The SEDs for all Class~I targets are given in Appendix~\ref{Sec:sed_ClassI}. The PACS flux densities are at or near the peak power point for these embedded sources. 

Among the Class~II targets, 75\% of the sample is detected in at least one of the {\it Herschel} channels, and the SEDs for the majority of these sources are given in Appendix~\ref{Sec:sed_ClassII}, Figures~\ref{sed1_multi}-\ref{sed2d}. Owing to to the difficulty in interpreting the SEDs for targets with unresolved companions, the SEDs for that category of target are plotted separately in Figure~\ref{sed1_multi}. The SEDs of the smaller number of undetected Class~II targets are shown in Figure~\ref{sed2d}.
Based on the {\it Herschel} PACS data, a subset of the detected Class~II targets are identified as candidate transition disk and candidate truncated disk objects; the SEDs for those targets are presented in Sections~\ref{Sec:TD} and \ref{Sec:Trunc}, respectively.

\begin{figure*}
\centering
\includegraphics[scale=0.6]{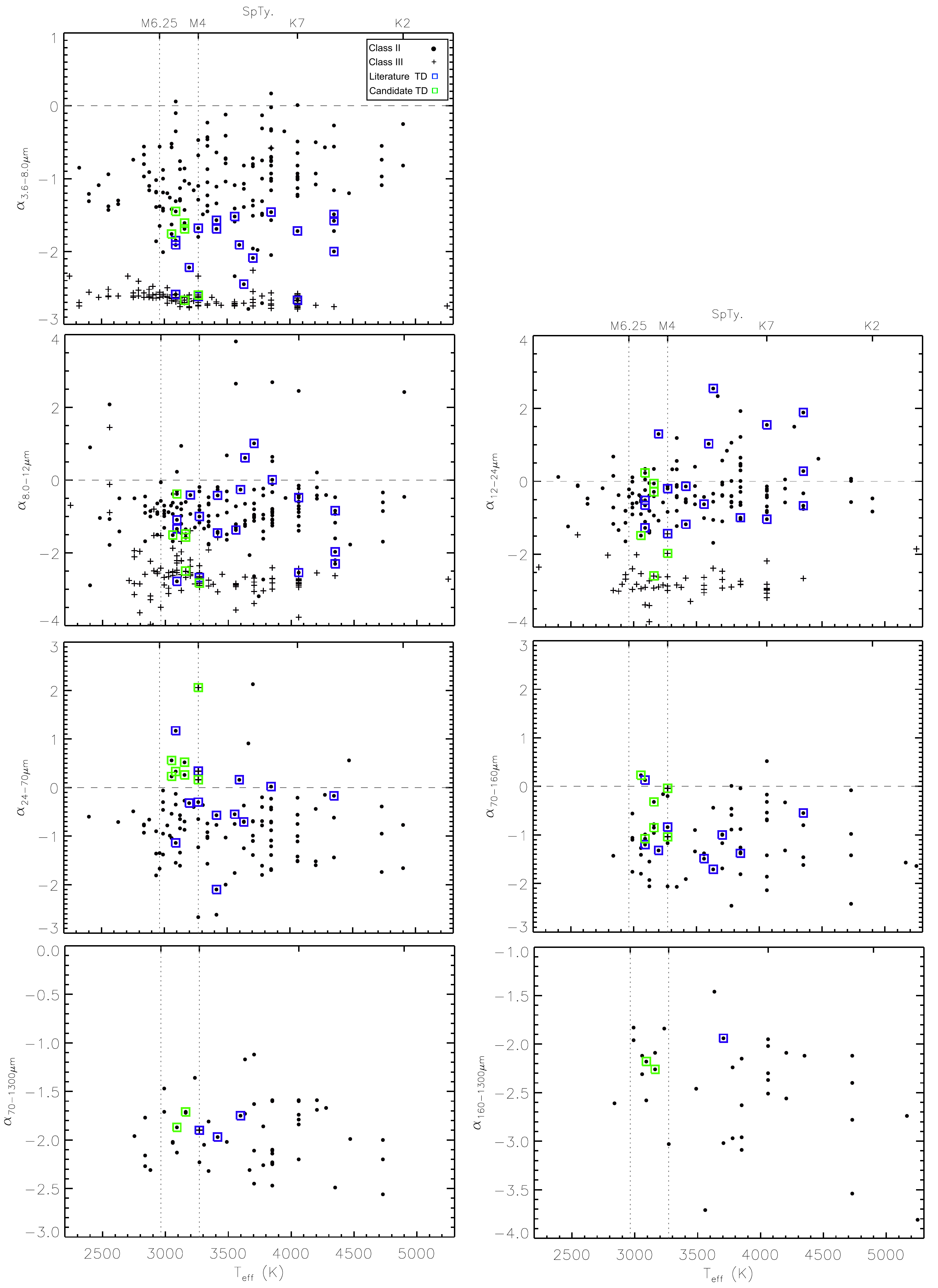}
\caption {Series of spectral index plots shown as a function of stellar effective temperature, merging the known Taurus Class~II (filled circles) and Class III (crosses) members with; mid-IR emission (IRAC, MIPS and {\it WISE}), far-IR emission (PACS and MIPS), and 1.3 mm emission (for those targets not with no reported 1.3 mm observations, the nearest submm/mm observation was scaled to 1.3~mm assuming {\it F}$_{\nu}$~$\propto$~$\nu^{2}$). Symbols leftwards of the vertical dotted lines, plotted at $T_{{\rm eff}}$=2963~K and $T_{{\rm eff}}$=3270~K, indicate the substellar ($\leq$M6.25) members and members of the TBOSS sample ($\leq$M4), respectively. Blue squares enclose the targets identified in the literature with transition disks and green squares enclose the candidate transition disks objects identified within this paper.
}
\label{index_panel}
\end{figure*}

In order to compare the shapes of the SEDs for different spectral types of the detected Class~II targets, the median SED for each spectral type is shown in Figure~\ref{medianSED}, and the median SEDs are scaled by the {\it J}-band flux density. As a result of the characteristic SED shape of Class~II objects with edge-on disks (similar in appearance to Class~I objects), targets with known edge-on disks -- IRAS~04248+2612, J04381486+2611399, J04442713+2512164, and ZZ Tau-IRS \citep{Padgett:1999,Andrews:2008,Luhman:2007,Bouy:2008,White:2004} -- are not included in the calculated median SEDs. Overall, the relative flux density at longer wavelengths declines as a function of spectral type, though there is no discontinuity at the stellar/substellar boundary. 

Only four Class~III targets have {\it Herschel} PACS detections, and the SEDs for these four sources are discussed further in Section~\ref{Sec:TD}. The remaining Class~III targets with {\it Herschel} upper limits are shown in Appendix~\ref{Sec:sed_ClassIII}.

\subsection{Candidate transition disks}
\label{Sec:TD}
Transition disks represent a more advanced stage of disk evolution in which material in the disk has dissipated, either marked by an inner hole (e.g., \citealp{Calvet:2005}), or a homologous depletion \citep{Currie:2009}. A number of physical processes have been suggested to clear disk material, including photoevaporation \citep{Alexander:2006}, disruption from companions \citep{Artymowicz:1994}, and grain growth \citep{Dullemond:2004}. The shape of the SED -- specifically spectral indices between pairs of wavelengths -- or spectral line features, or the absence of gas have been used to identify transition disks and later stage debris disks \citep{Muzerolle:2010,Kim:2013}. Within the TBOSS sample, 4 transition disks are known \citep{Currie:2011,Cieza:2012}, along with 2 debris/photoevaporative disks \citep{Currie:2011}. Details of the previously known TBOSS transition disk and debris disk objects are listed in Table \ref{table:TD}.  

\begin{figure}
\centering
\includegraphics[scale=0.65]{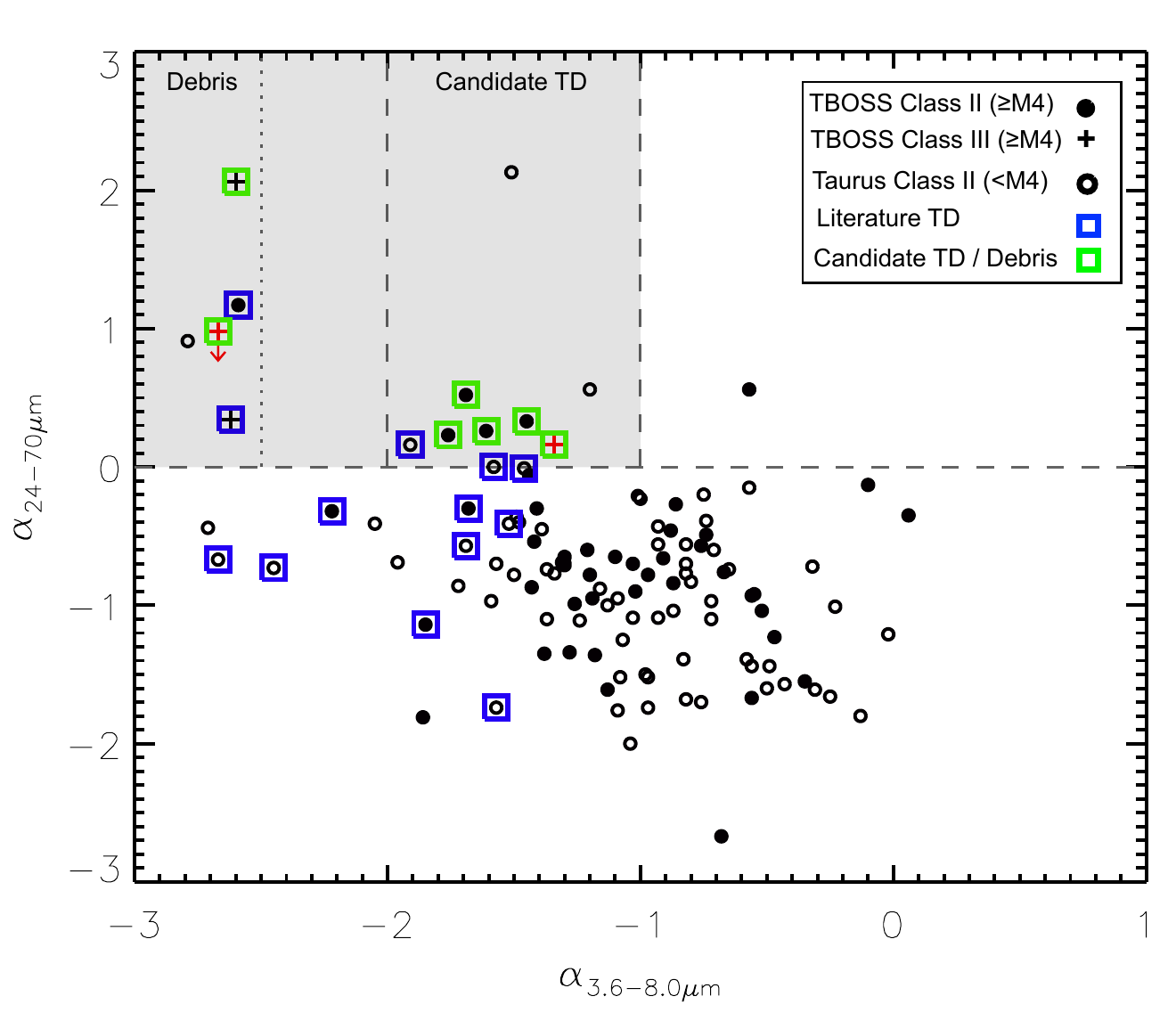}
\caption {Spectral index plot showing the Taurus Class~II and Class~III members with mid-IR and far-IR excess. The spectral indices; $\alpha_{3.6-8.0{\mu}m}$ and $\alpha_{24-70{\mu}m}$ are used to identify candidate targets with transition disks (TDs; points bounded by the horizontal and vertical dashed lines) and debris disks (points bounded by the horizontal dashed and vertical dotted lines). Targets identified in the literature with TDs, and the candidate TD / debris disks reported here, are enclosed by blue and green squares respectively. J04295422+1754041 (red cross) was unobserved with IRAC and MIPS, and the spectral indices are calculated from the {\it WISE} 3.4~$\mu$m and 12~$\mu$m flux densities, and the {\it WISE} 22~$\mu$m and PACS~70~$\mu$m flux densities. XEST~08-033 (red cross with downward arrow) was only detected with PACS at 160~$\mu$m, and the 24-70 spectral index shown is calculated from the 70~$\mu$m upper limit.
}
\label{IndexTD}
\end{figure}

\begin{figure*}
\centering
\includegraphics[scale=1.74]{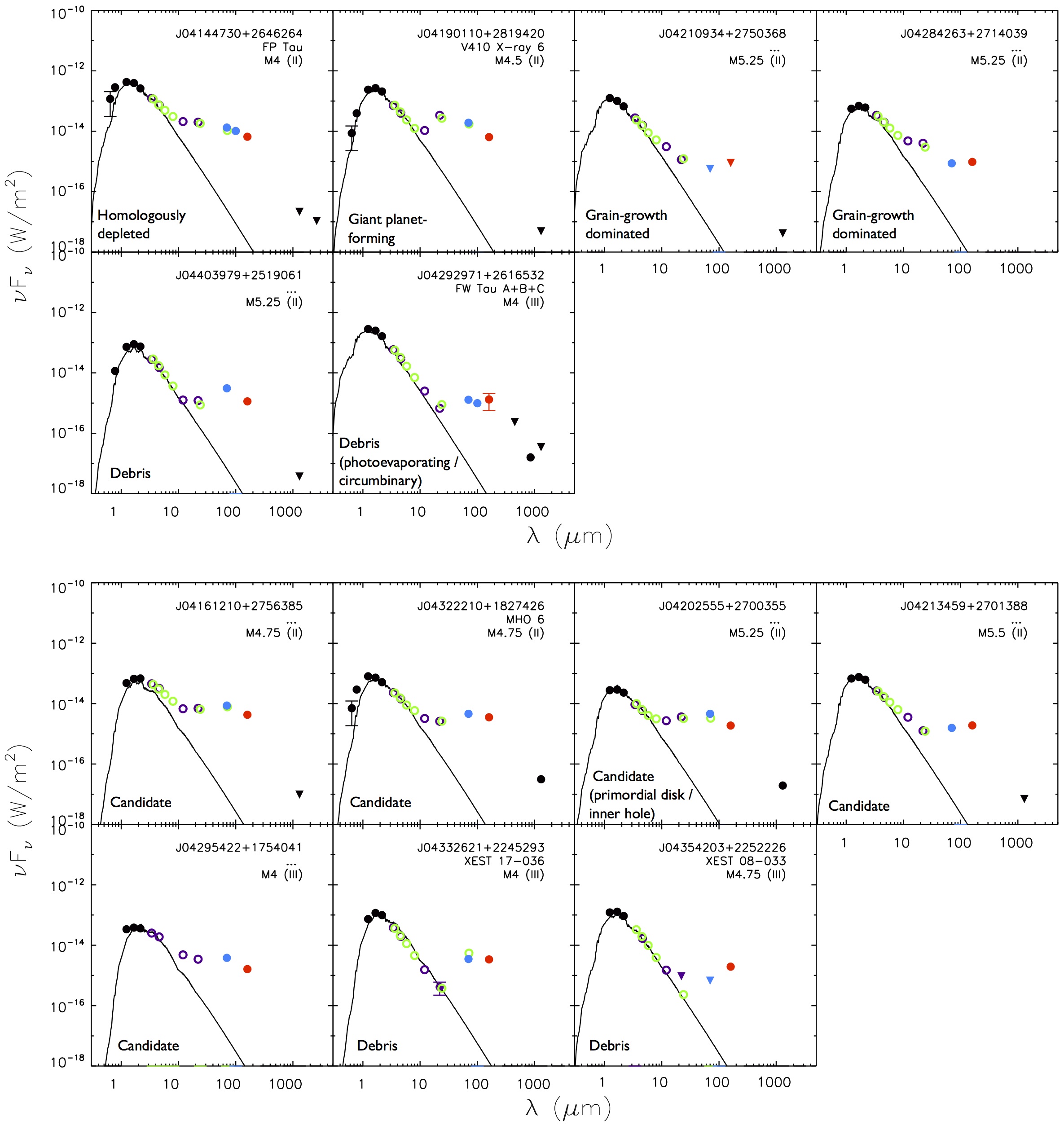}
\caption {SEDs showing the targets with transition disks (TD) previously reported in the literature ({\bf top panel}) and the new candidate targets with TDs identified in this paper ({\bf bottom panel}). Additionally, J04202555+2700355 for which the evolutionary status was  previously reported as inconclusive \citep{Currie:2011}, is identified as a candidate TD here.  For those targets identified in the literature the TD types, compiled from \citet{Currie:2011} and \citet{Cieza:2012}, are displayed on each SED. The candidate TDs are identified by the rising spectral index from 24-70 $\mu$m ($\alpha_{24-70{\mu}m}$>0). In addition to the positive $\alpha_{24-70{\mu}m}$ indices for XEST~17-036 and XEST~08-033, the $\alpha_{3.6-8.0{\mu}m}$ indices are typical for those of Class~III (debris disk) objects. The target name, spectral type, and spectral class are labeled in each SED. The observed broadband photometry (see also Appendix~\ref{Appendix:A}) is compiled from optical ({\it R}$_{C}$, {\it I}$_{C}$), and near-IR (2MASS; {\it JHK$_{S}$}) wavelengths (black points), the mid-IR (IRAC and {\it WISE}; green and purple open circles respectively), the far-IR  (MIPS, PACS Blue and Red channels; green open circles, blue and red points respectively) and submm-mm wavelengths (black points). 3$\sigma$ upper limits are represented by the downwards triangles. The best-fit atmospheric model are displayed for each target.
}
\label{sedTD}
\end{figure*}

A series of spectral indices ($\alpha = - \frac{d~\mathrm{log}({\nu}F_{\nu})}{ d~\mathrm{log}({\nu})}$) as a function of effective temperature calculated over wavelength ranges from the 3.6 - 8.0~$\mu$m IRAC bands to the 160 - 1300~$\mu$m bands are plotted in 7 panels in Figure~\ref{index_panel} for the full Taurus Class~II and Class~III population with detections. Over the 3.6 - 8.0~$\mu$m wavelength range, the known transition disk SEDs have steeply declining slopes and occupy the region at the extreme of the Class~II distribution and overlapping with Class~III members. The transition disks again have a distinct distribution in the SED power law index over the 24 - 70~$\mu$m range; the majority of the slopes are flat to rising. Based on the 24 - 70~$\mu$m spectral index, we have identified a new population of 7 additional transition disk candidates with rising slopes, and these new systems are noted in Table~\ref{table:TD}. Amongst these 7 transition disk candidates, J04202555+2700355 was previously investigated \citep{Currie:2011} and an inconclusive evolutionary status (primordial disk or transition disk with inner hole) was reported based on SED modeling results. XEST~17-036 and XEST~08-033 are considered to be debris disks based on their $\alpha_{3.6-8.0{\mu}m}$ values that are consistent with those of previously reported debris/photoevaporative disks in the literature. Furthermore, XEST~17-036 -- is remarkable in that it is among the few spatially resolved targets. Our selection criteria and classification for the targets that we identify to be candidate transition/debris disks is highlighted in the spectral index plot shown in Figure~\ref{IndexTD}. Based on conservative estimates of transition and debris disk objects reported in the literature, targets with $\alpha_{24-70{\mu}m}$~>0, and $-2{\la}~\alpha_{3.6-8.0{\mu}m}{\la}-1$ are identified as candidate transition disk objects, and targets with $\alpha_{3.6-8.0{\mu}m}{\la}-2.5$ are identified as debris disk objects. The SEDs of the known and candidate transition disks are shown in Figure~\ref{sedTD}. The 5 new transition disk candidates have spectral types ranging from M4-M5.5.

From the total number of transition disk objects, it is possible to determine the typical statistical lifetime for this evolutionary stage. The estimated transition disk lifetime of $\sim$0.45~Myr for targets of spectral types K5-M5 was calculated from the ratio of the number of transition disk objects to that of primordial disk objects, and assuming that the average primordial disk lifetime is $\sim$3~Myr \citep{Luhman:2010}. Following the same procedure we obtain a ratio of 9/52 (accounting for the 4 Class~III objects that are detected at 70~$\mu$m and/or 160~$\mu$m), and an estimated lifetime of $\sim$0.5 Myr for targets of spectral types M4-M9, similar to the K5-M5 value.

\subsection{Candidate truncated disks}
\label{Sec:Trunc}
\begin{figure}
\centering
\includegraphics[scale=0.3]{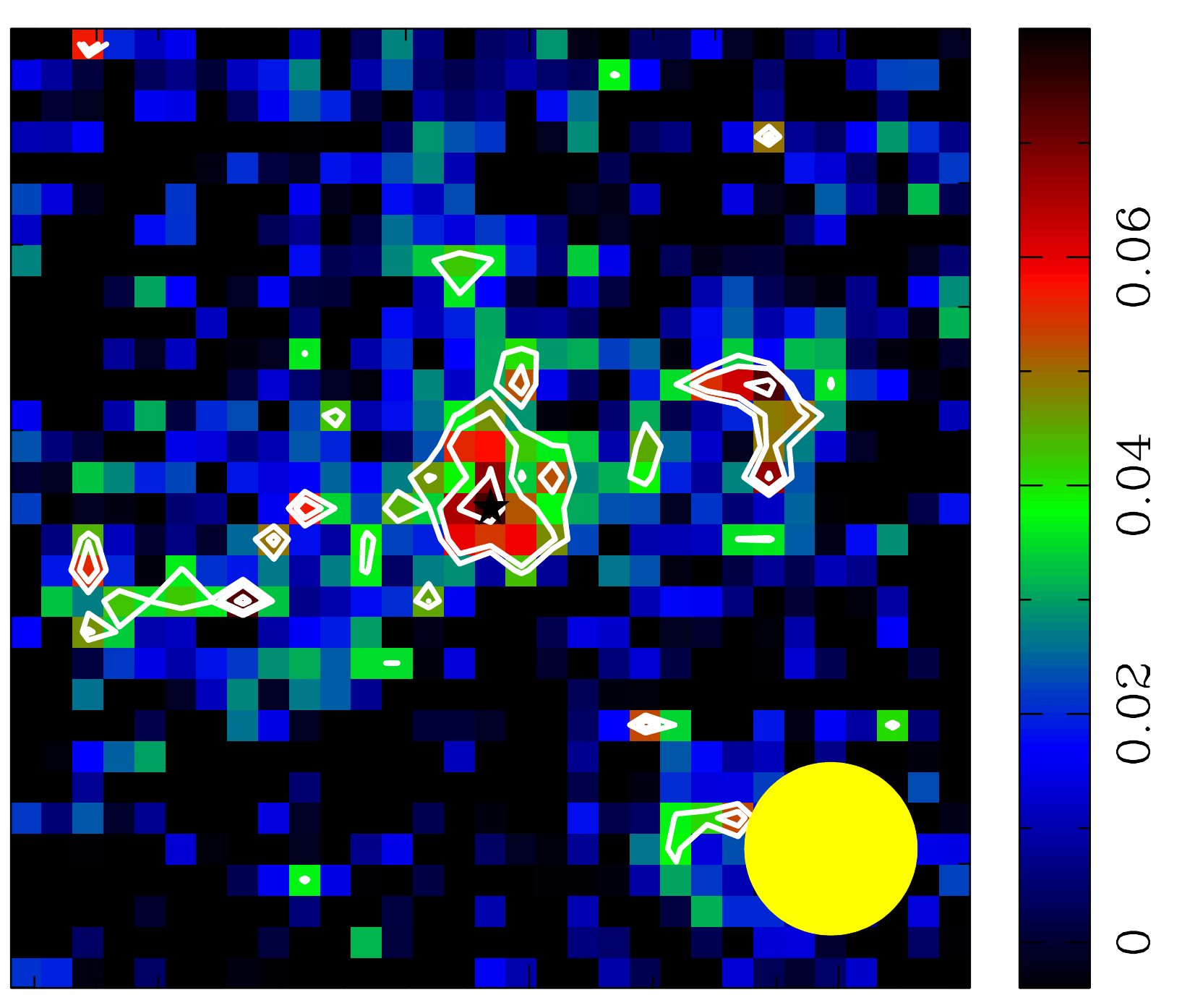}
\caption {PACS Blue channel (70~$\mu$m) combined map of the undetected Class II objects. The map is 30$\arcsec$ in size along each axis. The beam size is represented by the yellow circle in the bottom, right-hand corner and the black star marks the peak position in the map. Contour levels are plotted at 4, 5 and 6$\sigma$. The scale bar shows the intensity of the map in mJy/pixel.}
\label{ul}
\end{figure}

A total of 15 Class II targets with 24~$\mu$m flux densities were not detected with the {\it Herschel} 70~$\mu$m PACS maps, including one target not observed with {\it Herschel}, that is undetected at 70~$\mu$m with {\it Spitzer} MIPS. 
The non-detections are not simply the latest spectral type targets, but include examples extending to M4.25. To estimate the average flux density for this small subset of undetected Class~II members, all the maps were combined to form a single 70~$\mu$m map shown in Figure~\ref{ul}, and the flux density of the faint combination source was measured as 3.0~$\pm$~0.5~mJy. This exercise was repeated for the 160~$\mu$m maps in which a 3$\sigma$ upper limit of 14~mJy was measured in the combined map. Both the individual detection limit and these estimates of the average flux densities at 70~$\mu$m and 160~$\mu$m are plotted on the SEDs for the 17 Class II targets that are not detected at far-IR wavelengths with either {\it Herschel} PACS or {\it Spitzer} MIPS (presented in Appendix \ref{sed2d}). For 3 of the these targets -- CIDA~14, MHO~5 and J04322415+2251083 -- the individual limit alone indicates that the slope of the SED has a power law index ($\alpha_{24-70 {\mu}m}$)  of -2.0 or steeper. Considering the lower value of the combined flux from all the non-detected targets, 5 Class~II targets with 24~$\mu$m detections and 70~$\mu$m upper limits have slopes steeper than -2.0. 

\begin{figure*}
\centering
\includegraphics[scale=1.7]{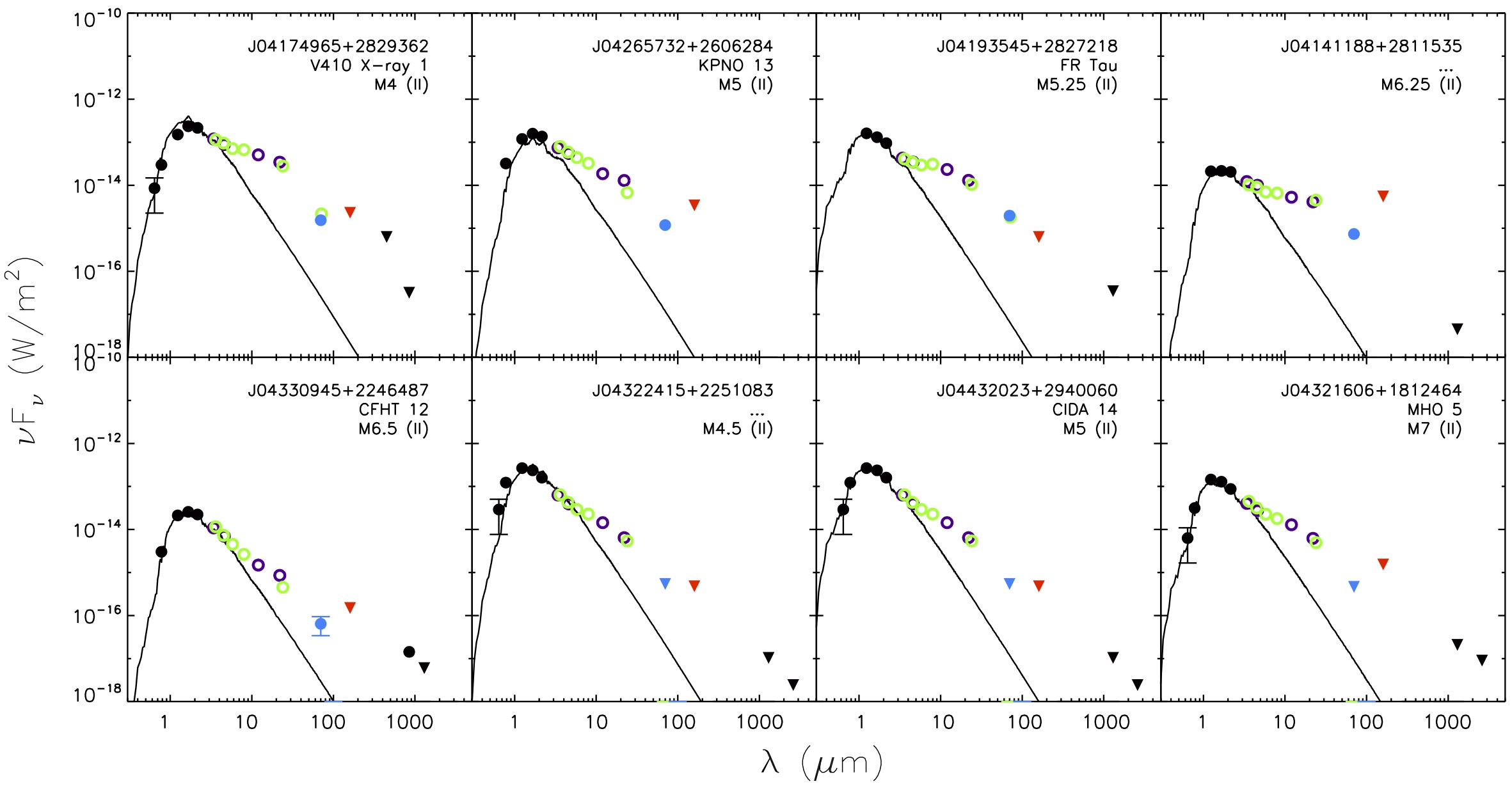}
\caption {SEDs of the eight candidate targets with truncated disks, identified from the steeply declining spectral index from 24-70 $\mu$m ($\alpha_{24-70{\mu}m}\mathrm{<-1.4}$). The target name, spectral type, and spectral class are labeled in each SED. The observed broadband photometry (see also Appendix~\ref{Appendix:A}) is compiled from optical ({\it R}$_{C}$, {\it I}$_{C}$), and near-IR (2MASS; {\it JHK$_{S}$}) wavelengths (black points), the mid-IR (IRAC and {\it WISE}; green and purple open circles respectively), the far-IR  (MIPS, PACS Blue and Red channels; green open circles, blue and red points respectively) and submm-mm wavelengths (black points). 3$\sigma$ upper limits are represented by the downwards triangles. The best-fit atmospheric model are displayed for each target.}
\label{sedTrunc}
\end{figure*}

One possible explanation for the shape of the Class II targets with exceptionally steep far-IR SED slopes is truncation of their disks. An example of a likely truncated disk in the $\epsilon$~Cha moving group around ET~Cha was identified from a similarly steep SED index ($\alpha_{24-70 {\mu}m}$ = -1.4) over wavelengths extending to the PACS bands combined with a non-detection of the CO(3-2) line \citep{Woitke:2011}. In addition to the 3 Class~II non-detected sources, there are 5 Class~II targets with 70 $\mu$m detections that also define a slope as steep as that of ET~Cha -- CFHT~12, FR~Tau, KPNO~13, J04141188+2811535, and V410~Xray~1 -- resulting in a total of 8 candidate truncated disks. The SEDs of these candidate truncated disk targets are presented in Figure~\ref{sedTrunc}. Follow-up high sensitivity CO spectral line observations of these steep slope SED targets could provide further evidence of truncated disks around these 8 Taurus targets.

\subsection{Impact of companions}

Many of the 154 M4+ TBOSS sample have been observed by infrared direct imaging from the ground \citep{Duchene:2004, Luhman:2009}, high resolution imaging employing space-based measurements with {\it HST} \citep{White:2001, Kraus:2006, Todorov:2014}, ground-based speckle imaging \citep{Kohler:1998, Konopacky:2007}, and aperture masking or AO imaging \citep{Kraus:2009, Kraus:2011}. The results of the PACS observations presented here therefore provide the opportunity to assess the similarity or difference in the frequency or typical flux density level of disks for targets with and without companions.
Whilst the comprehensive analysis of multiplicity and completeness for those TBOSS targets with existing imaging data will be addressed in a forthcoming coming paper, the subset of known and candidate transition disk and candidate truncated disk objects represent interesting cases in which to initially assess the multiplicity rate, since the inner holes of misidentified transition disks may be due to the presence of a companion that gravitationally truncates the inner radii of a circumbinary disk (e.g., \citealp{Ireland:2008}). For binary systems with semi-major axis {\it a}, the outer edges of the circumstellar disk(s) are expected to be truncated to disk radii of $\sim$$0.2-0.5~a$ \citep{Artymowicz:1994}. Details of the targets for which imaging observations have been investigated for companions are given in Table~\ref{table:Multiplicity}, along with the corresponding literature reference. The binary systems that were detected with the PACS Blue channel (70~$\mu$m or 100~$\mu$m) among the categories of transition, truncated, mixed pairs, and extended objects are given in Table~\ref{table:Multiplicity}.

Amongst the transition disk objects identified within the TBOSS sample, 10 of13 targets possess high resolution imaging data, from which a binary disk fraction of 40$_{-13}^{+16}$\%, and single disk fraction of 60$_{-16}^{+13}$\%  is calculated. Similarly, amongst the truncated disk objects identified within the TBOSS sample 5 of 8 targets possess high resolution imaging data, from which a binary disk fraction of 40$_{-16}^{+22}$\%, and single disk fraction of 60$_{-22}^{+16}$\% is calculated. There is no significant difference between the binary and single star disk fractions for the transition disk and truncated disk objects.

\subsection{Mixed pair systems}
\label{Sec:mix}
\begin{figure}
\centering
\includegraphics[scale=0.85]{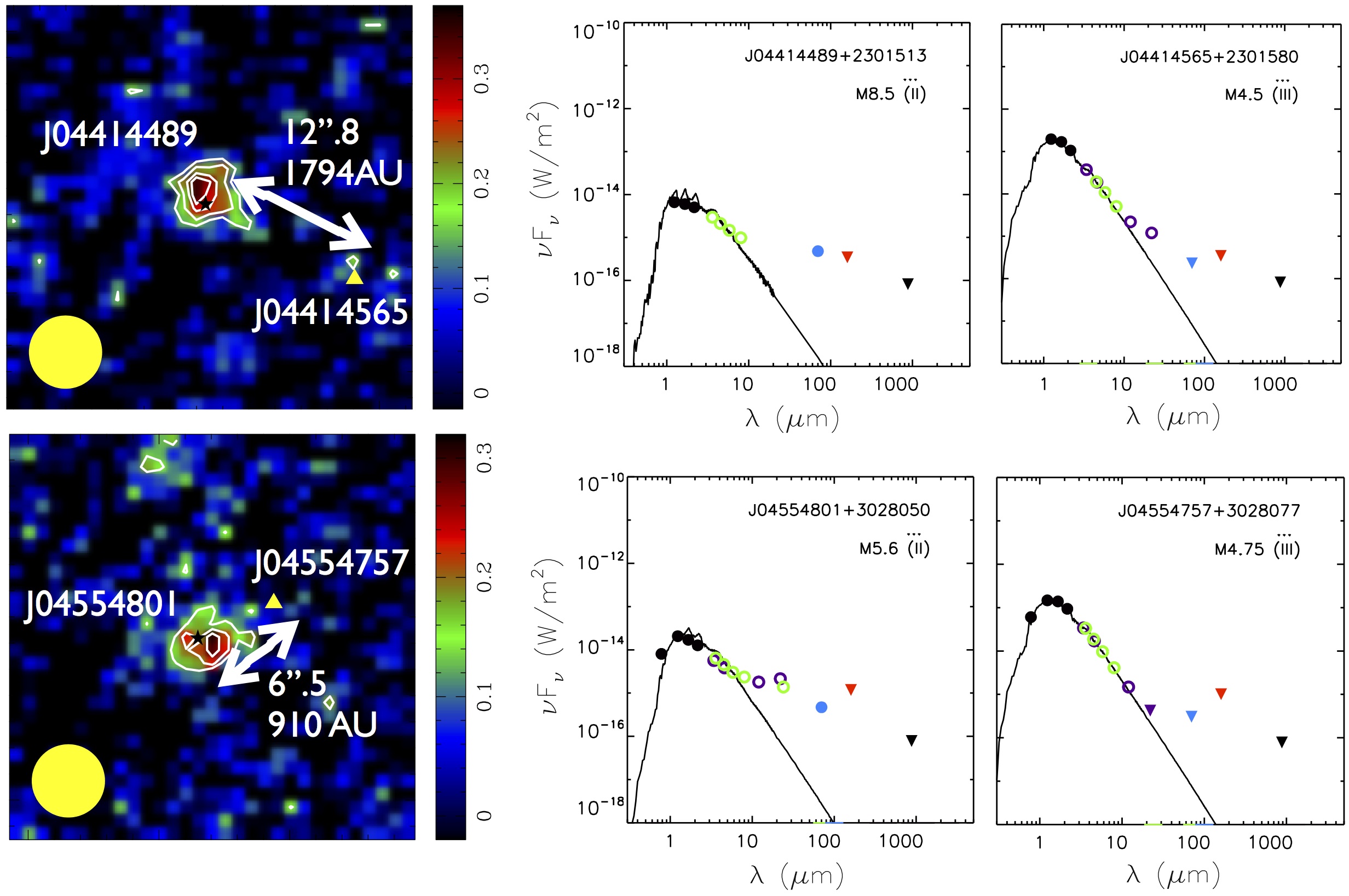}
\caption {PACS Blue channel (70 $\mu$m) maps {(\bf left)} for the two mixed pair systems identified within the TBOSS sample; J0441489+2301513 / J04414565+2301580 and J04554801+3028050 / J04554757+3028077, and the corresponding SEDs {(\bf right}). The black stars in the maps show the 2MASS positions of the detected Class~II secondaries (J0441489 and J04554801), and the yellow triangles mark the 2MASS positions of the undetected Class~III primaries (J04414565 and J04554757). Contours are plotted at 3, 4, 5 and 6$\sigma$ and the PACS 70~$\mu$m beam size is represented by the yellow circle in the bottom, left corner of each map. Target name, spectral type, and spectral class are displayed in the SEDs. Broadband photometry, compiled from optical-mm wavelengths and the underlying best-fit atmospheric models are displayed in the SEDs.
}
\label{mixed}
\end{figure}

Mixed pairs form an interesting subset of multiple systems comprised of one component with a disk and one component without a disk. 
Studies of spatially resolved binary systems in Taurus investigating inner disk signatures such as {\it H}$\alpha$ emission and near-IR excesses reveal that mixed pair systems are rare, representing $\sim$15-20\% of the population (e.g., \citealp{Prato:1997, Duchene:1999, Hartigan:2003, Monin:2007}). Furthermore, considering these inner disk diagnostics, the component with the disk is equally likely to be the primary or secondary \citep{Monin:2007,Daemgen:2012}.
With interferometric submm/mm observations sensitive to the bulk of the disks in Class II binaries in Taurus and Ophiuchus \citep{Jensen:2003,Patience:2008,Harris:2012}, 19 pairs have been spatially resolved, consisting of 14 systems with only a circumprimary disk, 3 systems of both circumprimary and circumsecondary disks, and 2 systems with only a circumsecondary disk. In each of the systems with only a circumsecondary disk in the submm/mm, the primaries are actually close binary pairs with separations of $\sim$5-45~AU \citep{Kraus:2011}. Among the TBOSS sample, two examples of Class~II~/~Class~III mixed systems are identified -- J04414565+2301580/J04414489+2301513 and J04554757+3028077/J04554801+3028050. The PACS maps and SEDs of these systems are shown in Figure~\ref{mixed}. In each case, the secondary is the disk-bearing component based on the far-IR emission from the {\it Herschel} data.
Based on high resolution imaging, both the primary and secondary of the J04414565/J04414489 system posses close companions, while neither the primary nor secondary of the J04554757/J04554801 system has a close companion \citep{Kraus:2011, Todorov:2014}. The angular separations of the two mixed pair systems, and the additional separations for the close companions of the J04414565/J04414489 system are listed in Table~\ref{table:Multiplicity}. In the J04414565/J04414489 system, the host of the secondary disk is a very low mass M8.5 ($\sim$20~{\it M}$_{\mathrm{Jup}}$) brown dwarf approaching the planetary mass regime.


\subsection{Candidate proto-brown dwarfs in Taurus}
Proto-brown dwarfs are cores of substellar mass, at the Class~0 or Class~I stage of low mass star-formation \citep{Adams:1987, Andre:1993, Pound:1993}. 
Observable diagnostics, revealed in characterizing the earliest stages of brown dwarf evolution, can be used to distinguish between the number of different brown dwarf formation scenarios such as embryo ejection \citep{Bate:2002}, disk fragmentation \citep{Stamatellos:2011} or photoionization from nearby massive stars \citep{Whitworth:2004}. It is also noteworthy that the first discovery of a pre-brown dwarf core -- a self-gravitating starless core of dust and gas with a mass in the brown-dwarf regime -- is located in the dense Ophiuchus L1688 cloud \citep{Andre:2012}, indicating that these objects can form in environmental conditions denser than Taurus. 
Given the divergence of possible formation models, it is important to identify the youngest population of proto-brown dwarfs. Among the Taurus population, only two Class 0 objects are known, and none have spectral types in the M4-L0 TBOSS range, however, there are several candidate proto-brown dwarfs among the Class~I TBOSS range, and each case is described below.

Three sources within the TBOSS sample may be considered as proto-brown dwarf candidates -- [GKH94]~41, IRAS~04191+1523~B, and L1521F-IRS. [GKH94]~41 has been identified to be a disk-dominated source based on IRS spectra \citep{Furlan:2011}, and is therefore at a more evolved stage than the Class~I classification that has been adopted here \citep{Luhman:2010}. The accretion rate for L1521F-IRS is estimated to be low and likely towards the end of its main (Class~0) accretion phase \citep{Bourke:2006}. IRAS~04191+1523~B is the secondary component of a 6$\farcs$1 binary \citep{Duchene:2004}, making the data at wavelengths longer than 8 $\mu$m contaminated by the primary emission. Given the quiescent phase of accretion and the bolometric temperature consistent with M6-M8 \citep{Luhman:2010}, L1521F-IRS and IRAS~04191+1523~B will remain substellar unless a large amount of material is added to the central source. From the SED of L1521F-IRS, this candidate proto-brown dwarf appears heavily embedded and the slope over the 160~$\mu$m to 1.3~mm range is $\alpha$=-1.8, which is markedly shallower than the other isolated Class~I TBOSS targets ($\alpha$ typically ranging from -3.4 to -2.7). The SED slope of L1521F-IRS is similar to ISM grains and may indicate that this target has only recently transitioned from a Class~0 to Class~I object.

\subsection{Comparison with other star-forming regions}
Comparison of the results of large-scale disk population studies performed in different star-forming regions and in young clusters can provide an indication of the impact of environmental factors and evolutionary trends. From IR observations over the {\it J}- to {\it L}-band of young populations spanning the 0.3-5 Myr age range, a typical disk lifetime of $\la$3~Myr was inferred from the decline in IR excess frequency \citep{Haisch:2001}. From longer wavelength {\it Spitzer} observations of Taurus extending to 24 $\mu$m, the disk fraction (defined by the  ratio of Class~II to Class~II+III members) as a function of central object mass was similar to Chameleon~I, but higher than IC~348, and the lower stellar density of Taurus and Chameleon was suggested as a possible explanation \citep{Luhman:2010}. Submillimeter population studies comparing Taurus and Ophiuchus measured similar disk fractions and average disk masses for both regions \citep{Andrews:2005,Andrews:2007}. These existing IR and submm population studies mainly focus on the earlier spectral type members outside the TBOSS sample range.

Given the unique sensitivity and scope of the TBOSS survey, it is difficult to compare the Taurus M4+ far-IR disk results with those of other star-forming regions. The most analogous study is a recent {\it Herschel} investigation of 43 brown dwarfs (taken as objects with M6+ spectral types) in the L1688 cloud of Ophiuchus, for which a disk fraction of $\sim$0.3 at 70~$\mu$m is revealed (Alves de Oliveira et al. 2013). Considering the M6+ targets in the TBOSS sample, the detected disk fraction is also $\sim$0.3 (22 of 67 targets), however the sensitivity of the TBOSS observations is much deeper. Restricting the TBOSS M6+ disks to those that could have been detected in the Ophiuchus study ($\ga$25 mJy) yields a lower disk fraction of $\sim$0.15, suggesting that the average disk flux density for a Taurus member at or below the stellar/substellar boundary is lower than for an equivalent object in Ophiuchus. 

\section{Summary}
\label{Sec:Conclusions}
Of the 154 total M4-L0 members of the Taurus star-forming region, we report new 70~$\mu$m and 160~$\mu$m  flux densities or upper limits for 134 targets and combine the results with other {\it Herschel} and {\it Spitzer} programs to compile the PACS measurements for 153 targets comprising the TBOSS sample. For the 148 TBOSS targets not contaminated by emission from unresolved earlier spectral type primaries, the detection rates at 70~$\mu$m with PACS and MIPS were calculated for the different evolutionary classes and spectral type ranges. Considering all M4-L0 TBOSS targets, 100\% of Class~I, 75\% of Class~II, and 4\% of Class~III TBOSS targets were detected. Dividing the Class~II targets into M4-M6 members above the substellar limit and M6.25+ brown dwarfs, the far-IR detection rate is significantly higher for the lowest mass stars relative to brown dwarfs. Although the disk frequencies are different, the distributions of disk-to-central-object far-IR flux density are indistinguishable above and below the substellar limit for disk targets, based on a K-S test.

To obtain a benchmark first census of far-IR disk emission across the full stellar and substellar population of any star-forming region, the TBOSS results were combined with {\it Spitzer} and {\it Herschel} surveys of the earlier spectral type members. The 70~$\mu$m flux density as a function of spectral type declines with a trend similar to the decline in flux density of the central object, and the range of flux density at a given spectral type span at least an order of magnitude. Using the radiative transfer code MCFOST to model the range of {\it Herschel} flux densities, the dominant parameters influencing the PACS bands was found to be the disk inclination, scale height, and flaring index. The majority of the detected Taurus Class~II population can be constrained with flared disks ($\beta$ = 1.125), with scale heights of 10 AU at a reference disk radius of 100~AU (equivalent to an opening angle of 5$\degr$), and with disk inclinations from face- to edge-on.

For all TBOSS targets, the SEDs were constructed by joining the far-IR data with results from optical to mid-IR surveys and submm/mm flux densities when available. Based on the spectral indices over the 24~$\mu$m to 70~$\mu$m range, 5 new candidate transition disks were identified. While the new and known transition disks have a distinct distribution in the slope over the 24~$\mu$m to 70~$\mu$m range, the spectral indices for these objects at longer wavelengths are indistinguishable from the values for the general Class~II population. Another set of 8 targets have very steep SED slopes over the 24~$\mu$m to 70~$\mu$m range, possibly indicating truncated disks such as has been suggested for the disk around ET~Cha \citep{Woitke:2011}. Two other unusual targets are disk-bearing secondary components of mixed systems. Finally, the overall Taurus substellar disk frequency is similar to that of the brown dwarfs in Ophiuchus, but the average disk flux density for a Taurus M6+ target is lower than for a comparable object in Ophiuchus.

\begin{longtab}
\begin{longtable}{c c c c r r}
\caption{\label{Table:sample} Properties of the known 154 M4-L0 Taurus members, ordered by target RA.} \\
\hline\hline
 2MASS & Other name & SED class$^{1}$ & SpTy. & Ref. & Notes  \\
\hline
\endfirsthead
\caption{continued.} \\
\hline\hline
 2MASS & Other name & SED class$^{1}$ & SpTy. & Ref. & Notes  \\
\hline
\endhead
\hline
\endfoot
J04034997+2620382	&	XEST 06-006	&	III	&	M5.25 $\pm$ 0.25	&	3	&		\\
J04131414+2819108	&	LkCa 1	&	III	&	M4 $\pm$ 1	&	16	&	b	\\
J04141188+2811535	&	 ... 	&	II	&	M6.25 $\pm$ 0.5	&	8	&		\\
J04141458+2827580	&	FN Tau	&	II	&	M5 $\pm$ 1	&	16	&		\\
J04141760+2806096	&	CIDA 1	&	II	&	M5.5 $\pm$ 0.5	&	10	&		\\
J04144730+2646264	&	FP Tau	&	II	&	M4 $\pm$ 1	&	16	&		\\
J04144739+2803055	&	XEST 20-066	&	III	&	M5.25 $\pm$ 0.25	&	3	&		\\
J04150515+2808462	&	CIDA 2	&	III	&	M4.5 $\pm$ 1	&	16	&	b	\\
J04151471+2800096	&	KPNO 1	&	III	&	M9 $\pm$ 0.5	&	7	&		\\
J04152409+2910434	&	 ... 	&	III	&	M7 $\pm$ 0.25	&	5	&		\\
J04155799+2746175	&	 ... 	&	II	&	M5 $\pm$ 0.25	&	3	&		\\
J04161210+2756385	&	 ... 	&	II	&	M4.75 $\pm$ 0.25	&	8	&		\\
J04161885+2752155	&	 ... 	&	III	&	M6.25 $\pm$ 0.25	&	5	&		\\
J04162725+2053091	&	 ... 	&	III	&	M5 $\pm$ 0.25	&	4	&		\\
J04163048+3037053	&	 ... 	&	III	&	M4.5 $\pm$ 0.25	&	8	&		\\
J04163911+2858491	&	 ... 	&	II	&	M5.5 $\pm$ 0.25	&	5	&		\\
J04174955+2813318	&	KPNO 10	&	II	&	M5.5 $\pm$ 0.5	&	7	&		\\
J04174965+2829362	&	V410 X-ray 1	&	II	&	M4 $\pm$ 0.5	&	17	&		\\
J04180796+2826036	&	V410 X-ray 3	&	III	&	M6.5 $\pm$ 0.5	&	10	&		\\
J04181710+2828419	&	V410 Anon 13	&	II	&	M5.75 $\pm$ 0.25	&	11	&		\\
J04183030+2743208	&	KPNO 11	&	III	&	M5.5 $\pm$ 0.25	&	9	&		\\
J04184023+2824245	&	V410 X-ray 4	&	III	&	M4 $\pm$ 1	&	14	&		\\
J04185115+2814332	&	KPNO 2	&	III	&	M6.75 $\pm$ 0.5	&	7	&		\\
J04185813+2812234	&	IRAS 04158+2805	&	I	&	M5.25 $\pm$ 0.25	&	5	&		\\
J04190110+2819420	&	V410 X-ray 6	&	II	&	M4.5 $\pm$ 2	&	14	&		\\
J04190126+2802487	&	KPNO 12	&	II	&	M9 $\pm$ 0.25	&	9	&		\\
J04190197+2822332	&	V410 X-ray 5a	&	III	&	M5.5 $\pm$ 0.5	&	15	&		\\
J04193545+2827218	&	FR Tau	&	II	&	M5.25 $\pm$ 0.25	&	3	&		\\
J04194657+2712552	&	$[$GKH94$]$ 41	&	I	&	M7.5 $\pm$ 1.5	&	3	&		\\
J04201611+2821325	&	 ... 	&	II	&	M6.5 $\pm$ 0.25	&	3	&		\\
J04202555+2700355	&	 ... 	&	II	&	M5.25 $\pm$ 0.25	&	8	&		\\
J04202583+2819237	&	IRAS 04173+2812	&	II	&	mid-M	&	3	&	f	\\
J04203918+2717317	&	XEST 16-045	&	III	&	M4.5 $\pm$ 0.25	&	3	&		\\
J04205273+1746415	&	J2-157	&	III	&	M5.5 $\pm$ 1	&	16	&		\\
J04210795+2702204	&	CFHT 19	&	II	&	M5.25 $\pm$ 0.5	&	7	&		\\
J04210934+2750368	&	 ... 	&	II	&	M5.25 $\pm$ 0.25	&	3	&		\\
J04213459+2701388	&	 ... 	&	II	&	M5.5 $\pm$ 0.25	&	8	&		\\
J04214013+2814224	&	XEST 21-026	&	III	&	M5.75 $\pm$ 0.25	&	3	&		\\
J04214631+2659296	&	CFHT 10	&	II	&	M6.25 $\pm$ 0.5	&	7	&		\\
J04215450+2652315	&	 ... 	&	III	&	M8.5 $\pm$ 0.25	&	5	&		\\
J04220007+1530248	&	IRAS 04191+1523 B	&	I	&	M6-M8	&	2	&	g	\\
J04221332+1934392	&	 ... 	&	III	&	M8 $\pm$ 0.25	&	5	&		\\
J04221644+2549118	&	CFHT 14	&	III	&	M7.75 $\pm$ 0.5	&	7	&		\\
J04222404+2646258	&	XEST 11-087	&	III	&	M4.75 $\pm$ 0.25	&	3	&		\\
J04230607+2801194	&	 ... 	&	II	&	M6 $\pm$ 0.25	&	6	&		\\
J04233539+2503026	&	FU Tau A	&	II	&	M7.25 $\pm$ 0.25	&	3	&	a	\\
J04233573+2502596	&	FU Tau B	&	II	&	M9.25 $\pm$ 0.25	&	3	&	a	\\
J04242090+2630511	&	 ... 	&	II	&	M6.5 $\pm$ 0.25	&	6	&		\\
J04242646+2649503	&	CFHT 9	&	II	&	M6.25 $\pm$ 0.5	&	7	&	c	\\
J04244506+2701447	&	J1-4423	&	III	&	M5 $\pm$ 1	&	16	&		\\
J04251550+2829275	&	 ... 	&	II	&	M5 $\pm$ 1	&	1	&	d, h	\\
J04262939+2624137	&	KPNO 3	&	II	&	M6 $\pm$ 0.25	&	11	&		\\
J04263055+2443558	&	 ... 	&	II	&	M8.75 $\pm$ 0.25	&	5	&		\\
J04265732+2606284	&	KPNO 13	&	II	&	M5 $\pm$ 0.25	&	9	&		\\
J04270739+2215037	&	 ... 	&	III	&	M6.75 $\pm$ 0.25	&	4	&		\\
J04272799+2612052	&	KPNO 4	&	III	&	M9.5 $\pm$ 0.5	&	7	&		\\
J04274538+2357243	&	CFHT 15	&	III	&	M8.25 $\pm$ 0.5	&	7	&		\\
J04275730+2619183	&	IRAS 04248+2612	&	I	&	M4.5 $\pm$ 0.25	&	5	&		\\
 ... 	&	L1521F-IRS	&	I	&	M6-M8	&	2	&	g	\\
J04284263+2714039	&	 ... 	&	II	&	M5.25 $\pm$ 0.25	&	8	&		\\
J04290068+2755033	&	 ... 	&	II	&	M8.25 $\pm$ 0.25	&	5	&		\\
J04292071+2633406	&	J1-507	&	III	&	M4 $\pm$ 1	&	16	&	b	\\
J04292165+2701259	&	IRAS 04263+2654	&	II	&	M6 $\pm$ 0.5	&	7	&		\\
J04292971+2616532	&	FW Tau A+B+C	&	III	&	M4 $\pm$ 1	&	16	&	b	\\
J04294568+2630468	&	KPNO 5	&	III	&	M7.5 $\pm$ 0.25	&	11	&		\\
J04295422+1754041	&	 ... 	&	III	&	M4 $\pm$ 0.25	&	5	&		\\
J04295950+2433078	&	CFHT 20	&	II	&	M5.5 $\pm$ 0.5	&	7	&		\\
J04300724+2608207	&	KPNO 6	&	II	&	M9 $\pm$ 0.5	&	7	&	c	\\
J04302365+2359129	&	CFHT 16	&	III	&	M8.5 $\pm$ 0.5	&	7	&		\\
J04305171+2441475	&	ZZ Tau IRS	&	II	&	M5.25 $\pm$ 0.5	&	7	&		\\
J04305718+2556394	&	KPNO 7	&	II	&	M8.25 $\pm$ 0.25	&	11	&	c	\\
J04311578+1820072	&	MHO 9	&	III	&	M5 $\pm$ 0.5	&	10	&		\\
J04311907+2335047	&	 ... 	&	III	&	M7.75 $\pm$ 0.25	&	5	&		\\
J04312382+2410529	&	V927 Tau A+B	&	III	&	M5.5 $\pm$ 1	&	16	&	b	\\
J04312405+1800215	&	MHO 4	&	III	&	M7 $\pm$ 0.5	&	10	&		\\
J04312669+2703188	&	CFHT 13	&	III	&	M7.25 $\pm$ 0.5	&	7	&		\\
J04313613+1813432	&	LkHa 358	&	I	&	M5.5 $\pm$ 1	&	16	&		\\
J04315844+2543299	&	J1-665	&	III	&	M5 $\pm$ 1	&	16	&	b	\\
J04320329+2528078	&	 ... 	&	III	&	M6.25 $\pm$ 0.25	&	5	&		\\
J04321606+1812464	&	MHO 5	&	II	&	M7 $\pm$ 0.5	&	10	&		\\
J04321786+2422149	&	CFHT 7	&	III	&	M6.5 $\pm$ 0.5	&	7	&	a	\\
J04322210+1827426	&	MHO 6	&	II	&	M4.75 $\pm$ 0.25	&	11	&		\\
J04322329+2403013	&	 ... 	&	III	&	M7.75 $\pm$ 0.25	&	5	&		\\
J04322415+2251083	&	 ... 	&	II	&	M4.5 $\pm$ 0.25	&	3	&		\\
J04322627+1827521	&	MHO 7	&	III	&	M5.25 $\pm$ 0.25	&	11	&		\\
J04323028+1731303	&	GG Tau Ba+Bb (Ba)	&	II	&	M6 $\pm$ 0.5	&	10	&	b	\\
	&	GG Tau Ba+Bb (Bb)	&	II	&	M7.5 $\pm$ 0.5	&	10	&	b	\\
J04324938+2253082	&	JH 112 B	&	II	&	M4.25 $\pm$ 0.25	&	3	&		\\
J04325026+2422115	&	 ... 	&	III	&	M7.5 $\pm$ 0.5	&	7	&		\\
J04325119+1730092	&	LH 0429+17	&	III	&	M8.25 $\pm$ 0.25	&	3	&		\\
J04330197+2421000	&	MHO 8	&	III	&	M6 $\pm$ 0.25	&	11	&		\\
J04330781+2616066	&	KPNO 14	&	III	&	M6 $\pm$ 0.25	&	9	&		\\
J04330945+2246487	&	CFHT 12	&	II	&	M6.5 $\pm$ 0.5	&	7	&	c	\\
J04332621+2245293	&	XEST 17-036	&	III	&	M4 $\pm$ 0.25	&	3	&		\\
J04334171+1750402	&	 ... 	&	II	&	M4 $\pm$ 0.5	&	3	&		\\
J04334291+2526470	&	 ... 	&	III	&	M8.75 $\pm$ 0.25	&	5	&		\\
J04334465+2615005	&	 ... 	&	II	&	M4.75 $\pm$ 0.25	&	3	&		\\
J04335245+2612548	&	 ... 	&	II	&	M8.5 $\pm$ 0.25	&	6	&		\\
J04335252+2256269	&	XEST 17-059	&	III	&	M5.75 $\pm$ 0.25	&	3	&		\\
J04341527+2250309	&	CFHT 1	&	III	&	M7 $\pm$ 0.5	&	12	&		\\
J04344544+2308027	&	 ... 	&	III	&	M5.25 $\pm$ 0.25	&	4	&		\\
J04350850+2311398	&	CFHT 11	&	III	&	M6.75 $\pm$ 0.5	&	7	&		\\
J04353536+2408266	&	IRAS 04325+2402 C	&	II	&	M6-M8	&	2	&	a, g	\\
J04354183+2234115	&	KPNO 8	&	III	&	M5.75 $\pm$ 0.25	&	11	&		\\
J04354203+2252226	&	XEST 08-033	&	III	&	M4.75 $\pm$ 0.25	&	3	&		\\
J04354526+2737130	&	 ... 	&	III	&	M9.25 $\pm$ 0.25	&	5	&		\\
J04355143+2249119	&	KPNO 9	&	III	&	M8.5 $\pm$ 0.25	&	11	&		\\
J04355209+2255039	&	XEST 08-047	&	III	&	M4.5 $\pm$ 0.25	&	3	&		\\
J04355286+2250585	&	XEST 08-049	&	III	&	M4.25 $\pm$ 0.25	&	3	&		\\
J04355760+2253574	&	 ... 	&	III	&	M6 $\pm$ 1	&	1	&	d, h	\\
J04361030+2159364	&	 ... 	&	II	&	M8.7 $\pm$ 0.25	&	5	&		\\
J04361038+2259560	&	CFHT 2	&	III	&	M7.5 $\pm$ 0.25	&	11	&		\\
J04362151+2351165	&	 ... 	&	II	&	M5.25 $\pm$ 0.25	&	6	&		\\
J04363893+2258119	&	CFHT 3	&	III	&	M7.75 $\pm$ 0.25	&	11	&		\\
J04373705+2331080	&	 ... 	&	III	&	L0 $\pm$ 0.5	&	3	&		\\
J04380083+2558572	&	ITG 2	&	III	&	M7.25 $\pm$ 0.25	&	8	&		\\
J04381486+2611399	&	 ... 	&	II	&	M7.25 $\pm$ 0.25	&	8	&		\\
J04381630+2326402	&	 ... 	&	III	&	M4.75 $\pm$ 0.25	&	4	&		\\
J04382134+2609137	&	GM Tau	&	II	&	M6.5 $\pm$ 0.5	&	10	&		\\
J04385859+2336351	&	 ... 	&	II	&	M4.25 $\pm$ 0.25	&	4	&		\\
J04385871+2323595	&	 ... 	&	III	&	M6.5 $\pm$ 0.25	&	4	&		\\
J04390163+2336029	&	 ... 	&	II	&	M6 $\pm$ 0.25	&	4	&		\\
J04390396+2544264	&	CFHT 6	&	II	&	M7.25 $\pm$ 0.5	&	7	&		\\
J04390637+2334179	&	 ... 	&	III	&	M7.5 $\pm$ 0.25	&	4	&		\\
J04393364+2359212	&	 ... 	&	II	&	M5 $\pm$ 0.25	&	6	&		\\
J04394488+2601527	&	ITG 15	&	II	&	M5 $\pm$ 0.25	&	6	&		\\
J04394748+2601407	&	CFHT 4	&	II	&	M7 $\pm$ 0.5	&	12	&		\\
J04400067+2358211	&	 ... 	&	II	&	M6 $\pm$ 0.25	&	6	&		\\
J04400174+2556292	&	CFHT 17	&	III	&	M5.75 $\pm$ 0.5	&	7	&		\\
J04403979+2519061	&	 ... 	&	II	&	M5.25 $\pm$ 0.25	&	8	&		\\
J04410424+2557561	&	Haro 6-32	&	III	&	M5 $\pm$ 0.25	&	8	&		\\
J04411078+2555116	&	ITG 34	&	II	&	M6.5 $\pm$ 0.5	&	7	&		\\
J04414489+2301513	&	 ... 	&	II	&	M8.5 $\pm$ 0.25	&	5	&		\\
J04414565+2301580	&	 ... 	&	III	&	M4.5 $\pm$ 0.5	&	3	&		\\
J04414825+2534304	&	 ... 	&	II	&	M7.75 $\pm$ 0.25	&	8	&		\\
J04422101+2520343	&	CIDA 7	&	II	&	M4.75 $\pm$ 0.25	&	5	&		\\
J04432023+2940060	&	CIDA 14	&	II	&	M5 $\pm$ 0.5	&	13	&		\\
J04442713+2512164	&	IRAS S04414+2506	&	II	&	M7.25 $\pm$ 0.25	&	8	&	d	\\
J04464260+2459034	&	RXJ 04467+2459	&	III	&	M4 $\pm$ 0.5	&	13	&		\\
J04484189+1703374	&	 ... 	&	III	&	M7 $\pm$ 0.25	&	5	&		\\
J04520668+3047175	&	IRAS 04489+3042	&	I	&	M4 $\pm$ 1	&	5	&		\\
J04552333+3027366	&	 ... 	&	III	&	M6.25 $\pm$ 0.25	&	8	&		\\
J04554046+3039057	&	 ... 	&	III	&	M5.25 $\pm$ 0.25	&	8	&		\\
J04554535+3019389	&	 ... 	&	II	&	M4.75 $\pm$ 0.25	&	8	&		\\
J04554757+3028077	&	 ... 	&	III	&	M4.75 $\pm$ 0.25	&	8	&		\\
J04554801+3028050	&	 ... 	&	II	&	M5.6 $\pm$ 0.25	&	8	&		\\
J04554820+3030160	&	XEST 26-052	&	III	&	M4.5 $\pm$ 0.25	&	3	&	e	\\
J04554969+3019400	&	 ... 	&	II	&	M6 $\pm$ 0.25	&	8	&		\\
J04555288+3006523	&	 ... 	&	III	&	M5.25 $\pm$ 0.25	&	8	&		\\
J04555605+3036209	&	XEST 26-062	&	II	&	M4 $\pm$ 0.25	&	3	&		\\
J04555636+3049374	&	 ... 	&	III	&	M5 $\pm$ 0.25	&	8	&		\\
J04574903+3015195	&	 ... 	&	III	&	M9.25 $\pm$ 0.25	&	8	&		\\
J05061674+2446102	&	CIDA 10	&	III	&	M4 $\pm$ 1	&	16	&		\\
J05064662+2104296	&	 ... 	&	III	&	M5.25 $\pm$ 0.25	&	4	&		\\
J05075496+2500156	&	CIDA 12	&	II	&	M4 $\pm$ 1	&	16	&		\\
\end{longtable} 

\tablefoot{ \tablefoottext{1}{SED classes for all targets are those reported in \citet{Luhman:2010}, excluding the additional two new members reported in \citet{Rebull:2010}, as indicated in the Notes column.} \\
{\bf a-c.} For those targets not observed with {\it Herschel} PACS under our program (OT1\_jpatienc\_1), observations were performed under the following program IDs: a. KPGT\_pandre\_1, b. KPOT\_bdent\_1, c. GT1\_pharve01\_2. \\
{\bf d.} Observed in the far-IR with {\it Spitzer} MIPS only. \\
{\bf e.} No far-IR observations exist for this target. \\
{\bf f.} For the mid-M classification reported in \citet{Luhman:2010} we adopt a spectral type of M4. \\
{\bf g.} For the M6-M8 classification reported in \citet{Luhman:2010} we adopt a spectral type of M7. \\
{\bf h.} SED class from \citet{Rebull:2010}. \\
%
}

\tablebib{The spectral types and uncertainties have been extracted from the following references: 
(1) \citet{Rebull:2010}; (2) \citet{Luhman:2010}; (3) \citet{Luhman:2009}; (4) \citet{Slesnick:2006}; (5) \citet{Luhman:2006}; (6) \citet{Luhman:2006b}; (7) \citet{Guieu:2006}; (8) \citet{Luhman:2004}; (9) \citet{Luhman:2003}; (10) \citet{White:2003}; (11) \citet{Briceno:2002}; (12) \citet{Martin:2001}; (13) \citet{Briceno:1999}; (14) \citet{Luhman:1998R}; (15) \citet{Briceno:1998}; (16) \citet{Kenyon:1995}; (17) \citet{Strom:1994}.
}

\end{longtab}


\small
\begin{longtab}
\begin{longtable}{c c c c c c c}
\caption{\label{Table:obsLog} Observing log for the TBOSS sample observed under this program (OT1\_jpatienc\_1), ordered by target RA. } \\
\hline\hline
 2MASS & Other name & Scan ID & Cross scan ID & UT date & Tot. duration (s) &T$_{\mathrm{int}}$ on-source (s) \\ 
\hline
\endfirsthead
\caption{continued.} \\
\hline\hline
 2MASS & Other name & Scan ID & Cross scan ID & UT date & Tot. duration (s) &T$_{\mathrm{int}}$ on-source (s)  \\
\hline
\endhead
\hline
\endfoot
J04034997+2620382	&	XEST 06-006	&	1342241458	&	1342241459	&	2012 Mar 15	&	904	&	180	\\
J04141188+2811535	&	 ... 	&	1342241918	&	1342241919	&	2012 Mar 19	&	904	&	180	\\
J04141458+2827580	&	FN Tau	&	1342241920	&	1342241921	&	2012 Mar 19	&	1129	&	240	\\
J04141760+2806096	&	CIDA 1	&	1342228441	&	1342228442	&	2011 Sep 07	&	1129	&	240	\\
J04144730+2646264	&	FP Tau	&	1342241884	&	1342241885	&	2012 Mar 19	&	904	&	180	\\
J04144739+2803055	&	XEST 20-066	&	1342241916	&	1342241917	&	2012 Mar 19	&	904	&	180	\\
J04151471+2800096	&	KPNO 1	&	1342241914	&	1342241915	&	2012 Mar 19	&	904	&	180	\\
J04152409+2910434	&	 ... 	&	1342241924	&	1342241925	&	2012 Mar 19	&	904	&	180	\\
J04155799+2746175	&	 ... 	&	1342241646	&	1342241645	&	2012 Mar 17	&	904	&	240	\\
J04161210+2756385	&	 ... 	&	1342241912	&	1342241913	&	2012 Mar 19	&	904	&	180	\\
J04161885+2752155	&	 ... 	&	1342241910	&	1342241911	&	2012 Mar 19	&	904	&	180	\\
J04162725+2053091	&	 ... 	&	1342241880	&	1342241881	&	2012 Mar 19	&	1129	&	240	\\
J04163048+3037053	&	 ... 	&	1342241641	&	1342241642	&	2012 Mar 17	&	904	&	180	\\
J04163911+2858491	&	 ... 	&	1342241922	&	1342241923	&	2012 Mar 19	&	904	&	180	\\
J04174955+2813318	&	KPNO 10	&	1342241908	&	1342241909	&	2012 Mar 19	&	904	&	180	\\
J04174965+2829362	&	V410 X-ray 1	&	1342241906	&	1342241907	&	2012 Mar 19	&	904	&	180	\\
J04180796+2826036	&	V410 X-ray 3	&	1342241904	&	1342241905	&	2012 Mar 19	&	904	&	180	\\
J04181710+2828419	&	V410 Anon 13	&	1342241902	&	1342241903	&	2012 Mar 19	&	904	&	180	\\
J04183030+2743208	&	KPNO 11	&	1342241886	&	1342241887	&	2012 Mar 19	&	904	&	180	\\
J04184023+2824245	&	V410 X-ray 4	&	1342241900	&	1342241901	&	2012 Mar 19	&	904	&	180	\\
J04185115+2814332	&	KPNO 2	&	1342241893	&	1342241892	&	2012 Mar 19	&	904	&	180	\\
J04185813+2812234	&	IRAS 04158+2805	&	1342241890	&	1342241891	&	2012 Mar 19	&	904	&	180	\\
J04190110+2819420	&	V410 X-ray 6	&	1342241894	&	1342241895	&	2012 Mar 19	&	904	&	180	\\
J04190126+2802487	&	KPNO 12	&	1342241888	&	1342241889	&	2012 Mar 19	&	904	&	180	\\
J04190197+2822332	&	V410 X-ray 5a	&	1342241896	&	1342241897	&	2012 Mar 19	&	904	&	180	\\
J04193545+2827218	&	FR Tau	&	1342241898	&	1342241899	&	2012 Mar 19	&	904	&	180	\\
J04194657+2712552	&	$[$GKH94$]$ 41	&	1342242045	&	1342242046	&	2012 Mar 20	&	904	&	180	\\
J04201611+2821325	&	 ... 	&	1342242056	&	1342242057	&	2012 Mar 20	&	904	&	180	\\
J04202555+2700355	&	 ... 	&	1342242041	&	1342242042	&	2012 Mar 20	&	904	&	180	\\
J04202583+2819237	&	IRAS 04173+2812	&	1342242054	&	1342242055	&	2012 Mar 20	&	904	&	180	\\
J04203918+2717317	&	XEST 16-045	&	1342242043	&	1342242044	&	2012 Mar 20	&	904	&	180	\\
J04205273+1746415	&	J2-157	&	1342241876	&	1342241877	&	2012 Mar 19	&	904	&	180	\\
J04210795+2702204	&	CFHT 19	&	1342242039	&	1342242040	&	2012 Mar 20	&	904	&	180	\\
J04210934+2750368	&	 ... 	&	1342242048	&	1342242049	&	2012 Mar 20	&	904	&	180	\\
J04213459+2701388	&	 ... 	&	1342242037	&	1342242038	&	2012 Mar 20	&	904	&	180	\\
J04214013+2814224	&	XEST 21-026	&	1342242052	&	1342242053	&	2012 Mar 20	&	904	&	180	\\
J04214631+2659296	&	CFHT 10	&	1342242035	&	1342242036	&	2012 Mar 20	&	904	&	180	\\
J04215450+2652315	&	 ... 	&	1342242033	&	1342242034	&	2012 Mar 20	&	904	&	180	\\
J04220007+1530248	&	IRAS 04191+1523 B	&	1342241874	&	1342241875	&	2012 Mar 19	&	904	&	180	\\
J04221332+1934392	&	 ... 	&	1342241878	&	1342241879	&	2012 Mar 19	&	904	&	180	\\
J04221644+2549118	&	CFHT 14	&	1342242025	&	1342242026	&	2012 Mar 20	&	904	&	180	\\
J04222404+2646258	&	XEST 11-087	&	1342242031	&	1342242032	&	2012 Mar 20	&	904	&	180	\\
J04230607+2801194	&	 ... 	&	1342242051	&	1342242050	&	2012 Mar 20	&	904	&	180	\\
J04242090+2630511	&	 ... 	&	1342242027	&	1342242028	&	2012 Mar 20	&	904	&	180	\\
J04244506+2701447	&	J1-4423	&	1342242029	&	1342242030	&	2012 Mar 20	&	904	&	180	\\
J04262939+2624137	&	KPNO 3	&	1342242060	&	1342242061	&	2012 Mar 20	&	904	&	180	\\
J04263055+2443558	&	 ... 	&	1342242023	&	1342242024	&	2012 Mar 20	&	904	&	180	\\
J04265732+2606284	&	KPNO 13	&	1342242062	&	1342242063	&	2012 Mar 20	&	904	&	180	\\
J04270739+2215037	&	 ... 	&	1342242005	&	1342242006	&	2012 Mar 20	&	904	&	180	\\
J04272799+2612052	&	KPNO 4	&	1342242065	&	1342242064	&	2012 Mar 20	&	904	&	180	\\
J04274538+2357243	&	CFHT 15	&	1342242021	&	1342242022	&	2012 Mar 20	&	904	&	180	\\
J04275730+2619183	&	IRAS 04248+2612	&	1342242066	&	1342242067	&	2012 Mar 20	&	904	&	180	\\
 ... 	&	L1521F-IRS	&	1342241643	&	1342241644	&	2012 Mar 17	&	1129	&	240	\\
J04284263+2714039	&	 ... 	&	1342243068	&	1342243069	&	2012 Mar 21	&	904	&	180	\\
J04290068+2755033	&	 ... 	&	1342243066	&	1342243067	&	2012 Mar 21	&	904	&	180	\\
J04292165+2701259	&	IRAS 04263+2654	&	1342243070	&	1342243071	&	2012 Mar 21	&	904	&	180	\\
J04294568+2630468	&	KPNO 5	&	1342243074	&	1342243075	&	2012 Mar 21	&	904	&	180	\\
J04295422+1754041	&	 ... 	&	1342241961	&	1342241962	&	2012 Mar 20	&	904	&	180	\\
J04295950+2433078	&	CFHT 20	&	1342242019	&	1342242020	&	2012 Mar 20	&	904	&	180	\\
J04302365+2359129	&	CFHT 16	&	1342242009	&	1342242010	&	2012 Mar 20	&	904	&	180	\\
J04305171+2441475	&	ZZ Tau IRS	&	1342242017	&	1342242018	&	2012 Mar 20	&	904	&	180	\\
J04311578+1820072	&	MHO 9	&	1342241967	&	1342241968	&	2012 Mar 20	&	904	&	180	\\
J04311907+2335047	&	 ... 	&	1342242007	&	1342242008	&	2012 Mar 20	&	904	&	180	\\
J04312405+1800215	&	MHO 4	&	1342241963	&	1342241964	&	2012 Mar 20	&	904	&	180	\\
J04312669+2703188	&	CFHT 13	&	1342243072	&	1342243073	&	2012 Mar 21	&	904	&	180	\\
J04320329+2528078	&	 ... 	&	1342243082	&	1342243083	&	2012 Mar 21	&	904	&	180	\\
J04321606+1812464	&	MHO 5	&	1342241965	&	1342241966	&	2012 Mar 20	&	904	&	180	\\
J04322210+1827426	&	MHO 6	&	1342241969	&	1342241970	&	2012 Mar 20	&	904	&	180	\\
J04322329+2403013	&	 ... 	&	1342242011	&	1342242012	&	2012 Mar 20	&	904	&	180	\\
J04322415+2251083	&	 ... 	&	1342242003	&	1342242004	&	2012 Mar 20	&	904	&	180	\\
J04322627+1827521	&	MHO 7	&	1342243094	&	1342243095	&	2012 Mar 21	&	904	&	180	\\
J04324938+2253082	&	JH 112 B	&	1342242001	&	1342242002	&	2012 Mar 20	&	904	&	180	\\
J04325026+2422115	&	 ... 	&	1342242015	&	1342242016	&	2012 Mar 20	&	904	&	180	\\
J04325119+1730092	&	LH 0429+17	&	1342241957	&	1342241958	&	2012 Mar 20	&	904	&	180	\\
J04330197+2421000	&	MHO 8	&	1342242013	&	1342242014	&	2012 Mar 20	&	904	&	180	\\
J04330781+2616066	&	KPNO 14	&	1342243076	&	1342243077	&	2012 Mar 21	&	904	&	180	\\
J04332621+2245293	&	XEST 17-036	&	1342241999	&	1342242000	&	2012 Mar 20	&	904	&	180	\\
J04334171+1750402	&	 ... 	&	1342241959	&	1342241960	&	2012 Mar 20	&	904	&	180	\\
J04334291+2526470	&	 ... 	&	1342243084	&	1342243085	&	2012 Mar 21	&	904	&	180	\\
J04334465+2615005	&	 ... 	&	1342243078	&	1342243079	&	2012 Mar 21	&	904	&	180	\\
J04335245+2612548	&	 ... 	&	1342243081	&	1342243080	&	2012 Mar 21	&	904	&	180	\\
J04335252+2256269	&	XEST 17-059	&	1342241997	&	1342241998	&	2012 Mar 20	&	904	&	180	\\
J04341527+2250309	&	CFHT 1	&	1342241995	&	1342241996	&	2012 Mar 20	&	904	&	180	\\
J04344544+2308027	&	 ... 	&	1342241992	&	1342241991	&	2012 Mar 20	&	904	&	180	\\
J04350850+2311398	&	CFHT 11	&	1342241993	&	1342241994	&	2012 Mar 20	&	904	&	180	\\
J04354183+2234115	&	KPNO 8	&	1342241977	&	1342241978	&	2012 Mar 20	&	904	&	180	\\
J04354203+2252226	&	XEST 08-033	&	1342241983	&	1342241984	&	2012 Mar 20	&	904	&	180	\\
J04354526+2737130	&	 ... 	&	1342243466	&	1342243467	&	2012 Mar 24	&	904	&	180	\\
J04355143+2249119	&	KPNO 9	&	1342241979	&	1342241980	&	2012 Mar 20	&	904	&	180	\\
J04355209+2255039	&	XEST 08-047	&	1342241989	&	1342241990	&	2012 Mar 20	&	904	&	180	\\
J04355286+2250585	&	XEST 08-049	&	1342241981	&	1342241982	&	2012 Mar 20	&	904	&	180	\\
J04361030+2159364	&	 ... 	&	1342241975	&	1342241976	&	2012 Mar 20	&	904	&	180	\\
J04361038+2259560	&	CFHT 2	&	1342241987	&	1342241988	&	2012 Mar 20	&	904	&	180	\\
J04362151+2351165	&	 ... 	&	1342242072	&	1342242073	&	2012 Mar 20	&	904	&	180	\\
J04363893+2258119	&	CFHT 3	&	1342241985	&	1342241986	&	2012 Mar 20	&	904	&	180	\\
J04373705+2331080	&	 ... 	&	1342243088	&	1342243089	&	2012 Mar 21	&	904	&	180	\\
J04380083+2558572	&	ITG 2	&	1342243458	&	1342243459	&	2012 Mar 24	&	904	&	180	\\
J04381486+2611399	&	 ... 	&	1342243462	&	1342243463	&	2012 Mar 24	&	904	&	180	\\
J04381630+2326402	&	 ... 	&	1342243090	&	1342243091	&	2012 Mar 21	&	904	&	180	\\
J04382134+2609137	&	GM Tau	&	1342243460	&	1342243461	&	2012 Mar 24	&	904	&	180	\\
J04385859+2336351	&	 ... 	&	1342243434	&	1342243435	&	2012 Mar 24	&	904	&	180	\\
J04385871+2323595	&	 ... 	&	1342243092	&	1342243093	&	2012 Mar 21	&	904	&	180	\\
J04390163+2336029	&	 ... 	&	1342243086	&	1342243087	&	2012 Mar 21	&	904	&	180	\\
J04390396+2544264	&	CFHT 6	&	1342243456	&	1342243457	&	2012 Mar 24	&	904	&	180	\\
J04390637+2334179	&	 ... 	&	1342243432	&	1342243433	&	2012 Mar 24	&	904	&	180	\\
J04393364+2359212	&	 ... 	&	1342243438	&	1342243439	&	2012 Mar 24	&	904	&	180	\\
J04394488+2601527	&	ITG 15	&	1342243453	&	1342243452	&	2012 Mar 24	&	904	&	180	\\
J04394748+2601407	&	CFHT 4	&	1342243454	&	1342243455	&	2012 Mar 24	&	904	&	180	\\
J04400067+2358211	&	 ... 	&	1342243436	&	1342243437	&	2012 Mar 24	&	904	&	180	\\
J04400174+2556292	&	CFHT 17	&	1342243450	&	1342243451	&	2012 Mar 24	&	904	&	180	\\
J04403979+2519061	&	 ... 	&	1342243442	&	1342243443	&	2012 Mar 24	&	904	&	180	\\
J04410424+2557561	&	Haro 6-32	&	1342243449	&	1342243448	&	2012 Mar 24	&	904	&	180	\\
J04411078+2555116	&	ITG 34	&	1342243446	&	1342243447	&	2012 Mar 24	&	904	&	180	\\
J04414489+2301513	&	 ... 	&	1342240750	&	1342240751	&	2012 Mar 07	&	904	&	180	\\
J04414565+2301580	&	 ... 	&	1342243430	&	1342243431	&	2012 Mar 24	&	904	&	180	\\
J04414825+2534304	&	 ... 	&	1342243444	&	1342243445	&	2012 Mar 24	&	904	&	180	\\
J04422101+2520343	&	CIDA 7	&	1342243440	&	1342243441	&	2012 Mar 24	&	904	&	180	\\
J04432023+2940060	&	CIDA 14	&	1342243472	&	1342243473	&	2012 Mar 24	&	904	&	180	\\
J04464260+2459034	&	RXJ 04467+2459	&	1342228864	&	1342228865	&	2011 Sep 18	&	904	&	180	\\
J04484189+1703374	&	 ... 	&	1342243428	&	1342243429	&	2012 Mar 24	&	904	&	180	\\
J04520668+3047175	&	IRAS 04489+3042	&	1342243492	&	1342243493	&	2012 Mar 24	&	904	&	180	\\
J04552333+3027366	&	 ... 	&	1342243484	&	1342243485	&	2012 Mar 24	&	904	&	180	\\
J04554046+3039057	&	 ... 	&	1342243488	&	1342243489	&	2012 Mar 24	&	904	&	180	\\
J04554535+3019389	&	 ... 	&	1342250330	&	1342250331	&	2012 Aug 26	&	904	&	180	\\
J04554757+3028077	&	 ... 	&	1342242762	&	1342242763	&	2012 Mar 30	&	1129	&	240	\\
J04554801+3028050	&	 ... 	&	1342243482	&	1342243483	&	2012 Mar 24	&	904	&	180	\\
J04554969+3019400	&	 ... 	&	1342243480	&	1342243481	&	2012 Mar 24	&	904	&	180	\\
J04555288+3006523	&	 ... 	&	1342243477	&	1342243476	&	2012 Mar 24	&	904	&	180	\\
J04555605+3036209	&	XEST 26-062	&	1342243486	&	1342243487	&	2012 Mar 24	&	904	&	180	\\
J04555636+3049374	&	 ... 	&	1342243490	&	1342243491	&	2012 Mar 24	&	904	&	180	\\
J04574903+3015195	&	 ... 	&	1342243478	&	1342243479	&	2012 Mar 24	&	904	&	180	\\
J05061674+2446102	&	CIDA 10	&	1342242686	&	1342242687	&	2012 Mar 29	&	904	&	180	\\
J05064662+2104296	&	 ... 	&	1342242684	&	1342242685	&	2012 Mar 29	&	904	&	180	\\
J05075496+2500156	&	CIDA 12	&	1342242688	&	1342242689	&	2012 Mar 29	&	904	&	180	\\

\end{longtable} 
\end{longtab}

{\small
\begin{table*}
\caption{Observing log for the TBOSS sample observed with {\it Herschel} PACS under other programs. Targets are ordered by RA.}             
\label{Table:obsLog_other}      
\centering          
\begin{tabular}{c c c c c c l}   
\hline\hline                             
 2MASS & Other name & Scan & Cross scan & UT  & Total  &  Program \\ 
   & & ID &  ID & date & duration (s) &  ID \\
\hline

J04131414+2819108	&	LkCa 1	&	1342216513	&	1342216514	&	2011 Mar 21	&	552	&	KPOT\_bdent\_1	\\
	&		&	(1342216515)	&	(1342216516)	&	2011 Mar 21	&	552	&	KPOT\_bdent\_1	\\
J04150515+2808462	&	CIDA 2	&	1342216529	&	1342216530	&	2011 Mar 21	&	552	&	KPOT\_bdent\_1	\\
	&		&	(1342216531)	&	(1342216532)	&	2011 Mar 21	&	552	&	KPOT\_bdent\_1	\\
J04233539+2503026	&	FU Tau A	&	(1342227304)	&	(1342227305)	&	2011 Aug 24	&	21222	&	KPGT\_pandre\_1	\\
J04233573+2502596	&	FU Tau B	&	(1342227304)	&	(1342227305)	&	2011 Aug 24	&	21222	&	KPGT\_pandre\_1	\\
J04242646+2649503	&	CFHT 9	&	1342227059	&	1342227060	&	2011 Aug 21	&	2690	&	GT1\_pharve01\_2	\\
J04292071+2633406	&	J1-507	&	1342227979	&	1342227980	&	2011 Sep 04	&	552	&	KPOT\_bdent\_1	\\
	&		&	(1342227981)	&	(1342227982)	&	2011 Sep 04	&	552	&	KPOT\_bdent\_1	\\
J04292971+2616532	&	FW Tau A+B+C	&	1342227987	&	1342227988	&	2011 Sep 04	&	552	&	KPOT\_bdent\_1	\\
	&		&	(1342227989)	&	(1342227990)	&	2011 Sep 04	&	552	&	KPOT\_bdent\_1	\\
J04300724+2608207	&	KPNO 6	&	1342227012	&	1342227013	&	2011 Aug 21	&	3140	&	GT1\_pharve01\_2	\\
J04305718+2556394	&	KPNO 7	&	1342227999	&	1342228000	&	2011 Sep 04	&	3140	&	GT1\_pharve01\_2	\\
J04312382+2410529	&	V927 Tau A+B	&	1342227055	&	1342227056	&	2011 Aug 21	&	552	&	KPOT\_bdent\_1	\\
	&		&	(1342227057)	&	(1342227058)	&	2011 Aug 21	&	552	&	KPOT\_bdent\_1	\\
J04313613+1813432	&	LkHa 358	&	1342228928	&	1342228929	&	2011 Sep 19	&	552	&	KPOT\_bdent\_1	\\
	&		&	(1342228930)	&	(1342228931)	&	2011 Sep 19	&	552	&	KPOT\_bdent\_1	\\
J04315844+2543299	&	J1-665	&	1342228001	&	1342228002	&	2011 Sep 04	&	552	&	KPOT\_bdent\_1	\\
	&		&	(1342228003)	&	(1342228004)	&	2011 Sep 04	&	552	&	KPOT\_bdent\_1	\\
J04321786+2422149	&	CFHT 7	&	(1342228005)	&	(1342228006)	&	2011 Sep 04	&	39224	&	KPGT\_pandre\_1	\\
J04323028+1731303	&	GG Tau Ba+Bb	&	1342228940	&	1342228941	&	2011 Sep 19	&	552	&	KPOT\_bdent\_1	\\
	&		&	(1342228942)	&	(1342228943)	&	2011 Sep 19	&	552	&	KPOT\_bdent\_1	\\
J04330945+2246487	&	CFHT 12	&	1342227013	&	1342227014	&	2011 Aug 21	&	3140	&	GT1\_pharve01\_2	\\
J04353536+2408266	&	IRAS 04325+2402 C	&	(1342228005)	&	(1342228006)	&	2011 Sep 04	&	39224	&	KPGT\_pandre\_1	\\

\hline
\end{tabular}

\tablefoot{The observations listed were performed with the PACS photometer Blue1 channel (70 $\mu$m) and the Red channel (160 $\mu$m). For the scan ID numbers within parenthesis, observations were performed with the PACS photometer Blue2 channel (100 $\mu$m) and the Red channel (160 $\mu$m).
}
\end{table*}
}

\begin{longtab}
\begin{longtable}{c c c c c c c c r}
\caption{\label{Table:sed} TBOSS stellar properties derived from the literature and evolutionary models in comparison with the best fit values from our SED fitting.} \\
\hline\hline
 2MASS & Other name & SpTy. & {\it A}$_V$ & {\it T}$\mathrm{_{eff}}$ & {\it R}$_{\mathrm{star}}$ & {\it A}$_V$ fit & {\it R}$_{\mathrm{star}}$ fit & Ref.  \\
    & &  & (mag) & (K) & (R$_{\sun}$) & (mag) & (R$_{\sun}$) & \\
\hline
\endfirsthead
\caption{continued.} \\
\hline\hline
 2MASS & Other name & SpTy. & {\it A}$_V$ & {\it T}$\mathrm{_{eff}}$ & {\it R}$_{\mathrm{star}}$ & {\it A}$_V$ fit & {\it R}$_{\mathrm{star}}$ fit & Ref.  \\
   & &  & (mag) & (K) &(R$_{\sun}$) & (mag) & (R$_{\sun}$) & \\
\hline
\endhead
\hline
\endfoot
J04034997+2620382	&	XEST 06-006	&	M5.25	&	0.0	&	3091	&	1.217	&	2.0	&	0.608	&	12	\\
J04131414+2819108	&	LkCa 1	&	M4	&	0.0	&	3270	&	1.652	&	0.2	&	1.817	&	6	\\
J04141188+2811535	&	 ... 	&	M6.25	&	1.0	&	2963	&	0.873	&	2.5	&	0.873	&	9	\\
J04141458+2827580	&	FN Tau	&	M5   	&	1.4	&	3125	&	1.300	&	0.0	&	2.600	&	6	\\
J04141760+2806096	&	CIDA 1	&	M5.5	&	5.5	&	3058	&	1.136	&	1.7	&	1.420	&	4	\\
J04144730+2646264	&	FP Tau	&	M4	&	0.2	&	3270	&	1.652	&	0.7	&	1.734	&	6	\\
J04144739+2803055	&	XEST 20-066	&	M5.25	&	0.0	&	3091	&	1.217	&	0.2	&	0.973	&	12	\\
J04150515+2808462	&	CIDA 2	&	M4.5	&	0.8	&	3198	&	1.472	&	0.3	&	1.620	&	6	\\
J04151471+2800096	&	KPNO 1	&	M9	&	0.4	&	2400	&	0.295	&	0.2	&	0.295	&	5	\\
J04152409+2910434	&	 ... 	&	M7	&	2.4	&	2880	&	0.641	&	0.8	&	0.320	&	10	\\
J04155799+2746175	&	 ... 	&	M5	&	0.0	&	3125	&	1.300	&	1.0	&	1.300	&	12	\\
J04161210+2756385	&	 ... 	&	M4.75	&	2.0	&	3161	&	1.385	&	2.0	&	1.385	&	9	\\
J04161885+2752155	&	 ... 	&	M6.25	&	1.2	&	2963	&	0.873	&	0.8	&	0.436	&	10	\\
J04162725+2053091	&	 ... 	&	M5	&	0.2	&	3125	&	1.300	&	0.5	&	1.365	&	4	\\
J04163048+3037053	&	 ... 	&	M4.5	&	0.7	&	3198	&	1.472	&	0.5	&	0.736	&	9	\\
J04163911+2858491	&	 ... 	&	M5.5	&	3.4	&	3058	&	1.136	&	0.3	&	0.455	&	10	\\
J04174955+2813318	&	KPNO 10	&	M5.5	&	0.0	&	3058	&	1.136	&	1.3	&	1.136	&	5	\\
J04174965+2829362	&	V410 X-ray 1	&	M4	&	1.0	&	3270	&	1.652	&	2.5	&	2.065	&	15	\\
J04180796+2826036	&	V410 X-ray 3	&	M6.5	&	1.5	&	2935	&	0.794	&	0.6	&	0.635	&	4	\\
J04181710+2828419	&	V410 Anon 13	&	M5.75	&	2.8	&	3024	&	1.054	&	2.8	&	0.422	&	3	\\
J04183030+2743208	&	KPNO 11	&	M5.5	&	0.0	&	3058	&	1.136	&	0.5	&	0.568	&	8	\\
J04184023+2824245	&	V410 X-ray 4	&	M4	&	18.9	&	3270	&	1.652	&	3.8	&	1.982	&	7	\\
J04185115+2814332	&	KPNO 2	&	M6.75	&	0.4	&	2908	&	0.708	&	1.0	&	0.354	&	5	\\
J04185813+2812234	&	IRAS 04158+2805	&	M5.25	&	1.8	&	3091	&	1.217	&	2.5	&	0.608	&	10	\\
J04190110+2819420	&	V410 X-ray 6	&	M4.5	&	0.9	&	3198	&	1.472	&	1.1	&	0.589	&	7	\\
J04190126+2802487	&	KPNO 12	&	M9	&	0.5	&	2400	&	0.295	&	0.7	&	0.413	&	8	\\
J04190197+2822332	&	V410 X-ray 5a	&	M5.5	&	2.5	&	3058	&	1.136	&	1.5	&	0.909	&	1	\\
J04193545+2827218	&	FR Tau	&	M5.25	&	0.0	&	3091	&	1.217	&	0.0	&	1.217	&	12	\\
J04194657+2712552	&	$[$GKH94$]$ 41	&	M7.5	&	27.0	&	2795	&	0.470	&	4.0	&	0.470	&	12	\\
J04201611+2821325	&	 ... 	&	M6.5	&	0.0	&	2935	&	0.794	&	1.0	&	0.397	&	12	\\
J04202555+2700355	&	 ... 	&	M5.25	&	2.0	&	3091	&	1.217	&	1.6	&	0.487	&	9	\\
J04202583+2819237	&	IRAS 04173+2812	&	mid-M	&	12.0	&	3270	&	1.652	&	6.0	&	1.734	&	4	\\
J04203918+2717317	&	XEST 16-045	&	M4.5	&	0.0	&	3198	&	1.472	&	0.8	&	1.620	&	12	\\
J04205273+1746415	&	J2-157	&	M5.5	&	0.0	&	3058	&	1.136	&	0.5	&	0.568	&	6	\\
J04210795+2702204	&	CFHT 19	&	M5.25	&	7.3	&	3091	&	1.217	&	5.1	&	1.521	&	5	\\
J04210934+2750368	&	 ... 	&	M5.25	&	0.0	&	3091	&	1.217	&	0.5	&	1.217	&	12	\\
J04213459+2701388	&	 ... 	&	M5.5	&	1.7	&	3058	&	1.136	&	0.9	&	1.136	&	9	\\
J04214013+2814224	&	XEST 21-026	&	M5.75	&	0.0	&	3024	&	1.054	&	0.2	&	0.527	&	12	\\
J04214631+2659296	&	CFHT 10	&	M6.25	&	3.6	&	2963	&	0.873	&	1.8	&	0.349	&	5	\\
J04215450+2652315	&	 ... 	&	M8.5	&	1.2	&	2555	&	0.348	&	1.3	&	0.522	&	10	\\
J04221332+1934392	&	 ... 	&	M8	&	0.0	&	2710	&	0.402	&	0.4	&	0.402	&	10	\\
J04221644+2549118	&	CFHT 14	&	M7.75	&	0.6	&	2753	&	0.422	&	1.2	&	0.422	&	5	\\
J04222404+2646258	&	XEST 11-087	&	M4.75	&	1.1	&	3161	&	1.385	&	0.3	&	1.385	&	12	\\
J04230607+2801194	&	 ... 	&	M6	&	0.0	&	2990	&	0.942	&	1.5	&	0.942	&	11	\\
J04242090+2630511	&	 ... 	&	M6.5	&	0.0	&	2935	&	0.794	&	0.5	&	0.397	&	11	\\
J04242646+2649503	&	CFHT 9	&	M6.25	&	0.9	&	2963	&	0.873	&	0.9	&	0.349	&	5	\\
J04244506+2701447	&	J1-4423	&	M5	&	1.0	&	3125	&	1.300	&	0.4	&	1.365	&	6	\\
J04251550+2829275	&	 ... 	&	M5	&	3.4	&	3125	&	1.300	&	0.3	&	1.300	&	14	\\
J04262939+2624137	&	KPNO 3	&	M6	&	1.6	&	2990	&	0.942	&	1.6	&	0.377	&	3	\\
J04263055+2443558	&	 ... 	&	M8.75	&	0.0	&	2478	&	0.321	&	0.9	&	0.321	&	10	\\
J04265732+2606284	&	KPNO 13	&	M5	&	2.5	&	3125	&	1.300	&	1.7	&	2.600	&	8	\\
J04270739+2215037	&	 ... 	&	M6.75	&	0.4	&	2908	&	0.708	&	0.5	&	0.849	&	4	\\
J04272799+2612052	&	KPNO 4	&	M9.5	&	2.5	&	2245	&	0.268	&	1.3	&	0.268	&	5	\\
J04274538+2357243	&	CFHT 15	&	M8.25	&	1.3	&	2633	&	0.375	&	0.5	&	0.563	&	5	\\
J04275730+2619183	&	IRAS 04248+2612	&	M4.5	&	3.9	&	3198	&	1.472	&	1.9	&	1.620	&	10	\\
 ... 	&	L1521F-IRS	&	M6-M8	&	...$^{a}$	&	2880	&	0.641	&	0.5	&	0.384	&	4	\\
J04284263+2714039	&	 ... 	&	M5.25	&	0.5	&	3091	&	1.217	&	1.3	&	1.217	&	9	\\
J04290068+2755033	&	 ... 	&	M8.25	&	0.0	&	2633	&	0.375	&	0.6	&	0.375	&	10	\\
J04292071+2633406	&	J1-507	&	M4   	&	0.8	&	3270	&	1.652	&	0.3	&	1.817	&	6	\\
J04292165+2701259	&	IRAS 04263+2654	&	M6	&	4.9	&	2990	&	0.942	&	0.4	&	1.884	&	5	\\
J04292971+2616532	&	FW Tau A+B+C	&	M4	&	0.4	&	3270	&	1.652	&	0.0	&	1.652	&	6	\\
J04294568+2630468	&	KPNO 5	&	M7.5	&	0.0	&	2795	&	0.470	&	0.3	&	0.470	&	3	\\
J04295422+1754041	&	 ... 	&	M4	&	0.0	&	3270	&	1.652	&	2.6	&	1.652	&	10	\\
J04295950+2433078	&	CFHT 20	&	M5.5	&	3.6	&	3058	&	1.136	&	2.5	&	2.273	&	5	\\
J04300724+2608207	&	KPNO 6	&	M9	&	0.9	&	2400	&	0.295	&	1.4	&	0.295	&	5	\\
J04302365+2359129	&	CFHT 16	&	M8.5	&	1.5	&	2555	&	0.348	&	0.6	&	0.522	&	5	\\
J04305171+2441475	&	ZZ Tau IRS	&	M5.25	&	2.4	&	3091	&	1.217	&	3.1	&	1.278	&	5	\\
J04305718+2556394	&	KPNO 7	&	M8.25	&	0.0	&	2633	&	0.375	&	1.5	&	0.375	&	3	\\
J04311578+1820072	&	MHO 9	&	M5	&	0.9	&	3125	&	1.300	&	0.3	&	1.300	&	4	\\
J04311907+2335047	&	 ... 	&	M7.75	&	0.6	&	2753	&	0.422	&	0.6	&	0.634	&	10	\\
J04312382+2410529	&	V927 Tau A+B	&	M5.5	&	0.4	&	3058	&	1.136	&	0.2	&	1.704	&	6	\\
J04312405+1800215	&	MHO 4	&	M7	&	1.3	&	2880	&	0.641	&	0.7	&	0.961	&	4	\\
J04312669+2703188	&	CFHT 13	&	M7.25	&	3.5	&	2838	&	0.552	&	1.4	&	0.441	&	5	\\
J04313613+1813432	&	LkHa 358	&	M5.5	&	13.6	&	3058	&	1.136	&	2.7	&	1.591	&	6	\\
J04315844+2543299	&	J1-665	&	M5	&	1.0	&	3125	&	1.300	&	0.4	&	1.560	&	6	\\
J04320329+2528078	&	 ... 	&	M6.25	&	0.0	&	2963	&	0.873	&	0.5	&	0.698	&	10	\\
J04321606+1812464	&	MHO 5	&	M7	&	1.2	&	2880	&	0.641	&	0.1	&	0.961	&	4	\\
J04321786+2422149	&	CFHT 7	&	M6.5	&	0.0	&	2935	&	0.794	&	0.3	&	1.191	&	5	\\
J04322210+1827426	&	MHO 6	&	M4.75	&	1.1	&	3161	&	1.385	&	1.4	&	1.385	&	3	\\
J04322329+2403013	&	 ... 	&	M7.75	&	0.0	&	2753	&	0.422	&	0.2	&	0.422	&	10	\\
J04322415+2251083	&	 ... 	&	M4.5	&	1.7	&	3198	&	1.472	&	0.9	&	0.589	&	12	\\
J04322627+1827521	&	MHO 7	&	M5.25	&	0.1	&	3091	&	1.217	&	0.2	&	1.217	&	3	\\
J04324938+2253082	&	JH 112 B	&	M4.25	&	3.1	&	3234	&	1.554	&	2.2	&	1.865	&	12	\\
J04325026+2422115	&	 ... 	&	M7.5	&	9.2	&	2795	&	0.470	&	2.8	&	0.705	&	5	\\
J04325119+1730092	&	LH 0429+17	&	M8.25	&	0.0	&	2633	&	0.375	&	0.2	&	0.563	&	12	\\
J04330197+2421000	&	MHO 8	&	M6	&	1.0	&	2990	&	0.942	&	1.0	&	1.413	&	3	\\
J04330781+2616066	&	KPNO 14	&	M6	&	3.0	&	2990	&	0.942	&	1.2	&	1.130	&	8	\\
J04330945+2246487	&	CFHT 12	&	M6.5	&	3.4	&	2935	&	0.794	&	1.7	&	0.794	&	5	\\
J04332621+2245293	&	XEST 17-036	&	M4	&	3.9	&	3270	&	1.652	&	1.6	&	1.652	&	12	\\
J04334171+1750402	&	 ... 	&	M4	&	0.3	&	3270	&	1.652	&	1.3	&	1.652	&	12	\\
J04334291+2526470	&	 ... 	&	M8.75	&	0.0	&	2478	&	0.321	&	1.1	&	0.321	&	10	\\
J04334465+2615005	&	 ... 	&	M4.75	&	3.0	&	3161	&	1.385	&	3.0	&	0.554	&	12	\\
J04335245+2612548	&	 ... 	&	M8.5	&	5.0	&	2555	&	0.348	&	0.5	&	0.522	&	4	\\
J04335252+2256269	&	XEST 17-059	&	M5.75	&	0.0	&	3024	&	1.054	&	0.2	&	1.582	&	12	\\
J04341527+2250309	&	CFHT 1	&	M7	&	3.1	&	2880	&	0.641	&	1.9	&	0.801	&	13	\\
J04344544+2308027	&	 ... 	&	M5.25	&	2.1	&	3091	&	1.217	&	0.9	&	0.608	&	4	\\
J04350850+2311398	&	CFHT 11	&	M6.75	&	0.0	&	2908	&	0.708	&	0.2	&	0.566	&	5	\\
J04354183+2234115	&	KPNO 8	&	M5.75	&	0.5	&	3024	&	1.054	&	0.3	&	0.527	&	3	\\
J04354203+2252226	&	XEST 08-033	&	M4.75	&	1.7	&	3161	&	1.385	&	0.7	&	1.385	&	12	\\
J04354526+2737130	&	 ... 	&	M9.25	&	0.0	&	2323	&	0.281	&	0.2	&	0.281	&	10	\\
J04355143+2249119	&	KPNO 9	&	M8.5	&	0.0	&	2555	&	0.348	&	0.7	&	0.522	&	3	\\
J04355209+2255039	&	XEST 08-047	&	M4.5	&	2.0	&	3198	&	1.472	&	0.6	&	1.472	&	12	\\
J04355286+2250585	&	XEST 08-049	&	M4.25	&	1.2	&	3234	&	1.554	&	0.4	&	1.554	&	12	\\
J04355760+2253574	&	 ... 	&	M6	&	0.0	&	2990	&	0.942	&	1.0	&	0.104	&	14	\\
J04361030+2159364	&	 ... 	&	M8.5	&	0.0	&	2555	&	0.348	&	0.1	&	0.522	&	10	\\
J04361038+2259560	&	CFHT 2	&	M7.5	&	2.0	&	2795	&	0.470	&	0.8	&	0.705	&	3	\\
J04362151+2351165	&	 ... 	&	M5.25	&	1.5	&	3091	&	1.217	&	1.0	&	0.487	&	11	\\
J04363893+2258119	&	CFHT 3	&	M7.75	&	1.0	&	2753	&	0.422	&	1.0	&	0.634	&	3	\\
J04373705+2331080	&	 ... 	&	L0	&	0.0	&	2090	&	0.240	&	0.7	&	0.100	&	12	\\
J04380083+2558572	&	ITG 2	&	M7.25	&	0.6	&	2838	&	0.552	&	0.2	&	0.827	&	9	\\
J04381486+2611399	&	 ... 	&	M7.25	&	0.0	&	2838	&	0.552	&	1.0	&	0.579	&	9	\\
J04381630+2326402	&	 ... 	&	M4.75	&	...$^{b}$	&	3161	&	1.385	&	0.3	&	0.693	&	4	\\
J04382134+2609137	&	GM Tau	&	M6.5	&	6.1	&	2935	&	0.794	&	0.6	&	0.794	&	4	\\
J04385859+2336351	&	 ... 	&	M4.25	&	0.0	&	3234	&	1.554	&	1.5	&	0.622	&	4	\\
J04385871+2323595	&	 ... 	&	M6.5	&	...$^{b}$	&	2935	&	0.794	&	0.2	&	0.635	&	4	\\
J04390163+2336029	&	 ... 	&	M6	&	1.8	&	2990	&	0.942	&	0.5	&	1.130	&	4	\\
J04390396+2544264	&	CFHT 6	&	M7.25	&	0.4	&	2838	&	0.552	&	0.5	&	0.552	&	5	\\
J04390637+2334179	&	 ... 	&	M7.5	&	...$^{b}$	&	2795	&	0.470	&	0.2	&	0.470	&	4	\\
J04393364+2359212	&	 ... 	&	M5	&	0.0	&	3125	&	1.300	&	1.0	&	1.300	&	11	\\
J04394488+2601527	&	ITG 15	&	M5	&	2.7	&	3125	&	1.300	&	0.5	&	2.600	&	11	\\
J04394748+2601407	&	CFHT 4	&	M7	&	10.7	&	2880	&	0.641	&	0.0	&	0.641	&	13	\\
J04400067+2358211	&	 ... 	&	M6	&	0.0	&	2990	&	0.942	&	0.5	&	0.377	&	11	\\
J04400174+2556292	&	CFHT 17	&	M5.75	&	6.5	&	3024	&	1.054	&	2.6	&	1.318	&	5	\\
J04403979+2519061	&	 ... 	&	M5.25	&	2.5	&	3091	&	1.217	&	1.2	&	1.217	&	9	\\
J04410424+2557561	&	Haro 6-32	&	M5	&	0.6	&	3125	&	1.300	&	0.4	&	0.520	&	9	\\
J04411078+2555116	&	ITG 34	&	M6.5	&	1.8	&	2935	&	0.794	&	1.8	&	1.191	&	5	\\
J04414489+2301513	&	 ... 	&	M8.5	&	0.0	&	2555	&	0.348	&	0.9	&	0.348	&	10	\\
J04414565+2301580	&	 ... 	&	M4.5	&	0.4	&	3198	&	1.472	&	0.2	&	1.472	&	12	\\
J04414825+2534304	&	 ... 	&	M7.75	&	1.0	&	2753	&	0.422	&	1.3	&	0.634	&	9	\\
J04422101+2520343	&	CIDA 7	&	M4.75	&	1.2	&	3161	&	1.385	&	1.2	&	1.385	&	10	\\
J04432023+2940060	&	CIDA 14	&	M5	&	0.3	&	3125	&	1.300	&	0.4	&	1.560	&	2	\\
J04442713+2512164	&	IRAS S04414+2506	&	M7.25	&	0.0	&	2838	&	0.552	&	0.1	&	0.552	&	9	\\
J04464260+2459034	&	RXJ 04467+2459	&	M4	&	1.6	&	3270	&	1.652	&	1.1	&	1.652	&	2	\\
J04484189+1703374	&	 ... 	&	M7	&	0.0	&	2880	&	0.641	&	0.8	&	0.320	&	10	\\
J04520668+3047175	&	IRAS 04489+3042	&	M4	&	4.8	&	3270	&	1.652	&	5.0	&	1.734	&	10	\\
J04552333+3027366	&	 ... 	&	M6.25	&	0.0	&	2963	&	0.873	&	1.0	&	0.436	&	9	\\
J04554046+3039057	&	 ... 	&	M5.25	&	0.2	&	3091	&	1.217	&	0.3	&	0.608	&	9	\\
J04554535+3019389	&	 ... 	&	M4.75	&	0.0	&	3161	&	1.385	&	1.3	&	1.385	&	9	\\
J04554757+3028077	&	 ... 	&	M4.75	&	0.0	&	3161	&	1.385	&	0.4	&	0.693	&	9	\\
J04554801+3028050	&	 ... 	&	M5.6	&	0.0	&	3044	&	1.103	&	1.0	&	0.441	&	9	\\
J04554820+3030160	&	XEST 26-052	&	M4.5	&	0.0	&	3198	&	1.472	&	0.3	&	0.736	&	12	\\
J04554969+3019400	&	 ... 	&	M6	&	0.0	&	2990	&	0.942	&	1.1	&	0.377	&	9	\\
J04555288+3006523	&	 ... 	&	M5.25	&	0.0	&	3091	&	1.217	&	0.2	&	0.608	&	9	\\
J04555605+3036209	&	XEST 26-062	&	M4	&	1.9	&	3270	&	1.652	&	0.0	&	1.652	&	12	\\
J04555636+3049374	&	 ... 	&	M5	&	0.4	&	3125	&	1.300	&	0.2	&	1.365	&	9	\\
J04574903+3015195	&	 ... 	&	M9.25	&	0.0	&	2323	&	0.281	&	0.2	&	0.100	&	9	\\
J05061674+2446102	&	CIDA 10	&	M4	&	0.5	&	3270	&	1.652	&	0.4	&	1.652	&	6	\\
J05064662+2104296	&	 ... 	&	M5.25	&	0.8	&	3091	&	1.217	&	0.5	&	0.608	&	4	\\
J05075496+2500156	&	CIDA 12	&	M4	&	0.8	&	3270	&	1.652	&	1.2	&	1.652	&	6	\\
\noalign{\smallskip} \hline \noalign{\smallskip} 
\multicolumn{9}{c}{Multiple systems with combined spectra} \\
\noalign{\smallskip} \hline \noalign{\smallskip} 
J04220043+1530212	&	IRAS 04191+1523 A	&	K6-M3.5	&	33.3	&	3705	&	2.544	&	 	&	  	&	4	\\
J04220007+1530248	&	IRAS 04191+1523 B	&	M6-M8	&	34.4	&	2880	&	0.641	&	 	&	 	&	4	\\
  	&	{\bf IRAS 04191+1523 AB + C}	&	  	&	 	&	  	&		&	2.7	&	3.312	&		\\
J04233539+2503026	&	FU Tau A	&	M7.25	&	2.0	&	2838	&	0.552	&	  	&	  	&	12	\\
J04233573+2502596	&	FU Tau B	&	M9.25	&	0.0	&	2323	&	0.276	&	  	&	  	&	12	\\
  	&	{\bf FU Tau A+B}	&	  	&	  	&	  	& 		&	0.0	&	9.113$^{c}$	&		\\
&	GG Tau Ba	&	M6	&	...  	&	2990	&	0.942	&	  	&	  	&		\\
&	GG Tau Bb	&	M7.5	&	...  	&	2795	&	0.470	&	 	&	  	&		\\
{\bf J04323028+1731303}	&	{\bf GG Tau Ba + Bb}	&	  	&	1.1	&	  	&	&	0.3	&	0.847	&	4	\\
J04353539+2408194	&	IRAS 04325+2402 AB	&	K6-M3.5	&	36.9	&	3705	&	2.544	&	  	&	  	&	4	\\
J04353536+2408266	&	IRAS 04325+2402 C	&	M6-M8	&	38.0	&	2880	&	0.641	&	  	&	  	&	4	\\
  	&	{\bf IRAS 04325+2402 AB + C}	&	  	&	  	&	  	&	3.185	&	3.7	&	4.457	&		\\

\end{longtable} 

\tablefoot{References refer to the extinctions compiled from the literature, as listed in the {\it A}$_V$ column. $^{a}$ For L1521F-IRS upper limits are reported in 2MASS at {\it JHK}. $^{b}$ For these systems the calculated the excesses between the 2MASS {\it JH} and the intrinsic colors \citep{Luhman:2010} are negative. $^{c}$ The best-fit spectrum of FU Tau A+B resulted in an extreme fit to the stellar radius. The system is noted to be an over luminous source in \citet{Luhman:2009b}.
}
\tablebib{Extinction values have been compiled from the following references: (1) \citet{Briceno:1998}; (2) \citet{Briceno:1999}; (3) \citet{Briceno:2002}; (4) This work; (5) \citet{Guieu:2006}; (6) \citet{Kenyon:1995}; (7) \citet{Luhman:1998R}; (8) \citet{Luhman:2003}; (9) \citet{Luhman:2004}; (10) \citet{Luhman:2006}; (11) \citet{Luhman:2006b}; (12) \citet{Luhman:2009}; (13) \citet{Martin:2001}; (14) \citet{Rebull:2010}; (15) \citet{Strom:1994}.
}

\end{longtab}


\begin{longtab}
\begin{longtable}{l c c c c c c r r}
\caption{\label{table:PACS} {\it Herschel} PACS photometry compilation. Targets are ordered by evolutionary class, spectral type, and RA.} \\      
\hline\hline                             
2MASS & Other name & Class & SpTy. & 70 $\mu$m & 100 $\mu$m & 160 $\mu$m & Ref. & Notes \\ 
& & & & (mJy) & (mJy) & (mJy) & & \\
\hline
\endfirsthead
\caption{continued.} \\
\hline\hline
2MASS & Other name & Class & SpTy. & 70 $\mu$m & 100 $\mu$m & 160 $\mu$m & Ref. & Notes \\ 
& & & & (mJy) & (mJy) & (mJy) & & \\
\hline
\endhead
\hline
\endfoot
J04520668+3047175	&	IRAS 04489+3042	&	I	&	M4	&	2151 $\pm$ 5	&		&	2103 $\pm$ 34	&		&		\\
J04275730+2619183	&	IRAS 04248+2612	&	I	&	M4.5	&	4531 $\pm$ 17	&		&	10093 $\pm$ 141	&		&		\\
J04185813+2812234	&	IRAS 04158+2805	&	I	&	M5.25	&	1089 $\pm$ 3	&		&	2953 $\pm$ 25	&		&		\\
J04313613+1813432	&	LkHa 358	&	I	&	M5.5	&	1300 $\pm$ 200	&	1320 $\pm$ 200	&	1530 $\pm$ 300	&	2	&		\\
J04220007+1530248	&	IRAS 04191+1523 B	&	I	&	M6-M8	&	(7002 $\pm$ 13)	&		&	(8884 $\pm$ 232)	&		&		\\
 ... 	&	L1521F-IRS	&	I	&	M6-M8	&	522 $\pm$ 4	&		&	3712 $\pm$ 52	&		&		\\
J04194657+2712552	&	$[$GKH94$]$ 41	&	I	&	M7.5	&	269 $\pm$ 5	&		&	279 $\pm$ 66	&		&		\\
J04144730+2646264	&	FP Tau	&	II	&	M4	&	307 $\pm$ 3	&		&	351 $\pm$ 11	&		&		\\
J04174965+2829362	&	V410 X-ray 1	&	II	&	M4	&	36 $\pm$ 3	&		&	$<$122	&		&		\\
J04202583+2819237	&	IRAS 04173+2812	&	II	&	mid-M	&	172 $\pm$ 3	&		&	72 $\pm$ 6	&		&		\\
J04334171+1750402	&	 ... 	&	II	&	M4	&	17 $\pm$ 3	&		&	$<$28	&		&		\\
J04555605+3036209	&	XEST 26-062	&	II	&	M4	&	330 $\pm$ 4	&		&	639 $\pm$ 39	&		&		\\
J05075496+2500156	&	CIDA 12	&	II	&	M4	&	51 $\pm$ 3	&		&	44 $\pm$ 6	&		&		\\
J04324938+2253082	&	JH 112 B	&	II	&	M4.25	&	$<$318	&		&	(284 $\pm$ 26)	&		&		\\
J04385859+2336351	&	 ... 	&	II	&	M4.25	&	38 $\pm$ 3	&		&	76 $\pm$ 13	&		&		\\
J04190110+2819420	&	V410 X-ray 6	&	II	&	M4.5	&	445 $\pm$ 4	&		&	342 $\pm$ 36	&		&		\\
J04322415+2251083	&	 ... 	&	II	&	M4.5	&	$<$9	&		&	$<$77	&		&		\\
J04161210+2756385	&	 ... 	&	II	&	M4.75	&	201 $\pm$ 3	&		&	228 $\pm$ 8	&		&		\\
J04322210+1827426	&	MHO 6	&	II	&	M4.75	&	107 $\pm$ 2	&		&	188 $\pm$ 7	&		&		\\
J04334465+2615005	&	 ... 	&	II	&	M4.75	&	149 $\pm$ 2	&		&	178 $\pm$ 22	&		&		\\
J04422101+2520343	&	CIDA 7	&	II	&	M4.75	&	330 $\pm$ 2	&		&	342 $\pm$ 19	&		&		\\
J04554535+3019389	&	 ... 	&	II	&	M4.75	&	22 $\pm$ 3	&		&	$<$30	&		&		\\
J04141458+2827580	&	FN Tau	&	II	&	M5   	&	1755 $\pm$ 4	&		&	816 $\pm$ 16	&		&		\\
J04155799+2746175	&	 ... 	&	II	&	M5	&	25 $\pm$ 3	&		&	$<$48	&		&		\\
J04265732+2606284	&	KPNO 13	&	II	&	M5	&	28 $\pm$ 4	&		&	$<$182	&		&		\\
J04393364+2359212	&	 ... 	&	II	&	M5	&	70 $\pm$ 1	&		&	44 $\pm$ 15	&		&		\\
J04394488+2601527	&	ITG 15	&	II	&	M5	&	272 $\pm$ 3	&		&	114 $\pm$ 28	&		&		\\
J04432023+2940060	&	CIDA 14	&	II	&	M5	&	$<$13	&		&	$<$25	&		&		\\
J04193545+2827218	&	FR Tau	&	II	&	M5.25	&	46 $\pm$ 3	&		&	$<$33	&		&		\\
J04202555+2700355	&	 ... 	&	II	&	M5.25	&	107 $\pm$ 3	&		&	100 $\pm$ 15	&		&		\\
J04210795+2702204	&	CFHT 19	&	II	&	M5.25	&	3277 $\pm$ 8	&		&	3076 $\pm$ 61	&		&		\\
J04210934+2750368	&	 ... 	&	II	&	M5.25	&	$<$13	&		&	$<$47	&		&		\\
J04284263+2714039	&	 ... 	&	II	&	M5.25	&	20 $\pm$ 2	&		&	51 $\pm$ 12	&		&		\\
J04305171+2441475	&	ZZ Tau IRS	&	II	&	M5.25	&	2901 $\pm$ 5	&		&	2922 $\pm$ 26	&		&		\\
J04362151+2351165	&	 ... 	&	II	&	M5.25	&	$<$15	&		&	$<$64	&		&		\\
J04403979+2519061	&	 ... 	&	II	&	M5.25	&	72 $\pm$ 2	&		&	61 $\pm$ 20	&		&		\\
J04141760+2806096	&	CIDA 1	&	II	&	M5.5	&	266 $\pm$ 2	&		&	212 $\pm$ 21	&		&		\\
J04163911+2858491	&	 ... 	&	II	&	M5.5	&	$<$9	&		&	$<$55	&		&		\\
J04174955+2813318	&	KPNO 10	&	II	&	M5.5	&	160 $\pm$ 2	&		&	82 $\pm$ 26	&		&		\\
J04213459+2701388	&	 ... 	&	II	&	M5.5	&	37 $\pm$ 2	&		&	101 $\pm$ 19	&		&		\\
J04295950+2433078	&	CFHT 20	&	II	&	M5.5	&	128 $\pm$ 4	&		&	91 $\pm$ 23	&		&		\\
J04554801+3028050	&	 ... 	&	II	&	M5.6	&	11 $\pm$ 3	&		&	$<$63	&		&		\\
J04181710+2828419	&	V410 Anon 13	&	II	&	M5.75	&	35 $\pm$ 2	&		&	$<$113	&		&		\\
J04230607+2801194	&	 ... 	&	II	&	M6	&	41 $\pm$ 3	&		&	38 $\pm$ 9	&		&		\\
J04262939+2624137	&	KPNO 3	&	II	&	M6	&	23 $\pm$ 4	&		&	33 $\pm$ 12	&		&		\\
J04292165+2701259	&	IRAS 04263+2654	&	II	&	M6	&	329 $\pm$ 3	&		&	176 $\pm$ 14	&		&		\\
J04323028+1731303	&	GG Tau Ba+Bb	&	II	&	M6+M7.5	&	210 $\pm$ 20	&	 ... 	&	 ... 	&	2	&	a	\\
J04390163+2336029	&	 ... 	&	II	&	M6	&	15 $\pm$ 3	&		&	$<$24	&		&		\\
J04400067+2358211	&	 ... 	&	II	&	M6	&	55 $\pm$ 2	&		&	52 $\pm$ 5	&		&		\\
J04554969+3019400	&	 ... 	&	II	&	M6	&	$<$5	&		&	$<$17	&		&		\\
J04141188+2811535	&	 ... 	&	II	&	M6.25	&	17 $\pm$ 5	&		&	$<$293	&		&		\\
J04214631+2659296	&	CFHT 10	&	II	&	M6.25	&	$<$10	&		&	$<$84	&		&		\\
J04242646+2649503	&	CFHT 9	&	II	&	M6.25	&	10 $\pm$ 1	&		&	$<$6	&	1	&		\\
J04201611+2821325	&	 ... 	&	II	&	M6.5	&	$<$7	&		&	$<$26	&		&		\\
J04242090+2630511	&	 ... 	&	II	&	M6.5	&	$<$9	&		&	$<$58	&		&		\\
J04330945+2246487	&	CFHT 12	&	II	&	M6.5	&	2 $\pm$ 1	&		&	$<$8	&	1	&		\\
J04382134+2609137	&	GM Tau	&	II	&	M6.5	&	36 $\pm$ 2	&		&	$<$35	&		&		\\
J04411078+2555116	&	ITG 34	&	II	&	M6.5	&	21 $\pm$ 4	&		&	$<$38	&		&		\\
J04321606+1812464	&	MHO 5	&	II	&	M7	&	$<$11	&		&	$<$82	&		&		\\
J04353536+2408266	&	IRAS 04325+2402 C	&	II	&	M6-M8	&		&	(16906 $\pm$ 26)	&		&		&	b, c	\\
J04394748+2601407	&	CFHT 4	&	II	&	M7	&	109 $\pm$ 5	&		&	$<$150	&		&		\\
J04233539+2503026	&	FU Tau A	&	II	&	M7.25	&		&	$<$39	&	$<$247	&		&	b	\\
J04381486+2611399	&	 ... 	&	II	&	M7.25	&	95 $\pm$ 2	&		&	67 $\pm$ 24	&		&		\\
J04390396+2544264	&	CFHT 6	&	II	&	M7.25	&	23 $\pm$ 3	&		&	$<$56	&		&		\\
J04414825+2534304	&	 ... 	&	II	&	M7.75	&	37 $\pm$ 3	&		&	$<$122	&		&		\\
J04290068+2755033	&	 ... 	&	II	&	M8.25	&	$<$8	&		&	$<$36	&		&		\\
J04305718+2556394	&	KPNO 7	&	II	&	M8.25	&	4 $\pm$ 1	&		&	$<$7	&	1	&		\\
J04335245+2612548	&	 ... 	&	II	&	M8.5	&	$<$11	&		&	$<$47	&		&		\\
J04361030+2159364	&	 ... 	&	II	&	M8.5	&	$<$7	&		&	$<$22	&		&		\\
J04414489+2301513	&	 ... 	&	II	&	M8.5	&	11 $\pm$ 2	&		&	$<$18	&		&		\\
J04263055+2443558	&	 ... 	&	II	&	M8.75	&	$<$12	&		&	$<$51	&		&		\\
J04190126+2802487	&	KPNO 12	&	II	&	M9	&	$<$6	&		&	$<$88	&		&		\\
J04300724+2608207	&	KPNO 6	&	II	&	M9	&	2 $\pm$ 1	&		&	$<$5	&	1	&		\\
J04233573+2502596	&	FU Tau B	&	II	&	M9.25	&		&	$<$39	&	$<$247	&		&	b	\\
J04131414+2819108	&	LkCa 1	&	III	&	M4	&	$<$9	&	$<$9	&	$<$21	&	2	&		\\
J04184023+2824245	&	V410 X-ray 4	&	III	&	M4	&	$<$23	&		&	$<$88	&		&		\\
J04292071+2633406	&	J1-507	&	III	&	M4	&	$<$8	&	$<$8	&	$<$21	&	2	&		\\
J04292971+2616532	&	FW Tau A+B+C	&	III	&	M4	&	30 $\pm$ 4	&	33 $\pm$ 4	&	70 $\pm$ 40	&	2	&		\\
J04295422+1754041	&	 ... 	&	III	&	M4	&	89 $\pm$ 3	&		&	86 $\pm$ 8	&		&		\\
J04332621+2245293	&	XEST 17-036	&	III	&	M4	&	82 $\pm$ 4	&		&	181 $\pm$ 27	&		&		\\
J04464260+2459034	&	RXJ 04467+2459	&	III	&	M4	&	$<$10	&		&	$<$35	&		&		\\
J05061674+2446102	&	CIDA 10	&	III	&	M4	&	$<$6	&		&	$<$42	&		&		\\
J04355286+2250585	&	XEST 08-049	&	III	&	M4.25	&	$<$6	&		&	$<$109	&		&		\\
J04150515+2808462	&	CIDA 2	&	III	&	M4.5	&	$<$9	&	$<$9	&	$<$22	&	2	&		\\
J04163048+3037053	&	 ... 	&	III	&	M4.5	&	$<$11	&		&	$<$28	&		&		\\
J04203918+2717317	&	XEST 16-045	&	III	&	M4.5	&	$<$5	&		&	$<$45	&		&		\\
J04355209+2255039	&	XEST 08-047	&	III	&	M4.5	&	$<$11	&		&	$<$1657	&		&		\\
J04414565+2301580	&	 ... 	&	III	&	M4.5	&	$<$5	&		&	$<$18	&		&		\\
J04222404+2646258	&	XEST 11-087	&	III	&	M4.75	&	$<$6	&		&	$<$44	&		&		\\
J04354203+2252226	&	XEST 08-033	&	III	&	M4.75	&	$<$16	&		&	104 $\pm$ 28	&		&		\\
J04381630+2326402	&	 ... 	&	III	&	M4.75	&	$<$6	&		&	$<$26	&		&		\\
J04554757+3028077	&	 ... 	&	III	&	M4.75	&	$<$7	&		&	$<$63	&		&		\\
J04162725+2053091	&	 ... 	&	III	&	M5	&	$<$9	&		&	$<$39	&		&		\\
J04244506+2701447	&	J1-4423	&	III	&	M5	&	$<$5	&		&	$<$41	&		&		\\
J04311578+1820072	&	MHO 9	&	III	&	M5	&	$<$8	&		&	$<$76	&		&		\\
J04315844+2543299	&	J1-665	&	III	&	M5	&	$<$9	&	$<$9	&	$<$25	&	2	&		\\
J04410424+2557561	&	Haro 6-32	&	III	&	M5	&	$<$8	&		&	$<$48	&		&		\\
J04555636+3049374	&	 ... 	&	III	&	M5	&	$<$5	&		&	$<$35	&		&		\\
J04034997+2620382	&	XEST 06-006	&	III	&	M5.25	&	$<$5	&		&	$<$20	&		&		\\
J04144739+2803055	&	XEST 20-066	&	III	&	M5.25	&	$<$6	&		&	$<$67	&		&		\\
J04322627+1827521	&	MHO 7	&	III	&	M5.25	&	$<$8	&		&	$<$23	&		&		\\
J04344544+2308027	&	 ... 	&	III	&	M5.25	&	$<$10	&		&	$<$48	&		&		\\
J04554046+3039057	&	 ... 	&	III	&	M5.25	&	$<$10	&		&	$<$73	&		&		\\
J04555288+3006523	&	 ... 	&	III	&	M5.25	&	$<$7	&		&	$<$16	&		&		\\
J05064662+2104296	&	 ... 	&	III	&	M5.25	&	$<$6	&		&	$<$22	&		&		\\
J04183030+2743208	&	KPNO 11	&	III	&	M5.5	&	$<$6	&		&	$<$55	&		&		\\
J04190197+2822332	&	V410 X-ray 5a	&	III	&	M5.5	&	$<$5	&		&	$<$75	&		&		\\
J04205273+1746415	&	J2-157	&	III	&	M5.5	&	$<$5	&		&	$<$27	&		&		\\
J04312382+2410529	&	V927 Tau A+B	&	III	&	M5.5	&	$<$9	&	$<$9	&	$<$32	&	2	&		\\
J04214013+2814224	&	XEST 21-026	&	III	&	M5.75	&	$<$8	&		&	$<$31	&		&		\\
J04335252+2256269	&	XEST 17-059	&	III	&	M5.75	&	$<$12	&		&	$<$27	&		&		\\
J04354183+2234115	&	KPNO 8	&	III	&	M5.75	&	$<$3	&		&	$<$31	&		&		\\
J04400174+2556292	&	CFHT 17	&	III	&	M5.75	&	$<$11	&		&	$<$159	&		&		\\
J04330197+2421000	&	MHO 8	&	III	&	M6	&	$<$4	&		&	$<$48	&		&		\\
J04330781+2616066	&	KPNO 14	&	III	&	M6	&	$<$6	&		&	$<$79	&		&		\\
J04161885+2752155	&	 ... 	&	III	&	M6.25	&	$<$6	&		&	$<$50	&		&		\\
J04320329+2528078	&	 ... 	&	III	&	M6.25	&	$<$9	&		&	$<$33	&		&		\\
J04552333+3027366	&	 ... 	&	III	&	M6.25	&	$<$7	&		&	$<$34	&		&		\\
J04180796+2826036	&	V410 X-ray 3	&	III	&	M6.5	&	$<$7	&		&	$<$86	&		&		\\
J04321786+2422149	&	CFHT 7	&	III	&	M6.5	&		&	$<$39	&	$<$322	&		&	b	\\
J04385871+2323595	&	 ... 	&	III	&	M6.5	&	$<$4	&		&	$<$30	&		&		\\
J04185115+2814332	&	KPNO 2	&	III	&	M6.75	&	$<$7	&		&	$<$74	&		&		\\
J04270739+2215037	&	 ... 	&	III	&	M6.75	&	$<$8	&		&	$<$27	&		&		\\
J04350850+2311398	&	CFHT 11	&	III	&	M6.75	&	$<$7	&		&	$<$56	&		&		\\
J04152409+2910434	&	 ... 	&	III	&	M7	&	$<$7	&		&	$<$29	&		&		\\
J04312405+1800215	&	MHO 4	&	III	&	M7	&	$<$11	&		&	$<$38	&		&		\\
J04341527+2250309	&	CFHT 1	&	III	&	M7	&	$<$7	&		&	$<$67	&		&		\\
J04484189+1703374	&	 ... 	&	III	&	M7	&	$<$6	&		&	$<$25	&		&		\\
J04312669+2703188	&	CFHT 13	&	III	&	M7.25	&	$<$8	&		&	$<$54	&		&		\\
J04380083+2558572	&	ITG 2	&	III	&	M7.25	&	$<$6	&		&	$<$69	&		&		\\
J04294568+2630468	&	KPNO 5	&	III	&	M7.5	&	$<$7	&		&	$<$37	&		&		\\
J04325026+2422115	&	 ... 	&	III	&	M7.5	&	$<$10	&		&	$<$82	&		&		\\
J04361038+2259560	&	CFHT 2	&	III	&	M7.5	&	$<$7	&		&	$<$82	&		&		\\
J04390637+2334179	&	 ... 	&	III	&	M7.5	&	$<$5	&		&	$<$44	&		&		\\
J04221644+2549118	&	CFHT 14	&	III	&	M7.75	&	$<$5	&		&	$<$26	&		&		\\
J04311907+2335047	&	 ... 	&	III	&	M7.75	&	$<$7	&		&	$<$24	&		&		\\
J04322329+2403013	&	 ... 	&	III	&	M7.75	&	$<$9	&		&	$<$35	&		&		\\
J04363893+2258119	&	CFHT 3	&	III	&	M7.75	&	$<$5	&		&	$<$50	&		&		\\
J04221332+1934392	&	 ... 	&	III	&	M8	&	$<$5	&		&	$<$45	&		&		\\
J04274538+2357243	&	CFHT 15	&	III	&	M8.25	&	$<$6	&		&	$<$47	&		&		\\
J04325119+1730092	&	LH 0429+17	&	III	&	M8.25	&	$<$7	&		&	$<$33	&		&		\\
J04215450+2652315	&	 ... 	&	III	&	M8.5	&	$<$5	&		&	$<$35	&		&		\\
J04302365+2359129	&	CFHT 16	&	III	&	M8.5	&	$<$6	&		&	$<$51	&		&		\\
J04355143+2249119	&	KPNO 9	&	III	&	M8.5	&	$<$7	&		&	$<$102	&		&		\\
J04334291+2526470	&	 ... 	&	III	&	M8.75	&	$<$6	&		&	$<$32	&		&		\\
J04151471+2800096	&	KPNO 1	&	III	&	M9	&	$<$3	&		&	$<$41	&		&		\\
J04354526+2737130	&	 ... 	&	III	&	M9.25	&	$<$8	&		&	$<$30	&		&		\\
J04574903+3015195	&	 ... 	&	III	&	M9.25	&	$<$9	&		&	$<$15	&		&		\\
J04272799+2612052	&	KPNO 4	&	III	&	M9.5	&	$<$7	&		&	$<$21	&		&		\\
J04373705+2331080	&	 ... 	&	III	&	L0	&	$<$5	&		&	$<$27	&		&		\\

\end{longtable} 

\tablefoot{Fluxes in reported in parenthesis are unresolved from a companion of earlier spectral type (>M4), not included within the TBOSS sample. Upper limits are reported at a 3$\sigma$ level. The errors listed are the 1$\sigma$ statistical measurement error. Uncertainties in the absolute flux calibration  at 70~$\mu$m, 100~$\mu$m and 160~$\mu$m are 2.64\%, 2.75\% and 4.15\% respectively. \\
{\bf a.} GG Tau Ba+Bb is unresolved from the primary system (GG Tau Aa+Ab) at 100 $\mu$m and 160 $\mu$m \citep{Howard:2013}. \\
{\bf b.} {Fluxes measured from the level 2.5 processed maps observed under the program KPGT\_pandre\_1.} \\
{\bf c.} {IRAS 04325+2402 falls outside of the coverage region from the KPGT\_pandre\_1 processed map at 160 $\mu$m.} \\
}
\tablebib{(1) \citet{Harvey:2012}; (2) \citet{Howard:2013}.
}

\end{longtab}


\begin{table*}
\caption{Detection Rates}             
\label{table:diskFraction}      
\centering          
\begin{tabular}{l c c c c c c}   
\hline\hline \noalign{\smallskip}    
Class /  & & \multicolumn{2}{c}{70 $\mu$m} & \multicolumn{2}{c}{160 $\mu$m} \\                  
SpTy & Total & Observed & Detection Rate & Observed & Detection Rate \\ 
 \noalign{\smallskip} 
\hline \hline
\noalign{\smallskip} 

\noalign{\smallskip} 
{\bf Class I}  & 33  &  27 & 93 $_{-8}^{+2}$\% (25) & 22 & 73 $_{-11}^{+7}$\% (16)  \\   
 \noalign{\smallskip} 
\hline \hline
\noalign{\smallskip} 

K0-K7	& 18	&	13	&	100 $_{-12}^{+0}$\% (13) 	&	10	&	80 $_{-17}^{+7}$\% (8)  \\
\noalign{\smallskip} 
M0-M3	& 8	& 	8	&	75 $_{-20}^{+9}$\% (6)  	& 	6	&	33 $_{-13}^{+22}$\% (2)  \\
\noalign{\smallskip} 
M4-M6	& 4	 &	4 & 100 $_{-31}^{+0}$\% (4) & 4 & 100 $_{-31}^{+0}$\% (4) 	 \\
\noalign{\smallskip} 
M6.25-L0	& 3	&	2	&	100 $_{-46}^{+0}$\% (2)  	&	2	&	100 $_{-46}^{+0}$\% (2)  \\

\noalign{\smallskip}  \hline \noalign{\smallskip}                              
{\bf Class II} & 178 &  156 & 84 $_{-3}^{+3}$\% (131)   & 138 & 52 $_{-4}^{+4}$\% (72)  \\   
 \noalign{\smallskip} 
\hline \hline
\noalign{\smallskip} 

K0-K7	& 37	&	32	&	97 $_{-7}^{+1}$\% (31) 	&	26	&	81 $_{-10}^{+5}$\% (21)  \\
\noalign{\smallskip} 
M0-M3	& 72	& 	60	&	87 $_{-6}^{+3}$\% (52)  	& 	48	&	50 $_{-7}^{+7}$\% (24)  \\
\noalign{\smallskip} 
M4-M6	& 44	 &	43 & 84 $_{-7}^{+4}$\% (36) & 41 & 63 $_{-8}^{+7}$\% (26) 	 \\
\noalign{\smallskip} 
M6.25-L0	& 25	&	21	&	57 $_{-10}^{+11}$\% (12)  	&	23	&	4 $_{-1}^{+9}$\% (1)  \\

\noalign{\smallskip}  \hline \noalign{\smallskip}                              
{\bf Class III}  & 128 & 117 & 4 $_{-1}^{+3}$\% (5) & 114 & 4 $_{-1}^{+3}$\% (5) \\   
 \noalign{\smallskip} 
\hline \hline
\noalign{\smallskip} 

K0-K7	& 24	&	22	&	5 $_{-1}^{+9}$\% (1) 	&	20	&	5 $_{-2}^{+10}$\% (1)  \\
\noalign{\smallskip} 
M0-M3	& 26	& 	20	&	5 $_{-2}^{+10}$\% (1)  	& 	18	&	6 $_{-2}^{+10}$\% (1)  \\
\noalign{\smallskip} 
M4-M6	& 43	 &	41 & 7 $_{-2}^{+6}$\% (3) & 41 & 7 $_{-2}^{+6}$\% (3) 	 \\
\noalign{\smallskip} 
M6.25-L0	& 35	&	34	&	0 $_{-0}^{+5}$\% (0)  	&	35	&	0 $_{-0}^{+5}$\% (0)  \\

\noalign{\smallskip}  \hline
\end{tabular}
\tablefoot{The detection rates for the K0-K7, and M0-M3 bins have been tabulated from the {\it Spitzer} MIPS observations reported in \citet{Rebull:2010} and the {\it Herschel} PACS observations reported in \citet{Howard:2013}. The detection rates for the M4-M6, and M6.25-L0 bins have been tabulated from the {\it Herschel} PACS 70 $\mu$m and 160 $\mu$m observations reported here and are supplemented with results from the literature \citep{Harvey:2012, Howard:2013}.
}

\end{table*}

\begin{table*}
\caption{ Multiplicity of the different categories of disk type objects identified within TBOSS sample. }             
\label{table:Multiplicity}      
\centering          
\begin{tabular}{l c c  c l l l r}     
\hline\hline \noalign{\smallskip}
 2MASS & Other & Class & SpTy.  & Companion  & $\rho$ & PACS &Ref. \\
& name &  &  &  name & ($\arcsec$) & Blue Det. & \\
(1) & (2) & (3) & (4) &  (5) &  (6) & (7) & (8) \\
   
\noalign{\smallskip} \hline\hline \noalign{\smallskip}
\multicolumn{8}{c}{Extended disk objects} \\
\noalign{\smallskip} \hline \noalign{\smallskip} 
J04220007+1530248	&	IRAS 04191+1523 B	&	I	&	M6-M8	&	IRAS 04191+1523 A	&	6.09	&	y-contam$^{\star}$	&	5	\\
J04275730+2619183	&	IRAS 04248+2612	&	I	&	M4.5	&	IRAS 04248+2612 B	&	4.55	&	y-contam	&	5	\\
J04353536+2408266	&	IRAS 04325+2402 C	&	II	&	M6-M8	&	IRAS 04325+2402 AB$^{a}$	&	8.15	&	y-contam$^{\star}$	&	5	\\

\noalign{\smallskip} \hline\hline \noalign{\smallskip}
\multicolumn{8}{c}{Transition disk objects} \\
\noalign{\smallskip} \hline \noalign{\smallskip}
J04144730+2646264	&	FP Tau	&	II	&	M4	&	...	&	...	&	y	&	9	\\
J04161210+2756385	&	...	&	II	&	M4.75	&	...	&	...	&	y	&	6	\\
J04190110+2819420	&	V410 X-ray 6	&	II	&	M4.5	&	...	&	...	&	y	&	4	\\
J04202555+2700355	&	...	&	II	&	M5.25	&	...	&	...	&	y	&	9	\\
J04210934+2750368	&	...	&	II	&	M5.25	&	J04210934+2750368 B	&	0.77	&	n	&	4	\\
J04213459+2701388	&	...	&	II	&	M5.5	&	...	&	...	&	y	&	6	\\
J04284263+2714039	&	...	&	II	&	M5.25	&		&	0.62	&	y	&	6	\\
J04292971+2616532	&	FW Tau A+B+C	&	III	&	M4   	&	FW Tau A - B	&	0.08	&	y-contam	&	13	\\
	&		&		&		&	FW Tau A - C	&	2.36	&	y-contam	&	11	\\
J04322210+1827426	&	MHO 6	&	II	&	M4.75	&	...	&	...	&	y	&	9	\\
J04403979+2519061	&	...	&	II	&	M5.25	&	J04403979+2519061 B	&	0.04	&	y	&	6	\\

\noalign{\smallskip} \hline\hline \noalign{\smallskip}
\multicolumn{8}{c}{Truncated disk objects} \\
\noalign{\smallskip} \hline \noalign{\smallskip}
J04141188+2811535	&	...	&	II	&	M6.25	&	V773 Tau	&	26.21	&	y	&	8	\\
	&		&		&		&	V773 Tau A(ab)	&	SB$^{b}$	&	y-contam$^{\star}$	&	2	\\
	&		&		&		&	V773 Tau A - B	&	0.12	&	y-contam$^{\star}$	&	3	\\
	&		&		&		&	V773 Tau A - C	&	0.26	&	y-contam$^{\star}$	&	3	\\
J04174965+2829362	&	V410 X-ray 1	&	II	&	M4	&	...	&	...	&	y	&	9	\\
J04265732+2606284	&	KPNO 13	&	II	&	M5	&	...	&	...	&	y	&	9	\\
J04321606+1812464	&	MHO 5	&	II	&	M7	&	...	&	...	&	n	&	7	\\
J04432023+2940060	&	CIDA 14	&	II	&	M5	&	J04432023+2940060 B	&	8.4$^{c}$	&	n	&	1	\\

\noalign{\smallskip} \hline\hline \noalign{\smallskip}
\multicolumn{8}{c}{Mixed pair systems$^{d}$} \\
\noalign{\smallskip} \hline \noalign{\smallskip} 
J04414565+2301580	&	...	&	III	&	M4.5	&	J04414565+2301580 Aa - Ab	&	0.23	&	n	&	10	\\
	&		&		&		&	J04414489+2301513	&	12.81	&	y	&	8	\\
	&		&		&		&	J04414489+2301513 Ba - Bb	&	0.10	&	y	&	12	\\
J04554757+3028077	&	...	&	III	&	M4.75	&	J04554801+3028050	&	6.31	&	n	&	8	\\

\noalign{\smallskip} \hline
\end{tabular}

\tablefoot{Col. (1) 2MASS identifier. Col. (2) Other name. Col. (3) Spectral evolutionary class. Col. (4) Spectral type.  Col. (5) Companion name. Col. (6) Projected companion separations. Col. (7) Targets with PACS Blue channel (70 $\mu$m or 100 $\mu$m) detections are designated "y", and targets with upper limits are designated "n". "y-contam" indicates detected emission that is contaminated due the companion listed, and "y-contam$^{\star}$" indicates contamination due to the listed companion that is of a spectral type earlier than that of the TBOSS sample (<M4). Col. (8) Reference of companion separation. 
\tablefoottext{a}{IRAS 04325+2402 AB is a speculated binary \citep{Hartmann:1999}.}
\tablefoottext{b}{Spectroscopic binary (SB).}
\tablefoottext{c}{{\it HST} WFC3 archive data was inspected for candidate companions.}
\tablefoottext{d}{For the mixed pair systems, the primary component name and properties are listed in columns 1-4.}
}
\tablebib{(1) This work; (2) \citet{Boden:2007}; (3) \citet{Boden:2012}; (4) \citet{Cieza:2012}; (5) \citet{Duchene:2004}; (6) \citet{Konopacky:2007}; 
(7) \citet{Kraus:2006}; (8) \citet{Kraus:2009}; (9) \citet{Kraus:2009b}; (10) \citet{Kraus:2011}; (11) \citet{Kraus:2014}; (12) \citet{Todorov:2014}; (13) \citet{White:2001}.
}
\end{table*}


\begin{table*}
\caption{Transition disks within the TBOSS sample}             
\label{table:TD}      
\centering          
\begin{tabular}{l c c c c c}   
\hline\hline                             
 2MASS & Other name & Class & SpTy. & Transition disk type & Ref. \\ 
\hline

J04144730+2646264	&	FP Tau	&	2	&	M4	&	Homologously depleted	&	1	\\
J04190110+2819420	&	V410 X-ray 6	&	2	&	M4.5	&	Giant planet-forming	&	2	\\
J04161210+2756385	&	 ... 	&	2	&	M4.75	&	Candidate	&	3	\\
J04322210+1827426	&	MHO 6	&	2	&	M4.75	&	Candidate	&	3	\\
J04202555+2700355	&	 ... 	&	2	&	M5.25	&	Candidate, (PD/IH)$^{a}$	&	3, (1)	\\
J04210934+2750368	&	 ... 	&	2	&	M5.25	&	Grain-growth dominated	&	2	\\
J04284263+2714039	&	 ... 	&	2	&	M5.25	&	Grain-growth dominated	&	2	\\
J04403979+2519061	&	 ... 	&	2	&	M5.25	&	Debris disk	&	2	\\
J04213459+2701388	&	 ... 	&	2	&	M5.5	&	Candidate	&	3	\\
J04292971+2616532	&	FW Tau A+B+C	&	3	&	M4   	&	Photoevaporating/Circumbinary 	&	2	\\
J04295422+1754041	&		&	3	&	M4	&	Candidate	&	3	\\
J04332621+2245293	&	XEST 17-036	&	3	&	M4	&	Candidate	&	3	\\
J04354203+2252226	&	XEST 08-033	&	3	&	M4.75	&	Candidate	&	3	\\

\hline
\end{tabular}
\tablefoot{
\tablefoottext{a}{The evolutionary status for this target, previously reported in the literature was identified as inconclusive, being either a primordial disk (PD) object or a transition disk object with an inner hole (IH).}
}
\tablebib{(1) \citet{Currie:2011}; (2) \citet{Cieza:2012}; (3) This work.
}
\end{table*}

\begin{acknowledgements}
J. Bulger and J. Patience gratefully acknowledge support to Exeter from the STFC (ST/F0071241/1, ST/H002707/1), support from the Faculty of the European Space Astronomy Centre (ESAC), and support for this work was provided through an award issued by JPL/Caltech. C. Pinte acknowledges funding from the European Commission's 7$^\mathrm{th}$ Framework Program (contract PERG06-GA-2009-256513) and from Agence Nationale pour la Recherche (ANR) of France under contract ANR-2010-JCJC-0504-01. We thank Dr. G. Matthews, Dr. F. M{\'e}nard, Dr. C. Pinte and collaborators from the GASPS team for providing some of their results prior to publication. We thank the staff at the NASA {\it Herschel} Science Center for their informative responses to queries regarding PACS data. This research has made use of the SIMBAD database, operated at CDS, Strasbourg, France. This publication makes use of data products from the Two Micron All Sky Survey, which is a joint project of the University of Massachusetts and the Infrared Processing and Analysis Center/California Institute of Technology, funded by the National Aeronautics and Space Administration and the National Science Foundation. This publication makes use of data products from the Wide-field Infrared Survey Explorer, which is a joint project of the University of California, Los Angeles, and the Jet Propulsion Laboratory/California Institute of Technology, funded by the National Aeronautics and Space Administration. HIPE is a joint development by the {\it Herschel} Science Ground Segment Consortium, consisting of ESA, the NASA Herschel Science Center, and the HIFI, PACS and SPIRE consortia. This publication makes use of observations made with the NASA/ESA Hubble Space Telescope, obtained from the data archive at the Space Telescope Science Institute. STScI is operated by the Association of Universities for Research in Astronomy, Inc. under NASA contract NAS 5-26555.
\end{acknowledgements}

\bibliographystyle{aa}
\bibliography{TBOSS_PACS}


\Online
\begin{appendix} 
\section{Compilation of literature flux densities}
\label{Appendix:A}
The compilation of flux densities from optical-millimeter wavelength of the known 154 M4-L0 Taurus members is given in Tables~\ref{table:Optical}-\ref{Table:submm}. \\
Table~\ref{table:Optical} lists the {\it R}$_{C}$, {\it I}$_{C}$, and 2MASS {\it JHK$_{S}$} flux densities. \\
Table~\ref{Table:midIR} lists the {\it Spitzer} IRAC, MIPS-1, and {\it WISE} All-Sky Data release flux densities.\\
Table~\ref{Table:MIPS} lists the {\it Herschel} PACS, and {\it Spitzer} MIPS-2 flux densities. \\
Table~\ref{Table:submm} lists the submillimeter and millimeter (350 $\mu$m - 2.6 mm) flux densities.

\small
\begin{longtab}
\onecolumn
\begin{landscape}
\begin{longtable}{l c l l l l l l}
\caption{\label{table:Optical} Optical and near-IR photometry compilation of the known 154 M4-L0 Taurus members, ordered by target RA. }\\
\hline\hline
2MASS & Other name & $R_{\mathrm{C}}^{a}$ & $I_{\mathrm{C}}^{a}$ & {\it J} & {\it H}  & $K_{\mathrm{S}}$ & Reference	\\ 
 & & (mJy) &  (mJy)  & (mJy) & (mJy)  & (mJy)  & 	\\ 
\hline
\endfirsthead
\caption{continued.}\\
\hline\hline
2MASS & Other name & $R_{\mathrm{C}}^{a}$ & $I_{\mathrm{C}}^{a}$ & {\it J} & {\it H}  & $K_{\mathrm{S}}$ & Reference	\\ 
 & & (mJy) &  (mJy)  & (mJy) & (mJy)  & (mJy)  & 	\\ 
\hline
\endhead
\hline
\endfoot
J04034997+2620382	&	XEST 06-006	&		&		&	7.8 $\pm$ 0.2	&	8.8 $\pm$ 0.3	&	7.7 $\pm$ 0.2	&		\\
J04131414+2819108	&	LkCa 1	&	28.88 $\pm$ 21.28	&	95.18 $\pm$ 26.3	&	222 $\pm$ 6	&	291 $\pm$ 7	&	237 $\pm$ 6	&	$[$9$]$; $[$9$]$	\\
J04141188+2811535	&	 ... 	&		&		&	8.7 $\pm$ 0.2	&	12.0 $\pm$ 0.3	&	14.7 $\pm$ 0.4	&		\\
J04141458+2827580	&	FN Tau	&	3.95 $\pm$ 2.91	&		&	260 $\pm$ 7	&	349 $\pm$ 13	&	353 $\pm$ 9	&	$[$9$]$; $[$$]$	\\
J04141760+2806096	&	CIDA 1	&		&	4.73 $\pm$ 1.31	&	32.5 $\pm$ 0.8	&	60 $\pm$ 2	&	75 $\pm$ 2	&	$[$$]$; $[$7$]$	\\
J04144730+2646264	&	FP Tau	&	25.15 $\pm$ 18.53	&	74.91 $\pm$ 20.7	&	175 $\pm$ 5	&	219 $\pm$ 6	&	188 $\pm$ 5	&	$[$9$]$; $[$9$]$	\\
J04144739+2803055	&	XEST 20-066	&		&		&	77 $\pm$ 2	&	88 $\pm$ 2	&	72 $\pm$ 2	&		\\
J04150515+2808462	&	CIDA 2	&		&		&	144 $\pm$ 4	&	175 $\pm$ 5	&	154 $\pm$ 4	&		\\
J04151471+2800096	&	KPNO 1	&		&	0.12 $\pm$ 0.02	&	1.45 $\pm$ 0.06	&	2.05 $\pm$ 0.08	&	2.07 $\pm$ 0.08	&	$[$$]$; $[$2, 5$]$	\\
J04152409+2910434	&	 ... 	&		&		&	5.4 $\pm$ 0.2	&	7.2 $\pm$ 0.2	&	7.6 $\pm$ 0.2	&		\\
J04155799+2746175	&	 ... 	&		&		&	32.0 $\pm$ 0.8	&	41 $\pm$ 1	&	41 $\pm$ 1	&		\\
J04161210+2756385	&	 ... 	&		&		&	19.7 $\pm$ 0.5	&	37 $\pm$ 1	&	49 $\pm$ 1	&		\\
J04161885+2752155	&	 ... 	&		&		&	15.3 $\pm$ 0.4	&	19.9 $\pm$ 0.6	&	19.2 $\pm$ 0.5	&		\\
J04162725+2053091	&	 ... 	&		&		&	24.1 $\pm$ 0.6	&	26.3 $\pm$ 0.7	&	23.9 $\pm$ 0.6	&		\\
J04163048+3037053	&	 ... 	&		&		&	5.7 $\pm$ 0.2	&	6.6 $\pm$ 0.2	&	6.0 $\pm$ 0.2	&		\\
J04163911+2858491	&	 ... 	&		&		&	13.1 $\pm$ 0.3	&	18.9 $\pm$ 0.5	&	20.5 $\pm$ 0.5	&		\\
J04174955+2813318	&	KPNO 10	&		&	6.26 $\pm$ 1.73	&	28.0 $\pm$ 0.7	&	36 $\pm$ 1	&	32.1 $\pm$ 0.8	&	$[$$]$; $[$2, 4$]$	\\
J04174965+2829362	&	V410 X-ray 1	&	1.82 $\pm$ 1.34	&	7.84 $\pm$ 2.17	&	62 $\pm$ 2	&	131 $\pm$ 4	&	155 $\pm$ 4	&	$[$7$]$; $[$7, 5$]$	\\
J04180796+2826036	&	V410 X-ray 3	&	0.81 $\pm$ 0.6	&	6.15 $\pm$ 1.13	&	38 $\pm$ 1	&	48 $\pm$ 1	&	44 $\pm$ 1	&	$[$7$]$; $[$5, 7$]$	\\
J04181710+2828419	&	V410 Anon 13	&	0.06 $\pm$ 0.03	&	0.7 $\pm$ 0.13	&	10.5 $\pm$ 0.3	&	22.2 $\pm$ 0.6	&	27.6 $\pm$ 0.7	&	$[$6, 7$]$; $[$5, 6, 7$]$	\\
J04183030+2743208	&	KPNO 11	&		&	8.29 $\pm$ 2.29	&	28.1 $\pm$ 0.7	&	32 $\pm$ 1	&	26.4 $\pm$ 0.7	&	$[$$]$; $[$4$]$	\\
J04184023+2824245	&	V410 X-ray 4	&		&	0.03 $\pm$ 0.01	&	5.6 $\pm$ 0.2	&	42 $\pm$ 1	&	89 $\pm$ 2	&	$[$$]$; $[$7$]$	\\
J04185115+2814332	&	KPNO 2	&		&	0.64 $\pm$ 0.12	&	4.3 $\pm$ 0.1	&	5.2 $\pm$ 0.2	&	5.3 $\pm$ 0.1	&	$[$$]$; $[$2, 5$]$	\\
J04185813+2812234	&	IRAS 04158+2805	&	0.11 $\pm$ 0.08	&	0.68 $\pm$ 0.19	&	4.9 $\pm$ 0.2	&	11.8 $\pm$ 0.4	&	22.6 $\pm$ 0.6	&	$[$7$]$; $[$7, 5$]$	\\
J04190110+2819420	&	V410 X-ray 6	&	1.84 $\pm$ 1.35	&	10.34 $\pm$ 2.86	&	98 $\pm$ 3	&	148 $\pm$ 4	&	149 $\pm$ 4	&	$[$7$]$; $[$7, 5$]$	\\
J04190126+2802487	&	KPNO 12	&		&	0.03 $\pm$ 0.01	&	0.48 $\pm$ 0.04	&	0.66 $\pm$ 0.06	&	0.71 $\pm$ 0.06	&	$[$$]$; $[$4$]$	\\
J04190197+2822332	&	V410 X-ray 5a	&	0.22 $\pm$ 0.16	&	1.81 $\pm$ 0.5	&	25.5 $\pm$ 0.7	&	50 $\pm$ 1	&	58 $\pm$ 1	&	$[$7$]$; $[$7, 5$]$	\\
J04193545+2827218	&	FR Tau	&		&		&	66 $\pm$ 2	&	73 $\pm$ 2	&	68 $\pm$ 2	&		\\
J04194657+2712552	&	$[$GKH94$]$ 41	&		&		&	<0.1	&	2.50 $\pm$ 0.10	&	11.7 $\pm$ 0.3	&		\\
J04201611+2821325	&	 ... 	&		&		&	4.9 $\pm$ 0.2	&	5.8 $\pm$ 0.2	&	6.4 $\pm$ 0.2	&		\\
J04202555+2700355	&	 ... 	&		&		&	11.4 $\pm$ 0.3	&	16.1 $\pm$ 0.5	&	16.6 $\pm$ 0.5	&		\\
J04202583+2819237	&	IRAS 04173+2812	&		&		&	1.18 $\pm$ 0.05	&	4.4 $\pm$ 0.1	&	13.7 $\pm$ 0.4	&		\\
J04203918+2717317	&	XEST 16-045	&		&		&	101 $\pm$ 3	&	116 $\pm$ 4	&	100 $\pm$ 3	&		\\
J04205273+1746415	&	J2-157	&		&		&	35.9 $\pm$ 0.9	&	39 $\pm$ 1	&	32.6 $\pm$ 0.8	&		\\
J04210795+2702204	&	CFHT 19	&		&	0.62 $\pm$ 0.17	&	4.6 $\pm$ 0.1	&	15.3 $\pm$ 0.5	&	40 $\pm$ 1	&	$[$$]$; $[$2$]$	\\
J04210934+2750368	&	 ... 	&		&		&	51 $\pm$ 1	&	56 $\pm$ 2	&	48 $\pm$ 1	&		\\
J04213459+2701388	&	 ... 	&		&		&	27.8 $\pm$ 0.8	&	42 $\pm$ 1	&	45 $\pm$ 1	&		\\
J04214013+2814224	&	XEST 21-026	&		&		&	27.0 $\pm$ 0.8	&	29.9 $\pm$ 1.0	&	25.8 $\pm$ 0.7	&		\\
J04214631+2659296	&	CFHT 10	&		&	0.46 $\pm$ 0.13	&	4.7 $\pm$ 0.1	&	8.3 $\pm$ 0.3	&	9.3 $\pm$ 0.2	&	$[$$]$; $[$2$]$	\\
J04215450+2652315	&	 ... 	&		&		&	0.97 $\pm$ 0.04	&	1.62 $\pm$ 0.07	&	1.83 $\pm$ 0.08	&		\\
J04220007+1530248	&	IRAS 04191+1523 B	&		&		&	0.4 $\pm$ 0.2	&	12 $\pm$ 1	&	21 $\pm$ 2	&		\\
J04221332+1934392	&	 ... 	&		&		&	11.4 $\pm$ 0.3	&	15.4 $\pm$ 0.4	&	16.4 $\pm$ 0.4	&		\\
J04221644+2549118	&	CFHT 14	&		&	1.44 $\pm$ 0.4	&	9.5 $\pm$ 0.3	&	11.6 $\pm$ 0.4	&	11.2 $\pm$ 0.3	&	$[$$]$; $[$2$]$	\\
J04222404+2646258	&	XEST 11-087	&		&		&	59 $\pm$ 1	&	86 $\pm$ 3	&	82 $\pm$ 2	&		\\
J04230607+2801194	&	 ... 	&		&		&	20.2 $\pm$ 0.5	&	23.2 $\pm$ 0.7	&	22.0 $\pm$ 0.5	&		\\
J04233539+2503026	&	FU Tau A	&		&		&	78 $\pm$ 2	&	108 $\pm$ 3	&	124 $\pm$ 4	&		\\
J04233573+2502596	&	FU Tau B	&		&		&	1.5 $\pm$ 0.2	&	2.8 $\pm$ 0.2	&	3.1 $\pm$ 0.3	&		\\
J04242090+2630511	&	 ... 	&		&		&	6.4 $\pm$ 0.2	&	7.7 $\pm$ 0.2	&	7.1 $\pm$ 0.2	&		\\
J04242646+2649503	&	CFHT 9	&		&	1.85 $\pm$ 0.51	&	11.2 $\pm$ 0.3	&	13.6 $\pm$ 0.4	&	13.1 $\pm$ 0.3	&	$[$$]$; $[$2$]$	\\
J04244506+2701447	&	J1-4423	&		&		&	46 $\pm$ 1	&	53 $\pm$ 2	&	44 $\pm$ 1	&		\\
J04251550+2829275	&	 ... 	&		&		&	50 $\pm$ 1	&	53 $\pm$ 2	&	49 $\pm$ 1	&		\\
J04262939+2624137	&	KPNO 3	&		&	1.25 $\pm$ 0.35	&	7.5 $\pm$ 0.2	&	10.2 $\pm$ 0.3	&	9.8 $\pm$ 0.3	&	$[$$]$; $[$5$]$	\\
J04263055+2443558	&	 ... 	&		&		&	2.17 $\pm$ 0.06	&	2.7 $\pm$ 0.1	&	2.9 $\pm$ 0.1	&		\\
J04265732+2606284	&	KPNO 13	&		&	8.44 $\pm$ 2.33	&	49 $\pm$ 1	&	87 $\pm$ 2	&	98 $\pm$ 3	&	$[$$]$; $[$4$]$	\\
J04270739+2215037	&	 ... 	&		&		&	19.6 $\pm$ 0.5	&	22.4 $\pm$ 0.6	&	20.2 $\pm$ 0.5	&		\\
J04272799+2612052	&	KPNO 4	&	0.02 $\pm$ 0.01	&	0.08 $\pm$ 0.01	&	1.60 $\pm$ 0.06	&	2.5 $\pm$ 0.1	&	3.2 $\pm$ 0.1	&	$[$1$]$; $[$2, 5$]$	\\
J04274538+2357243	&	CFHT 15	&		&	0.17 $\pm$ 0.05	&	1.70 $\pm$ 0.07	&	2.07 $\pm$ 0.10	&	2.2 $\pm$ 0.1	&	$[$$]$; $[$2$]$	\\
J04275730+2619183	&	IRAS 04248+2612	&		&		&	36 $\pm$ 1	&	75 $\pm$ 2	&	85 $\pm$ 3	&		\\
 ... 	&	L1521F-IRS	&		&		&	<0.05	&	<0.4	&	0.74 $\pm$ 0.08	&		\\
J04284263+2714039	&	 ... 	&		&		&	22.9 $\pm$ 0.6	&	38.3 $\pm$ 0.9	&	44 $\pm$ 1	&		\\
J04290068+2755033	&	 ... 	&		&		&	3.9 $\pm$ 0.1	&	4.8 $\pm$ 0.2	&	4.8 $\pm$ 0.1	&		\\
J04292071+2633406	&	J1-507	&		&		&	188 $\pm$ 5	&	237 $\pm$ 7	&	203 $\pm$ 5	&		\\
J04292165+2701259	&	IRAS 04263+2654	&		&	3.59 $\pm$ 0.99	&	76 $\pm$ 2	&	163 $\pm$ 5	&	216 $\pm$ 5	&	$[$$]$; $[$2$]$	\\
J04292971+2616532	&	FW Tau A+B+C	&		&		&	117 $\pm$ 3	&	138 $\pm$ 4	&	117 $\pm$ 3	&		\\
J04294568+2630468	&	KPNO 5	&	0.07 $\pm$ 0.05	&	2.37 $\pm$ 0.65	&	14.0 $\pm$ 0.4	&	17.5 $\pm$ 0.5	&	16.2 $\pm$ 0.4	&	$[$1$]$; $[$5$]$	\\
J04295422+1754041	&	 ... 	&		&		&	13.9 $\pm$ 0.4	&	21.2 $\pm$ 0.5	&	26.0 $\pm$ 0.6	&		\\
J04295950+2433078	&	CFHT 20	&		&	2.33 $\pm$ 0.64	&	33.8 $\pm$ 1.0	&	63 $\pm$ 2	&	79 $\pm$ 2	&	$[$$]$; $[$2$]$	\\
J04300724+2608207	&	KPNO 6	&	0.02 $\pm$ 0.01	&	0.18 $\pm$ 0.03	&	1.60 $\pm$ 0.05	&	2.15 $\pm$ 0.09	&	2.23 $\pm$ 0.09	&	$[$1$]$; $[$2, 5$]$	\\
J04302365+2359129	&	CFHT 16	&		&	0.17 $\pm$ 0.05	&	1.65 $\pm$ 0.06	&	2.05 $\pm$ 0.09	&	2.2 $\pm$ 0.1	&	$[$$]$; $[$2$]$	\\
J04305171+2441475	&	ZZ Tau IRS	&		&	1.06 $\pm$ 0.29	&	11.6 $\pm$ 0.3	&	27.3 $\pm$ 0.8	&	50 $\pm$ 1	&	$[$$]$; $[$2$]$	\\
J04305718+2556394	&	KPNO 7	&		&	0.35 $\pm$ 0.1	&	2.48 $\pm$ 0.09	&	3.01 $\pm$ 0.10	&	3.3 $\pm$ 0.1	&	$[$$]$; $[$5$]$	\\
J04311578+1820072	&	MHO 9	&	3.02 $\pm$ 2.23	&	16.85 $\pm$ 4.66	&	52 $\pm$ 1	&	62 $\pm$ 2	&	50 $\pm$ 1	&	$[$7$]$; $[$7, 5$]$	\\
J04311907+2335047	&	 ... 	&		&		&	6.3 $\pm$ 0.2	&	8.4 $\pm$ 0.3	&	8.8 $\pm$ 0.2	&		\\
J04312382+2410529	&	V927 Tau A+B	&	14.21 $\pm$ 10.47	&	66.46 $\pm$ 18.36	&	205 $\pm$ 5	&	243 $\pm$ 8	&	207 $\pm$ 5	&	$[$9$]$; $[$9$]$	\\
J04312405+1800215	&	MHO 4	&	0.66 $\pm$ 0.48	&	4.77 $\pm$ 1.32	&	35 $\pm$ 1	&	44 $\pm$ 2	&	40 $\pm$ 1	&	$[$7$]$; $[$7, 5$]$	\\
J04312669+2703188	&	CFHT 13	&		&	0.18 $\pm$ 0.05	&	1.87 $\pm$ 0.07	&	2.7 $\pm$ 0.1	&	2.78 $\pm$ 0.10	&	$[$$]$; $[$2$]$	\\
J04313613+1813432	&	LkHa 358	&		&	0.97 $\pm$ 0.27	&	12.2 $\pm$ 0.4	&	44 $\pm$ 1	&	89 $\pm$ 2	&	$[$$]$; $[$7$]$	\\
J04315844+2543299	&	J1-665	&		&		&	93 $\pm$ 2	&	120 $\pm$ 3	&	100 $\pm$ 3	&		\\
J04320329+2528078	&	 ... 	&		&		&	32.8 $\pm$ 0.9	&	37 $\pm$ 1	&	34.3 $\pm$ 0.9	&		\\
J04321606+1812464	&	MHO 5	&	1.34 $\pm$ 0.99	&	8.29 $\pm$ 2.29	&	59 $\pm$ 2	&	71 $\pm$ 2	&	63 $\pm$ 2	&	$[$7$]$; $[$7, 5$]$	\\
J04321786+2422149	&	CFHT 7	&	0.69 $\pm$ 0.51	&	5.74 $\pm$ 1.58	&	38.6 $\pm$ 1.0	&	49 $\pm$ 1	&	47 $\pm$ 1	&	$[$1$]$; $[$2$]$	\\
J04322210+1827426	&	MHO 6	&	1.5 $\pm$ 1.11	&	7.7 $\pm$ 2.13	&	33 $\pm$ 1	&	40 $\pm$ 2	&	37 $\pm$ 1	&	$[$7$]$; $[$7$]$	\\
J04322329+2403013	&	 ... 	&		&		&	18.6 $\pm$ 0.5	&	21.6 $\pm$ 0.6	&	19.6 $\pm$ 0.5	&		\\
J04322415+2251083	&	 ... 	&		&		&	36 $\pm$ 1	&	48 $\pm$ 2	&	41 $\pm$ 1	&		\\
J04322627+1827521	&	MHO 7	&	2.89 $\pm$ 2.13	&	13.63 $\pm$ 3.77	&	57 $\pm$ 2	&	73 $\pm$ 3	&	57 $\pm$ 2	&	$[$7$]$; $[$7, 5$]$	\\
J04323028+1731303	&	GG Tau Ba+Bb	&		&		&	60 $\pm$ 2	&	71 $\pm$ 2	&	68 $\pm$ 2	&		\\
J04324938+2253082	&	JH 112 B	&		&		&	58 $\pm$ 2	&	122 $\pm$ 5	&	140 $\pm$ 4	&		\\
J04325026+2422115	&	 ... 	&		&	0.08 $\pm$ 0.02	&	4.2 $\pm$ 0.1	&	13.2 $\pm$ 0.4	&	20.5 $\pm$ 0.6	&	$[$$]$; $[$2$]$	\\
J04325119+1730092	&	LH 0429+17	&		&		&	2.12 $\pm$ 0.07	&	2.6 $\pm$ 0.1	&	2.52 $\pm$ 0.10	&		\\
J04330197+2421000	&	MHO 8	&	1.6 $\pm$ 1.18	&	9.01 $\pm$ 2.49	&	72 $\pm$ 2	&	90 $\pm$ 3	&	86 $\pm$ 2	&	$[$7$]$; $[$7, 5$]$	\\
J04330781+2616066	&	KPNO 14	&		&	2.6 $\pm$ 0.72	&	27.5 $\pm$ 0.7	&	49 $\pm$ 1	&	52 $\pm$ 1	&	$[$$]$; $[$4$]$	\\
J04330945+2246487	&	CFHT 12	&		&	0.8 $\pm$ 0.22	&	8.7 $\pm$ 0.2	&	14.2 $\pm$ 0.4	&	16.1 $\pm$ 0.4	&	$[$$]$; $[$2$]$	\\
J04332621+2245293	&	XEST 17-036	&		&		&	30.3 $\pm$ 0.8	&	65 $\pm$ 2	&	72 $\pm$ 2	&		\\
J04334171+1750402	&	 ... 	&		&		&	49 $\pm$ 1	&	57 $\pm$ 2	&	50 $\pm$ 1	&		\\
J04334291+2526470	&	 ... 	&		&		&	2.22 $\pm$ 0.08	&	3.0 $\pm$ 0.1	&	3.1 $\pm$ 0.1	&		\\
J04334465+2615005	&	 ... 	&		&		&	35.2 $\pm$ 0.9	&	72 $\pm$ 2	&	84 $\pm$ 2	&		\\
J04335245+2612548	&	 ... 	&		&		&	0.76 $\pm$ 0.05	&	1.49 $\pm$ 0.09	&	1.69 $\pm$ 0.08	&		\\
J04335252+2256269	&	XEST 17-059	&		&		&	128 $\pm$ 3	&	167 $\pm$ 5	&	151 $\pm$ 4	&		\\
J04341527+2250309	&	CFHT 1	&	0.02 $\pm$ 0.01	&	0.28 $\pm$ 0.08	&	5.1 $\pm$ 0.2	&	9.9 $\pm$ 0.3	&	12.1 $\pm$ 0.3	&	$[$6$]$; $[$6$]$	\\
J04344544+2308027	&	 ... 	&		&		&	12.0 $\pm$ 0.3	&	15.9 $\pm$ 0.4	&	13.9 $\pm$ 0.4	&		\\
J04350850+2311398	&	CFHT 11	&		&	2.85 $\pm$ 0.79	&	15.5 $\pm$ 0.4	&	17.2 $\pm$ 0.5	&	15.4 $\pm$ 0.4	&	$[$$]$; $[$2$]$	\\
J04353536+2408266	&	IRAS 04325+2402 C	&		&		&	0.6 $\pm$ 0.2	&	26 $\pm$ 4	&	78 $\pm$ 4	&		\\
J04354183+2234115	&	KPNO 8	&		&	2.33 $\pm$ 0.64	&	10.6 $\pm$ 0.3	&	11.6 $\pm$ 0.3	&	10.7 $\pm$ 0.3	&	$[$$]$; $[$5$]$	\\
J04354203+2252226	&	XEST 08-033	&		&		&	50 $\pm$ 1	&	71 $\pm$ 2	&	67 $\pm$ 2	&		\\
J04354526+2737130	&	 ... 	&		&		&	1.57 $\pm$ 0.07	&	2.06 $\pm$ 0.09	&	2.18 $\pm$ 0.08	&		\\
J04355143+2249119	&	KPNO 9	&		&	0.08 $\pm$ 0.02	&	1.03 $\pm$ 0.05	&	1.40 $\pm$ 0.06	&	1.41 $\pm$ 0.07	&	$[$$]$; $[$5$]$	\\
J04355209+2255039	&	XEST 08-047	&		&		&	48 $\pm$ 1	&	83 $\pm$ 2	&	79 $\pm$ 2	&		\\
J04355286+2250585	&	XEST 08-049	&		&		&	64 $\pm$ 2	&	92 $\pm$ 3	&	84 $\pm$ 2	&		\\
J04355760+2253574	&	 ... 	&		&		&	1.32 $\pm$ 0.05	&	2.14 $\pm$ 0.07	&	2.33 $\pm$ 0.08	&		\\
J04361030+2159364	&	 ... 	&		&		&	1.81 $\pm$ 0.07	&	2.32 $\pm$ 0.07	&	2.32 $\pm$ 0.09	&		\\
J04361038+2259560	&	CFHT 2	&	0.04 $\pm$ 0.03	&	0.48 $\pm$ 0.13	&	5.0 $\pm$ 0.1	&	8.0 $\pm$ 0.2	&	9.0 $\pm$ 0.2	&	$[$6$]$; $[$5, 6$]$	\\
J04362151+2351165	&	 ... 	&		&		&	8.6 $\pm$ 0.3	&	9.9 $\pm$ 0.3	&	8.4 $\pm$ 0.2	&		\\
J04363893+2258119	&	CFHT 3	&	0.03 $\pm$ 0.02	&	0.45 $\pm$ 0.12	&	5.2 $\pm$ 0.2	&	7.3 $\pm$ 0.2	&	7.5 $\pm$ 0.2	&	$[$6$]$; $[$5, 6$]$	\\
J04373705+2331080	&	 ... 	&		&		&	0.18 $\pm$ 0.03	&	0.36 $\pm$ 0.05	&	0.45 $\pm$ 0.06	&		\\
J04380083+2558572	&	ITG 2	&	0.03 $\pm$ 0.02	&	3.33 $\pm$ 0.92	&	39 $\pm$ 1	&	58 $\pm$ 1	&	61 $\pm$ 2	&	$[$1$]$; $[$3$]$	\\
J04381486+2611399	&	 ... 	&	0.02 $\pm$ 0.02	&	0.19 $\pm$ 0.05	&	1.36 $\pm$ 0.07	&	2.3 $\pm$ 0.1	&	4.3 $\pm$ 0.1	&	$[$1$]$; $[$3$]$	\\
J04381630+2326402	&	 ... 	&		&		&	30.2 $\pm$ 0.8	&	32.7 $\pm$ 0.9	&	27.5 $\pm$ 0.7	&		\\
J04382134+2609137	&	GM Tau	&	1.76 $\pm$ 1.29	&	2.46 $\pm$ 0.68	&	12.0 $\pm$ 0.3	&	23.8 $\pm$ 0.6	&	37.3 $\pm$ 0.9	&	$[$9$]$; $[$3$]$	\\
J04385859+2336351	&	 ... 	&		&		&	25.9 $\pm$ 0.7	&	29.2 $\pm$ 0.8	&	25.7 $\pm$ 0.6	&		\\
J04385871+2323595	&	 ... 	&		&		&	16.1 $\pm$ 0.4	&	17.2 $\pm$ 0.5	&	15.3 $\pm$ 0.4	&		\\
J04390163+2336029	&	 ... 	&		&		&	47 $\pm$ 1	&	59 $\pm$ 2	&	56 $\pm$ 1	&		\\
J04390396+2544264	&	CFHT 6	&	0.13 $\pm$ 0.1	&	1.59 $\pm$ 0.44	&	13.9 $\pm$ 0.4	&	18.8 $\pm$ 0.5	&	18.9 $\pm$ 0.5	&	$[$$]$; $[$2, 3$]$	\\
J04390637+2334179	&	 ... 	&		&		&	23.2 $\pm$ 0.6	&	25.0 $\pm$ 0.7	&	22.2 $\pm$ 0.6	&		\\
J04393364+2359212	&	 ... 	&		&		&	38 $\pm$ 1	&	49 $\pm$ 1	&	52 $\pm$ 1	&		\\
J04394488+2601527	&	ITG 15	&		&		&	88 $\pm$ 2	&	160 $\pm$ 4	&	175 $\pm$ 4	&		\\
J04394748+2601407	&	CFHT 4	&	0.1 $\pm$ 0.07	&	1.25 $\pm$ 0.34	&	21.6 $\pm$ 0.6	&	40 $\pm$ 1	&	49 $\pm$ 1	&	$[$6$]$; $[$3, 6$]$	\\
J04400067+2358211	&	 ... 	&		&		&	17.0 $\pm$ 0.5	&	18.3 $\pm$ 0.5	&	17.0 $\pm$ 0.5	&		\\
J04400174+2556292	&	CFHT 17	&		&	0.23 $\pm$ 0.06	&	8.2 $\pm$ 0.2	&	22.6 $\pm$ 0.6	&	33.0 $\pm$ 0.9	&	$[$$]$; $[$2$]$	\\
J04403979+2519061	&	 ... 	&		&	3.07 $\pm$ 0.85	&	29.8 $\pm$ 0.8	&	49 $\pm$ 2	&	54 $\pm$ 1	&	$[$$]$; $[$3$]$	\\
J04410424+2557561	&	Haro 6-32	&		&	15.94 $\pm$ 4.4	&	67 $\pm$ 2	&	80 $\pm$ 2	&	70 $\pm$ 2	&	$[$$]$; $[$3$]$	\\
J04411078+2555116	&	ITG 34	&	0.06 $\pm$ 0.04	&	0.75 $\pm$ 0.21	&	8.4 $\pm$ 0.2	&	14.5 $\pm$ 0.4	&	17.6 $\pm$ 0.5	&	$[$1$]$; $[$2, 3$]$	\\
J04414489+2301513	&	 ... 	&		&		&	2.72 $\pm$ 0.08	&	3.3 $\pm$ 0.1	&	3.6 $\pm$ 0.1	&		\\
J04414565+2301580	&	 ... 	&		&		&	81 $\pm$ 2	&	94 $\pm$ 3	&	76 $\pm$ 2	&		\\
J04414825+2534304	&	 ... 	&		&	0.39 $\pm$ 0.11	&	5.1 $\pm$ 0.2	&	7.8 $\pm$ 0.2	&	8.6 $\pm$ 0.2	&	$[$$]$; $[$3$]$	\\
J04422101+2520343	&	CIDA 7	&		&	9.69 $\pm$ 2.68	&	44 $\pm$ 1	&	60 $\pm$ 2	&	57 $\pm$ 1	&	$[$$]$; $[$3$]$	\\
J04432023+2940060	&	CIDA 14	&	6.2 $\pm$ 4.57	&	32.4 $\pm$ 8.95	&	110 $\pm$ 3	&	131 $\pm$ 4	&	115 $\pm$ 3	&	$[$8$]$; $[$8$]$	\\
J04442713+2512164	&	IRAS S04414+2506	&		&		&	21.1 $\pm$ 0.6	&	29.3 $\pm$ 0.7	&	33.1 $\pm$ 0.8	&		\\
J04464260+2459034	&	RXJ 04467+2459	&	2.68 $\pm$ 1.98	&	14.54 $\pm$ 4.02	&	50 $\pm$ 1	&	55 $\pm$ 1	&	49 $\pm$ 1	&	$[$8$]$; $[$8$]$	\\
J04484189+1703374	&	 ... 	&		&		&	6.2 $\pm$ 0.2	&	6.9 $\pm$ 0.2	&	6.7 $\pm$ 0.2	&		\\
J04520668+3047175	&	IRAS 04489+3042	&		&	0.18 $\pm$ 0.05	&	2.71 $\pm$ 0.09	&	15.9 $\pm$ 0.4	&	47 $\pm$ 1	&	$[$$]$; $[$3$]$	\\
J04552333+3027366	&	 ... 	&		&	1.54 $\pm$ 0.42	&	9.5 $\pm$ 0.3	&	11.4 $\pm$ 0.3	&	10.9 $\pm$ 0.3	&	$[$$]$; $[$3$]$	\\
J04554046+3039057	&	 ... 	&		&	3.33 $\pm$ 0.92	&	13.1 $\pm$ 0.3	&	15.2 $\pm$ 0.4	&	13.0 $\pm$ 0.3	&	$[$$]$; $[$3$]$	\\
J04554535+3019389	&	 ... 	&		&	13.51 $\pm$ 3.73	&	42 $\pm$ 1	&	50 $\pm$ 1	&	44 $\pm$ 1	&	$[$$]$; $[$3$]$	\\
J04554757+3028077	&	 ... 	&		&	15.94 $\pm$ 4.4	&	61 $\pm$ 2	&	77 $\pm$ 2	&	68 $\pm$ 2	&	$[$$]$; $[$3$]$	\\
J04554801+3028050	&	 ... 	&		&	2.12 $\pm$ 0.59	&	8.5 $\pm$ 0.3	&	9.4 $\pm$ 0.3	&	9.2 $\pm$ 0.3	&	$[$$]$; $[$3$]$	\\
J04554820+3030160	&	XEST 26-052	&		&		&	27.9 $\pm$ 0.7	&	33.2 $\pm$ 0.9	&	27.8 $\pm$ 0.7	&		\\
J04554969+3019400	&	 ... 	&		&	2.5 $\pm$ 0.69	&	12.0 $\pm$ 0.3	&	13.1 $\pm$ 0.4	&	12.0 $\pm$ 0.3	&	$[$$]$; $[$3$]$	\\
J04555288+3006523	&	 ... 	&		&	8.44 $\pm$ 2.33	&	35.0 $\pm$ 0.9	&	39.5 $\pm$ 1.0	&	33.9 $\pm$ 0.9	&	$[$$]$; $[$3$]$	\\
J04555605+3036209	&	XEST 26-062	&		&		&	103 $\pm$ 3	&	140 $\pm$ 4	&	131 $\pm$ 3	&		\\
J04555636+3049374	&	 ... 	&		&	95.18 $\pm$ 26.3	&	25.2 $\pm$ 0.6	&	28.3 $\pm$ 0.7	&	24.4 $\pm$ 0.6	&	$[$$]$; $[$3$]$	\\
J04574903+3015195	&	 ... 	&		&	95.18 $\pm$ 26.3	&	0.78 $\pm$ 0.06	&	0.91 $\pm$ 0.08	&	1.07 $\pm$ 0.08	&	$[$$]$; $[$3$]$	\\
J05061674+2446102	&	CIDA 10	&		&		&	77 $\pm$ 2	&	94 $\pm$ 3	&	79 $\pm$ 2	&		\\
J05064662+2104296	&	 ... 	&		&		&	24.1 $\pm$ 0.7	&	28.0 $\pm$ 0.8	&	23.9 $\pm$ 0.6	&		\\
J05075496+2500156	&	CIDA 12	&		&		&	43 $\pm$ 1	&	53 $\pm$ 1	&	46 $\pm$ 1	&		\\

\end{longtable}
\tablefoot{The optical $R_{\mathrm{C}}$ and  $I_{\mathrm{C}}$ flux densities are taken from those reported in the literature, and the median flux density is listed for those targets of which multiple observations with the same filter are reported in the literature. 2MASS $JHK_{\mathrm{S}}$ flux densities are taken from the 2MASS point source catalog \citep{Cutri:2003}. \\
\tablefoottext{a}{Uncertainties of 0.8~mag and 0.3~mag were adopted for the $R_{\mathrm{C}}$ and $I_{\mathrm{C}}$ photometry measurements, respectively.}
}
\tablebib{$R_{\mathrm{C}}$ references are listed in the first square parenthesis. $I_{\mathrm{C}}$ references are listed second square parenthesis. \\ 
(1) \citet{Guieu:2007}; (2) \citet{Guieu:2006}; (3) \citet{Luhman:2004}; (4) \citet{Luhman:2003};
(5) \citet{Briceno:2002}; (6) \citet{Martin:2001}; (7) \citet{Luhman:2000}; (8) \citet{Briceno:1999}; (9) \citet{Kenyon:1995}.
}
\end{landscape}
\twocolumn
\end{longtab}
\begin{longtab}
\onecolumn
\begin{landscape}
\begin{longtable}{l c l l l l l l l l}
\caption{\label{Table:midIR} {\it Spitzer} IRAC and {\it WISE} All-Sky Data release photometry compilation of the known 154 M4-L0 Taurus members, ordered by target RA.}\\
\hline\hline
2MASS & Other name & {\it WISE}-1  &  IRAC-1& IRAC-2	&	{\it WISE}-2	&	IRAC-3	&	IRAC-4 &	{\it WISE}-3	&	{\it WISE}-4	\\ 
& & (mJy) & (mJy) & (mJy) & (mJy) & (mJy) & (mJy) & (mJy) & (mJy) \\
\hline
\endfirsthead
\caption{continued.}\\
\hline\hline
2MASS & Other name & {\it WISE}-1  &  IRAC-1& IRAC-2	&	{\it WISE}-2	&	IRAC-3	&	IRAC-4 &	{\it WISE}-3	&	{\it WISE}-4	\\ 
& & (mJy) & (mJy) & (mJy) & (mJy) & (mJy) & (mJy) & (mJy) & (mJy) \\
\hline
\endhead
\hline
\endfoot
J04034997+2620382	&	XEST 06-006	&	4.4 $\pm$ 0.1	&		&		&	3.07 $\pm$ 0.08	&		&		&	0.4 $\pm$ 0.2	&	$<$2.1	\\
J04131414+2819108	&	LkCa 1	&	116 $\pm$ 3	&	112 $\pm$ 2	&	74 $\pm$ 1	&	71 $\pm$ 2	&	50 $\pm$ 1	&	27.8 $\pm$ 0.7	&	13.4 $\pm$ 0.4	&	2 $\pm$ 1	\\
J04141188+2811535	&	 ... 	&	13.6 $\pm$ 0.4	&	12.1 $\pm$ 0.2	&	13.5 $\pm$ 0.3	&	15.4 $\pm$ 0.4	&	13.3 $\pm$ 0.3	&	17.2 $\pm$ 0.4	&	20.3 $\pm$ 0.5	&	30 $\pm$ 1	\\
J04141458+2827580	&	FN Tau	&	316 $\pm$ 9	&	280 $\pm$ 5	&	275 $\pm$ 5	&	336 $\pm$ 8	&	267 $\pm$ 7	&	339 $\pm$ 8	&	716 $\pm$ 15	&	1422 $\pm$ 34	\\
J04141760+2806096	&	CIDA 1	&	75 $\pm$ 2	&	106 $\pm$ 2	&	117 $\pm$ 2	&	93 $\pm$ 2	&	116 $\pm$ 3	&	156 $\pm$ 4	&	197 $\pm$ 4	&	326 $\pm$ 7	\\
J04144730+2646264	&	FP Tau	&	138 $\pm$ 3	&	138 $\pm$ 3	&	112 $\pm$ 2	&	114 $\pm$ 3	&	94 $\pm$ 2	&	81 $\pm$ 2	&	81 $\pm$ 2	&	146 $\pm$ 4	\\
J04144739+2803055	&	XEST 20-066	&	38.5 $\pm$ 1.0	&	37.2 $\pm$ 0.7	&	26.8 $\pm$ 0.5	&	26.4 $\pm$ 0.6	&	17.5 $\pm$ 0.4	&	10.1 $\pm$ 0.2	&	5.6 $\pm$ 0.2	&	$<$3.9	\\
J04150515+2808462	&	CIDA 2	&	82 $\pm$ 2	&	83 $\pm$ 2	&	58 $\pm$ 1	&	55 $\pm$ 1	&	39 $\pm$ 1	&	21.5 $\pm$ 0.5	&	11.7 $\pm$ 0.3	&	$<$4.9	\\
J04151471+2800096	&	KPNO 1	&	1.25 $\pm$ 0.04	&	1.49 $\pm$ 0.03	&	1.08 $\pm$ 0.02	&	0.93 $\pm$ 0.03	&	0.67 $\pm$ 0.02	&	0.43 $\pm$ 0.02	&	$<$0.5	&	$<$1.9	\\
J04152409+2910434	&	 ... 	&	4.6 $\pm$ 0.1	&	5.2 $\pm$ 0.1	&	3.72 $\pm$ 0.07	&	3.48 $\pm$ 0.09	&	2.55 $\pm$ 0.06	&	1.56 $\pm$ 0.04	&	0.5 $\pm$ 0.1	&	$<$1.9	\\
J04155799+2746175	&	 ... 	&	35.9 $\pm$ 0.9	&	36.7 $\pm$ 0.9	&	32.1 $\pm$ 0.8	&	31.8 $\pm$ 0.7	&	27.8 $\pm$ 0.9	&	29.4 $\pm$ 0.9	&	31.8 $\pm$ 0.7	&	39 $\pm$ 2	\\
J04161210+2756385	&	 ... 	&	51 $\pm$ 1	&	51 $\pm$ 1	&	46 $\pm$ 1	&	49 $\pm$ 1	&	38 $\pm$ 1	&	31 $\pm$ 1	&	26.1 $\pm$ 0.6	&	51 $\pm$ 2	\\
J04161885+2752155	&	 ... 	&	11.9 $\pm$ 0.3	&	12 $\pm$ 0.3	&	9.1 $\pm$ 0.2	&	8.7 $\pm$ 0.2	&	6.4 $\pm$ 0.2	&	3.5 $\pm$ 0.1	&	2.0 $\pm$ 0.2	&	2 $\pm$ 1	\\
J04162725+2053091	&	 ... 	&	13.1 $\pm$ 0.3	&	13.8 $\pm$ 0.3	&	9.7 $\pm$ 0.2	&	9.1 $\pm$ 0.2	&	6.3 $\pm$ 0.2	&	3.8 $\pm$ 0.1	&	2.0 $\pm$ 0.2	&	$<$2.2	\\
J04163048+3037053	&	 ... 	&	3.24 $\pm$ 0.09	&	3.41 $\pm$ 0.08	&	2.44 $\pm$ 0.06	&	2.32 $\pm$ 0.06	&	1.57 $\pm$ 0.05	&	0.83 $\pm$ 0.03	&	0.7 $\pm$ 0.1	&	$<$3.1	\\
J04163911+2858491	&	 ... 	&	17.5 $\pm$ 0.5	&	18.4 $\pm$ 0.4	&	16.8 $\pm$ 0.4	&	15.8 $\pm$ 0.4	&	13.6 $\pm$ 0.4	&	11.2 $\pm$ 0.3	&	8.6 $\pm$ 0.3	&	9 $\pm$ 1	\\
J04174955+2813318	&	KPNO 10	&	8.8 $\pm$ 0.2	&	13.6 $\pm$ 0.2	&	13.6 $\pm$ 0.3	&	10.4 $\pm$ 0.3	&	13.3 $\pm$ 0.3	&	19.2 $\pm$ 0.5	&	19.7 $\pm$ 0.4	&	32 $\pm$ 2	\\
J04174965+2829362	&	V410 X-ray 1	&	133 $\pm$ 3	&	136 $\pm$ 3	&	143 $\pm$ 3	&	136 $\pm$ 3	&	136 $\pm$ 3	&	175 $\pm$ 4	&	196 $\pm$ 4	&	253 $\pm$ 8	\\
J04180796+2826036	&	V410 X-ray 3	&	25.7 $\pm$ 0.7	&	28.4 $\pm$ 0.5	&	20.1 $\pm$ 0.4	&	19.2 $\pm$ 0.5	&	13.1 $\pm$ 0.3	&	7.7 $\pm$ 0.2	&	4.3 $\pm$ 0.2	&	$<$2.8	\\
J04181710+2828419	&	V410 Anon 13	&	19.5 $\pm$ 0.5	&	22.9 $\pm$ 0.4	&	20.3 $\pm$ 0.4	&	20.1 $\pm$ 0.5	&	18.8 $\pm$ 0.5	&	19.5 $\pm$ 0.5	&	20.5 $\pm$ 0.5	&	31 $\pm$ 1	\\
J04183030+2743208	&	KPNO 11	&	14.9 $\pm$ 0.4	&	15.3 $\pm$ 0.4	&	10.9 $\pm$ 0.3	&	10.5 $\pm$ 0.3	&	6.9 $\pm$ 0.2	&	4.1 $\pm$ 0.1	&	2.3 $\pm$ 0.2	&	$<$3.3	\\
J04184023+2824245	&	V410 X-ray 4	&	64 $\pm$ 2	&	82 $\pm$ 2	&	65 $\pm$ 1	&	62 $\pm$ 1	&	45 $\pm$ 1	&	28.1 $\pm$ 0.7	&	13.3 $\pm$ 0.4	&	8 $\pm$ 1	\\
J04185115+2814332	&	KPNO 2	&	3.40 $\pm$ 0.09	&	3.7 $\pm$ 0.07	&	2.65 $\pm$ 0.05	&	2.54 $\pm$ 0.07	&	1.82 $\pm$ 0.05	&	1.14 $\pm$ 0.03	&	1.2 $\pm$ 0.2	&	3 $\pm$ 1	\\
J04185813+2812234	&	IRAS 04158+2805	&	51 $\pm$ 1	&	57 $\pm$ 1	&	75 $\pm$ 1	&	81 $\pm$ 2	&	91 $\pm$ 2	&	115 $\pm$ 3	&	163 $\pm$ 3	&	551 $\pm$ 13	\\
J04190110+2819420	&	V410 X-ray 6	&	79 $\pm$ 2	&	86 $\pm$ 2	&	66 $\pm$ 1	&	60 $\pm$ 1	&	45 $\pm$ 1	&	32.5 $\pm$ 0.8	&	40.8 $\pm$ 0.9	&	248 $\pm$ 7	\\
J04190126+2802487	&	KPNO 12	&	0.61 $\pm$ 0.02	&	0.75 $\pm$ 0.01	&	0.71 $\pm$ 0.02	&	0.70 $\pm$ 0.03	&	0.66 $\pm$ 0.02	&	0.59 $\pm$ 0.02	&	1.2 $\pm$ 0.2	&	$<$2.7	\\
J04190197+2822332	&	V410 X-ray 5a	&	35.2 $\pm$ 0.9	&	39 $\pm$ 0.8	&	27.5 $\pm$ 0.5	&	28.2 $\pm$ 0.6	&	18.6 $\pm$ 0.5	&	10.9 $\pm$ 0.3	&	6.1 $\pm$ 0.2	&	3 $\pm$ 1	\\
J04193545+2827218	&	FR Tau	&	48 $\pm$ 1	&	48.4 $\pm$ 0.9	&	52 $\pm$ 1	&	53 $\pm$ 1	&	56 $\pm$ 1	&	81 $\pm$ 2	&	90 $\pm$ 2	&	95 $\pm$ 4	\\
J04194657+2712552	&	$[$GKH94$]$ 41	&	15.0 $\pm$ 0.4	&	27.3 $\pm$ 0.5	&	34.2 $\pm$ 0.7	&	30.9 $\pm$ 0.7	&	41.8 $\pm$ 0.9	&	37.5 $\pm$ 0.9	&	46 $\pm$ 1	&	196 $\pm$ 5	\\
J04201611+2821325	&	 ... 	&	4.9 $\pm$ 0.1	&	5.3 $\pm$ 0.1	&	4.9 $\pm$ 0.1	&	4.5 $\pm$ 0.1	&	4.1 $\pm$ 0.1	&	3.9 $\pm$ 0.1	&	4.1 $\pm$ 0.2	&	$<$3.4	\\
J04202555+2700355	&	 ... 	&	10.4 $\pm$ 0.3	&	11.8 $\pm$ 0.2	&	9.5 $\pm$ 0.2	&	9.0 $\pm$ 0.2	&	7.7 $\pm$ 0.2	&	8.2 $\pm$ 0.2	&	10.5 $\pm$ 0.3	&	27 $\pm$ 2	\\
J04202583+2819237	&	IRAS 04173+2812	&	51 $\pm$ 1	&	60 $\pm$ 1	&	78 $\pm$ 2	&	76 $\pm$ 2	&	79 $\pm$ 2	&	91 $\pm$ 3	&	119 $\pm$ 2	&	205 $\pm$ 6	\\
J04203918+2717317	&	XEST 16-045	&	56 $\pm$ 1	&	50 $\pm$ 1	&	30.1 $\pm$ 0.7	&	35.5 $\pm$ 0.8	&	21.5 $\pm$ 0.5	&	12.3 $\pm$ 0.4	&	6.8 $\pm$ 0.3	&	$<$4.2	\\
J04205273+1746415	&	J2-157	&	18.2 $\pm$ 0.5	&	18.7 $\pm$ 0.4	&	13.0 $\pm$ 0.3	&	12.4 $\pm$ 0.3	&	8.6 $\pm$ 0.3	&	5.0 $\pm$ 0.2	&	3.6 $\pm$ 0.2	&	3.3 $\pm$ 0.9	\\
J04210795+2702204	&	CFHT 19	&	194 $\pm$ 5	&	205 $\pm$ 4	&	302 $\pm$ 6	&	309 $\pm$ 8	&	343 $\pm$ 8	&	479 $\pm$ 11	&	624 $\pm$ 12	&	1539 $\pm$ 35	\\
J04210934+2750368	&	 ... 	&	30.9 $\pm$ 0.8	&	28.1 $\pm$ 0.7	&	23.3 $\pm$ 0.5	&	24.4 $\pm$ 0.6	&	16.9 $\pm$ 0.5	&	13.6 $\pm$ 0.4	&	11.9 $\pm$ 0.4	&	8 $\pm$ 2	\\
J04213459+2701388	&	 ... 	&	29.4 $\pm$ 0.8	&	30.5 $\pm$ 0.7	&	25.7 $\pm$ 0.6	&	24.8 $\pm$ 0.6	&	21.1 $\pm$ 0.7	&	16.6 $\pm$ 0.5	&	13.6 $\pm$ 0.4	&	9 $\pm$ 2	\\
J04214013+2814224	&	XEST 21-026	&	14.6 $\pm$ 0.4	&	15.7 $\pm$ 0.4	&	10.5 $\pm$ 0.2	&	9.9 $\pm$ 0.3	&	7.0 $\pm$ 0.2	&	4.0 $\pm$ 0.1	&	2.2 $\pm$ 0.3	&	$<$7.6	\\
J04214631+2659296	&	CFHT 10	&	6.2 $\pm$ 0.2	&	6.9 $\pm$ 0.2	&	5.7 $\pm$ 0.1	&	5.5 $\pm$ 0.1	&	4.5 $\pm$ 0.1	&	4.1 $\pm$ 0.1	&	5.9 $\pm$ 0.3	&	7 $\pm$ 2	\\
J04215450+2652315	&	 ... 	&	1.21 $\pm$ 0.04	&	1.52 $\pm$ 0.03	&	1.07 $\pm$ 0.02	&	1.08 $\pm$ 0.04	&	0.75 $\pm$ 0.02	&	0.45 $\pm$ 0.02	&	0.6 $\pm$ 0.2	&	$<$5.3	\\
J04220007+1530248	&	IRAS 04191+1523 B	&	3.7 $\pm$ 0.2	&	3.75 $\pm$ 0.09	&	6.0 $\pm$ 0.1	&	8.4 $\pm$ 0.5	&	6.0 $\pm$ 0.2	&	7.8 $\pm$ 0.3	&	43 $\pm$ 2	&	546 $\pm$ 24	\\
J04221332+1934392	&	 ... 	&	9.7 $\pm$ 0.3	&	11 $\pm$ 0.3	&	7.9 $\pm$ 0.2	&	7.6 $\pm$ 0.2	&	5.4 $\pm$ 0.2	&	3.8 $\pm$ 0.1	&	1.9 $\pm$ 0.2	&	5 $\pm$ 1	\\
J04221644+2549118	&	CFHT 14	&	6.8 $\pm$ 0.2	&	7.5 $\pm$ 0.1	&	5.5 $\pm$ 0.1	&	5.1 $\pm$ 0.1	&	3.67 $\pm$ 0.09	&	2.07 $\pm$ 0.05	&	0.9 $\pm$ 0.2	&	$<$3.2	\\
J04222404+2646258	&	XEST 11-087	&	44 $\pm$ 1	&	47 $\pm$ 1	&	32.4 $\pm$ 0.8	&	31.5 $\pm$ 0.7	&	21.9 $\pm$ 0.7	&	12.0 $\pm$ 0.4	&	6.7 $\pm$ 0.3	&	$<$5.4	\\
J04230607+2801194	&	 ... 	&	14.8 $\pm$ 0.4	&	16.3 $\pm$ 0.3	&	14.7 $\pm$ 0.3	&	13.6 $\pm$ 0.3	&	12.0 $\pm$ 0.3	&	11.8 $\pm$ 0.4	&	11.9 $\pm$ 0.6	&	22 $\pm$ 4	\\
J04233539+2503026	&	FU Tau A	&	111 $\pm$ 3	&	130 $\pm$ 3	&	139 $\pm$ 3	&	125 $\pm$ 3	&	134 $\pm$ 4	&	152 $\pm$ 4	&	117 $\pm$ 2	&	121 $\pm$ 4	\\
J04233573+2502596	&	FU Tau B	&	111 $\pm$ 3	&	2.7 $\pm$ 0.3	&	3.1 $\pm$ 0.2	&	125 $\pm$ 3	&	3.0 $\pm$ 0.3	&	3.0 $\pm$ 0.2	&	117 $\pm$ 2	&	121 $\pm$ 4	\\
J04242090+2630511	&	 ... 	&	4.9 $\pm$ 0.1	&	5.4 $\pm$ 0.1	&	4.9 $\pm$ 0.1	&	4.4 $\pm$ 0.1	&	4.6 $\pm$ 0.1	&	4.6 $\pm$ 0.1	&		&		\\
J04242646+2649503	&	CFHT 9	&	8.7 $\pm$ 0.2	&	10.1 $\pm$ 0.2	&	8.6 $\pm$ 0.2	&	7.5 $\pm$ 0.2	&	7.4 $\pm$ 0.2	&	7.5 $\pm$ 0.2	&		&		\\
J04244506+2701447	&	J1-4423	&	23.5 $\pm$ 0.6	&	23.8 $\pm$ 0.5	&	16.0 $\pm$ 0.3	&	15.4 $\pm$ 0.4	&	11.0 $\pm$ 0.3	&	6.2 $\pm$ 0.1	&	3.9 $\pm$ 0.5	&	$<$7.0	\\
J04251550+2829275	&	 ... 	&	29.2 $\pm$ 0.8	&	31 $\pm$ 2	&	22 $\pm$ 1	&	20.8 $\pm$ 0.5	&	15.0 $\pm$ 0.7	&	8.5 $\pm$ 0.4	&	4.4 $\pm$ 0.3	&	$<$3.8	\\
J04262939+2624137	&	KPNO 3	&	7.4 $\pm$ 0.2	&	7.9 $\pm$ 0.2	&	7.5 $\pm$ 0.1	&	7.2 $\pm$ 0.2	&	7.4 $\pm$ 0.2	&	8.7 $\pm$ 0.2	&	9.8 $\pm$ 0.4	&	12 $\pm$ 2	\\
J04263055+2443558	&	 ... 	&	2.20 $\pm$ 0.06	&	2.63 $\pm$ 0.05	&	2.42 $\pm$ 0.05	&	2.37 $\pm$ 0.09	&	2.24 $\pm$ 0.05	&	2.44 $\pm$ 0.06	&	2.4 $\pm$ 0.6	&	$<$6.0	\\
J04265732+2606284	&	KPNO 13	&	83 $\pm$ 2	&	94 $\pm$ 2	&	88 $\pm$ 2	&	85 $\pm$ 2	&	84 $\pm$ 2	&	85 $\pm$ 2	&	72 $\pm$ 2	&	95 $\pm$ 3	\\
J04270739+2215037	&	 ... 	&	11.6 $\pm$ 0.3	&	12.3 $\pm$ 0.3	&	8.8 $\pm$ 0.2	&	8.4 $\pm$ 0.2	&	5.8 $\pm$ 0.2	&	3.3 $\pm$ 0.1	&	1.9 $\pm$ 0.3	&	$<$2.7	\\
J04272799+2612052	&	KPNO 4	&	2.35 $\pm$ 0.06	&	2.68 $\pm$ 0.05	&	2.09 $\pm$ 0.04	&	2.04 $\pm$ 0.06	&	1.47 $\pm$ 0.04	&	0.92 $\pm$ 0.03	&	1.0 $\pm$ 0.2	&	$<$2.8	\\
J04274538+2357243	&	CFHT 15	&	1.30 $\pm$ 0.04	&	1.47 $\pm$ 0.04	&	1.08 $\pm$ 0.03	&	1.07 $\pm$ 0.05	&	0.74 $\pm$ 0.03	&	0.41 $\pm$ 0.03	&	$<$0.9	&	$<$5.2	\\
J04275730+2619183	&	IRAS 04248+2612	&	36.7 $\pm$ 0.9	&	64 $\pm$ 3	&	71 $\pm$ 3	&	55 $\pm$ 1	&	68 $\pm$ 2	&	105 $\pm$ 3	&	205 $\pm$ 4	&	1005 $\pm$ 20	\\
 ... 	&	L1521F-IRS	&	0.21 $\pm$ 0.01	&	0.66 $\pm$ 0.04	&	0.98 $\pm$ 0.05	&	0.79 $\pm$ 0.03	&	1.0 $\pm$ 0.1	&	0.91 $\pm$ 0.08	&	$<$0.8	&	21 $\pm$ 1	\\
J04284263+2714039	&	 ... 	&	37.8 $\pm$ 0.9	&	37.4 $\pm$ 0.9	&	29.8 $\pm$ 0.7	&	30.7 $\pm$ 0.7	&	24.2 $\pm$ 0.8	&	19.1 $\pm$ 0.6	&	18.4 $\pm$ 0.5	&	29 $\pm$ 2	\\
J04290068+2755033	&	 ... 	&	3.14 $\pm$ 0.08	&	3.54 $\pm$ 0.08	&	2.98 $\pm$ 0.07	&	2.92 $\pm$ 0.08	&	2.68 $\pm$ 0.08	&	2.68 $\pm$ 0.08	&	3.3 $\pm$ 0.2	&	$<$4.6	\\
J04292071+2633406	&	J1-507	&	109 $\pm$ 3	&	109 $\pm$ 2	&	72 $\pm$ 1	&	70 $\pm$ 2	&	48 $\pm$ 1	&	27.3 $\pm$ 0.7	&	14.6 $\pm$ 0.4	&	6 $\pm$ 1	\\
J04292165+2701259	&	IRAS 04263+2654	&	153 $\pm$ 4	&	166 $\pm$ 4	&	158 $\pm$ 4	&	154 $\pm$ 4	&	142 $\pm$ 4	&	143 $\pm$ 4	&	153 $\pm$ 3	&	339 $\pm$ 10	\\
J04292971+2616532	&	FW Tau A+B+C	&	65 $\pm$ 2	&	67 $\pm$ 1	&	48.6 $\pm$ 0.9	&	46 $\pm$ 1	&	31.7 $\pm$ 0.7	&	18.4 $\pm$ 0.4	&	9.6 $\pm$ 0.3	&	5 $\pm$ 1	\\
J04294568+2630468	&	KPNO 5	&	10.0 $\pm$ 0.3	&	10.7 $\pm$ 0.2	&	7.7 $\pm$ 0.1	&	7.3 $\pm$ 0.2	&	5.3 $\pm$ 0.1	&	2.98 $\pm$ 0.07	&	2.1 $\pm$ 0.2	&	$<$2.8	\\
J04295422+1754041	&	 ... 	&	28.3 $\pm$ 0.7	&		&		&	29.1 $\pm$ 0.7	&		&		&	18.5 $\pm$ 0.5	&	25 $\pm$ 2	\\
J04295950+2433078	&	CFHT 20	&	73 $\pm$ 2	&	69 $\pm$ 2	&	68 $\pm$ 2	&	72 $\pm$ 2	&	55 $\pm$ 1	&	49 $\pm$ 2	&	57 $\pm$ 1	&	93 $\pm$ 3	\\
J04300724+2608207	&	KPNO 6	&	1.48 $\pm$ 0.04	&	1.64 $\pm$ 0.03	&	1.43 $\pm$ 0.03	&	1.36 $\pm$ 0.05	&	1.25 $\pm$ 0.04	&	1.38 $\pm$ 0.04	&	0.7 $\pm$ 0.3	&	$<$3.3	\\
J04302365+2359129	&	CFHT 16	&	1.24 $\pm$ 0.04	&	1.46 $\pm$ 0.03	&	1.02 $\pm$ 0.02	&	0.95 $\pm$ 0.04	&	0.71 $\pm$ 0.03	&	0.40 $\pm$ 0.02	&	0.4 $\pm$ 0.2	&	3 $\pm$ 1	\\
J04305171+2441475	&	ZZ Tau IRS	&	78 $\pm$ 2	&	161 $\pm$ 3	&	209 $\pm$ 4	&	238 $\pm$ 6	&	248 $\pm$ 6	&	330 $\pm$ 8	&	529 $\pm$ 10	&	1436 $\pm$ 30	\\
J04305718+2556394	&	KPNO 7	&	2.27 $\pm$ 0.06	&	2.59 $\pm$ 0.05	&	2.27 $\pm$ 0.04	&	2.20 $\pm$ 0.07	&	1.88 $\pm$ 0.05	&	2.04 $\pm$ 0.05	&	1.7 $\pm$ 0.2	&	3 $\pm$ 1	\\
J04311578+1820072	&	MHO 9	&	26.7 $\pm$ 0.7	&	27.1 $\pm$ 0.6	&	18.9 $\pm$ 0.4	&	18.1 $\pm$ 0.5	&	13.1 $\pm$ 0.4	&	7.1 $\pm$ 0.2	&	4.9 $\pm$ 0.3	&	$<$5.6	\\
J04311907+2335047	&	 ... 	&	5.5 $\pm$ 0.2	&	6.1 $\pm$ 0.1	&	4.4 $\pm$ 0.1	&	4.3 $\pm$ 0.1	&	2.81 $\pm$ 0.09	&	1.68 $\pm$ 0.07	&	1.1 $\pm$ 0.2	&	$<$3.6	\\
J04312382+2410529	&	V927 Tau A+B	&	109 $\pm$ 3	&	110 $\pm$ 2	&	77 $\pm$ 1	&	72 $\pm$ 2	&	51 $\pm$ 1	&	28.7 $\pm$ 0.7	&	15.4 $\pm$ 0.4	&	5 $\pm$ 1	\\
J04312405+1800215	&	MHO 4	&	22.4 $\pm$ 0.6	&	25.2 $\pm$ 0.6	&	18.1 $\pm$ 0.4	&	16.4 $\pm$ 0.4	&	12.2 $\pm$ 0.4	&	6.9 $\pm$ 0.2	&	4.3 $\pm$ 0.3	&	$<$3.7	\\
J04312669+2703188	&	CFHT 13	&	1.86 $\pm$ 0.05	&	2.01 $\pm$ 0.04	&	1.51 $\pm$ 0.04	&	1.41 $\pm$ 0.05	&	0.95 $\pm$ 0.03	&	0.55 $\pm$ 0.02	&	$<$0.5	&	$<$3.4	\\
J04313613+1813432	&	LkHa 358	&	159 $\pm$ 4	&	117 $\pm$ 3	&	147 $\pm$ 3	&	234 $\pm$ 6	&	163 $\pm$ 5	&	174 $\pm$ 5	&	302 $\pm$ 6	&	1146 $\pm$ 33	\\
J04315844+2543299	&	J1-665	&	51 $\pm$ 1	&	53 $\pm$ 1	&	36.0 $\pm$ 0.7	&	34.4 $\pm$ 0.9	&	23.5 $\pm$ 0.6	&	13.1 $\pm$ 0.3	&	6.0 $\pm$ 0.4	&	$<$5.2	\\
J04320329+2528078	&	 ... 	&	20.1 $\pm$ 0.5	&	21.9 $\pm$ 0.5	&	15.8 $\pm$ 0.4	&	14.7 $\pm$ 0.3	&	10.6 $\pm$ 0.3	&	6.0 $\pm$ 0.2	&	2.3 $\pm$ 0.3	&	$<$4.0	\\
J04321606+1812464	&	MHO 5	&	45 $\pm$ 1	&	54 $\pm$ 1	&	47 $\pm$ 1	&	42 $\pm$ 1	&	43 $\pm$ 1	&	47 $\pm$ 1	&	50 $\pm$ 1	&	46 $\pm$ 3	\\
J04321786+2422149	&	CFHT 7	&	27.4 $\pm$ 0.7	&	29.8 $\pm$ 0.6	&	21.6 $\pm$ 0.4	&	20.7 $\pm$ 0.5	&	14.7 $\pm$ 0.4	&	8.4 $\pm$ 0.2	&	4.6 $\pm$ 0.3	&	$<$5.6	\\
J04322210+1827426	&	MHO 6	&	25.7 $\pm$ 0.7	&	26.7 $\pm$ 0.5	&	22.8 $\pm$ 0.5	&	21.9 $\pm$ 0.5	&	17.1 $\pm$ 0.4	&	15.4 $\pm$ 0.5	&	12.4 $\pm$ 0.4	&	19 $\pm$ 2	\\
J04322329+2403013	&	 ... 	&	12.0 $\pm$ 0.3	&	12 $\pm$ 0.3	&	8.8 $\pm$ 0.2	&	8.5 $\pm$ 0.2	&	5.3 $\pm$ 0.2	&	3.4 $\pm$ 0.1	&	1.7 $\pm$ 0.3	&	$<$3.8	\\
J04322415+2251083	&	 ... 	&	26.8 $\pm$ 0.7	&	28.1 $\pm$ 0.6	&	25.0 $\pm$ 0.6	&	24.4 $\pm$ 0.6	&	21.7 $\pm$ 0.5	&	26.6 $\pm$ 0.8	&	27.3 $\pm$ 0.7	&	28 $\pm$ 2	\\
J04322627+1827521	&	MHO 7	&	30.5 $\pm$ 0.8	&	31.7 $\pm$ 0.6	&	21.8 $\pm$ 0.5	&	21.1 $\pm$ 0.5	&	14.3 $\pm$ 0.4	&	8.2 $\pm$ 0.3	&	4.7 $\pm$ 0.3	&	$<$5.4	\\
J04323028+1731303	&	GG Tau Ba+Bb	&	775 $\pm$ 28	&	59 $\pm$ 1	&	54 $\pm$ 1	&		&	50 $\pm$ 2	&	59 $\pm$ 2	&		&		\\
J04324938+2253082	&	JH 112 B	&	104 $\pm$ 4	&	113 $\pm$ 2	&	96 $\pm$ 2	&	96 $\pm$ 3	&	91 $\pm$ 2	&	99 $\pm$ 3	&	88 $\pm$ 3	&	113 $\pm$ 10	\\
J04325026+2422115	&	 ... 	&	14.3 $\pm$ 0.4	&	19.1 $\pm$ 0.3	&	15.1 $\pm$ 0.3	&	13.8 $\pm$ 0.3	&	10.7 $\pm$ 0.2	&	6.2 $\pm$ 0.1	&	2.3 $\pm$ 0.3	&	$<$3.9	\\
J04325119+1730092	&	LH 0429+17	&	1.55 $\pm$ 0.04	&		&	1.24 $\pm$ 0.03	&	1.12 $\pm$ 0.05	&		&	0.52 $\pm$ 0.04	&	$<$0.5	&	$<$5.3	\\
J04330197+2421000	&	MHO 8	&	50 $\pm$ 1	&	54 $\pm$ 1	&	39.0 $\pm$ 0.7	&	37.5 $\pm$ 0.9	&	25.9 $\pm$ 0.5	&	15.1 $\pm$ 0.3	&	8.0 $\pm$ 0.3	&	3 $\pm$ 1	\\
J04330781+2616066	&	KPNO 14	&	31.1 $\pm$ 0.8	&	35.4 $\pm$ 0.7	&	25.5 $\pm$ 0.5	&	24.1 $\pm$ 0.6	&	16.8 $\pm$ 0.4	&	9.8 $\pm$ 0.2	&	4.8 $\pm$ 0.3	&	$<$3.4	\\
J04330945+2246487	&	CFHT 12	&	12.2 $\pm$ 0.3	&	13.7 $\pm$ 0.2	&	10.9 $\pm$ 0.2	&	10.7 $\pm$ 0.3	&	8.6 $\pm$ 0.2	&	6.9 $\pm$ 0.1	&	5.7 $\pm$ 0.2	&	6 $\pm$ 1	\\
J04332621+2245293	&	XEST 17-036	&	42 $\pm$ 1	&	43.1 $\pm$ 0.8	&	30.8 $\pm$ 0.6	&	30.1 $\pm$ 0.7	&	21.6 $\pm$ 0.5	&	12.0 $\pm$ 0.3	&	6.0 $\pm$ 0.3	&	3 $\pm$ 1	\\
J04334171+1750402	&	 ... 	&	29.3 $\pm$ 0.8	&	28.9 $\pm$ 0.7	&	21.8 $\pm$ 0.5	&	22.0 $\pm$ 0.5	&	17.2 $\pm$ 0.5	&	15.3 $\pm$ 0.5	&	14.3 $\pm$ 0.4	&	23 $\pm$ 2	\\
J04334291+2526470	&	 ... 	&	1.96 $\pm$ 0.06	&	2.36 $\pm$ 0.06	&	1.67 $\pm$ 0.03	&	1.52 $\pm$ 0.05	&	1.18 $\pm$ 0.05	&	0.65 $\pm$ 0.02	&	$<$0.5	&	$<$3.2	\\
J04334465+2615005	&	 ... 	&	66 $\pm$ 2	&	90 $\pm$ 2	&	91 $\pm$ 2	&	74 $\pm$ 2	&	88 $\pm$ 2	&	88 $\pm$ 2	&	70 $\pm$ 1	&	106 $\pm$ 3	\\
J04335245+2612548	&	 ... 	&	1.27 $\pm$ 0.04	&	1.57 $\pm$ 0.03	&	1.54 $\pm$ 0.03	&	1.49 $\pm$ 0.05	&	1.45 $\pm$ 0.04	&	1.65 $\pm$ 0.04	&	1.6 $\pm$ 0.2	&	$<$4.1	\\
J04335252+2256269	&	XEST 17-059	&	83 $\pm$ 2	&	86 $\pm$ 2	&	63 $\pm$ 1	&	60 $\pm$ 1	&	42 $\pm$ 1	&	23.7 $\pm$ 0.6	&	12.6 $\pm$ 0.4	&	6 $\pm$ 1	\\
J04341527+2250309	&	CFHT 1	&	7.8 $\pm$ 0.2	&	9.3 $\pm$ 0.2	&	6.8 $\pm$ 0.1	&	6.5 $\pm$ 0.2	&	4.8 $\pm$ 0.1	&	2.62 $\pm$ 0.06	&	1.1 $\pm$ 0.2	&	$<$4.4	\\
J04344544+2308027	&	 ... 	&	8.3 $\pm$ 0.2	&	9 $\pm$ 0.2	&	6.0 $\pm$ 0.1	&	6.0 $\pm$ 0.2	&	4.2 $\pm$ 0.1	&	2.40 $\pm$ 0.07	&	1.6 $\pm$ 0.2	&	$<$3.2	\\
J04350850+2311398	&	CFHT 11	&	9.1 $\pm$ 0.2	&	9.4 $\pm$ 0.2	&	6.8 $\pm$ 0.2	&	6.6 $\pm$ 0.2	&	4.5 $\pm$ 0.1	&	2.63 $\pm$ 0.08	&	2.2 $\pm$ 0.2	&	3 $\pm$ 1	\\
J04353536+2408266	&	IRAS 04325+2402 C	&	36.4 $\pm$ 0.9	&	1.1 $\pm$ 0.2	&	1.2 $\pm$ 0.1	&	56 $\pm$ 1	&	1.0 $\pm$ 0.1	&	1.0 $\pm$ 0.1	&	84 $\pm$ 2	&	1893 $\pm$ 45	\\
J04354183+2234115	&	KPNO 8	&	6.1 $\pm$ 0.2	&	6.4 $\pm$ 0.1	&	4.66 $\pm$ 0.09	&	4.3 $\pm$ 0.1	&	3.10 $\pm$ 0.08	&	1.75 $\pm$ 0.05	&	0.8 $\pm$ 0.2	&	$<$4.1	\\
J04354203+2252226	&	XEST 08-033	&		&	38.8 $\pm$ 0.8	&	29.0 $\pm$ 0.7	&	25.9 $\pm$ 0.7	&	18.8 $\pm$ 0.5	&	10.3 $\pm$ 0.3	&	5.8 $\pm$ 0.2	&	$<$7.0	\\
J04354526+2737130	&	 ... 	&	1.30 $\pm$ 0.04	&	1.59 $\pm$ 0.04	&	1.09 $\pm$ 0.03	&	1.01 $\pm$ 0.03	&	0.80 $\pm$ 0.03	&	0.39 $\pm$ 0.03	&	$<$0.3	&	$<$2.7	\\
J04355143+2249119	&	KPNO 9	&	0.87 $\pm$ 0.03	&	1.01 $\pm$ 0.02	&	0.72 $\pm$ 0.02	&	0.69 $\pm$ 0.03	&	0.46 $\pm$ 0.02	&	0.28 $\pm$ 0.02	&	0.7 $\pm$ 0.2	&	6 $\pm$ 1	\\
J04355209+2255039	&	XEST 08-047	&	42 $\pm$ 1	&	43.3 $\pm$ 0.8	&	30.2 $\pm$ 0.6	&	28.6 $\pm$ 0.7	&	20.4 $\pm$ 0.4	&	11.6 $\pm$ 0.3	&	6.2 $\pm$ 0.2	&	8 $\pm$ 1	\\
J04355286+2250585	&	XEST 08-049	&	48 $\pm$ 1	&	47.9 $\pm$ 0.9	&	34.2 $\pm$ 0.7	&	31.8 $\pm$ 0.7	&	22.3 $\pm$ 0.5	&	12.6 $\pm$ 0.3	&	6.2 $\pm$ 0.2	&	6 $\pm$ 1	\\
J04355760+2253574	&	 ... 	&	1.50 $\pm$ 0.04	&	1.68 $\pm$ 0.08	&	1.28 $\pm$ 0.07	&	1.57 $\pm$ 0.05	&	1.23 $\pm$ 0.07	&	0.75 $\pm$ 0.06	&	0.6 $\pm$ 0.2	&	14 $\pm$ 1	\\
J04361030+2159364	&	 ... 	&	1.47 $\pm$ 0.04	&	1.81 $\pm$ 0.04	&	1.52 $\pm$ 0.03	&	1.44 $\pm$ 0.05	&	1.31 $\pm$ 0.04	&	1.29 $\pm$ 0.03	&	1.0 $\pm$ 0.2	&	3 $\pm$ 1	\\
J04361038+2259560	&	CFHT 2	&	6.0 $\pm$ 0.2	&	6.5 $\pm$ 0.1	&	5.04 $\pm$ 0.09	&	4.6 $\pm$ 0.1	&	3.26 $\pm$ 0.07	&	1.92 $\pm$ 0.04	&	0.9 $\pm$ 0.2	&	4 $\pm$ 1	\\
J04362151+2351165	&	 ... 	&	5.1 $\pm$ 0.1	&	5 $\pm$ 0.1	&	4.6 $\pm$ 0.1	&	4.5 $\pm$ 0.1	&	4.1 $\pm$ 0.1	&	3.9 $\pm$ 0.1	&	3.3 $\pm$ 0.2	&	4 $\pm$ 1	\\
J04363893+2258119	&	CFHT 3	&	4.9 $\pm$ 0.1	&	5.5 $\pm$ 0.1	&	3.93 $\pm$ 0.07	&	3.8 $\pm$ 0.1	&	2.73 $\pm$ 0.06	&	1.57 $\pm$ 0.04	&	1.1 $\pm$ 0.2	&	$<$3.2	\\
J04373705+2331080	&	 ... 	&	0.50 $\pm$ 0.02	&	0.61 $\pm$ 0.01	&	0.50 $\pm$ 0.02	&	0.49 $\pm$ 0.02	&	0.42 $\pm$ 0.02	&	0.31 $\pm$ 0.02	&	0.3 $\pm$ 0.1	&	$<$2.1	\\
J04380083+2558572	&	ITG 2	&	38.2 $\pm$ 1.0	&	42 $\pm$ 1	&	30.7 $\pm$ 0.7	&	29.8 $\pm$ 0.7	&	21.3 $\pm$ 0.7	&	11.9 $\pm$ 0.4	&	6.5 $\pm$ 0.2	&	$<$4.4	\\
J04381486+2611399	&	 ... 	&	11.4 $\pm$ 0.3	&	13.6 $\pm$ 0.3	&	15.8 $\pm$ 0.4	&	16.3 $\pm$ 0.4	&	16.5 $\pm$ 0.5	&	17.7 $\pm$ 0.5	&	21.9 $\pm$ 0.5	&	80 $\pm$ 3	\\
J04381630+2326402	&	 ... 	&	15.1 $\pm$ 0.4	&		&		&	10.5 $\pm$ 0.3	&		&		&	2.6 $\pm$ 0.2	&	3 $\pm$ 1	\\
J04382134+2609137	&	GM Tau	&	49 $\pm$ 1	&	52 $\pm$ 1	&	52 $\pm$ 1	&	51 $\pm$ 1	&	45 $\pm$ 1	&	45 $\pm$ 1	&	46.2 $\pm$ 0.9	&	63 $\pm$ 2	\\
J04385859+2336351	&	 ... 	&	16.2 $\pm$ 0.4	&	17.9 $\pm$ 0.4	&		&	14.4 $\pm$ 0.4	&	13.3 $\pm$ 0.4	&		&	14.8 $\pm$ 0.4	&	22 $\pm$ 1	\\
J04385871+2323595	&	 ... 	&	8.8 $\pm$ 0.2	&		&		&	6.3 $\pm$ 0.2	&		&		&	2.3 $\pm$ 0.2	&	$<$3.6	\\
J04390163+2336029	&	 ... 	&	34.8 $\pm$ 0.9	&	36.4 $\pm$ 0.9	&		&	28.0 $\pm$ 0.7	&	24.7 $\pm$ 0.8	&		&	20.7 $\pm$ 0.5	&	26 $\pm$ 1	\\
J04390396+2544264	&	CFHT 6	&	12.6 $\pm$ 0.3	&	13.7 $\pm$ 0.3	&	12.8 $\pm$ 0.3	&	12.2 $\pm$ 0.3	&	11.2 $\pm$ 0.4	&	14.1 $\pm$ 0.4	&	14.6 $\pm$ 0.4	&	21 $\pm$ 1	\\
J04390637+2334179	&	 ... 	&	13.7 $\pm$ 0.4	&	13.6 $\pm$ 0.3	&		&	9.7 $\pm$ 0.2	&	6.1 $\pm$ 0.2	&		&	2.4 $\pm$ 0.2	&	$<$3.5	\\
J04393364+2359212	&	 ... 	&	38.2 $\pm$ 1.0	&	40.2 $\pm$ 0.9	&	39.0 $\pm$ 0.9	&	37.6 $\pm$ 0.9	&	34 $\pm$ 1	&	44 $\pm$ 1	&	51 $\pm$ 1	&	67 $\pm$ 2	\\
J04394488+2601527	&	ITG 15	&	135 $\pm$ 4	&	119 $\pm$ 2	&	101 $\pm$ 2	&	134 $\pm$ 3	&	95 $\pm$ 2	&	94 $\pm$ 2	&	95 $\pm$ 2	&	195 $\pm$ 6	\\
J04394748+2601407	&	CFHT 4	&	36.6 $\pm$ 0.9	&	44.9 $\pm$ 0.9	&	42.7 $\pm$ 0.8	&	44 $\pm$ 1	&	42 $\pm$ 1	&	48 $\pm$ 1	&	48 $\pm$ 1	&	73 $\pm$ 2	\\
J04400067+2358211	&	 ... 	&	10.6 $\pm$ 0.3	&	12.3 $\pm$ 0.3	&	10.2 $\pm$ 0.2	&	9.4 $\pm$ 0.2	&	8.3 $\pm$ 0.3	&	8.6 $\pm$ 0.3	&	8.9 $\pm$ 0.3	&	20 $\pm$ 1	\\
J04400174+2556292	&	CFHT 17	&	21.0 $\pm$ 0.5	&	25.6 $\pm$ 0.5	&	19.5 $\pm$ 0.5	&	18.6 $\pm$ 0.5	&	13.5 $\pm$ 0.3	&	7.7 $\pm$ 0.2	&	3.4 $\pm$ 0.2	&	$<$3.9	\\
J04403979+2519061	&	 ... 	&	30.8 $\pm$ 0.8	&	34.1 $\pm$ 0.8	&	25.7 $\pm$ 0.6	&	23.0 $\pm$ 0.6	&	16.3 $\pm$ 0.5	&	9.6 $\pm$ 0.3	&	4.9 $\pm$ 0.2	&	9 $\pm$ 1	\\
J04410424+2557561	&	Haro 6-32	&	39 $\pm$ 1	&	41.4 $\pm$ 0.7	&	28.7 $\pm$ 0.5	&	27.5 $\pm$ 0.6	&	18.7 $\pm$ 0.4	&	10.6 $\pm$ 0.2	&	8.1 $\pm$ 0.4	&	7 $\pm$ 1	\\
J04411078+2555116	&	ITG 34	&	12.4 $\pm$ 0.3	&	14.1 $\pm$ 0.3	&	14.2 $\pm$ 0.2	&	13.0 $\pm$ 0.3	&	12.6 $\pm$ 0.3	&	13.9 $\pm$ 0.3	&	16.0 $\pm$ 0.4	&	20 $\pm$ 1	\\
J04414489+2301513	&	 ... 	&	42 $\pm$ 1	&	3.5 $\pm$ 0.08	&	3.18 $\pm$ 0.07	&	28.9 $\pm$ 0.7	&	2.81 $\pm$ 0.09	&	2.58 $\pm$ 0.08	&	8.4 $\pm$ 0.3	&	9 $\pm$ 1	\\
J04414565+2301580	&	 ... 	&	42 $\pm$ 1	&		&	29.0 $\pm$ 0.7	&	28.9 $\pm$ 0.7	&	20.5 $\pm$ 0.6	&	13.3 $\pm$ 0.4	&	8.4 $\pm$ 0.3	&	9 $\pm$ 1	\\
J04414825+2534304	&	 ... 	&	7.1 $\pm$ 0.2	&	7.9 $\pm$ 0.2	&	8.1 $\pm$ 0.2	&	8.3 $\pm$ 0.2	&	7.5 $\pm$ 0.2	&	9.7 $\pm$ 0.2	&	11.8 $\pm$ 0.3	&	25 $\pm$ 2	\\
J04422101+2520343	&	CIDA 7	&	42 $\pm$ 1	&	43.9 $\pm$ 0.9	&	39.5 $\pm$ 0.8	&	41 $\pm$ 1	&	40 $\pm$ 1	&	49 $\pm$ 1	&	57 $\pm$ 1	&	152 $\pm$ 4	\\
J04432023+2940060	&	CIDA 14	&	71 $\pm$ 2	&	76 $\pm$ 1	&	64 $\pm$ 1	&	60 $\pm$ 1	&	56 $\pm$ 1	&	60 $\pm$ 1	&	55 $\pm$ 1	&	47 $\pm$ 2	\\
J04442713+2512164	&	IRAS S04414+2506	&	29.7 $\pm$ 0.8	&	39.3 $\pm$ 0.8	&	39.3 $\pm$ 0.8	&	31.3 $\pm$ 0.8	&	42 $\pm$ 1	&	56 $\pm$ 1	&	63 $\pm$ 1	&	159 $\pm$ 5	\\
J04464260+2459034	&	RXJ 04467+2459	&	26.9 $\pm$ 0.7	&	27.5 $\pm$ 0.5	&	19.3 $\pm$ 0.4	&	18.6 $\pm$ 0.5	&	12.4 $\pm$ 0.3	&	7.1 $\pm$ 0.2	&	4.2 $\pm$ 0.2	&	$<$3.1	\\
J04484189+1703374	&	 ... 	&	4.0 $\pm$ 0.1	&	4.2 $\pm$ 0.1	&	3.01 $\pm$ 0.07	&	2.91 $\pm$ 0.08	&	2.02 $\pm$ 0.06	&	1.15 $\pm$ 0.04	&	0.5 $\pm$ 0.2	&	$<$2.1	\\
J04520668+3047175	&	IRAS 04489+3042	&	91 $\pm$ 2	&	120 $\pm$ 3	&	152 $\pm$ 4	&	164 $\pm$ 4	&	172 $\pm$ 5	&	221 $\pm$ 7	&	283 $\pm$ 6	&	861 $\pm$ 20	\\
J04552333+3027366	&	 ... 	&	6.5 $\pm$ 0.2	&	7 $\pm$ 0.1	&	5.0 $\pm$ 0.1	&	4.8 $\pm$ 0.1	&	3.35 $\pm$ 0.07	&	1.91 $\pm$ 0.06	&	1.6 $\pm$ 0.1	&	$<$3.5	\\
J04554046+3039057	&	 ... 	&	7.3 $\pm$ 0.2	&	7.4 $\pm$ 0.1	&	5.4 $\pm$ 0.1	&	5.1 $\pm$ 0.1	&	3.57 $\pm$ 0.09	&	2.09 $\pm$ 0.05	&	1.3 $\pm$ 0.2	&	3 $\pm$ 1	\\
J04554535+3019389	&	 ... 	&	26.6 $\pm$ 0.7	&	29.3 $\pm$ 0.6	&	25.5 $\pm$ 0.5	&	23.4 $\pm$ 0.6	&	21.3 $\pm$ 0.5	&	20.7 $\pm$ 0.4	&	18.1 $\pm$ 0.4	&	20 $\pm$ 1	\\
J04554757+3028077	&	 ... 	&	38.0 $\pm$ 1.0	&	39.5 $\pm$ 0.7	&	28.1 $\pm$ 0.5	&	26.2 $\pm$ 0.6	&	18.4 $\pm$ 0.4	&	10.8 $\pm$ 0.3	&	5.7 $\pm$ 0.3	&	$<$3.1	\\
J04554801+3028050	&	 ... 	&	6.3 $\pm$ 0.3	&	7.6 $\pm$ 0.1	&	6.9 $\pm$ 0.1	&	5.9 $\pm$ 0.3	&	5.8 $\pm$ 0.1	&	6.2 $\pm$ 0.1	&	6.9 $\pm$ 0.3	&	16 $\pm$ 1	\\
J04554820+3030160	&	XEST 26-052	&	15.0 $\pm$ 0.4	&	15.4 $\pm$ 0.3	&	10.9 $\pm$ 0.2	&	10.2 $\pm$ 0.2	&	7.0 $\pm$ 0.2	&	4.3 $\pm$ 0.1	&	2.5 $\pm$ 0.2	&	8 $\pm$ 1	\\
J04554969+3019400	&	 ... 	&	7.4 $\pm$ 0.2	&	8 $\pm$ 0.2	&	6.2 $\pm$ 0.1	&	5.9 $\pm$ 0.1	&	4.8 $\pm$ 0.2	&	3.60 $\pm$ 0.09	&	3.2 $\pm$ 0.2	&	5 $\pm$ 1	\\
J04555288+3006523	&	 ... 	&	18.7 $\pm$ 0.5	&	19.6 $\pm$ 0.5	&	13.5 $\pm$ 0.3	&	12.7 $\pm$ 0.3	&	9.1 $\pm$ 0.3	&	5.2 $\pm$ 0.2	&	2.9 $\pm$ 0.2	&	$<$2.7	\\
J04555605+3036209	&	XEST 26-062	&	84 $\pm$ 2	&	94 $\pm$ 2	&	81 $\pm$ 1	&	73 $\pm$ 2	&	73 $\pm$ 2	&	87 $\pm$ 2	&	119 $\pm$ 2	&	272 $\pm$ 7	\\
J04555636+3049374	&	 ... 	&	13.8 $\pm$ 0.4	&	14.1 $\pm$ 0.3	&	9.9 $\pm$ 0.2	&	9.5 $\pm$ 0.2	&	6.7 $\pm$ 0.2	&	3.8 $\pm$ 0.1	&	2.4 $\pm$ 0.2	&	$<$2.9	\\
J04574903+3015195	&	 ... 	&	0.61 $\pm$ 0.02	&	0.79 $\pm$ 0.02	&	0.56 $\pm$ 0.01	&	0.49 $\pm$ 0.02	&	0.40 $\pm$ 0.01	&	0.20 $\pm$ 0.01	&	$<$0.3	&	$<$2.3	\\
J05061674+2446102	&	CIDA 10	&	42 $\pm$ 1	&	41 $\pm$ 1	&	29.8 $\pm$ 0.7	&	27.5 $\pm$ 0.7	&	19.3 $\pm$ 0.6	&	11.2 $\pm$ 0.3	&	6.2 $\pm$ 0.2	&	3 $\pm$ 1	\\
J05064662+2104296	&	 ... 	&	13.7 $\pm$ 0.4	&	14.3 $\pm$ 0.3	&	10.0 $\pm$ 0.2	&	9.4 $\pm$ 0.2	&	6.7 $\pm$ 0.2	&	3.8 $\pm$ 0.1	&	2.4 $\pm$ 0.2	&	2 $\pm$ 1	\\
J05075496+2500156	&	CIDA 12	&	27.0 $\pm$ 0.7	&	26.6 $\pm$ 0.6	&	20.8 $\pm$ 0.5	&	21.0 $\pm$ 0.5	&	18.1 $\pm$ 0.6	&	21.1 $\pm$ 0.7	&	22.6 $\pm$ 0.5	&	35 $\pm$ 2	\\

\end{longtable}
\tablefoot{{\it Spitzer} IRAC flux densities are those reported in \citet{Luhman:2010}, excluding 2MASS~J04251550+2829275 and 2MASS~J04355760+2253574, for which the IRAC flux densities are taken from those reported in \citet{Rebull:2010}. The weighted average flux density is reported for targets with mulit-epoch IRAC observations. {\it WISE} flux densities are taken from the {\it WISE} All-Sky Catalog \citep{Wright:2010}.
}
\end{landscape}
\twocolumn
\end{longtab}

\begin{longtab}
\onecolumn
\begin{landscape}
\begin{longtable}{l c l l l l l l l}
\caption{\label{Table:MIPS} {\it Spitzer} MIPS and {\it Herschel} PACS photometry compilation of the known 154 M4-L0 Taurus members, ordered by target RA.}\\
\hline\hline
2MASS & Other name	&	MIPS-1 &	MIPS-2	&	PACS-1	&	PACS-2	&	PACS-3	&	Reference & Notes	\\  
& & (mJy) & (mJy) & (mJy) & (mJy) & (mJy) \\
\hline
\endfirsthead
\caption{continued.}\\
\hline\hline
2MASS & Other name	&	MIPS-1 &	MIPS-2	&	PACS-1	&	PACS-2	&	PACS-3	&	Reference & Notes	\\  
& & (mJy) & (mJy) & (mJy) & (mJy) & (mJy) \\
\hline
\endhead
\hline
\endfoot
J04034997+2620382	&	XEST 06-006	&		&		&	$<$5	&		&	$<$20	&	$[$$]$; $[$1$]$	&		\\
J04131414+2819108	&	LkCa 1	&	3.5 $\pm$ 0.3	&	$<$235	&	$<$9	&	$<$9	&	$<$21	&	$[$2$]$; $[$3$]$	&		\\
J04141188+2811535	&	 ... 	&	36 $\pm$ 2	&	$<$301	&	17 $\pm$ 5	&		&	$<$293	&	$[$2$]$; $[$1$]$	&		\\
J04141458+2827580	&	FN Tau	&	1105 $\pm$ 62	&	979 $\pm$ 199	&	1755 $\pm$ 4	&		&	816 $\pm$ 16	&	$[$2$]$; $[$1$]$	&		\\
J04141760+2806096	&	CIDA 1	&	278 $\pm$ 16	&	239 $\pm$ 49	&	266 $\pm$ 2	&		&	212 $\pm$ 21	&	$[$2$]$; $[$1$]$	&		\\
J04144730+2646264	&	FP Tau	&	143 $\pm$ 8	&	253 $\pm$ 51	&	307 $\pm$ 3	&		&	351 $\pm$ 11	&	$[$2$]$; $[$1$]$	&		\\
J04144739+2803055	&	XEST 20-066	&	1.4 $\pm$ 0.2	&		&	$<$6	&		&	$<$67	&	$[$1$]$; $[$1$]$	&		\\
J04150515+2808462	&	CIDA 2	&	3.0 $\pm$ 0.3	&	$<$233	&	$<$9	&	$<$9	&	$<$22	&	$[$2$]$; $[$3$]$	&		\\
J04151471+2800096	&	KPNO 1	&	$<$0.4	&	$<$340	&	$<$3	&		&	$<$41	&	$[$2$]$; $[$1$]$	&		\\
J04152409+2910434	&	 ... 	&	$<$0.7	&	$<$272	&	$<$7	&		&	$<$29	&	$[$2$]$; $[$1$]$	&		\\
J04155799+2746175	&	 ... 	&	36 $\pm$ 2	&	$<$253	&	25 $\pm$ 3	&		&	$<$48	&	$[$2$]$; $[$1$]$	&		\\
J04161210+2756385	&	 ... 	&	51 $\pm$ 3	&	187 $\pm$ 38	&	201 $\pm$ 3	&		&	228 $\pm$ 8	&	$[$2$]$; $[$1$]$	&		\\
J04161885+2752155	&	 ... 	&	$<$0.8	&	$<$244	&	$<$6	&		&	$<$50	&	$[$2$]$; $[$1$]$	&		\\
J04162725+2053091	&	 ... 	&		&		&	$<$9	&		&	$<$39	&	$[$$]$; $[$1$]$	&		\\
J04163048+3037053	&	 ... 	&		&		&	$<$11	&		&	$<$28	&	$[$$]$; $[$1$]$	&		\\
J04163911+2858491	&	 ... 	&	9.3 $\pm$ 0.6	&	$<$248	&	$<$9	&		&	$<$55	&	$[$2$]$; $[$1$]$	&		\\
J04174955+2813318	&	KPNO 10	&	30 $\pm$ 2	&	133 $\pm$ 27	&	160 $\pm$ 2	&		&	82 $\pm$ 26	&	$[$2$]$; $[$1$]$	&		\\
J04174965+2829362	&	V410 X-ray 1	&	221 $\pm$ 12	&	51 $\pm$ 10	&	36 $\pm$ 3	&		&	$<$122	&	$[$2$]$; $[$1$]$	&		\\
J04180796+2826036	&	V410 X-ray 3	&	1.4 $\pm$ 0.3	&	$<$150	&	$<$7	&		&	$<$86	&	$[$2$]$; $[$1$]$	&		\\
J04181710+2828419	&	V410 Anon 13	&	28 $\pm$ 2	&		&	35 $\pm$ 2	&		&	$<$113	&	$[$2$]$; $[$1$]$	&		\\
J04183030+2743208	&	KPNO 11	&	$<$0.7	&	$<$203	&	$<$6	&		&	$<$55	&	$[$2$]$; $[$1$]$	&		\\
J04184023+2824245	&	V410 X-ray 4	&	4.2 $\pm$ 0.4	&	$<$178	&	$<$23	&		&	$<$88	&	$[$2$]$; $[$1$]$	&		\\
J04185115+2814332	&	KPNO 2	&	$<$1.0	&	$<$180	&	$<$7	&		&	$<$74	&	$[$2$]$; $[$1$]$	&		\\
J04185813+2812234	&	IRAS 04158+2805	&	580 $\pm$ 33	&	830 $\pm$ 169	&	1089 $\pm$ 3	&		&	2953 $\pm$ 25	&	$[$2$]$; $[$1$]$	&		\\
J04190110+2819420	&	V410 X-ray 6	&	213 $\pm$ 12	&	412 $\pm$ 84	&	445 $\pm$ 4	&		&	342 $\pm$ 36	&	$[$2$]$; $[$1$]$	&		\\
J04190126+2802487	&	KPNO 12	&	$<$0.6	&	$<$157	&	$<$6	&		&	$<$88	&	$[$2$]$; $[$1$]$	&		\\
J04190197+2822332	&	V410 X-ray 5a	&	2.0 $\pm$ 0.2	&	$<$180	&	$<$5	&		&	$<$75	&	$[$2$]$; $[$1$]$	&		\\
J04193545+2827218	&	FR Tau	&	83 $\pm$ 5	&	43 $\pm$ 9	&	46 $\pm$ 3	&		&	$<$33	&	$[$2$]$; $[$1$]$	&		\\
J04194657+2712552	&	$[$GKH94$]$ 41	&	172 $\pm$ 10	&	$<$267	&	269 $\pm$ 5	&		&	279 $\pm$ 66	&	$[$2$]$; $[$1$]$	&		\\
J04201611+2821325	&	 ... 	&	3.7 $\pm$ 0.3	&	$<$290	&	$<$7	&		&	$<$26	&	$[$2$]$; $[$1$]$	&		\\
J04202555+2700355	&	 ... 	&	25 $\pm$ 1	&	79 $\pm$ 16	&	107 $\pm$ 3	&		&	100 $\pm$ 15	&	$[$2$]$; $[$1$]$	&		\\
J04202583+2819237	&	IRAS 04173+2812	&	221 $\pm$ 12	&	133 $\pm$ 27	&	172 $\pm$ 3	&		&	72 $\pm$ 6	&	$[$2$]$; $[$1$]$	&		\\
J04203918+2717317	&	XEST 16-045	&	2.1 $\pm$ 0.2	&	$<$195	&	$<$5	&		&	$<$45	&	$[$2$]$; $[$1$]$	&		\\
J04205273+1746415	&	J2-157	&		&		&	$<$5	&		&	$<$27	&	$[$$]$; $[$1$]$	&		\\
J04210795+2702204	&	CFHT 19	&	1627 $\pm$ 92	&	2307 $\pm$ 469	&	3277 $\pm$ 8	&		&	3076 $\pm$ 61	&	$[$2$]$; $[$1$]$	&		\\
J04210934+2750368	&	 ... 	&	9.7 $\pm$ 0.6	&	$<$190	&	$<$13	&		&	$<$47	&	$[$2$]$; $[$1$]$	&		\\
J04213459+2701388	&	 ... 	&	9.6 $\pm$ 0.6	&	$<$172	&	37 $\pm$ 2	&		&	101 $\pm$ 19	&	$[$2$]$; $[$1$]$	&		\\
J04214013+2814224	&	XEST 21-026	&		&		&	$<$8	&		&	$<$31	&	$[$$]$; $[$1$]$	&		\\
J04214631+2659296	&	CFHT 10	&	8.9 $\pm$ 0.6	&	$<$205	&	$<$10	&		&	$<$84	&	$[$2$]$; $[$1$]$	&		\\
J04215450+2652315	&	 ... 	&	0.5 $\pm$ 0.1	&	$<$169	&	$<$5	&		&	$<$35	&	$[$2$]$; $[$1$]$	&		\\
J04220007+1530248	&	IRAS 04191+1523 B	&		&		&	(7002 $\pm$ 13)	&		&	(8884 $\pm$ 232)	&	$[$$]$; $[$1$]$	&	d	\\
J04221332+1934392	&	 ... 	&		&		&	$<$5	&		&	$<$45	&	$[$$]$; $[$1$]$	&		\\
J04221644+2549118	&	CFHT 14	&	$<$1.1	&	$<$267	&	$<$5	&		&	$<$26	&	$[$2$]$; $[$1$]$	&		\\
J04222404+2646258	&	XEST 11-087	&	1.7 $\pm$ 0.2	&	$<$185	&	$<$6	&		&	$<$44	&	$[$2$]$; $[$1$]$	&		\\
J04230607+2801194	&	 ... 	&	19 $\pm$ 1	&	$<$431	&	41 $\pm$ 3	&		&	38 $\pm$ 9	&	$[$2$]$; $[$1$]$	&		\\
J04233539+2503026	&	FU Tau A	&	104 $\pm$ 6	&	86 $\pm$ 17	&		&	$<$39	&	$<$247	&	$[$2$]$; $[$1$]$	&	a	\\
J04233573+2502596	&	FU Tau B	&		&		&		&	$<$39	&	$<$247	&	$[$$]$; $[$1$]$	&	a	\\
J04242090+2630511	&	 ... 	&	7.4 $\pm$ 0.5	&	$<$275	&	$<$9	&		&	$<$58	&	$[$2$]$; $[$1$]$	&		\\
J04242646+2649503	&	CFHT 9	&	13.9 $\pm$ 0.9	&	$<$383	&	10 $\pm$ 1	&		&	$<$6	&	$[$2$]$; $[$2$]$	&		\\
J04244506+2701447	&	J1-4423	&	$<$1.1	&	$<$296	&	$<$5	&		&	$<$41	&	$[$2$]$; $[$1$]$	&		\\
J04251550+2829275	&	 ... 	&	1.4 $\pm$ 0.3	&	$<$369	&		&		&		&	$[$2$]$; $[$$]$	&		\\
J04262939+2624137	&	KPNO 3	&	12.9 $\pm$ 0.8	&	$<$285	&	23 $\pm$ 4	&		&	33 $\pm$ 12	&	$[$2$]$; $[$1$]$	&		\\
J04263055+2443558	&	 ... 	&	2.0 $\pm$ 0.3	&	$<$285	&	$<$12	&		&	$<$51	&	$[$2$]$; $[$1$]$	&		\\
J04265732+2606284	&	KPNO 13	&	53 $\pm$ 3	&	$<$330	&	28 $\pm$ 4	&		&	$<$182	&	$[$2$]$; $[$1$]$	&		\\
J04270739+2215037	&	 ... 	&		&		&	$<$8	&		&	$<$27	&	$[$$]$; $[$1$]$	&		\\
J04272799+2612052	&	KPNO 4	&	0.4 $\pm$ 0.1	&	$<$296	&	$<$7	&		&	$<$21	&	$[$2$]$; $[$1$]$	&		\\
J04274538+2357243	&	CFHT 15	&	$<$0.4	&	$<$293	&	$<$6	&		&	$<$47	&	$[$2$]$; $[$1$]$	&		\\
J04275730+2619183	&	IRAS 04248+2612	&	886 $\pm$ 50	&	3155 $\pm$ 641	&	5208 $\pm$ 19	&		&	10093 $\pm$ 141	&	$[$2$]$; $[$1$]$	&		\\
 ... 	&	L1521F-IRS	&	25 $\pm$ 1	&	460 $\pm$ 94	&	631 $\pm$ 1	&		&	3712 $\pm$ 52	&	$[$2$]$; $[$1$]$	&		\\
J04284263+2714039	&	 ... 	&	24 $\pm$ 1	&	$<$346	&	20 $\pm$ 2	&		&	51 $\pm$ 12	&	$[$2$]$; $[$1$]$	&		\\
J04290068+2755033	&	 ... 	&	4.3 $\pm$ 0.3	&	$<$346	&	$<$8	&		&	$<$36	&	$[$2$]$; $[$1$]$	&		\\
J04292071+2633406	&	J1-507	&	3.5 $\pm$ 0.4	&	$<$299	&	$<$8	&	$<$8	&	$<$21	&	$[$2$]$; $[$3$]$	&		\\
J04292165+2701259	&	IRAS 04263+2654	&	310 $\pm$ 18	&	280 $\pm$ 57	&	329 $\pm$ 3	&		&	176 $\pm$ 14	&	$[$2$]$; $[$1$]$	&		\\
J04292971+2616532	&	FW Tau A+B+C	&	7.0 $\pm$ 0.5	&	$<$290	&	19 $\pm$ 0.01	&	33 $\pm$ 4	&	70 $\pm$ 40	&	$[$2$]$; $[$3$]$	&		\\
J04294568+2630468	&	KPNO 5	&	$<$0.9	&	$<$340	&	$<$7	&		&	$<$37	&	$[$2$]$; $[$1$]$	&		\\
J04295422+1754041	&	 ... 	&		&		&	89 $\pm$ 3	&		&	86 $\pm$ 8	&	$[$$]$; $[$1$]$	&		\\
J04295950+2433078	&	CFHT 20	&	78 $\pm$ 4	&	117 $\pm$ 24	&	128 $\pm$ 4	&		&	91 $\pm$ 23	&	$[$2$]$; $[$1$]$	&		\\
J04300724+2608207	&	KPNO 6	&	1.5 $\pm$ 0.3	&	$<$307	&	2 $\pm$ 1	&		&	$<$5	&	$[$2$]$; $[$2$]$	&		\\
J04302365+2359129	&	CFHT 16	&	$<$0.4	&	$<$310	&	$<$6	&		&	$<$51	&	$[$2$]$; $[$1$]$	&		\\
J04305171+2441475	&	ZZ Tau IRS	&	1126 $\pm$ 64	&	1936 $\pm$ 393	&	2901 $\pm$ 5	&		&	2922 $\pm$ 26	&	$[$2$]$; $[$1$]$	&		\\
J04305718+2556394	&	KPNO 7	&	2.6 $\pm$ 0.3	&	$<$251	&	4 $\pm$ 1	&		&	$<$7	&	$[$2$]$; $[$2$]$	&		\\
J04311578+1820072	&	MHO 9	&	0.9 $\pm$ 0.1	&		&	$<$8	&		&	$<$76	&	$[$1$]$; $[$1$]$	&		\\
J04311907+2335047	&	 ... 	&	$<$0.4	&	$<$352	&	$<$7	&		&	$<$24	&	$[$2$]$; $[$1$]$	&		\\
J04312382+2410529	&	V927 Tau A+B	&	3.8 $\pm$ 0.4	&	$<$336	&	$<$9	&	$<$9	&	$<$32	&	$[$2$]$; $[$3$]$	&		\\
J04312405+1800215	&	MHO 4	&	1.0 $\pm$ 0.1	&		&	$<$11	&		&	$<$38	&	$[$1$]$; $[$1$]$	&		\\
J04312669+2703188	&	CFHT 13	&	0.4 $\pm$ 0.1	&	$<$416	&	$<$8	&		&	$<$54	&	$[$2$]$; $[$1$]$	&		\\
J04313613+1813432	&	LkHa 358	&	678 $\pm$ 21	&		&	1440 $\pm$ 0.2	&	1320 $\pm$ 200	&	1530 $\pm$ 300	&	$[$1$]$; $[$3$]$	&		\\
J04315844+2543299	&	J1-665	&	1.7 $\pm$ 0.3	&	$<$288	&	$<$9	&	$<$9	&	$<$25	&	$[$2$]$; $[$$]$	&		\\
J04320329+2528078	&	 ... 	&	$<$1.0	&	$<$301	&	$<$9	&		&	$<$33	&	$[$2$]$; $[$1$]$	&		\\
J04321606+1812464	&	MHO 5	&	39 $\pm$ 1	&		&	$<$11	&		&	$<$82	&	$[$1$]$; $[$1$]$	&		\\
J04321786+2422149	&	CFHT 7	&	1.4 $\pm$ 0.4	&	$<$352	&		&	$<$39	&	$<$322	&	$[$2$]$; $[$1$]$	&	a	\\
J04322210+1827426	&	MHO 6	&	20.7 $\pm$ 0.6	&		&	107 $\pm$ 2	&		&	188 $\pm$ 7	&	$[$1$]$; $[$1$]$	&		\\
J04322329+2403013	&	 ... 	&	$<$0.8	&	$<$336	&	$<$9	&		&	$<$35	&	$[$2$]$; $[$1$]$	&		\\
J04322415+2251083	&	 ... 	&	26 $\pm$ 1	&	$<$208	&	$<$9	&		&	$<$77	&	$[$2$]$; $[$1$]$	&		\\
J04322627+1827521	&	MHO 7	&	0.8 $\pm$ 0.2	&		&	$<$8	&		&	$<$23	&	$[$1$]$; $[$1$]$	&		\\
J04323028+1731303	&	GG Tau Ba+Bb	&		&		&	210 $\pm$ 20	&	 ... 	&	 ... 	&	$[$$]$; $[$3$]$	&	b	\\
J04324938+2253082	&	JH 112 B	&		&		&	$<$318	&		&	(284 $\pm$ 26)	&	$[$$]$; $[$1$]$	&	c	\\
J04325026+2422115	&	 ... 	&	1.1 $\pm$ 0.3	&	$<$267	&	$<$10	&		&	$<$82	&	$[$2$]$; $[$1$]$	&		\\
J04325119+1730092	&	LH 0429+17	&		&		&	$<$7	&		&	$<$33	&	$[$$]$; $[$1$]$	&		\\
J04330197+2421000	&	MHO 8	&	1.9 $\pm$ 0.3	&	$<$346	&	$<$4	&		&	$<$48	&	$[$2$]$; $[$1$]$	&		\\
J04330781+2616066	&	KPNO 14	&	1.7 $\pm$ 0.2	&	$<$207	&	$<$6	&		&	$<$79	&	$[$2$]$; $[$1$]$	&		\\
J04330945+2246487	&	CFHT 12	&	3.6 $\pm$ 0.3	&	$<$267	&	2 $\pm$ 1	&		&	$<$8	&	$[$2$]$; $[$2$]$	&		\\
J04332621+2245293	&	XEST 17-036	&	3.0 $\pm$ 0.3	&	130 $\pm$ 26	&	108 $\pm$ 4	&		&	181 $\pm$ 27	&	$[$2$]$; $[$1$]$	&		\\
J04334171+1750402	&	 ... 	&		&		&	17 $\pm$ 3	&		&	$<$28	&	$[$$]$; $[$1$]$	&		\\
J04334291+2526470	&	 ... 	&	$<$0.3	&	$<$208	&	$<$6	&		&	$<$32	&	$[$2$]$; $[$1$]$	&		\\
J04334465+2615005	&	 ... 	&	108 $\pm$ 6	&	142 $\pm$ 29	&	149 $\pm$ 2	&		&	178 $\pm$ 22	&	$[$2$]$; $[$1$]$	&		\\
J04335245+2612548	&	 ... 	&	3.0 $\pm$ 0.3	&	$<$290	&	$<$11	&		&	$<$47	&	$[$2$]$; $[$1$]$	&		\\
J04335252+2256269	&	XEST 17-059	&	3.4 $\pm$ 0.3	&	$<$212	&	$<$12	&		&	$<$27	&	$[$2$]$; $[$1$]$	&		\\
J04341527+2250309	&	CFHT 1	&	$<$0.8	&	$<$310	&	$<$7	&		&	$<$67	&	$[$2$]$; $[$1$]$	&		\\
J04344544+2308027	&	 ... 	&		&		&	$<$10	&		&	$<$48	&	$[$$]$; $[$1$]$	&		\\
J04350850+2311398	&	CFHT 11	&	0.6 $\pm$ 0.1	&	$<$553	&	$<$7	&		&	$<$56	&	$[$2$]$; $[$1$]$	&		\\
J04353536+2408266	&	IRAS 04325+2402 C	&		&		&		&	(16906 $\pm$ 26)	&		&	$[$$]$; $[$1$]$	&	a, d, e	\\
J04354183+2234115	&	KPNO 8	&	$<$0.4	&	$<$277	&	$<$3	&		&	$<$31	&	$[$2$]$; $[$1$]$	&		\\
J04354203+2252226	&	XEST 08-033	&	1.8 $\pm$ 0.2	&	$<$290	&	$<$16	&		&	104 $\pm$ 28	&	$[$2$]$; $[$1$]$	&		\\
J04354526+2737130	&	 ... 	&	$<$0.4	&	$<$201	&	$<$8	&		&	$<$30	&	$[$2$]$; $[$1$]$	&		\\
J04355143+2249119	&	KPNO 9	&	$<$0.3	&	$<$352	&	$<$7	&		&	$<$102	&	$[$2$]$; $[$1$]$	&		\\
J04355209+2255039	&	XEST 08-047	&		&	$<$491	&	$<$11	&		&	$<$1657	&	$[$$]$; $[$1$]$	&		\\
J04355286+2250585	&	XEST 08-049	&	1.6 $\pm$ 0.2	&	$<$330	&	$<$6	&		&	$<$109	&	$[$2$]$; $[$1$]$	&		\\
J04355760+2253574	&	 ... 	&	10.1 $\pm$ 0.6	&	$<$346	&		&		&		&	$[$2$]$; $[$$]$	&		\\
J04361030+2159364	&	 ... 	&	1.8 $\pm$ 0.3	&	$<$290	&	$<$7	&		&	$<$22	&	$[$2$]$; $[$1$]$	&		\\
J04361038+2259560	&	CFHT 2	&	$<$0.4	&	$<$199	&	$<$7	&		&	$<$82	&	$[$2$]$; $[$1$]$	&		\\
J04362151+2351165	&	 ... 	&	5.7 $\pm$ 0.4	&	$<$293	&	$<$15	&		&	$<$64	&	$[$2$]$; $[$1$]$	&		\\
J04363893+2258119	&	CFHT 3	&	$<$2.7	&	$<$277	&	$<$5	&		&	$<$50	&	$[$2$]$; $[$1$]$	&		\\
J04373705+2331080	&	 ... 	&		&		&	$<$5	&		&	$<$27	&	$[$$]$; $[$1$]$	&		\\
J04380083+2558572	&	ITG 2	&	1.5 $\pm$ 0.3	&	$<$321	&	$<$6	&		&	$<$69	&	$[$2$]$; $[$1$]$	&		\\
J04381486+2611399	&	 ... 	&	73 $\pm$ 4	&	$<$372	&	95 $\pm$ 2	&		&	67 $\pm$ 24	&	$[$2$]$; $[$1$]$	&		\\
J04381630+2326402	&	 ... 	&		&		&	$<$6	&		&	$<$26	&	$[$$]$; $[$1$]$	&		\\
J04382134+2609137	&	GM Tau	&	53 $\pm$ 3	&	$<$318	&	36 $\pm$ 2	&		&	$<$35	&	$[$2$]$; $[$1$]$	&		\\
J04385859+2336351	&	 ... 	&	20 $\pm$ 1	&	$<$356	&	38 $\pm$ 3	&		&	76 $\pm$ 13	&	$[$2$]$; $[$1$]$	&		\\
J04385871+2323595	&	 ... 	&		&		&	$<$4	&		&	$<$30	&	$[$$]$; $[$1$]$	&		\\
J04390163+2336029	&	 ... 	&	22 $\pm$ 1	&	$<$94	&	15 $\pm$ 3	&		&	$<$24	&	$[$2$]$; $[$1$]$	&		\\
J04390396+2544264	&	CFHT 6	&	18 $\pm$ 1	&	$<$505	&	23 $\pm$ 3	&		&	$<$56	&	$[$2$]$; $[$1$]$	&		\\
J04390637+2334179	&	 ... 	&	$<$1.3	&		&	$<$5	&		&	$<$44	&	$[$2$]$; $[$1$]$	&		\\
J04393364+2359212	&	 ... 	&	59 $\pm$ 3	&	$<$333	&	70 $\pm$ 1	&		&	44 $\pm$ 15	&	$[$2$]$; $[$1$]$	&		\\
J04394488+2601527	&	ITG 15	&	187 $\pm$ 11	&	231 $\pm$ 47	&	272 $\pm$ 3	&		&	114 $\pm$ 28	&	$[$2$]$; $[$1$]$	&		\\
J04394748+2601407	&	CFHT 4	&	75 $\pm$ 4	&	$<$336	&	109 $\pm$ 5	&		&	$<$150	&	$[$2$]$; $[$1$]$	&		\\
J04400067+2358211	&	 ... 	&	20 $\pm$ 1	&	$<$379	&	55 $\pm$ 2	&		&	52 $\pm$ 5	&	$[$2$]$; $[$1$]$	&		\\
J04400174+2556292	&	CFHT 17	&	1.6 $\pm$ 0.3	&	$<$394	&	$<$11	&		&	$<$159	&	$[$2$]$; $[$1$]$	&		\\
J04403979+2519061	&	 ... 	&	6.8 $\pm$ 0.4	&	$<$310	&	72 $\pm$ 2	&		&	61 $\pm$ 20	&	$[$2$]$; $[$1$]$	&		\\
J04410424+2557561	&	Haro 6-32	&	1.0 $\pm$ 0.3	&	$<$408	&	$<$8	&		&	$<$48	&	$[$2$]$; $[$1$]$	&		\\
J04411078+2555116	&	ITG 34	&	18 $\pm$ 1	&	$<$394	&	21 $\pm$ 4	&		&	$<$38	&	$[$2$]$; $[$1$]$	&		\\
J04414489+2301513	&	 ... 	&		&		&	11 $\pm$ 2	&		&	$<$18	&	$[$$]$; $[$1$]$	&		\\
J04414565+2301580	&	 ... 	&		&		&	$<$5	&		&	$<$18	&	$[$$]$; $[$1$]$	&		\\
J04414825+2534304	&	 ... 	&	21 $\pm$ 1	&	$<$304	&	37 $\pm$ 3	&		&	$<$122	&	$[$2$]$; $[$1$]$	&		\\
J04422101+2520343	&	CIDA 7	&	150 $\pm$ 8	&	275 $\pm$ 56	&	330 $\pm$ 2	&		&	342 $\pm$ 19	&	$[$2$]$; $[$1$]$	&		\\
J04432023+2940060	&	CIDA 14	&	43 $\pm$ 2	&		&	$<$13	&		&	$<$25	&	$[$1$]$; $[$1$]$	&		\\
J04442713+2512164	&	IRAS S04414+2506	&	143 $\pm$ 8	&	154 $\pm$ 31	&		&		&		&	$[$2$]$; $[$$]$	&		\\
J04464260+2459034	&	RXJ 04467+2459	&	1.1 $\pm$ 0.3	&	$<$321	&	$<$10	&		&	$<$35	&	$[$2$]$; $[$1$]$	&		\\
J04484189+1703374	&	 ... 	&		&		&	$<$6	&		&	$<$25	&	$[$$]$; $[$1$]$	&		\\
J04520668+3047175	&	IRAS 04489+3042	&		&		&	2151 $\pm$ 5	&		&	2103 $\pm$ 34	&	$[$$]$; $[$1$]$	&		\\
J04552333+3027366	&	 ... 	&		&		&	$<$7	&		&	$<$34	&	$[$$]$; $[$1$]$	&		\\
J04554046+3039057	&	 ... 	&		&		&	$<$10	&		&	$<$73	&	$[$$]$; $[$1$]$	&		\\
J04554535+3019389	&	 ... 	&	18.9 $\pm$ 0.6	&		&	22 $\pm$ 3	&		&	$<$30	&	$[$1$]$; $[$1$]$	&		\\
J04554757+3028077	&	 ... 	&		&		&	$<$7	&		&	$<$63	&	$[$$]$; $[$1$]$	&		\\
J04554801+3028050	&	 ... 	&	10.9 $\pm$ 0.4	&		&	11 $\pm$ 3	&		&	$<$63	&	$[$1$]$; $[$1$]$	&		\\
J04554820+3030160	&	XEST 26-052	&		&		&		&		&		&		&	f	\\
J04554969+3019400	&	 ... 	&	3.9 $\pm$ 0.2	&		&	$<$5	&		&	$<$17	&	$[$1$]$; $[$1$]$	&		\\
J04555288+3006523	&	 ... 	&		&		&	$<$7	&		&	$<$16	&	$[$$]$; $[$1$]$	&		\\
J04555605+3036209	&	XEST 26-062	&	226 $\pm$ 7	&		&	330 $\pm$ 4	&		&	639 $\pm$ 39	&	$[$1$]$; $[$1$]$	&		\\
J04555636+3049374	&	 ... 	&		&		&	$<$5	&		&	$<$35	&	$[$$]$; $[$1$]$	&		\\
J04574903+3015195	&	 ... 	&		&		&	$<$9	&		&	$<$15	&	$[$$]$; $[$1$]$	&		\\
J05061674+2446102	&	CIDA 10	&		&		&	$<$6	&		&	$<$42	&	$[$$]$; $[$1$]$	&		\\
J05064662+2104296	&	 ... 	&		&		&	$<$6	&		&	$<$22	&	$[$$]$; $[$1$]$	&		\\
J05075496+2500156	&	CIDA 12	&		&		&	51 $\pm$ 3	&		&	44 $\pm$ 6	&	$[$$]$; $[$1$]$	&		\\

\end{longtable}
\tablefoot{Upper limits are reported at a 3$\sigma$ level. The errors listed are the 1$\sigma$ statistical measurement errors. Uncertainties in the absolute flux calibration with {\it Spitzer} MIPS at 24~$\mu$m and 70~$\mu$m are 4\% and 7\% respectively. Uncertainties in the absolute flux calibration with {\it Herschel} PACS at 70~$\mu$m, 100~$\mu$m and 160~$\mu$m are 2.64\%, 2.75\% and 4.15\% respectively. Flux densities reported in parenthesis indicates the known multiple systems that is unresolved at the observed wavelength. \\
{\bf a.} {\it Herschel} PACS fluxes measured from the level 2.5 processed maps observed under the program KPGT\_pandre\_1. \\
{\bf b.} GG Tau Ba+Bb is unresolved from the primary system (GG Tau Aa+Ab) at 100 $\mu$m and 160 $\mu$m \citep{Howard:2013}. \\
{\bf c.} Multiple system of which the target listed is resolved with {\it Herschel} PACS in the Blue1 and/or Blue2 channel(s) but unresolved from its companion in the Red channel. \\
{\bf d.} Multiple system of which the target listed is the later type secondary. The {\it Herschel} PACS flux densities reported are contaminated from the primary companion. \\
{\bf e.} IRAS 04325+2402 falls outside of the coverage region from the KPGT\_pandre\_1 processed map at 160 $\mu$m. \\
{\bf f.} No far-IR observation exists for this target. \\
}
\tablebib{{\it Spitzer} MIPS-1 references are listed in the first square parenthesis. MIPS-2 measurements are those reported in \citet{Rebull:2010}. {\it Herschel} PACS references are listed second square parenthesis. \\ 
{\it Spitzer} MIPS: (1) \citet{Luhman:2010}; (2) \citet{Rebull:2010}. \\
{\it Herschel} PACS: (1) This work; (2) \citet{Harvey:2012}; (3) \citet{Howard:2013}.
}
\end{landscape}
\twocolumn
\end{longtab}

\begin{longtab}
\onecolumn
\begin{landscape}
\begin{longtable}{l c l l l l l l l r r}
\caption{\label{Table:submm} Submillimeter and and millimeter photometry compilation of the known 154 M4-L0 Taurus members, ordered by target RA.} \\
\hline\hline
2MASS & Other name & 350 ${\mu}$m & 450 ${\mu}$m  & 850 ${\mu}$m & 880 ${\mu}$m & 1.2 mm & 1.3 mm & 2.6 mm & Reference &  Notes. \\
& & (mJy) & (mJy) & (mJy) & (mJy) & (mJy) & (mJy) & (mJy) \\
\hline
\endfirsthead
\caption{continued.}\\
\hline\hline
2MASS & Other name & 350 ${\mu}$m & 450 ${\mu}$m  & 850 ${\mu}$m & 880 ${\mu}$m & 1.2 mm & 1.3 mm & 2.6 mm & Reference &  Notes. \\
& & (mJy) & (mJy) & (mJy) & (mJy) & (mJy) & (mJy) & (mJy) \\
\hline
\endhead
\hline
\endfoot
J04034997+2620382	&	XEST 06-006	&		&		&		&		&		&		&		&		&		\\
J04131414+2819108	&	LkCa 1	&		&	$<$89	&	$<$8	&		&		&	$<$14	&		&	$[$1$]$; $[$4$]$	&		\\
J04141188+2811535	&	 ... 	&		&		&		&		&		&	$<$2	&		&	$[$$]$; $[$8$]$	&		\\
J04141458+2827580	&	FN Tau	&		&		&		&		&		&	$<$18	&	$<$5	&	$[$$]$; $[$7$]$	&		\\
J04141760+2806096	&	CIDA 1	&		&		&		&		&		&	14 $\pm$ 3	&	$<$8	&	$[$$]$; $[$7$]$	&		\\
J04144730+2646264	&	FP Tau	&		&		&		&		&		&	$<$9	&	$<$9	&	$[$$]$; $[$7$]$	&		\\
J04144739+2803055	&	XEST 20-066	&		&		&		&		&		&		&		&		&		\\
J04150515+2808462	&	CIDA 2	&		&	$<$165	&	$<$14	&		&		&		&		&	$[$1$]$; $[$$]$	&		\\
J04151471+2800096	&	KPNO 1	&		&		&		&		&		&	$<$2	&		&	$[$$]$; $[$8$]$	&		\\
J04152409+2910434	&	 ... 	&		&		&		&		&		&		&		&		&		\\
J04155799+2746175	&	 ... 	&		&		&		&		&		&		&		&		&		\\
J04161210+2756385	&	 ... 	&		&		&		&		&		&	$<$4	&		&	$[$$]$; $[$1$]$	&		\\
J04161885+2752155	&	 ... 	&		&		&		&		&		&		&		&		&		\\
J04162725+2053091	&	 ... 	&		&		&		&		&		&		&		&		&		\\
J04163048+3037053	&	 ... 	&		&		&		&		&		&		&		&		&		\\
J04163911+2858491	&	 ... 	&		&		&		&		&		&		&		&		&		\\
J04174955+2813318	&	KPNO 10	&		&		&		&		&		&	8 $\pm$ 1	&		&	$[$$]$; $[$1$]$	&		\\
J04174965+2829362	&	V410 X-ray 1	&		&	$<$94	&	$<$9	&		&		&		&		&	$[$1$]$; $[$$]$	&		\\
J04180796+2826036	&	V410 X-ray 3	&		&		&		&		&		&		&		&		&		\\
J04181710+2828419	&	V410 Anon 13	&		&		&		&		&		&	$<$9	&	$<$7	&	$[$$]$; $[$7$]$	&		\\
J04183030+2743208	&	KPNO 11	&		&		&		&		&		&		&		&		&		\\
J04184023+2824245	&	V410 X-ray 4	&		&		&		&		&		&		&		&		&		\\
J04185115+2814332	&	KPNO 2	&		&		&		&		&		&	1.8 $\pm$ 0.8	&		&	$[$$]$; $[$8$]$	&		\\
J04185813+2812234	&	IRAS 04158+2805	&		&	1600 $\pm$ 400	&	440 $\pm$ 40	&		&		&	110 $\pm$ 5	&		&	$[$2$]$; $[$3$]$	&		\\
J04190110+2819420	&	V410 X-ray 6	&		&		&		&		&		&	$<$2	&		&	$[$$]$; $[$1$]$	&		\\
J04190126+2802487	&	KPNO 12	&		&		&		&		&		&	$<$2	&		&	$[$$]$; $[$8$]$	&		\\
J04190197+2822332	&	V410 X-ray 5a	&		&		&		&		&		&		&		&		&		\\
J04193545+2827218	&	FR Tau	&		&		&		&		&		&	$<$15	&		&	$[$$]$; $[$4$]$	&		\\
J04194657+2712552	&	[GKH94] 41	&		&		&		&		&		&		&		&		&		\\
J04201611+2821325	&	 ... 	&		&		&		&		&		&	$<$2	&		&	$[$$]$; $[$1$]$	&		\\
J04202555+2700355	&	 ... 	&		&		&		&		&		&	8 $\pm$ 1	&		&	$[$$]$; $[$1$]$	&		\\
J04202583+2819237	&	IRAS 04173+2812	&		&		&		&		&		&	$<$2	&		&	$[$$]$; $[$1$]$	&		\\
J04203918+2717317	&	XEST 16-045	&		&		&		&		&		&		&		&		&		\\
J04205273+1746415	&	J2-157	&		&		&		&		&		&	$<$14	&		&	$[$$]$; $[$4$]$	&		\\
J04210795+2702204	&	CFHT 19	&		&		&		&		&		&	$<$3	&		&	$[$$]$; $[$1$]$	&		\\
J04210934+2750368	&	 ... 	&		&		&		&		&		&	$<$2	&		&	$[$$]$; $[$1$]$	&		\\
J04213459+2701388	&	 ... 	&		&		&		&		&		&	$<$3	&		&	$[$$]$; $[$1$]$	&		\\
J04214013+2814224	&	XEST 21-026	&		&		&		&		&		&		&		&		&		\\
J04214631+2659296	&	CFHT 10	&		&		&		&		&		&	$<$4	&		&	$[$$]$; $[$1$]$	&		\\
J04215450+2652315	&	 ... 	&		&		&		&		&		&		&		&		&		\\
J04220007+1530248	&	IRAS 04191+1523 B	&		&	1800 $\pm$ 220	&	440 $\pm$ 20	&		&		&	110 $\pm$ 7	&		&	$[$5$]$; $[$3$]$	&	a	\\
J04221332+1934392	&	 ... 	&		&		&		&		&		&		&		&		&		\\
J04221644+2549118	&	CFHT 14	&		&		&		&		&		&		&		&		&		\\
J04222404+2646258	&	XEST 11-087	&		&		&		&		&		&		&		&		&		\\
J04230607+2801194	&	 ... 	&		&		&		&		&		&	5 $\pm$ 1	&		&	$[$$]$; $[$1$]$	&		\\
J04233539+2503026	&	FU Tau A	&		&		&		&		&		&	$<$2	&		&	$[$$]$; $[$1$]$	&		\\
J04233573+2502596	&	FU Tau B	&		&		&		&		&		&	$<$2	&		&	$[$$]$; $[$1$]$	&		\\
J04242090+2630511	&	 ... 	&		&		&		&		&		&	$<$1	&		&	$[$$]$; $[$1$]$	&		\\
J04242646+2649503	&	CFHT 9	&		&		&		&		&		&	$<$3	&		&	$[$$]$; $[$1$]$	&		\\
J04244506+2701447	&	J1-4423	&		&	$<$52	&	$<$8	&		&		&	$<$11	&		&	$[$1$]$; $[$4$]$	&		\\
J04251550+2829275	&	 ... 	&		&		&		&		&		&		&		&		&		\\
J04262939+2624137	&	KPNO 3	&		&		&		&		&		&	6 $\pm$ 1	&		&	$[$$]$; $[$1$]$	&		\\
J04263055+2443558	&	 ... 	&		&		&		&		&		&		&		&		&		\\
J04265732+2606284	&	KPNO 13	&		&		&		&		&		&		&		&		&		\\
J04270739+2215037	&	 ... 	&		&		&		&		&		&		&		&		&		\\
J04272799+2612052	&	KPNO 4	&		&		&		&		&		&	$<$3	&		&	$[$$]$; $[$8$]$	&		\\
J04274538+2357243	&	CFHT 15	&		&		&		&		&		&		&		&		&		\\
J04275730+2619183	&	IRAS 04248+2612	&	1178 $\pm$ 30	&	2960 $\pm$ 1024	&	560 $\pm$ 178	&		&		&	60 $\pm$ 7	&		&	$[$9$]$; $[$3$]$	&	c	\\
 ... 	&	L1521F-IRS	&	4600 $\pm$ 700	&	7000 $\pm$ 1800	&	1500 $\pm$ 500	&		&	600 $\pm$ 150	&		&		&	$[$3$]$; $[$$]$	&	b	\\
J04284263+2714039	&	 ... 	&		&		&		&		&		&		&		&		&		\\
J04290068+2755033	&	 ... 	&		&		&		&		&		&	$<$2	&		&	$[$$]$; $[$1$]$	&		\\
J04292071+2633406	&	J1-507	&		&	$<$52	&	$<$6	&		&		&	$<$14	&		&	$[$1$]$; $[$4$]$	&		\\
J04292165+2701259	&	IRAS 04263+2654	&		&		&		&		&		&	$<$4	&		&	$[$$]$; $[$1$]$	&		\\
J04292971+2616532	&	FW Tau A+B+C	&		&	$<$35	&	5 $\pm$ 1	&		&		&	$<$15	&		&	$[$1$]$; $[$2$]$	&		\\
J04294568+2630468	&	KPNO 5	&		&		&		&		&		&	$<$2	&		&	$[$$]$; $[$8$]$	&		\\
J04295422+1754041	&	 ... 	&		&		&		&		&		&		&		&		&		\\
J04295950+2433078	&	CFHT 20	&		&		&		&		&		&	$<$3	&		&	$[$$]$; $[$1$]$	&		\\
J04300724+2608207	&	KPNO 6	&		&		&		&		&		&	$<$2	&		&	$[$$]$; $[$8$]$	&		\\
J04302365+2359129	&	CFHT 16	&		&		&		&		&		&		&		&		&		\\
J04305171+2441475	&	ZZ Tau IRS	&		&		&		&		&		&	106 $\pm$ 2	&		&	$[$$]$; $[$1$]$	&		\\
J04305718+2556394	&	KPNO 7	&		&		&		&		&		&	$<$3	&		&	$[$$]$; $[$8$]$	&		\\
J04311578+1820072	&	MHO 9	&		&		&		&		&		&		&		&		&		\\
J04311907+2335047	&	 ... 	&		&		&		&		&		&		&		&		&		\\
J04312382+2410529	&	V927 Tau A+B	&		&	$<$1030	&	$<$10	&		&		&	$<$20	&		&	$[$1$]$; $[$2$]$	&		\\
J04312405+1800215	&	MHO 4	&		&		&		&		&		&	$<$10	&	$<$8	&	$[$$]$; $[$7$]$	&		\\
J04312669+2703188	&	CFHT 13	&		&		&		&		&		&		&		&		&		\\
J04313613+1813432	&	LkHa 358	&		&		&		&		&		&	17 $\pm$ 1	&	4 $\pm$ 0.6	&	$[$$]$; $[$4, 7$]$	&		\\
J04315844+2543299	&	J1-665	&		&		&		&		&		&		&		&		&		\\
J04320329+2528078	&	 ... 	&		&		&		&		&		&		&		&		&		\\
J04321606+1812464	&	MHO 5	&		&		&		&		&		&	$<$9	&	$<$8	&	$[$$]$; $[$7$]$	&		\\
J04321786+2422149	&	CFHT 7	&		&		&		&		&		&		&		&		&		\\
J04322210+1827426	&	MHO 6	&		&		&		&		&		&	14 $\pm$ 2	&		&	$[$$]$; $[$1$]$	&		\\
J04322329+2403013	&	 ... 	&		&		&		&		&		&		&		&		&		\\
J04322415+2251083	&	 ... 	&		&		&		&		&		&	$<$6	&		&	$[$$]$; $[$1$]$	&		\\
J04322627+1827521	&	MHO 7	&		&		&		&		&		&		&		&		&		\\
J04323028+1731303	&	GG Tau Ba+Bb	&		&		&		&		&		&	$<$5	&		&	$[$6$]$; $[$$]$	&		\\
J04324938+2253082	&	JH 112 B	&		&		&		&		&		&		&		&		&		\\
J04325026+2422115	&	 ... 	&		&		&		&		&		&		&		&		&		\\
J04325119+1730092	&	LH 0429+17	&		&		&		&		&		&		&		&		&		\\
J04330197+2421000	&	MHO 8	&		&		&		&		&		&		&		&		&		\\
J04330781+2616066	&	KPNO 14	&		&		&		&		&		&		&		&		&		\\
J04330945+2246487	&	CFHT 12	&		&		&	4 $\pm$ 1	&		&		&	$<$3	&		&	$[$8$]$; $[$1$]$	&		\\
J04332621+2245293	&	XEST 17-036	&		&		&		&		&		&		&		&		&		\\
J04334171+1750402	&	 ... 	&		&		&		&		&		&	6 $\pm$ 2	&		&	$[$$]$; $[$1$]$	&		\\
J04334291+2526470	&	 ... 	&		&		&		&		&		&		&		&		&		\\
J04334465+2615005	&	 ... 	&		&		&		&		&		&	18 $\pm$ 2	&		&	$[$$]$; $[$1$]$	&		\\
J04335245+2612548	&	 ... 	&		&		&		&		&		&		&		&		&		\\
J04335252+2256269	&	XEST 17-059	&		&		&		&		&		&		&		&		&		\\
J04341527+2250309	&	CFHT 1	&		&		&		&		&		&	$<$3	&		&	$[$$]$; $[$8$]$	&		\\
J04344544+2308027	&	 ... 	&		&		&		&		&		&		&		&		&		\\
J04350850+2311398	&	CFHT 11	&		&		&		&		&		&		&		&		&		\\
J04353536+2408266	&	IRAS 04325+2402 C	&		&	606 $\pm$ 185	&	186 $\pm$ 11	&		&		&	110 $\pm$ 7	&		&	$[$1$]$; $[$3$]$	&	a	\\
J04354183+2234115	&	KPNO 8	&		&		&		&		&		&		&		&		&		\\
J04354203+2252226	&	XEST 08-033	&		&		&		&		&		&		&		&		&		\\
J04354526+2737130	&	 ... 	&		&		&		&		&		&		&		&		&		\\
J04355143+2249119	&	KPNO 9	&		&		&		&		&		&	$<$2	&		&	$[$$]$; $[$8$]$	&		\\
J04355209+2255039	&	XEST 08-047	&		&		&		&		&		&		&		&		&		\\
J04355286+2250585	&	XEST 08-049	&		&		&		&		&		&		&		&		&		\\
J04355760+2253574	&	 ... 	&		&		&		&		&		&		&		&		&		\\
J04361030+2159364	&	 ... 	&		&		&		&		&		&		&		&		&		\\
J04361038+2259560	&	CFHT 2	&		&		&	$<$6	&		&		&	$<$2	&		&	$[$7$]$; $[$7$]$	&		\\
J04362151+2351165	&	 ... 	&		&		&		&		&		&	$<$5	&		&	$[$$]$; $[$1$]$	&		\\
J04363893+2258119	&	CFHT 3	&		&		&		&		&		&	$<$2	&		&	$[$$]$; $[$8$]$	&		\\
J04373705+2331080	&	 ... 	&		&		&		&		&		&		&		&		&		\\
J04380083+2558572	&	ITG 2	&		&		&		&		&		&		&		&		&		\\
J04381486+2611399	&	 ... 	&		&		&		&		&		&	2.3 $\pm$ 0.8	&		&	$[$$]$; $[$8$]$	&		\\
J04381630+2326402	&	 ... 	&		&		&		&		&		&		&		&		&		\\
J04382134+2609137	&	GM Tau	&		&		&	$<$3	&		&		&	$<$5	&	$<$2	&	$[$8$]$; $[$7$]$	&		\\
J04385859+2336351	&	 ... 	&		&		&		&		&		&	13 $\pm$ 2	&		&	$[$$]$; $[$1$]$	&		\\
J04385871+2323595	&	 ... 	&		&		&		&		&		&		&		&		&		\\
J04390163+2336029	&	 ... 	&		&		&		&		&		&	$<$7	&		&	$[$$]$; $[$1$]$	&		\\
J04390396+2544264	&	CFHT 6	&		&		&		&		&		&	2.4 $\pm$ 0.4	&		&	$[$$]$; $[$5$]$	&		\\
J04390637+2334179	&	 ... 	&		&		&		&		&		&		&		&		&		\\
J04393364+2359212	&	 ... 	&		&		&		&		&		&	$<$6	&		&	$[$$]$; $[$1$]$	&		\\
J04394488+2601527	&	ITG 15	&		&		&		&		&		&	$<$6	&		&	$[$$]$; $[$1$]$	&		\\
J04394748+2601407	&	CFHT 4	&		&		&	11 $\pm$ 2	&		&		&	2.4 $\pm$ 0.8	&		&	$[$7$]$; $[$7$]$	&		\\
J04400067+2358211	&	 ... 	&		&		&		&		&		&	$<$6	&		&	$[$$]$; $[$1$]$	&		\\
J04400174+2556292	&	CFHT 17	&		&		&		&		&		&		&		&		&		\\
J04403979+2519061	&	 ... 	&		&		&		&		&		&	$<$2	&		&	$[$$]$; $[$1$]$	&		\\
J04410424+2557561	&	Haro 6-32	&		&		&		&		&		&		&		&		&		\\
J04411078+2555116	&	ITG 34	&		&		&		&		&		&	$<$5	&		&	$[$$]$; $[$1$]$	&		\\
J04414489+2301513	&	 ... 	&		&		&		&	$<$24	&		&		&		&	$[$6$]$; $[$$]$	&		\\
J04414565+2301580	&	 ... 	&		&		&		&	$<$24	&		&		&		&	$[$6$]$; $[$$]$	&		\\
J04414825+2534304	&	 ... 	&		&		&		&		&		&	2.2 $\pm$ 0.4	&		&	$[$$]$; $[$5$]$	&		\\
J04422101+2520343	&	CIDA 7	&		&	990 $\pm$ 330	&	38 $\pm$ 8	&		&		&	$<$19	&	$<$5	&	$[$1$]$; $[$7$]$	&	d	\\
J04432023+2940060	&	CIDA 14	&		&		&		&		&		&	$<$5	&	$<$2	&	$[$$]$; $[$7$]$	&		\\
J04442713+2512164	&	IRAS S04414+2506	&		&	36 $\pm$ 15	&	10 $\pm$ 1	&		&		&	5.2 $\pm$ 0.3	&		&	$[$4, 8$]$; $[$6$]$	&		\\
J04464260+2459034	&	RXJ 04467+2459	&		&		&		&		&		&	$<$5	&	$<$2	&	$[$$]$; $[$7$]$	&		\\
J04484189+1703374	&	 ... 	&		&		&		&		&		&		&		&		&		\\
J04520668+3047175	&	IRAS 04489+3042	&		&		&		&		&		&	15 $\pm$ 4	&		&	$[$$]$; $[$3$]$	&		\\
J04552333+3027366	&	 ... 	&		&		&		&		&		&	$<$3	&		&	$[$$]$; $[$8$]$	&		\\
J04554046+3039057	&	 ... 	&		&		&		&		&		&		&		&		&		\\
J04554535+3019389	&	 ... 	&		&		&		&		&		&	$<$5	&		&	$[$$]$; $[$1$]$	&		\\
J04554757+3028077	&	 ... 	&		&		&		&	$<$23	&		&		&		&	$[$6$]$; $[$$]$	&		\\
J04554801+3028050	&	 ... 	&		&		&		&	$<$23	&		&		&		&	$[$6$]$; $[$$]$	&		\\
J04554820+3030160	&	XEST 26-052	&		&		&		&		&		&		&		&		&		\\
J04554969+3019400	&	 ... 	&		&		&		&		&		&	$<$3	&		&	$[$$]$; $[$1$]$	&		\\
J04555288+3006523	&	 ... 	&		&		&		&		&		&		&		&		&		\\
J04555605+3036209	&	XEST 26-062	&		&		&		&		&		&	9 $\pm$ 2	&		&	$[$$]$; $[$1$]$	&		\\
J04555636+3049374	&	 ... 	&		&		&		&		&		&		&		&		&		\\
J04574903+3015195	&	 ... 	&		&		&		&		&		&	$<$2	&		&	$[$$]$; $[$8$]$	&		\\
J05061674+2446102	&	CIDA 10	&		&	$<$94	&	$<$11	&		&		&		&		&	$[$1$]$; $[$$]$	&		\\
J05064662+2104296	&	 ... 	&		&		&		&		&		&		&		&		&		\\
J05075496+2500156	&	CIDA 12	&		&	$<$88	&	$<$7	&		&		&	$<$5	&	$<$2	&	$[$1$]$; $[$7$]$	&		\\

\end{longtable}
\tablefoot{Upper limits are reported at a 3$\sigma$ level. The errors listed are the 1$\sigma$ statistical measurement errors. The typical quoted, absolute flux calibration uncertainties at 350~$\mu$m and 450~$\mu$m are $\sim$25\%; at 850~$\mu$m is $\sim$10\%, and with SCUBA-2 at 850~$\mu$m is $\sim$5\% \citep{Mohanty:2013}. The reported calibration uncertainties at millimeter wavelengths are $\sim$10-20\%. \\
{\bf a.} Multiple system of which the flux densities are unresolved. \\
{\bf b.} The listed sub-millimeter and millimeter uncertainties include both the statistical measurement errors and uncertainties in the absolute calibration. \\
{\bf c.} The 450 $\mu$m and 850 $\mu$m uncertainties include both the statistical measurement errors and uncertainties in the absolute calibration. \\
{\bf d.} An anomalous continuum slope is reported in \citet{Andrews:2005} for this member. \\
}
\tablebib{References for observations at 350 - 1200 $\mu$m are listed in the first square parenthesis. References for observations at 1.3 mm and/or 2.6 mm are listed in the second square parenthesis. \\
350 - 1200 $\mu$m: (1) \citet{Andrews:2005}; (2) \citet{Andrews:2008}; (3) \citet{Bourke:2006}; (4) \citet{Bouy:2008}; (5) \citet{DiFrancesco:2008}; (6) \citet{Harris:2012}; (7)\citet{Klein:2003}; (8) \citet{Mohanty:2013}; (9) \citet{Young:2003}. \\
1.3 - 2.6 mm:  (1) \citet{Andrews:2013}; (2) \citet{Beckwith:1990}; (3) \citet{Motte:2001}; (4) \citet{Osterloh:1995}; (5) \citet{PhanBao:2011}; (6) \citet{Ricci:2013}; (7) \citet{Schaefer:2009}; (8) \citet{Scholz:2006}.
}
\end{landscape}
\twocolumn
\end{longtab}


\section{{\itshape {\bfseries Herschel}} PACS maps}
\label{Sec:maps}
\begin{figure*}
\centering
\includegraphics[scale=0.28]{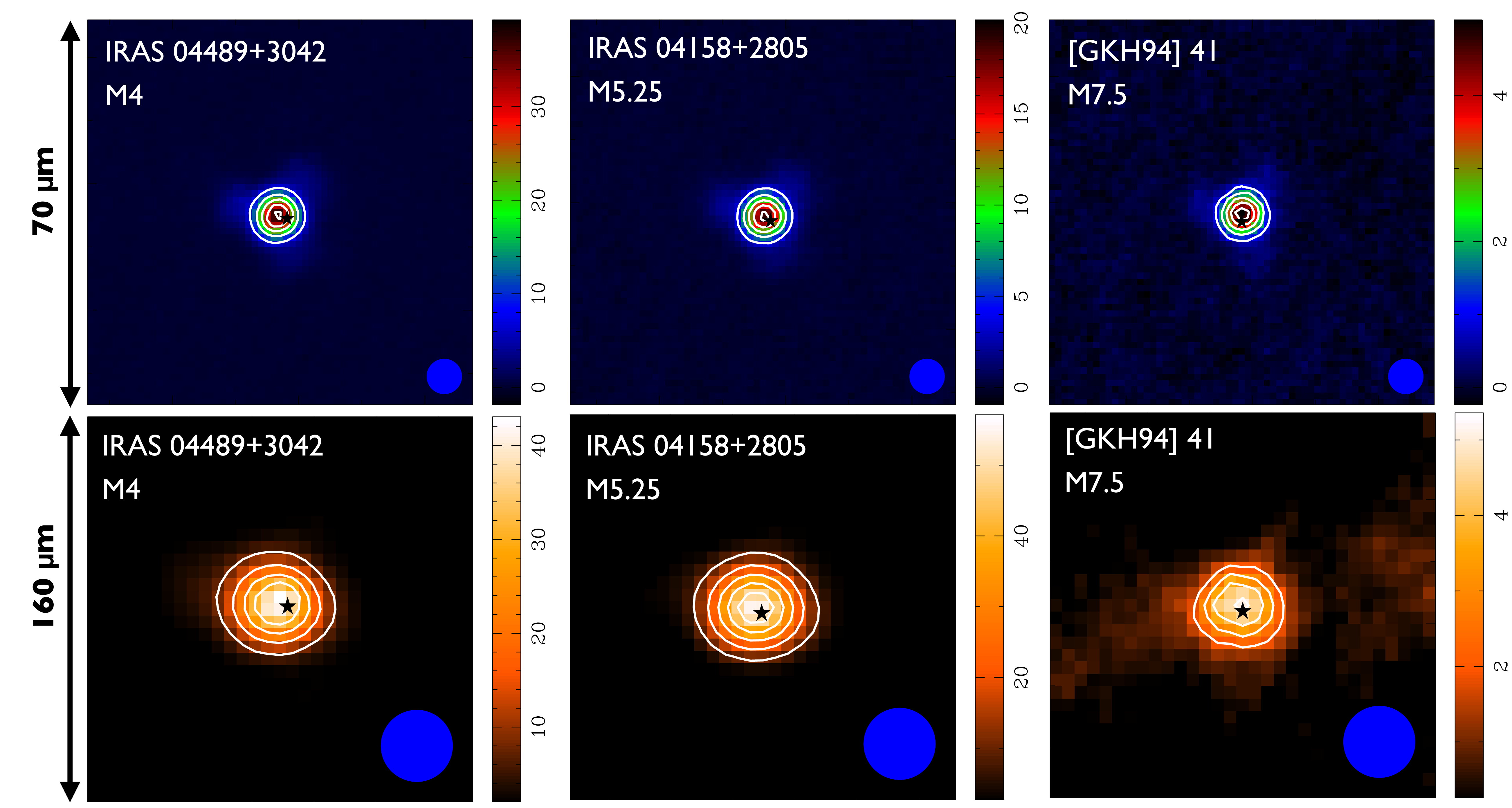}
\caption {Class~I objects with PACS Blue \& Red channel detections.}
\label{mapsClass1}
\end{figure*}

\begin{figure*}
\centering
\includegraphics[scale=0.28]{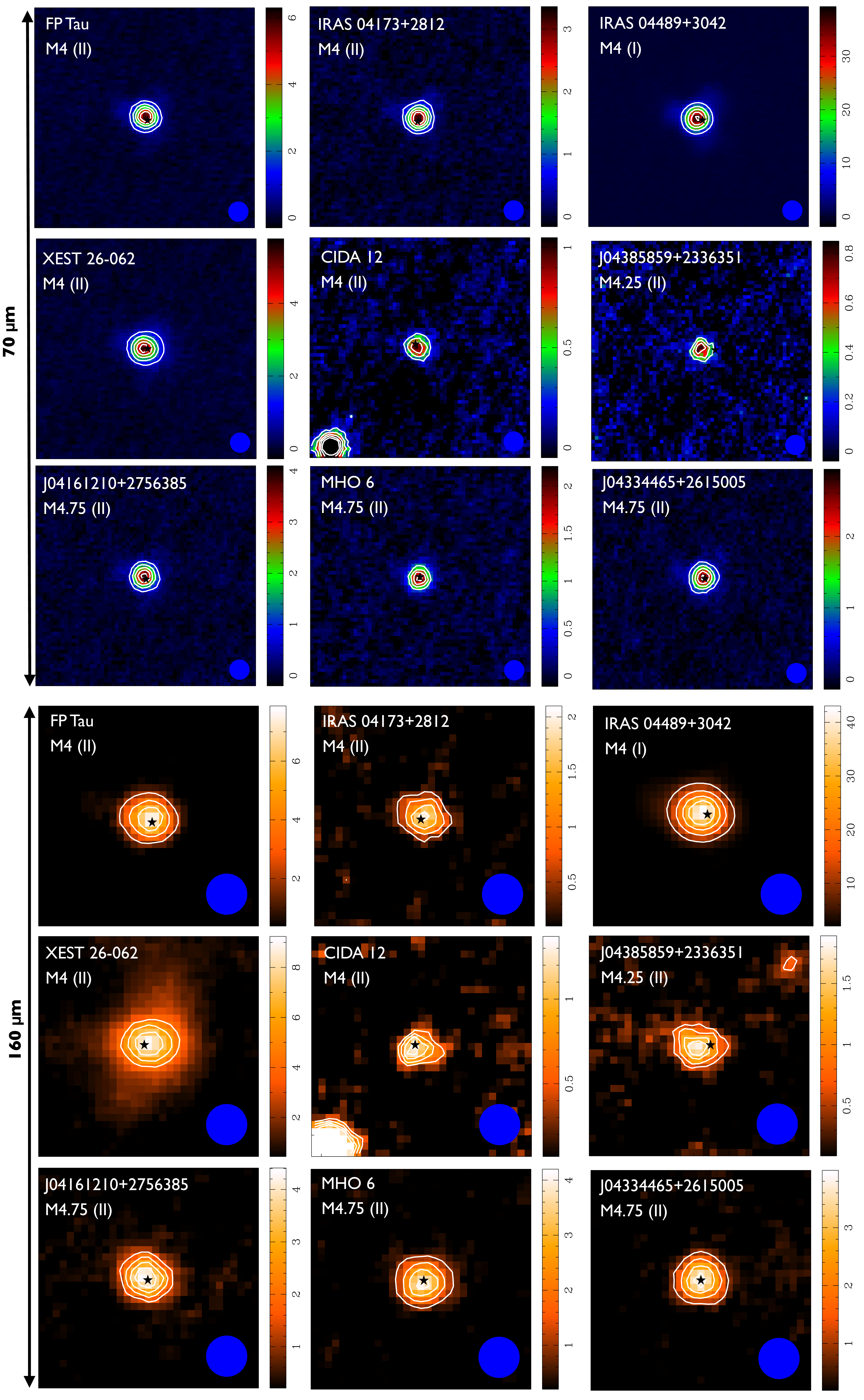}
\caption {Class~II objects with PACS Blue \& Red channel detections.}
\label{maps2a}
\end{figure*}

\begin{figure*}
\centering
\includegraphics[scale=0.28]{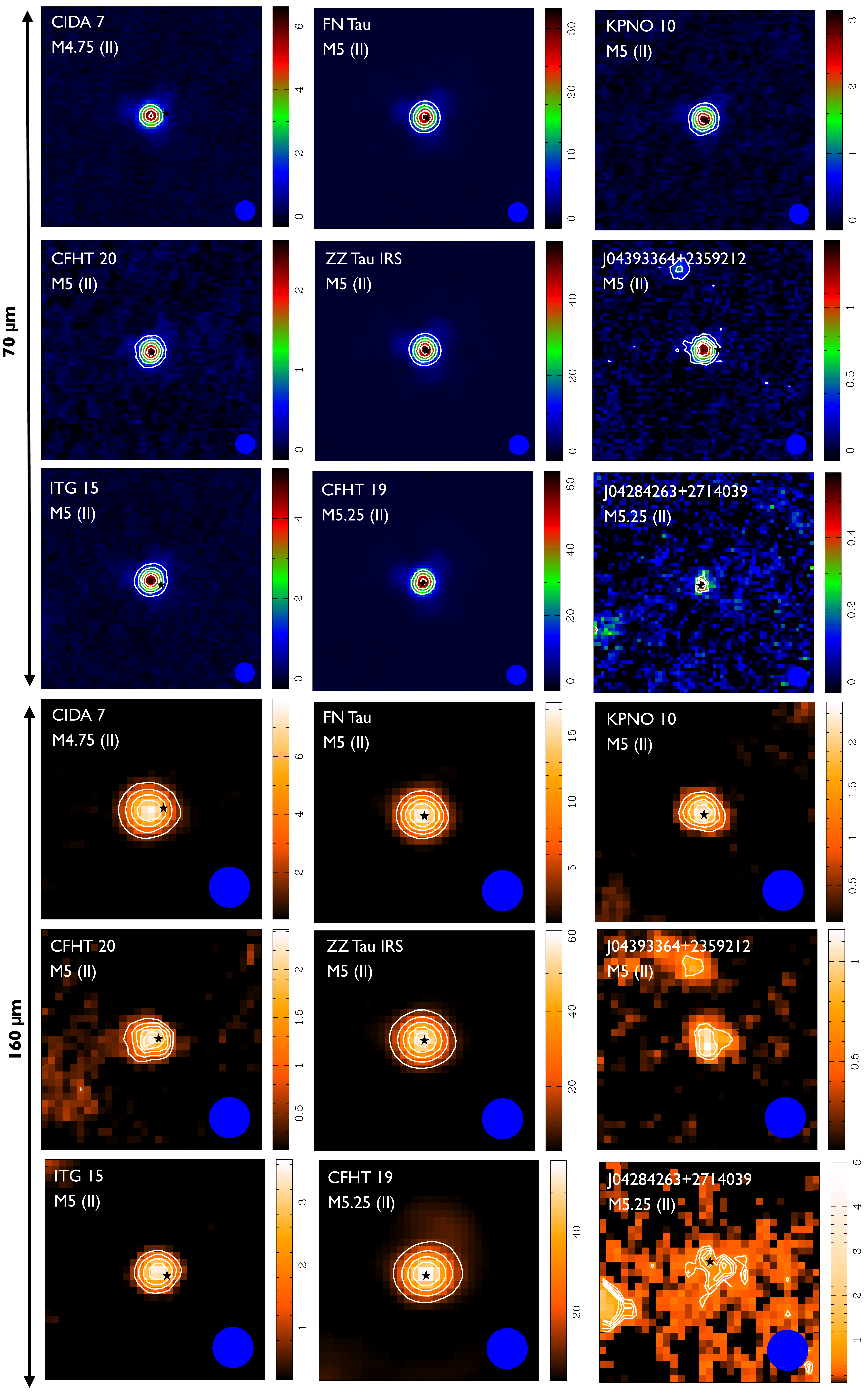}
\caption {Class~II objects with PACS Blue \& Red channel detections.}
\label{maps2b}
\end{figure*}

\begin{figure*}
\centering
\includegraphics[scale=0.28]{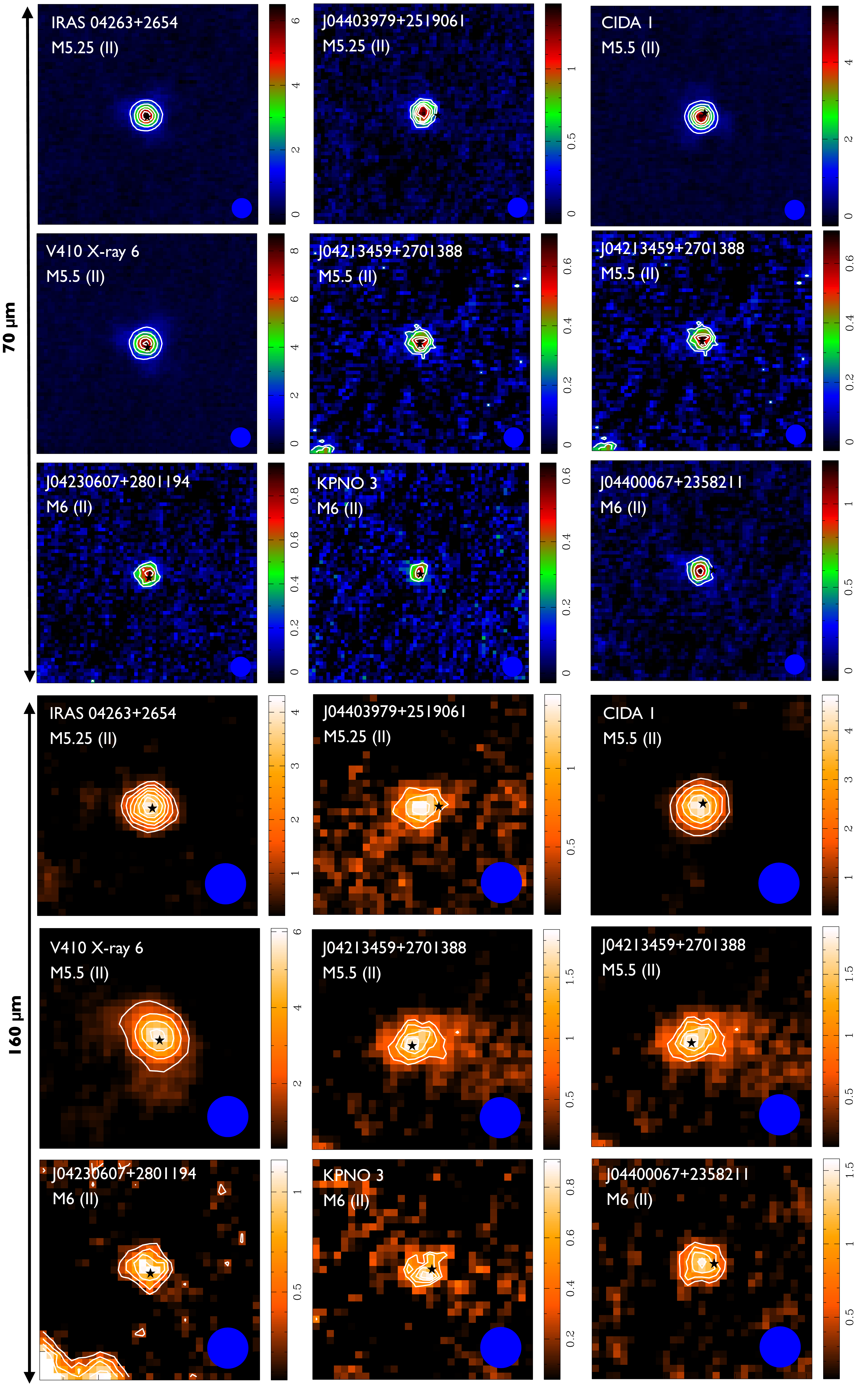}
\caption {Class II objects with PACS Blue \& Red channel detections.}
\label{maps2c}
\end{figure*}

\begin{figure*}
\centering
\includegraphics[scale=0.28]{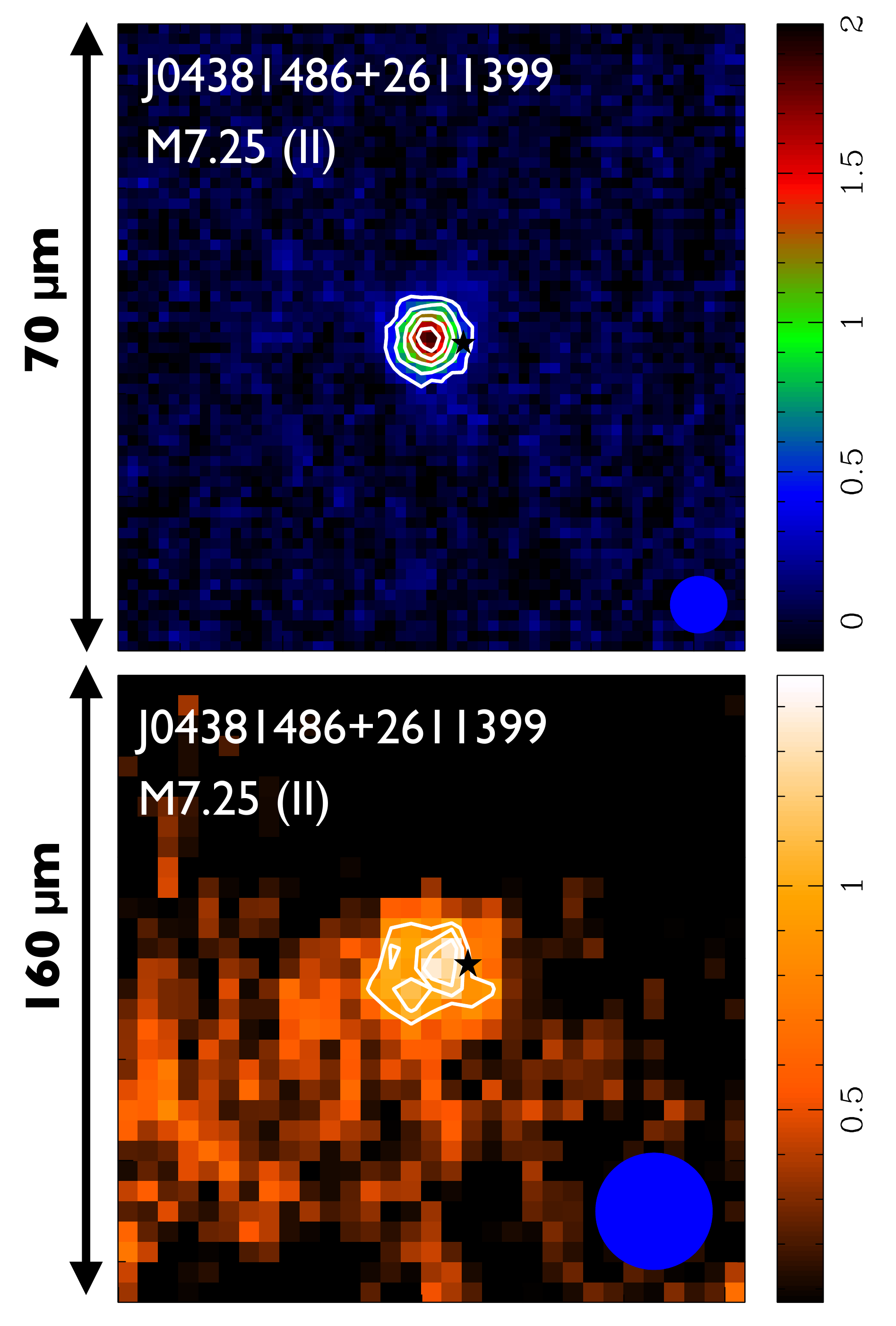}
\caption {Class~II objects with PACS Blue \& Red channel detections.}
\label{maps2d}
\end{figure*}

\begin{figure*}
\centering
\includegraphics[scale=0.28]{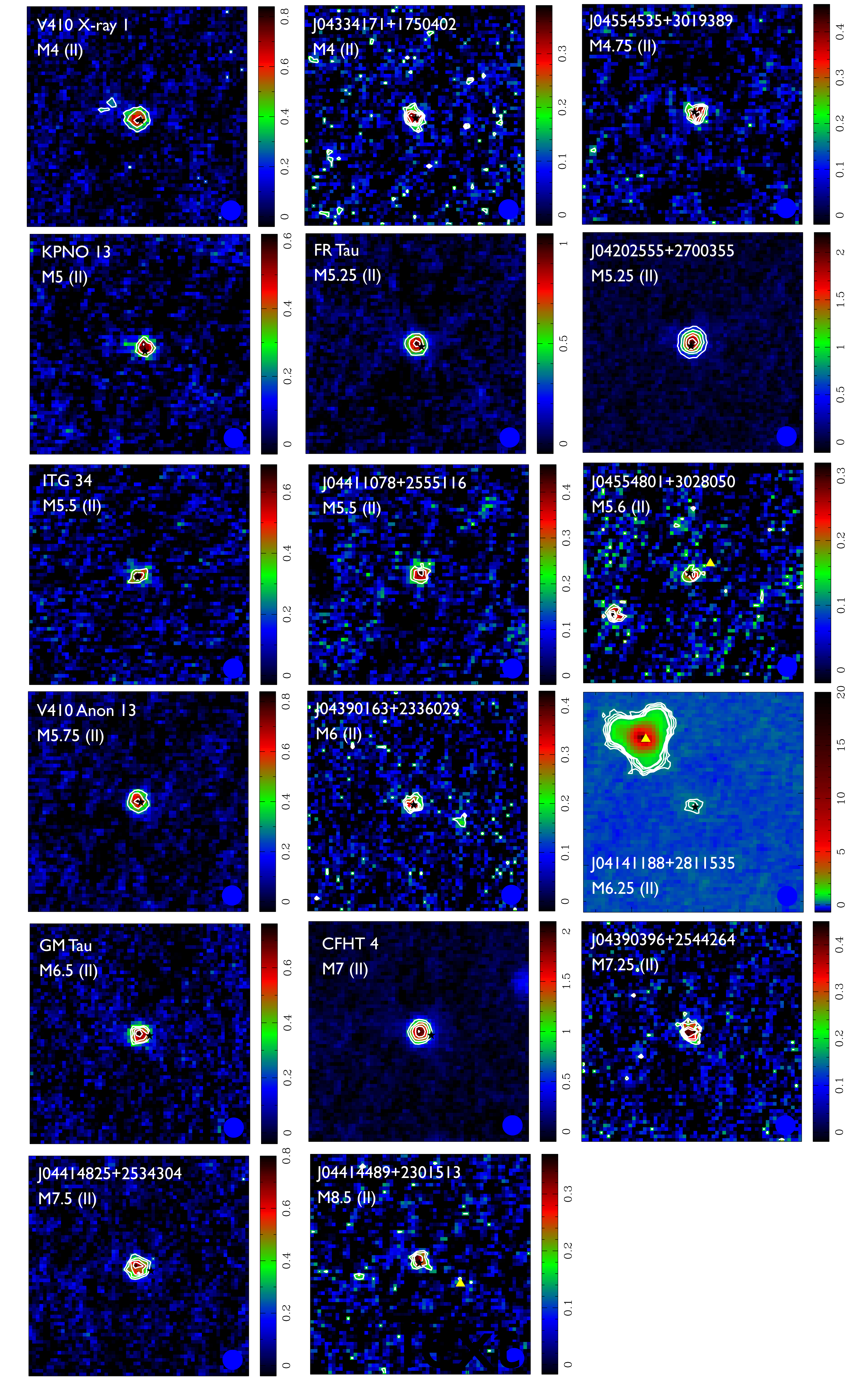}
\caption {Class~II objects with PACS Blue channel detections only.}
\label{maps2e}
\end{figure*}

\begin{figure*}
\centering
\includegraphics[scale=0.28]{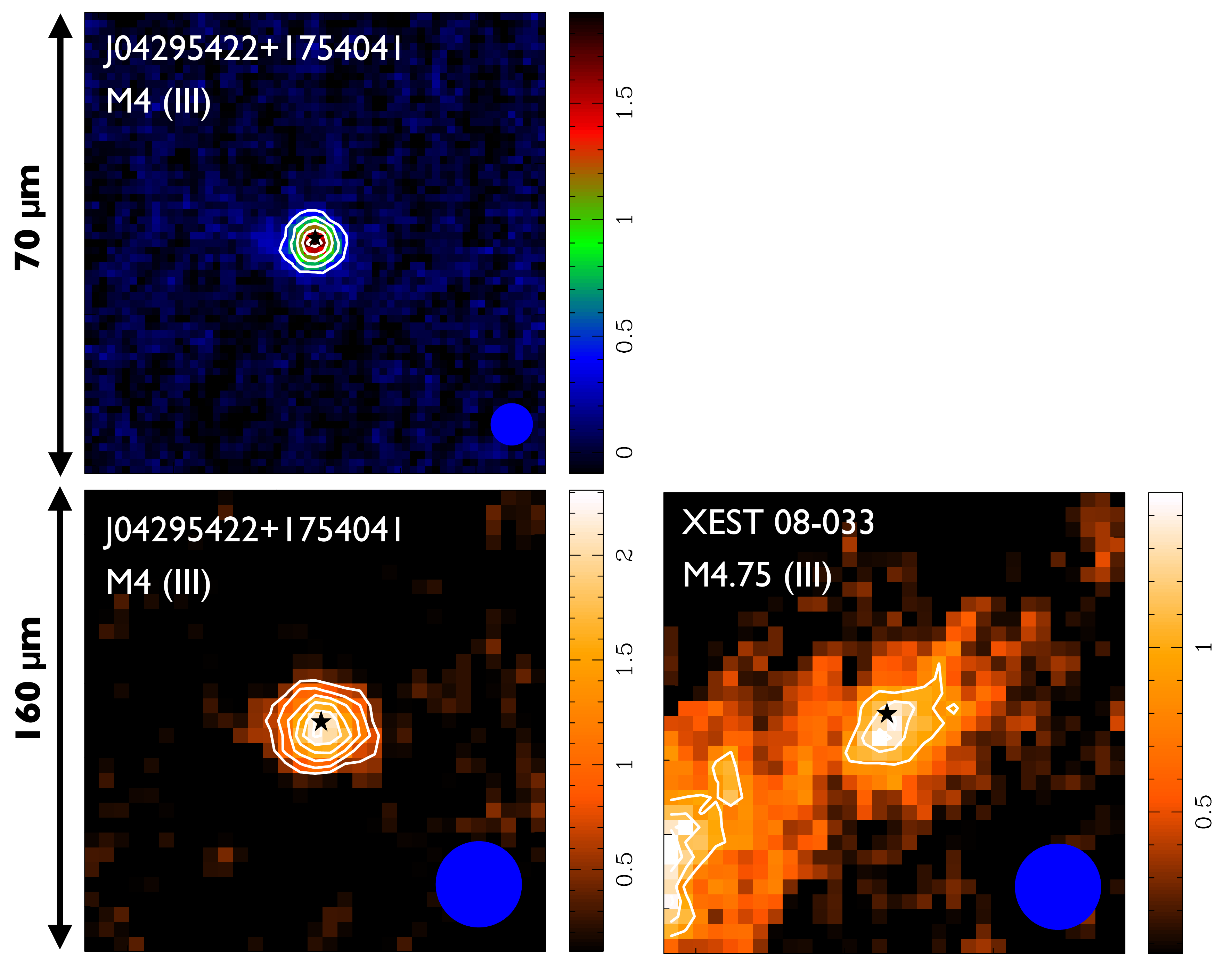}
\caption {Class~III objects with PACS Blue and/or Red channel detections.}
\label{maps3}
\end{figure*}

The {\it Herschel} PACS maps of the detected targets observed under this program shown in Figures~\ref{mapsClass1}-\ref{maps3}, and are displayed in order of spectral class and type. Target name and spectral type are labeled in the top left-hand corner of each map. Both the Blue channel (70~$\mu$m) and Red channel (160~$\mu$m) maps are 60$\arcsec$ in size along each axis. In each map the black star indicates the 2MASS position of the target. The 2MASS positions of known multiple companions are indicted by yellow triangles in the Blue channel maps, and green triangles in the Red channel maps. The scale bar shows the intensity of each map in units of mJy/pixel. Contour levels begin at 3$\sigma$ in all maps and are plotted at intervals depending on the source brightness. The PACS beam size is represented by the blue circle in the lower, right-hand corner of each map.



\section{SEDs of the TBOSS sample -- Class I objects}
\label{Sec:sed_ClassI}
\begin{figure*}
\centering
\includegraphics[scale=1.7]{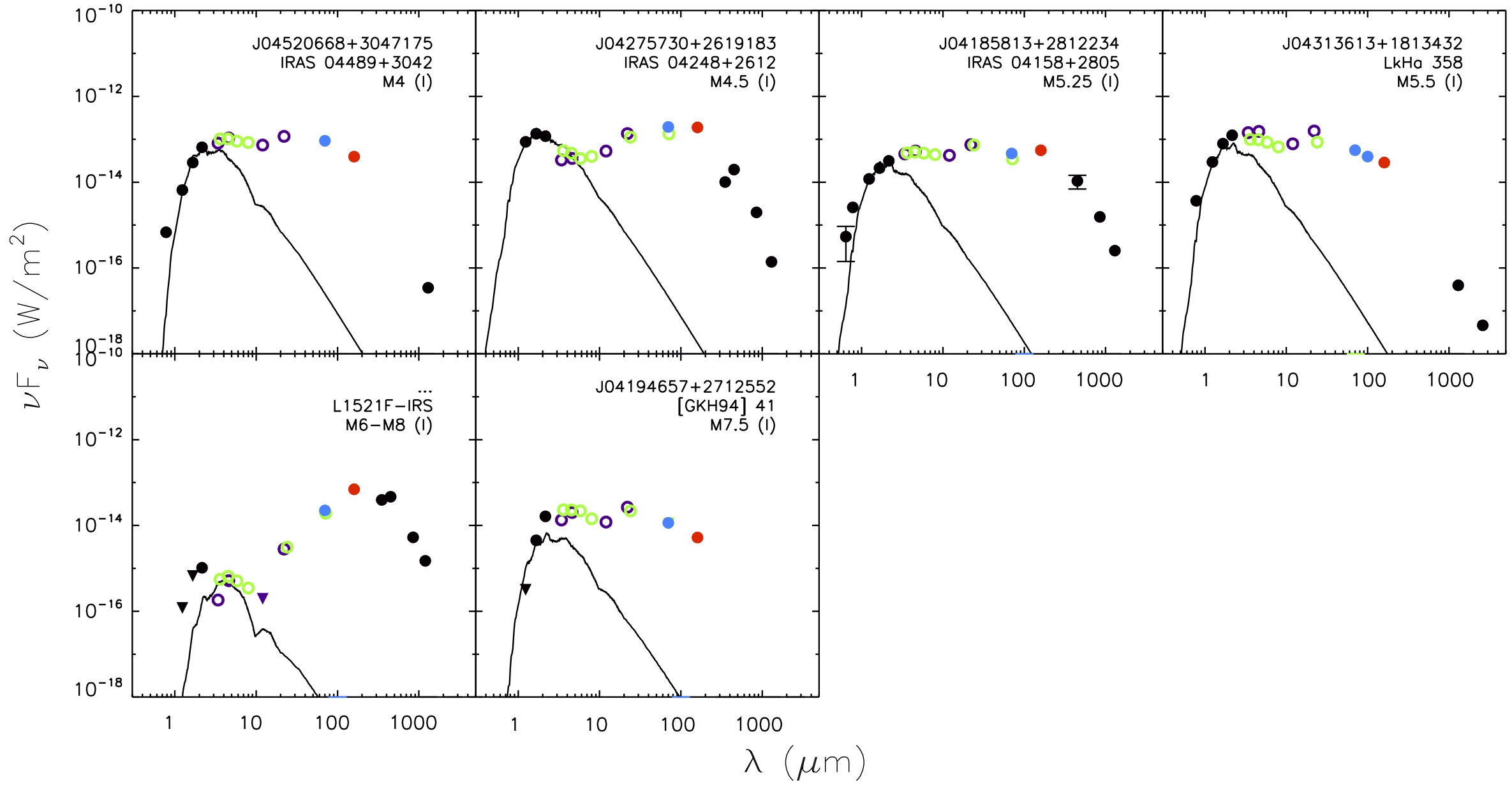}
\caption {SEDs of the TBOSS of the Class~I objects. The target name, spectral type, and spectral class are labeled in each SED. The observed broadband photometry is compiled from optical ({\it R}$_{C}$, {\it I}$_{C}$), and near-IR (2MASS; {\it JHK$_{S}$}) wavelengths (black points), the mid-IR (IRAC and {\it WISE}; green and purple open circles respectively), the far-IR  (MIPS, PACS Blue and Red channels; green open circles, blue and red points respectively) and submm-mm wavelengths (black points). 3$\sigma$ upper limits are represented by the downwards triangles. The best-fit atmospheric model are displayed for each target.
}
\label{sed1}
\end{figure*}

The SEDs for Class~I objects of the TBOSS sample are shown in Figure~\ref{sed1}. As IRAS~04191+1523~B is unresolved from IRAS~04191+1523~A with {\it Herschel} PACS, the SED(s) for this category of system(s) is shown in Figure~\ref{sed1_multi}. 

\section{SEDs of the TBOSS sample -- Class II objects}
\label{Sec:sed_ClassII}
\begin{figure*}
\centering
\includegraphics[scale=1.7]{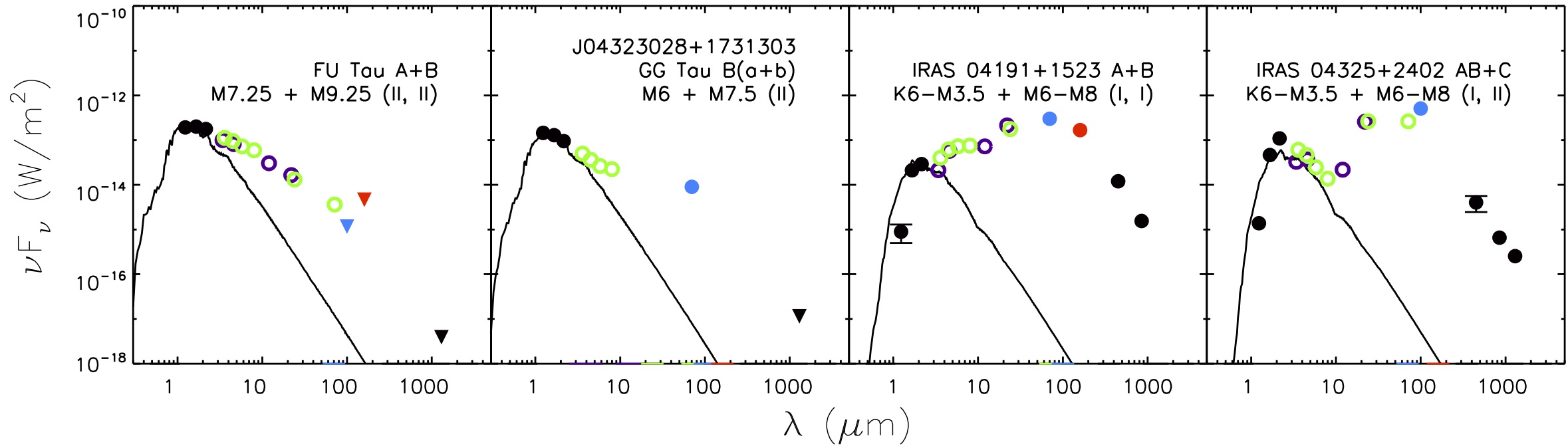}
\caption {SEDs of the four known TBOSS Class~I and/or Class~II multiple systems for which the system broadband photometry and the combined atmospheric models of each system component are shown. The system name, component spectral types, and spectral classes are labeled in each SED. The observed broadband photometry is compiled from optical ({\it R}$_{C}$, {\it I}$_{C}$), and near-IR (2MASS; {\it JHK$_{S}$}) wavelengths (black points), the mid-IR (IRAC and {\it WISE}; green and purple open circles respectively), the far-IR  (MIPS, PACS Blue and Red channels; green open circles, blue and red points respectively) and submm-mm wavelengths (black points). 3$\sigma$ upper limits are represented by the downwards triangles. The best-fit atmospheric model are displayed for each target.
}
\label{sed1_multi}
\end{figure*}

\begin{figure*}
\centering
\includegraphics[scale=1.7]{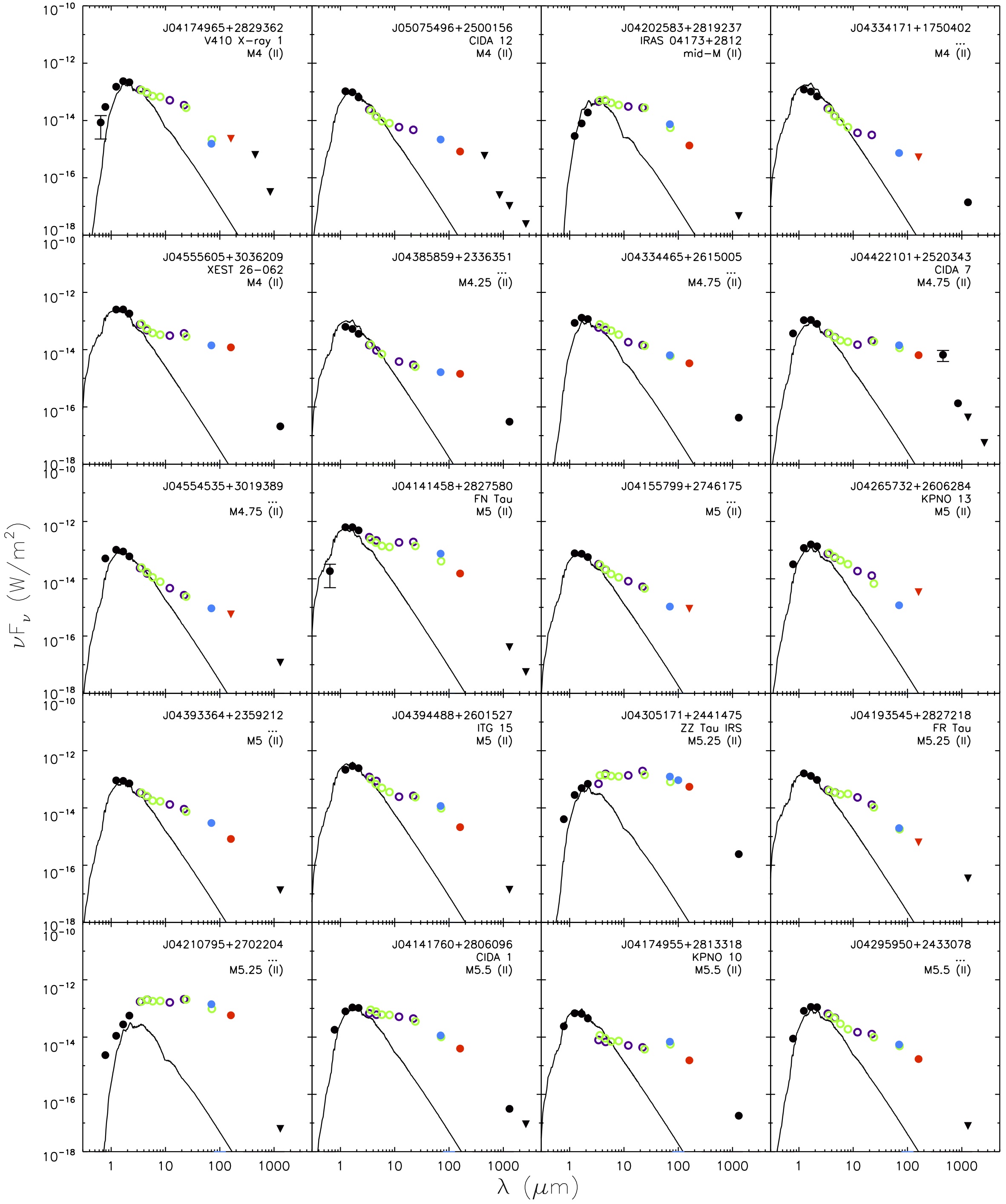}
\caption {SEDs of the detected Class~II objects with spectral types M4-M5.5. The target name, spectral type, and spectral class are labeled in each SED. The observed broadband photometry is compiled from optical ({\it R}$_{C}$, {\it I}$_{C}$), and near-IR (2MASS; {\it JHK$_{S}$}) wavelengths (black points), the mid-IR (IRAC and {\it WISE}; green and purple open circles respectively), the far-IR  (MIPS, PACS Blue and Red channels; green open circles, blue and red points respectively) and submm-mm wavelengths (black points). 3$\sigma$ upper limits are represented by the downwards triangles. The best-fit atmospheric model are displayed for each target.
}
\label{sed2a}
\end{figure*}

\begin{figure*}
\centering
\includegraphics[scale=1.7]{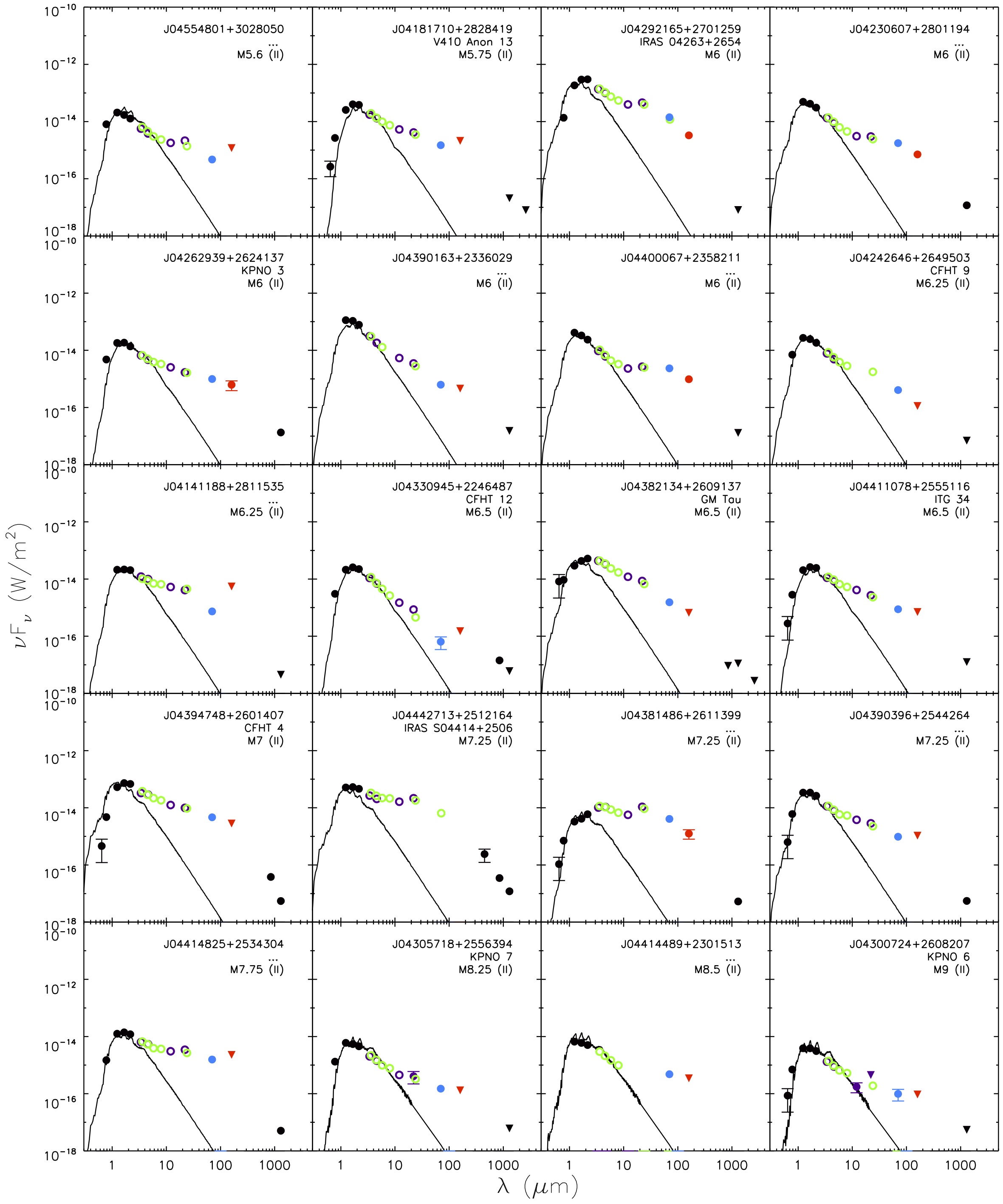}
\caption {SEDs of the detected Class~II objects with spectral types M5.6-M9. The target name, spectral type, and spectral class are labeled in each SED. The observed broadband photometry is compiled from optical ({\it R}$_{C}$, {\it I}$_{C}$), and near-IR (2MASS; {\it JHK$_{S}$}) wavelengths (black points), the mid-IR (IRAC and {\it WISE}; green and purple open circles respectively), the far-IR  (MIPS, PACS Blue and Red channels; green open circles, blue and red points respectively) and submm-mm wavelengths (black points). 3$\sigma$ upper limits are represented by the downwards triangles. The best-fit atmospheric model are displayed for each target.}
\label{sed2b}
\end{figure*}

\begin{figure*}
\centering
\includegraphics[scale=1.7]{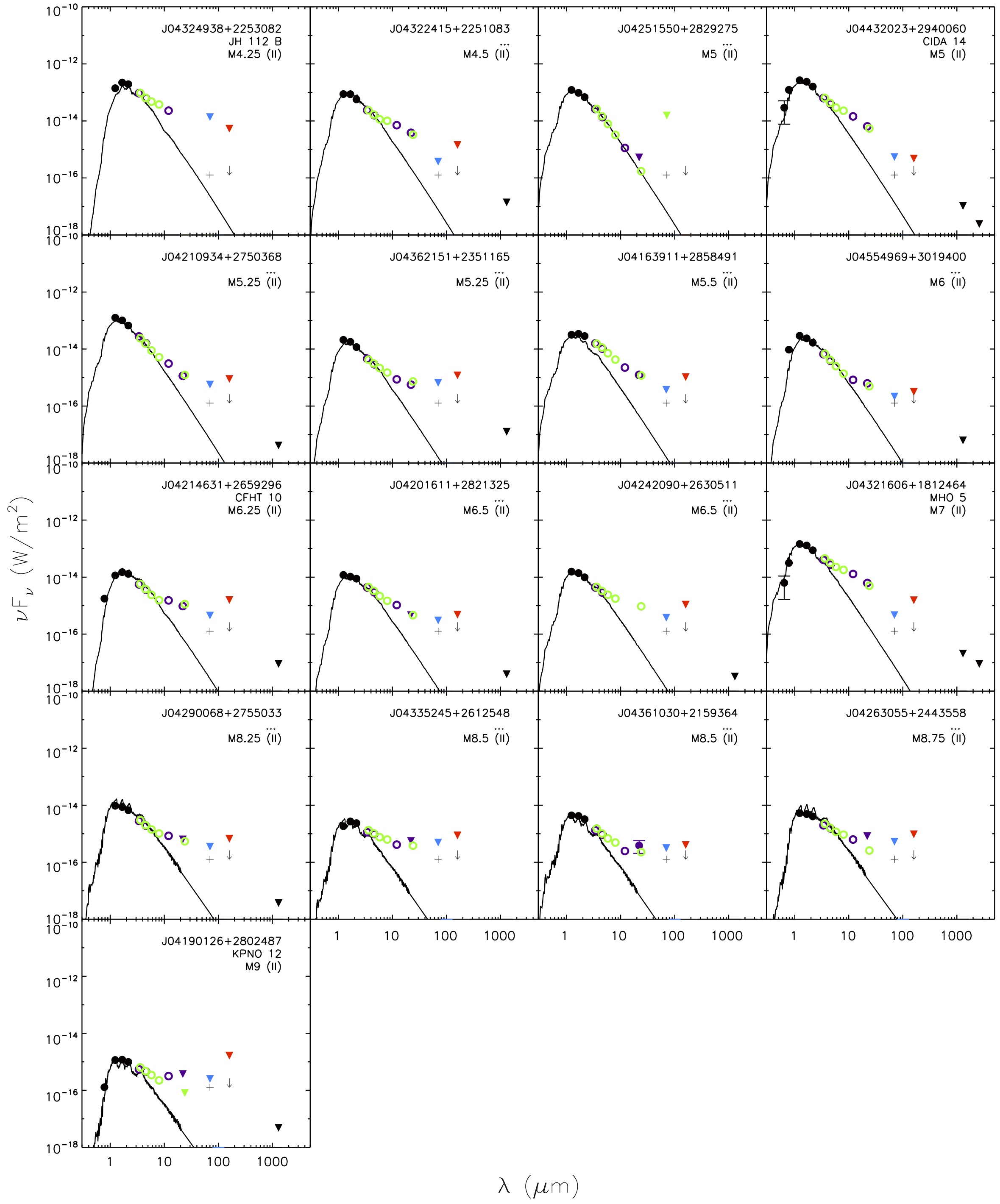}
\caption {SEDs of the undetected Class~II objects with spectral types spanning M4.25-M9. The target name, spectral type, and spectral class are labeled in each SED. The observed broadband photometry is compiled from optical ({\it R}$_{C}$, {\it I}$_{C}$), and near-IR (2MASS; {\it JHK$_{S}$}) wavelengths (black points), the mid-IR (IRAC and {\it WISE}; green and purple open circles respectively), the far-IR  (MIPS, PACS Blue and Red channels; green open circles, blue and red points respectively) and submm-mm wavelengths (black points). 3$\sigma$ upper limits are represented by the downwards triangles. The best-fit atmospheric model are displayed for each target. The gray cross shows level of the artificial detection generated from combined Class~II (70~$\mu$m) upper limit maps, and similarly the gray downward arrow shows the 160~$\mu$m upper limit from the combined maps.}
\label{sed2d}
\end{figure*}

The additional SEDs for the Class~II objects of the TBOSS sample are shown in Figures~\ref{sed1_multi}-\ref{sed2d}. Figure~\ref{sed1_multi} shows the targets that are in multiple systems of which are unresolved with {\it Herschel} PACS. Figures~\ref{sed2a} and \ref{sed2b} shows the SEDs the Class~II objects that are detected with {\it Herschel} PACS. Finally, Figure~\ref{sed2d} shows the SEDs of the Class~II objects that were undetected with {\it Herschel} PACS.

\section{SEDs of the TBOSS sample -- Class III objects}
\label{Sec:sed_ClassIII}
\begin{figure*}
\centering
\includegraphics[scale=1.7]{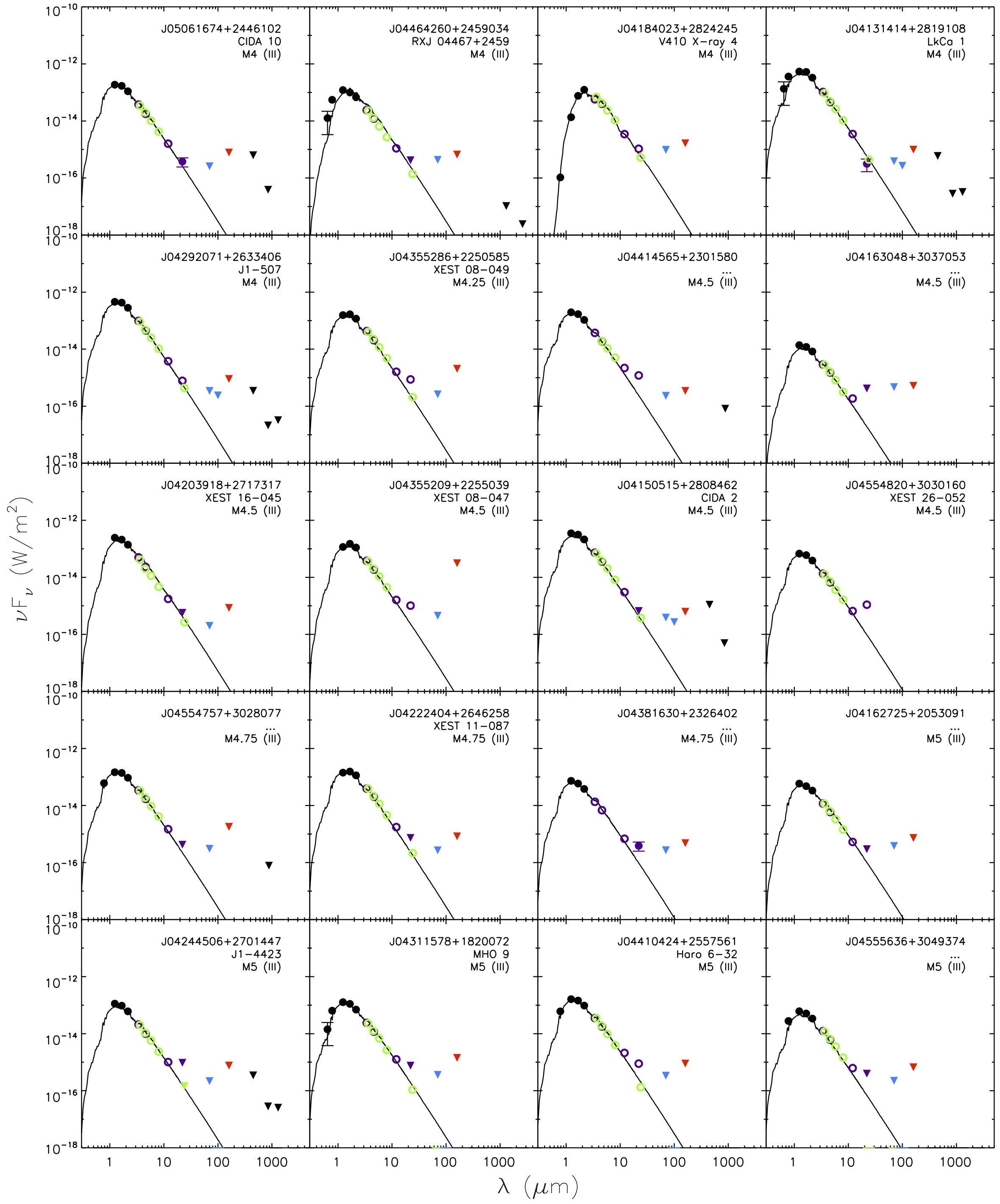}
\caption {SEDs of the undetected Class~III objects with spectral types spanning M4-M5. Additionally, the known M4.5 member -- XEST 26-052 -- for which no far-IR observations exist, is also included within the panel. The target name, spectral type, and spectral class are labeled in each SED. The observed broadband photometry is compiled from optical ({\it R}$_{C}$, {\it I}$_{C}$), and near-IR (2MASS; {\it JHK$_{S}$}) wavelengths (black points), the mid-IR (IRAC and {\it WISE}; green and purple open circles respectively), the far-IR  (MIPS, PACS Blue and Red channels; green open circles, blue and red points respectively) and submm-mm wavelengths (black points). 3$\sigma$ upper limits are represented by the downwards triangles. The best-fit atmospheric model are displayed for each target.}
\label{sed3a}
\end{figure*}

\begin{figure*}
\centering
\includegraphics[scale=1.7]{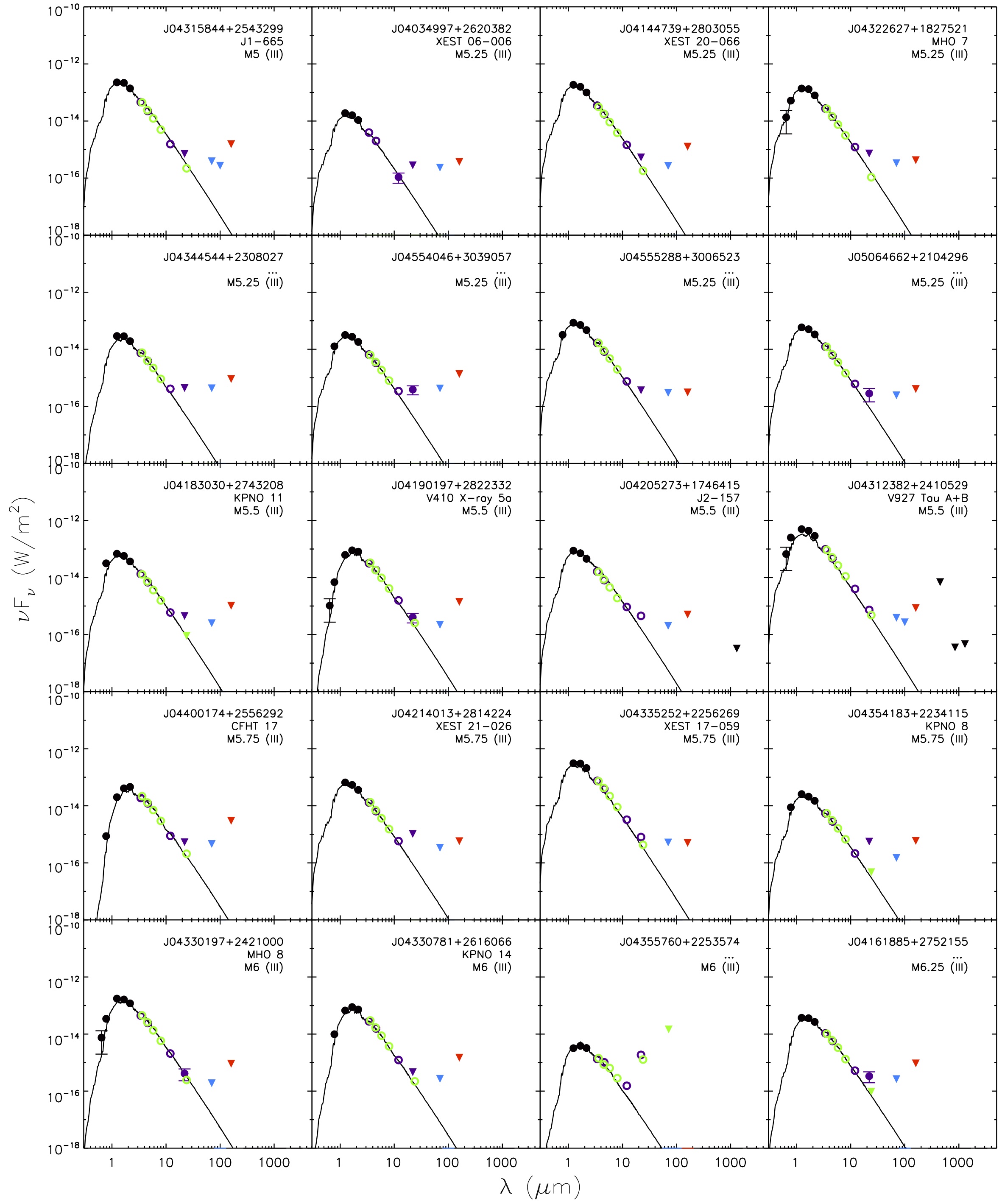}
\caption {SEDs of the undetected Class~III objects with spectral types spanning M5-M6.25. The target name, spectral type, and spectral class are labeled in each SED. The observed broadband photometry is compiled from optical ({\it R}$_{C}$, {\it I}$_{C}$), and near-IR (2MASS; {\it JHK$_{S}$}) wavelengths (black points), the mid-IR (IRAC and {\it WISE}; green and purple open circles respectively), the far-IR  (MIPS, PACS Blue and Red channels; green open circles, blue and red points respectively) and submm-mm wavelengths (black points). 3$\sigma$ upper limits are represented by the downwards triangles. The best-fit atmospheric model are displayed for each target.}
\label{sed3b}
\end{figure*}

\begin{figure*}
\centering
\includegraphics[scale=1.7]{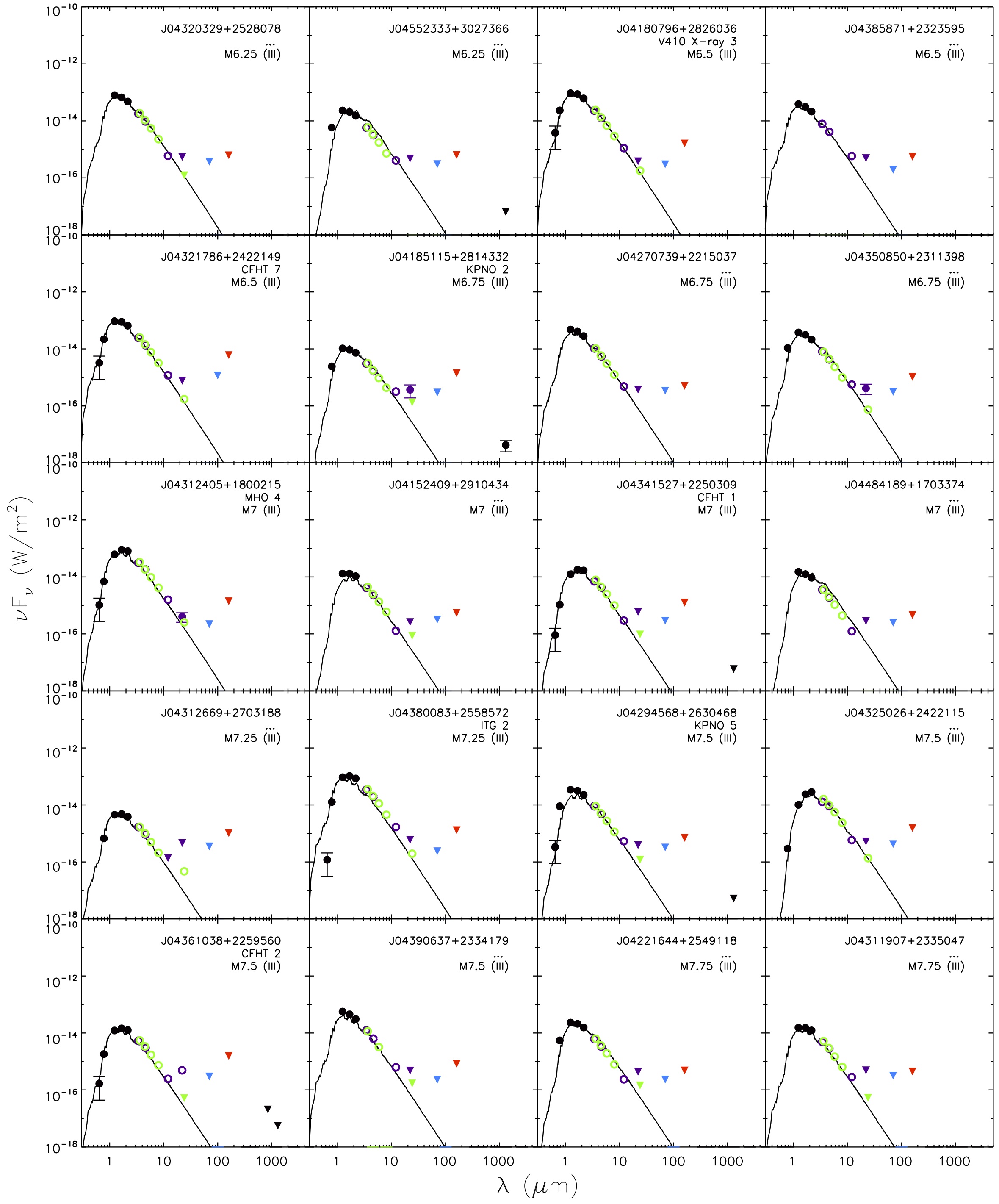}
\caption {SEDs of the undetected Class~III objects with spectral types spanning M6.25-M7.75. The target name, spectral type, and spectral class are labeled in each SED. The observed broadband photometry is compiled from optical ({\it R}$_{C}$, {\it I}$_{C}$), and near-IR (2MASS; {\it JHK$_{S}$}) wavelengths (black points), the mid-IR (IRAC and {\it WISE}; green and purple open circles respectively), the far-IR  (MIPS, PACS Blue and Red channels; green open circles, blue and red points respectively) and submm-mm wavelengths (black points). 3$\sigma$ upper limits are represented by the downwards triangles. The best-fit atmospheric model are displayed for each target.}
\label{sed3c}
\end{figure*}

\begin{figure*}
\centering
\includegraphics[scale=1.7]{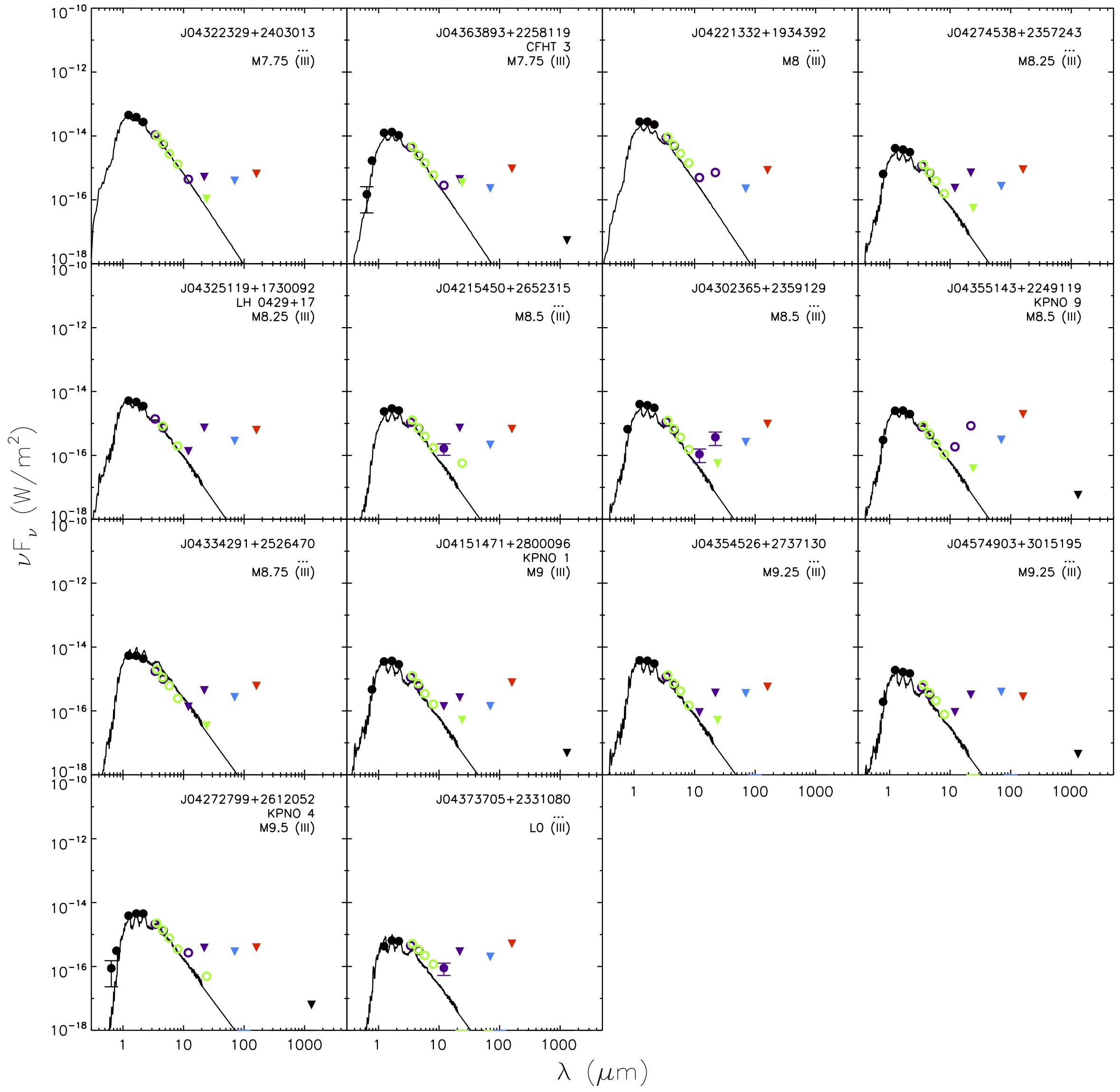}
\caption {SEDs of the undetected Class~III objects with spectral types spanning M7.75-L0. The target name, spectral type, and spectral class are labeled in each SED. The observed broadband photometry is compiled from optical ({\it R}$_{C}$, {\it I}$_{C}$), and near-IR (2MASS; {\it JHK$_{S}$}) wavelengths (black points), the mid-IR (IRAC and {\it WISE}; green and purple open circles respectively), the far-IR  (MIPS, PACS Blue and Red channels; green open circles, blue and red points respectively) and submm-mm wavelengths (black points). 3$\sigma$ upper limits are represented by the downwards triangles. The best-fit atmospheric model are displayed for each target.}
\label{sed3d}
\end{figure*}

The SEDs for the Class~III objects of the TBOSS sample that are undetected with {\it Herschel} PACS are shown in Figures~\ref{sed3a}-\ref{sed3d}.

\end{appendix}
\end{document}